\documentclass[sigconf]{acmart}
\acmConference[XXX]{XXX}{XXX}{XXX}

\usepackage{listings}  
\usepackage{multirow}
\usepackage{xspace}
\usepackage[textsize=tiny]{todonotes}
\usepackage{multicol}
\usepackage{stfloats}
\usepackage{placeins}
\usepackage{rotating}
\usepackage{hyperref}
\usepackage{booktabs}
\usepackage[export]{adjustbox}

\hypersetup{
   colorlinks,%
   citecolor=black,%
   filecolor=black,%
   linkcolor=black,%
   urlcolor=gray
}

\usepackage{mdframed}

\newcommand{\code}[1]{{\lstinline{#1}}\xspace}
\newcommand{\smallcode}[1]{{\lstinline[basicstyle=\scriptsize]{#1}}\xspace}

\usepackage{tcolorbox}
\newenvironment{takeaway}{
\vspace{.5em}
\begin{tcolorbox}[colback=blue!5!white,colframe=blue!5!white,arc=0mm,grow to left by=1.5mm,left=0mm,grow to right by=1.5mm,right=0mm,top=0mm,bottom=0mm]
\small
}
{
\end{tcolorbox}
}

\newcommand{\ToolName}{\mbox{\textit{LLMorpheus}}\xspace}
\newcommand{\StrykerJS}{\mbox{\textit{StrykerJS}}\xspace}

\definecolor{KWColor}{rgb}{0.5,0,0.67}
\definecolor{CommentColor}{rgb}{0.15,0.5,0.15}
\definecolor{sh_comment}{rgb}{0.12, 0.38, 0.18}
\definecolor{sh_keyword}{rgb}{0.37, 0.08, 0.25}  
\definecolor{sh_string}{rgb}{0.06, 0.10, 0.98} 
\definecolor{DarkOrchid3}{rgb}{0.6, 0.196, 0.8} 
\definecolor{ChangeHighlightColor}{rgb}{0.345, 0.361, 0.788} 

\lstdefinelanguage{JavaScript}[]{Java}{
   morekeywords={var,class,function, async, await, undefined, let, form, button, div, useState, number, textarea, then, <PLACEHOLDER>, in, of} 
} 

\lstdefinestyle{Eclipse}{
  xleftmargin=0pt,
  language = JavaScript,
  basicstyle=\sffamily\footnotesize,
  stringstyle=\color{sh_string},
  keywordstyle = \color{sh_keyword}\bfseries,  
  lineskip=-0.0em,
  commentstyle=\color{sh_comment}\itshape,  
  escapeinside={/*@}{@*/},
  numbersep=5pt,
  captionpos=b,
  xleftmargin=0.4cm, xrightmargin=0.5cm,
   morekeywords={invokestatic,invokeinterface,invokevirtual,invokespecial,then, <PLACEHOLDER>,in, of},
}

\newcommand{\WRAP}[1]{\begin{minipage}[t]{0.18\textwidth}{\scriptsize #1}\end{minipage}}

\makeatletter
\DeclareRobustCommand\onedot{\futurelet\@let@token\@onedot}
\def\@onedot{\ifx\@let@token.\else.\null\fi\xspace}

\def\etal{\emph{et al}\onedot}
\makeatother

\newcommand{\CodeLlamaThirteen}{\mbox{\textit{codellama-13b-instruct}}\xspace}
\newcommand{\CodeLlamaThirtyFour}{\mbox{\textit{codellama-34b-instruct}}\xspace}
\newcommand{\Mixtral}{\mbox{\textit{mixtral-8x7b-instruct}}\xspace}
\newcommand{\GPTFouroMini}{\mbox{\textit{gpt-4o-mini}}\xspace}
\newcommand{\GPTFouro}{\mbox{\textit{gpt-4o}}\xspace}
\newcommand{\LlamaThreeThree}{\mbox{\textit{llama-3.3-70b-instruct}}\xspace}

\newcommand{\CodeLlamaThirteenShort}{\begin{minipage}{1.5cm}\textit{codellama-\\13b-instruct\\}\end{minipage}\xspace}
\newcommand{\CodeLlamaThirtyFourShort}{\begin{minipage}{1.5cm}\textit{codellama-\\34b-instruct\\}\end{minipage}\xspace}
\newcommand{\MixtralShort}{\begin{minipage}{1.5cm}\textit{mixtral-8x7b-\\instruct\\}\end{minipage}\xspace}

\newcommand{\NrBenchmarks}{\mbox{13}\xspace}

\newcommand{\NrPromptsMin}{\mbox{42}\xspace}
\newcommand{\NrPromptsMax}{\mbox{1,051}\xspace}
\newcommand{\FullMutantsMin}{\mbox{89}\xspace}
\newcommand{\FullMutantsMax}{\mbox{2,035}\xspace}
\newcommand{\FullKilledMutantsMin}{\mbox{23}\xspace}
\newcommand{\FullKilledMutantsMax}{\mbox{725}\xspace}
\newcommand{\FullSurvivingMutantsMin}{\mbox{3}\xspace}
\newcommand{\FullSurvivingMutantsMax}{\mbox{1,792}\xspace}
\newcommand{\FullTimeoutMutantsMin}{\mbox{0}\xspace}
\newcommand{\FullTimeoutMutantsMax}{\mbox{85}\xspace}

\newcommand{\MinLLMorpheusTimeInSeconds}{\mbox{430.53}\xspace}
\newcommand{\MinLLMorpheusTimeInMinutes}{\mbox{7}\xspace}
\newcommand{\MaxLLMorpheusTimeInSeconds}{\mbox{5,241.46}\xspace}
\newcommand{\MaxLLMorpheusTimeInMinutes}{\mbox{87}\xspace}

\newcommand{\MinPrecomputedStrykerTimeInSeconds}{\mbox{155.24}\xspace}
\newcommand{\MaxPrecomputedStrykerTimeInSeconds}{\mbox{14,034.67}\xspace}
\newcommand{\MinPrecomputedStrykerTimeInMinutes}{\mbox{2.5}\xspace}
\newcommand{\MaxPrecomputedStrykerTimeInMinutes}{\mbox{234}\xspace}

\newcommand{\NrPromptTokensMin}{\mbox{24,655}\xspace}
\newcommand{\NrPromptTokensMax}{\mbox{2,127,655}\xspace}
\newcommand{\NrCompletionTokensMin}{\mbox{9,134}\xspace}
\newcommand{\NrCompletionTokensMax}{\mbox{220,215}\xspace}
\newcommand{\TotalPromptTokens}{\mbox{5,841,112}\xspace}
\newcommand{\TotalCompletionTokens}{\mbox{721,984}\xspace}

\newcommand{\MinPercentageOfCommonMutants}{\mbox{89.29\%}\xspace}
\newcommand{\MaxPercentageOfCommonMutants}{\mbox{98.89\%}\xspace}

\newcommand{\NrSubjectApplications}{\mbox{13}\xspace}

\newcommand{\Candidates}{\begin{turn}{90}{\bf \#candidates}\end{turn}}
\newcommand{\Invalid}{\begin{turn}{90}{\bf \#invalid}\end{turn}}
\newcommand{\Identical}{\begin{turn}{90}{\bf \#identical}\end{turn}}
\newcommand{\Duplicate}{\begin{turn}{90}{\bf \#duplicate}\end{turn}}
\newcommand{\Total}{\begin{turn}{90}{\bf \#mutants}\end{turn}}
\newcommand{\Killed}{\begin{turn}{90}{\bf \#killed}\end{turn}}
\newcommand{\Survived}{\begin{turn}{90}{\bf \#survived}\end{turn}}
\newcommand{\Timeout}{\begin{turn}{90}{\bf \#timeout}\end{turn}}
\newcommand{\Time}{\begin{turn}{90}{\bf time (sec)}\end{turn}}
\newcommand{\MutScore}{\begin{turn}{90}{\bf mut. score}\end{turn}}

\newcommand{\Equiv}{\begin{turn}{90}{\bf equiv}\end{turn}}

\newcommand{\NotEquiv}{\begin{turn}{90}{\bf not equiv}\end{turn}}
\newcommand{\Unknown}{\begin{turn}{90}{\bf unknown}\end{turn}}

\newcommand{\Rotate}[1]{\begin{turn}{90}#1\end{turn}}

\newcommand{\ComplexJSDistinctZero}{1,217\xspace}
\newcommand{\ComplexJSDistinctTwentyFive}{2,354\xspace}
\newcommand{\ComplexJSDistinctFifty}{3,196\xspace}
\newcommand{\ComplexJSDistinctOne}{4,200\xspace}
\newcommand{\ComplexJSCommonZero}{1,181\xspace}
\newcommand{\ComplexJSCommonZeroPercentage}{97.04\%\xspace}
\newcommand{\ComplexJSCommonTwentyFive}{447\xspace}
\newcommand{\ComplexJSCommonTwentyFivePercentage}{18.99\%\xspace}
\newcommand{\ComplexJSCommonFifty}{205\xspace}
\newcommand{\ComplexJSCommonFiftyPercentage}{6.41\%\xspace}
\newcommand{\ComplexJSCommonOne}{17\xspace}
\newcommand{\ComplexJSCommonOnePercentage}{0.4\%\xspace}

\newcommand{\CodeLlamaThirteenCommonMutantsMinPercentage}{96.15\%\xspace}
\newcommand{\CodeLlamaThirteenCommonMutantsMaxPercentage}{100\%\xspace}
\newcommand{\MixtralCommonMutantsMinPercentage}{34.22\%\xspace}
\newcommand{\MixtralCommonMutantsMaxPercentage}{50\%\xspace}
\newcommand{\LlamaThreeThreeCommonMutantsMinPercentage}{28.26\%\xspace}
\newcommand{\LlamaThreeThreeCommonMutantsMaxPercentage}{58.94\%\xspace}
\newcommand{\GPTFouroMiniCommonMutantsMinPercentage}{28.26\%\xspace}
\newcommand{\GPTFouroMiniCommonMutantsMaxPercentage}{58.94\%\xspace}

\newcommand{\NrMutantsInEquivalenceCaseStudy}{517\xspace} 
\newcommand{\NrNotEquivalentMutants}{403\xspace} 
\newcommand{\PercentageNotEquivalentMutants}{78\%\xspace}
\newcommand{\NrEquivalentMutants}{105\xspace}
\newcommand{\PercentageEquivalentMutants}{20\%\xspace}
\newcommand{\NrUnknownMutants}{9\xspace}
\newcommand{\PercentageUnknownMutants}{2\%\xspace}

\newcommand{\StrykerNrMutantsInEquivalenceCaseStudy}{430\xspace} 
\newcommand{\StrykerNrNotEquivalentMutants}{395\xspace}
\newcommand{\StrykerPercentageNotEquivalentMutants}{92\%\xspace}
\newcommand{\StrykerNrEquivalentMutants}{5\xspace}
\newcommand{\StykerPercentageEquivalentMutants}{1\%\xspace}
\newcommand{\StrykerNrUnknownMutants}{30\xspace}
\newcommand{\StrykerPercentageUnknownMutants}{7\%\xspace}

\newcommand{\ChangedText}[1]{#1}

\newenvironment{revision*}{%
\color{black}
}{\ignorespacesafterend}

\mdfdefinestyle{revisionframe}{%
	linecolor=ChangeHighlightColor,linewidth=1pt,innermargin=-0pc,outermargin=-0pc,innerleftmargin=0pc,innerrightmargin=0pc,
	apptotikzsetting={%
		\tikzset{mdfbackground/.append style={fill=white,fill opacity=0}}}
}

\newcommand{\TotalMutantCandidates}{9,967\xspace}
\newcommand{\NrInvalidMutantCandidates}{2,894\xspace}
\newcommand{\NrIdenticalMutantCandidates}{156\xspace}
\newcommand{\NrDuplicateMutantCandidates}{205\xspace}
\newcommand{\PercentageInvalidMutantCandidates}{29.0\%\xspace}
\newcommand{\PercentageIdenticalMutantCandidates}{1.6\%\xspace}
\newcommand{\PercentageDuplicateMutantCandidates}{2.1\%\xspace}

\newcommand{\TotalMutants}{6,712\xspace}
\newcommand{\NrKilledMutants}{3,237\xspace}
\newcommand{\NrSurvivingMutants}{3,155\xspace}
\newcommand{\NrTimedOutMutants}{320\xspace}
\newcommand{\PercentageKilledMutants}{48.2\%\xspace}
\newcommand{\PercentageSurvivingMutants}{47.0\%\xspace}
\newcommand{\PercentageTimedOutMutants}{4.8\%\xspace}
\newcommand{\StrykerJSNrSurvivingMutants}{1,956\xspace}

\newcommand{\StrykerComplexJSNrMutants}{1,302\xspace}
\newcommand{\StrykerComplexJSNrSurvivingMutants}{539\xspace}
\newcommand{\LLMorpheusComplexJSNrMutants}{1,199\xspace}
\newcommand{\LLMorpheusComplexJSNrSurvivingMutants}{473\xspace}

\newcommand{\StrykerQNrMutants}{1,058\xspace}
\newcommand{\StrykerQNrSurvivingMutants}{927\xspace}
\newcommand{\LLMorpheusQNrMutants}{2,035\xspace}
\newcommand{\LLMorpheusQNrSurvivingMutants}{1,792\xspace}  
  
\newcommand{\TotalMutantsOneMutation}{2,333\xspace}
\newcommand{\TotalMutantsBasic}{1,326\xspace}

\newcommand{\CodeLlamaThirteenTotalMutantCandidates}{8,088\xspace}
\newcommand{\CodeLlamaThirteenIdentical}{922\xspace}
\newcommand{\GPTFouroMiniInvalid}{3,703\xspace}

\newcommand{\MixtralInvalid}{2,540\xspace}

\newcommand{\MixtralTotalMutants}{5,402\xspace}
\newcommand{\LlamaThreeThreeTotalMutants}{6,823\xspace}
\newcommand{\LlamaThreeThreeSurvivingMutants}{3,423\xspace}

\begin{document}

\title{LLMorpheus: Mutation Testing using Large Language Models}
\author{Frank Tip}
\email{f.tip@northeastern.edu}
\affiliation{
  \institution{Northeastern University}
  \city{Boston, MA}
  \country{USA}
}
\author{Jonathan Bell}
\email{j.bell@northeastern.edu}
\affiliation{
  \institution{Northeastern University}
  \city{Boston, MA}
  \country{USA}
}
\author{Max Sch\"afer}
\email{max@xbow.com}
\affiliation{
  \institution{XBOW}
  \city{Oxford}
  \country{UK}
}

\begin{abstract}
  In mutation testing, the quality of a test suite is evaluated by introducing
  faults into a program and determining whether the program's tests detect them.   
  Most existing approaches for mutation testing involve the application of a fixed
  set of mutation operators, e.g., replacing a ``+'' with a ``-'',
  or removing a function's body.
  However, certain types of real-world bugs cannot easily be simulated by such approaches,
  limiting their effectiveness. 
  This paper presents a technique for mutation testing where placeholders are introduced
  at designated locations in a program's source code and where a  Large Language Model (LLM) is prompted 
  to ask what they could be replaced with.
  The technique is implemented in \ToolName, a mutation testing tool for JavaScript, and evaluated 
  on \NrBenchmarks subject packages, considering several variations on the
  prompting strategy, and using several LLMs.
  We find \ToolName to be capable of producing mutants that resemble
  existing bugs that cannot be produced by \StrykerJS, a state-of-the-art mutation testing
  tool. Moreover, we report on the running time, cost, and number of mutants produced by \ToolName,
  demonstrating its practicality. 
\end{abstract}

\maketitle
\thispagestyle{plain}
\pagestyle{plain}

\pagestyle{plain} 
\thispagestyle{plain}

\section{Introduction}
Mutation testing is an approach for evaluating the adequacy of a test suite and
is increasingly adopted in industrial settings~\cite{Petrovic21DoesMutation,Petrovic22Practical,Petrovic23PleaseFix}.
With mutation testing, an automated tool repeatedly injects a small modification to the system under test and executes the test suite on this mutated code.
Mutation testing is premised on the \emph{competent programmer hypothesis}, which posits that most buggy programs are quite close to being correct and that complex faults are \emph{coupled} with simpler faults~\cite{DeMillo78Hints},
i.e., a test that is strong enough to detect a simple fault should also be able to detect a more complex one.
Hence, mutation analysis tools typically apply a relatively small set of mutation operators: replacing constants, replacing 
operators, modifying branch conditions, and deleting statements.
Studies have shown that, given two test suites for the same system under test, the one that detects more mutants (even using only these limited mutation operators) is likely to also detect more real faults~\cite{Just14AreMutants,Laurent22Revisiting}.

However, not \emph{all} real faults are coupled to mutants due to the limited set of mutation operators.
For example, a fault resulting from calling the wrong method on an object is unlikely to be coupled to a mutant, as state-of-the-art mutation tools do 
not implement a ``change method call'' operator.
%
%
While a far wider range of mutation operators has been explored in the literature~\cite{Jia11Analysis,Ghiduk17Higher},  state-of-the-practice tools like Pitest~\cite{Coles16PIT,Coles24ArcMutate}, Major~\cite{Just14Major} and Stryker \cite{Stryker} 
typically do not implement them because of the implementation effort required and, especially, the increased cost of mutation analysis.  
\begin{revision*}
Each additional mutation operator will result in more mutants that must be run and analyzed, and since each mutant must be evaluated in isolation, this may dramatically increase the time needed for developers that use the tool.
\end{revision*}
Furthermore, some mutation operators might not be worthwhile to run, as noted in documentation from the developer of Pitest: ``Although pitest provides a number of other operators, they are not enabled by default as they may provide a poorer experience''~\cite{Coles24ArcMutateExtendedOperators}. 
An alternative approach for generating mutants is to use a dataset of real faults to train a machine learning model to learn how to inject mutants~\cite{Tufano19Learning,Tian23Learning,Patra21Semantic}.
However, the need for developers to train a model for their project is an impediment to adoption of such techniques.

\ChangedText{
Our approach, \ToolName, can be viewed as a generalization of rule-based mutation techniques \cite{Coles16PIT,Coles24ArcMutate,Just14Major,Stryker}
in which the location of mutations is determined using a set of predefined rules, and where an LLM is asked to suggest a diversity of mutations that
introduce buggy behavior at those locations. To this end, \ToolName repeatedly prompts an LLM to inject faults at designated locations into a code fragment
using prompts that include:}
  (i)  general background on mutation testing 
  (ii) (parts of) a source file in which a single code fragment is replaced with the word ``PLACEHOLDER'',
  (iii) the original code fragment that was replaced by the placeholder, and  
  (iv) a request to replace the placeholder with a buggy code fragment that has different behavior than the original code.
After discarding syntactically invalid suggestions, we use \StrykerJS, a state-of-the-art mutation testing tool for JavaScript that we modified 
to apply the mutations suggested by \ToolName instead of applying its standard mutators, classify mutants as killed, surviving, or timed out, and generate an interactive
web site for inspecting the results.  

\begin{sloppypar}
The effectiveness of our approach hinges on the assumption that LLMs can understand the surrounding context of the code fragment represented by a PLACEHOLDER 
well enough to suggest syntactically valid and realistic buggy code fragments. To determine whether this assumption holds, we
evaluate \ToolName on \NrSubjectApplications subject applications written in JavaScript and TypeScript and measure  
how many mutants are generated and how they are classified (killed, survived, timed-out) using \ChangedText{four ``open'' LLMs for 
which the training process is documented (Meta's \CodeLlamaThirtyFour, \CodeLlamaThirteen, \LlamaThreeThree and Mistral's \Mixtral) and one proprietary LLM (OpenAI's \GPTFouroMini}). 
We manually examine a subset of the surviving mutants 
to determine whether they are equivalent to the original source code or if they represent behavioral changes and contrast the results against mutants 
generated using \StrykerJS's standard mutators.  The cost of \ToolName is assessed by measuring its running time and the number of tokens used for prompts and completions. 
We also report on experiments with alternative prompts that omit parts of the information encoded in default prompts and 
with different ``temperature'' settings of an LLM.  
\end{sloppypar}

\ChangedText{
To investigate \ToolName's ability to produce mutants that resemble real-world faults, we conducted a detailed case study involving four real-world bugs from the
Bugs.js suite \cite{Gyimesi19BugsJS}. In this case study, we used \ToolName to mutate the \textit{fixed} version of a program  near the location of the fix,
executed the program's tests for each of these mutants and compared the test outcomes against those of the buggy version.
In each case, \ToolName produced either a mutant that was identical to a bug that was originally observed, or one that caused the same test failures 
as a bug that was originally observed. Hence, we} find \ToolName to be capable of generating a diverse set of mutants, some of which resemble real-world bugs that cannot be created by 
\StrykerJS' standard mutation operators.

\begin{sloppypar}
For surviving mutants generated using \CodeLlamaThirtyFour, we
find that the majority  (\PercentageNotEquivalentMutants) reflect behavioral differences and
\PercentageEquivalentMutants are equivalent to the original code. 
Using the \CodeLlamaThirtyFour and \CodeLlamaThirteen models, results are generally stable at temperature 0.0 when experiments are repeated, 
but the use of higher temperatures yields more variable results. For \Mixtral, \LlamaThreeThree, and \GPTFouroMini models, there is already 
significant variability at temperature 0.
The default template generally produces the largest number of mutants and surviving mutants, and removing different fragments of this prompt
degrades the results to varying degrees. 
The \LlamaThreeThree and \CodeLlamaThirtyFour LLMs generally produce the largest number of mutants and surviving mutants, 
but \ToolName is still effective when \CodeLlamaThirteen, \Mixtral, and \GPTFouroMini are used.
\end{sloppypar}

In summary, the contributions of this paper are:
\begin{enumerate}
  \item
    A technique for mutation testing in which placeholders are introduced at designated locations in a 
    program's source code, and where an LLM is prompted to ask what they could be replaced with.
  \item
    An implementation of this technique in \ToolName, a practical mutation testing tool for JavaScript.
  \item
    An empirical evaluation of \ToolName on \NrSubjectApplications subject applications, demonstrating its practicality
    and comparing it to a standard approach to mutation testing based on mutation operators.
\end{enumerate}

The remainder of this paper is organized as follows.
Section~\ref{sec:Motivation} presents motivating examples that illustrate the potential of LLM-based mutation techniques to introduce faults resembling real bugs.
In Section~\ref{sec:Approach}, an overview of our approach is presented.
Section~\ref{sec:Evaluation} presents an evaluation of \ToolName and Section~\ref{sec:Threats} covers threats to validity.
Related work is discussed in Section~\ref{sec:Related}.
Lastly, Section~\ref{sec:Conclusions} presents conclusions and directions for future work.

\section{Background and Motivation}
  \label{sec:Motivation}


In this section, we study a few bugs that do not correspond to mutation operators supported by state-of-the-art
mutation testing tools but that are similar to mutations \ToolName could suggest.

\paragraph{Example 1.}

\begin{figure}
\centering
\hspace*{-5mm}
\begin{tabular}{c}
\includegraphics[width=0.4855\textwidth]{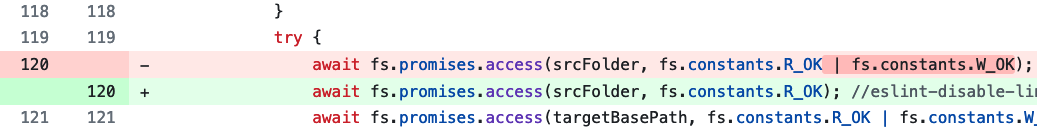}
\\
{\bf (a)}
\\
\includegraphics[width=0.485\textwidth]{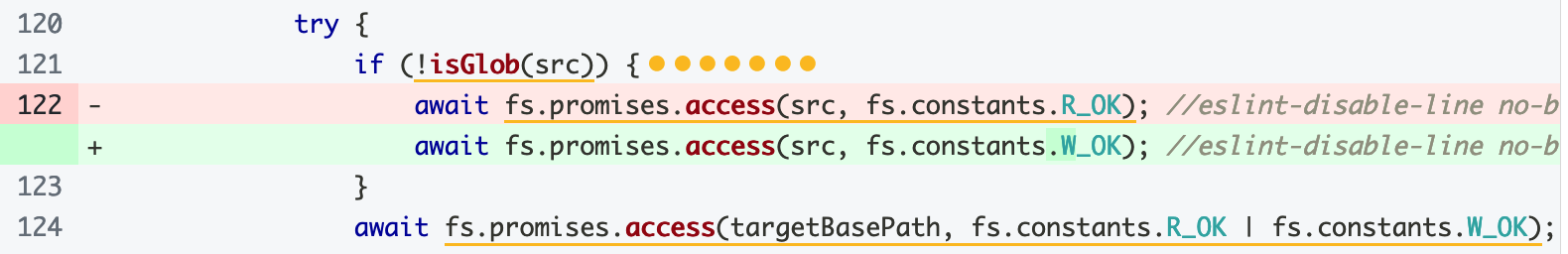}
\\
{\bf (b)}
\\
\includegraphics[width=0.485\textwidth]{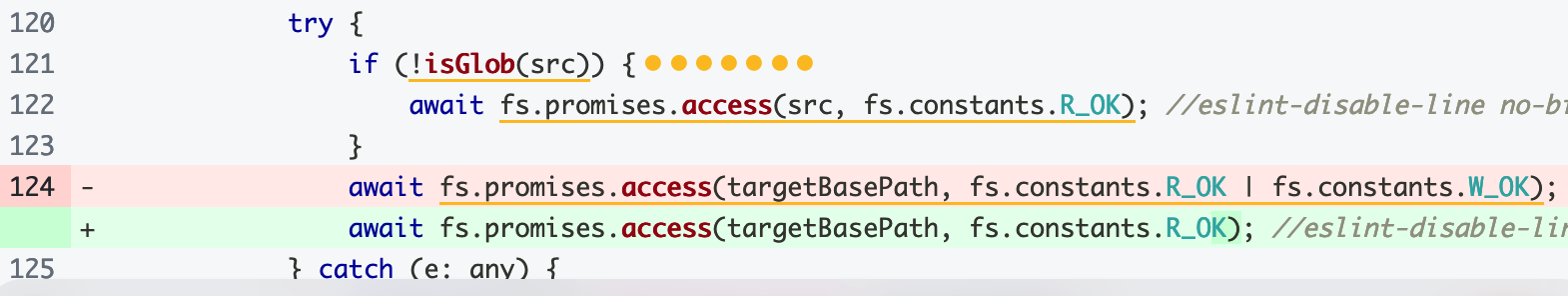}
\\
{\bf (c)}
\end{tabular}
\caption{
  (a) Fix for a bug reported in issue \#36 in \code{zip-a-folder}.
  (b) A mutation suggested by \ToolName at the same line that involves replacing read-access with write-access.
  (c) A mutation suggested by \ToolName elsewhere in the same file that mirrors the change made by the developer.  
}
\label{fig:ZipAFolderExample}
\end{figure}

Zip-a-folder \cite{ZipAFolder} is a library for compressing folders.
On January 31, 2022, a user observed that the library required write-access for source-folders unnecessarily
and opened issue \#36, requesting that this access be removed.  The developer applied the fix
shown in Figure~\ref{fig:ZipAFolderExample}(a) on the same day, which involves replacing a binary bitwise-or
expression with one of its operands. 

\ToolName can suggest mutations that involve \textit{changing or introducing} references to functions, variables, and properties.  
Figure~\ref{fig:ZipAFolderExample}(b) and (c) show two mutations that \ToolName suggests for this project and
that could result in bugs similar to the one described above:
   part (b) shows a mutation at the same line where the bug was located that involves replacing read-access with write-access, and 
   part (c) shows a mutation at a nearby location that mirrors the change made by the developer.  

The state-of-the-art \StrykerJS tool is unable to suggest either of these mutations 
because (i) it does not support the 
mutation of  bitwise operator expressions such as \code{fs.constants.R_OK | fs.constants.W_OK} unless they 
appear as part of a control-flow predicate, nor (ii) mutations that involve replacing a binary 
expression with one of its operands. 
While adding support for mutating bitwise operator expressions would be straightforward, 
concerns have been expressed that adding more mutation operators to traditional mutation testing tools might result in 
too many mutants and degraded performance~\cite{Just15Higher,Just17Inferring,Coles24ArcMutateExtendedOperators}. 
More significantly,  \StrykerJS does not introduce or modify
property access expressions and has very limited support for replacing an expression with a different expression%
\footnote{
  In particular, \StrykerJS  only replaces control-flow predicates in \code{if}-statements and loops with boolean constants,  
  string literals with the value \code{"Stryker was here"}, and object literals with an empty object literal.
}. 

\begin{figure}
\centering
\begin{tabular}[t]{c}
\begin{minipage}{0.45\textwidth}
\includegraphics[width=\textwidth]{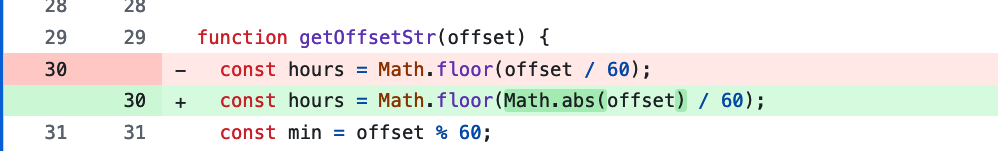}
\end{minipage}
\\
{\bf (a)}
\\
\begin{minipage}{0.45\textwidth}
\includegraphics[width=\textwidth]{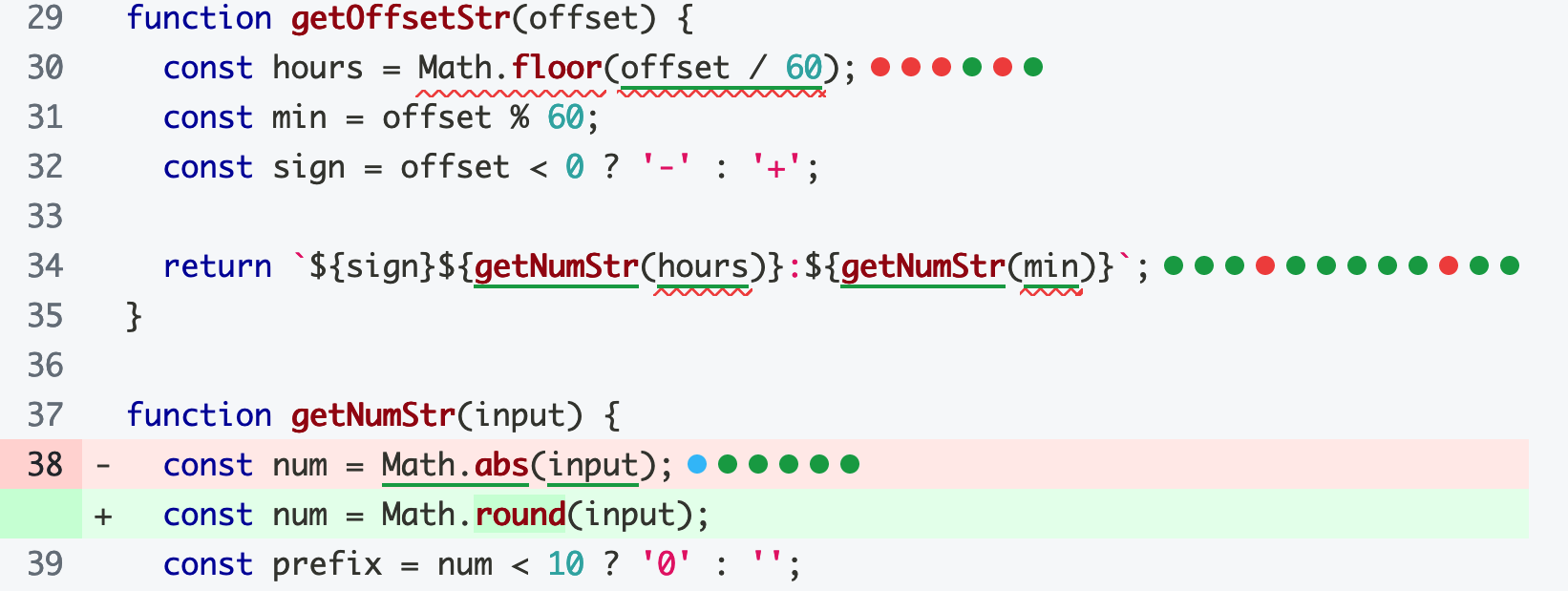}
\end{minipage}
\\
{\bf (b)}
\end{tabular}
\caption{
  (a) Fix for a bug reported in issue \#60 in \code{countries-and-timezones}.
  (b) A mutation suggested by \ToolName elsewhere in the same file.  
}
\label{fig:CountriesAndTimezonesExample}
\end{figure}

\paragraph{Example 2.}

Countries-and-timezones \cite{CountriesAndTimezones} is a library for working with countries and timezones.
In October 2023, a user reported a bug in function \code{getOffsetStr}, stating that it
produces incorrect results when invoked with negative values. The developer
proposed a simple fix that involves inserting a call to \code{Math.abs} to convert the
argument value to a non-negative number, and a variation on this fix was quickly adopted by the
developer, as shown in Figure~\ref{fig:CountriesAndTimezonesExample}(a). 

This bug-fix involves the introduction of a function call, so to \textit{introduce}
bugs like this one, a mutation testing tool would have to remove function calls or
change the function being invoked.
\StrykerJS only supports a very limited set of 20 mutations to function calls%
\footnote{
 See \url{https://stryker-mutator.io/docs/mutation-testing-elements/supported-mutators/}.
}, such as replacing calls to \code{String.startsWith} with call to \code{String.endsWith}
and removing a call to \code{Array.slice}. 
While one could extend \StrykerJS with a mutator that removes calls to \code{Math.abs}, many other function calls
could be handled similarly, and adding mutators for all of them would yield an overwhelmingly 
large number of mutants. 
Many such candidate functions would not be good choices for mutation, either because the function in
question is not a function that a developer inadvertently might have selected, or because it would
lead to syntactically invalid code. 

\ToolName suggests mutations that involve introducing and replacing function calls.  
Figure~\ref{fig:CountriesAndTimezonesExample}(b) shows a mutation that 
\ToolName suggested elsewhere in the same source file that involves replacing a call to
\code{Math.abs} with a call to \code{Math.round}, which could, in principle, introduce a bug 
like the one in Figure~\ref{fig:CountriesAndTimezonesExample}(a). Moreover, since LLMs are
trained to generate code that resembles code written by developers, it is likely that the
mutants produced by \ToolName involve the use of functions that a developer might have chosen.

\begin{figure}
\centering
\hspace*{-3mm}
\begin{tabular}[t]{c}
\begin{minipage}[t]{0.45\textwidth}
\includegraphics[width=\textwidth]{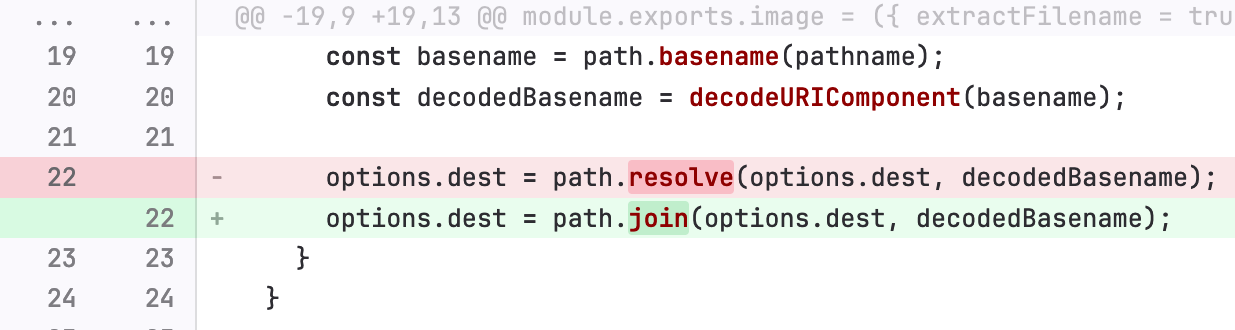}
\end{minipage}
\\
{\bf (a)}
\\
\begin{minipage}[t]{0.45\textwidth}
\includegraphics[width=\textwidth]{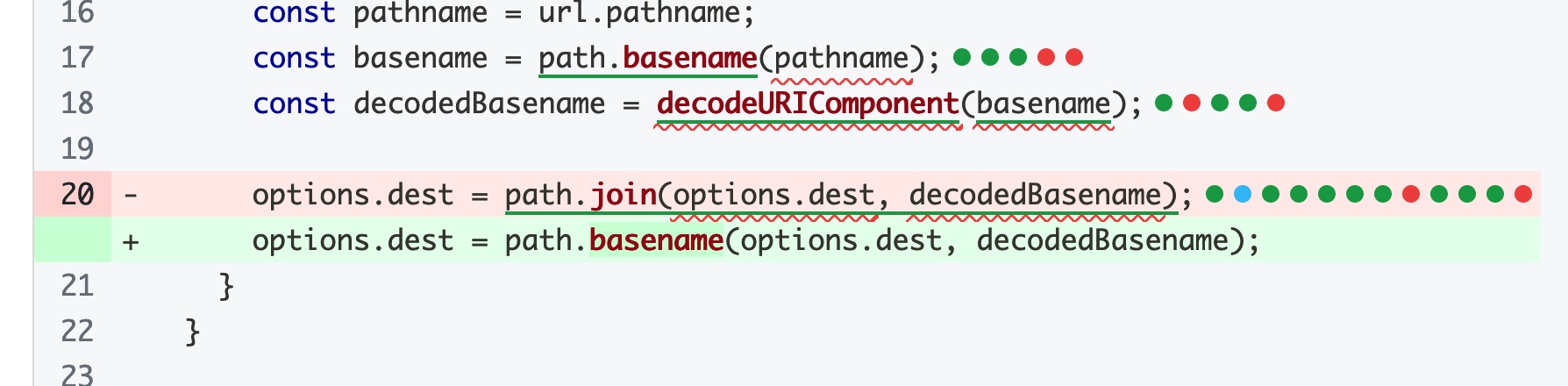}
\end{minipage}
\\
{\bf (b)}
\end{tabular}
\caption{
  (a) Fix for a bug reported in issue \#27 in \code{image-downloader}.
  (b) A mutation suggested by \ToolName at the same location that similarly involves
  calling a different function.
}
\label{fig:ImageDownloaderExample}
\end{figure}
\begin{figure}
\centering
\begin{tabular}{c}
\includegraphics[width=0.4\textwidth]{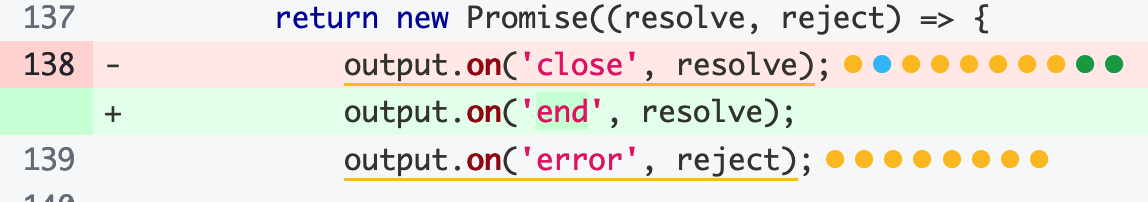}
\end{tabular}
\caption{
  A mutation suggested by \ToolName that involves associating an event listener
  with the \code{end} event instead of with the \code{close} event.
}
\label{fig:ZipAFolderEventNameMutation}
\end{figure}

\paragraph{Example 3.}

 \textit{image-downloader} is a module for downloading images.
In February 2022, a user opened issue~\#27, entitled ``If the directory name in dest: contains a dot \code{"."} then the download fails.'',
providing an example illustrating the problem. The developers soon responded with a fix, shown in Figure~\ref{fig:ImageDownloaderExample}(a),
that involves replacing a call to \code{path.resolve} with a call to \code{path.join}. 
While \ToolName does not produce a mutant that re-introduces this bug exactly, it does produce several at the same 
location%
\footnote{
  The line numbers have shifted slightly as the code has evolved since the bug report.
} that similarly replace the invoked function, including the one shown in Figure~\ref{fig:ImageDownloaderExample}(b). 
As mentioned, \StrykerJS has very limited support for mutations that involve calling different functions and
so it cannot suggest mutations like the one shown in  Figure~\ref{fig:ImageDownloaderExample}(b).

\paragraph{Example 4.}
 
\begin{sloppypar} 
Figure~\ref{fig:ZipAFolderEventNameMutation} shows another mutant produced by \ToolName for \textit{zip-a-folder}. Here,
the mutation involves changing the name of the event with which an event listener is associated. Such errors often
cause ``dead listeners'', i.e., situations where an event handler is never executed because it is associated
with the wrong event.  Dead listeners are quite common in JavaScript, where the use of string values to identify events 
precludes static checking, and previous research has focused on static analysis \cite{DBLP:conf/oopsla/MadsenTL15} and 
statistical methods \cite{DBLP:journals/tse/ArtecaST23} for detecting such errors.
\end{sloppypar} 
 
%
%
%

\paragraph{Discussion}

The above examples illustrate just a few of the kinds of mutations that \ToolName may produce.
Other mutations that it may suggest include:
  replacing a reference to a variable with a reference to a different variable,
  adding or removing arguments in function calls, and
  modifying object literals by adding or removing property-value pairs.

In practice, the number of such mutations is effectively infinite, so
an approach based on exhaustively applying a fixed set of mutation operators is unlikely to be practical.
\ToolName' LLM-based approach leverages the collective wisdom of programmers who 
wrote the code on which the LLM was trained to develop mutations.  As a result, suggested changes 
are likely to refer only to variables and functions that are in scope, and are likely to be type-correct.

\section{Approach}
  \label{sec:Approach}


\begin{figure*}[bht!]
\centering
\begin{tabular}{c}
\includegraphics[width=0.75\textwidth]{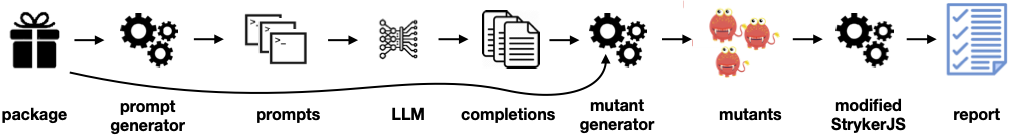}
\end{tabular}
\caption{Overview of approach. 
}
\label{fig:Overview}
\end{figure*}

\ToolName is capable of producing interesting mutants without requiring any training on a subject project, which
is a key distinction  compared to existing work that builds models of real bugs to generate mutants~\cite{Tufano19Learning,Tian23Learning,Patra21Semantic}.
This is accomplished by querying an LLM with a prompt that includes part of an application's source code
in which a code fragment is replaced with the text ``\code{<PLACEHOLDER>}''. Additional information provided in the prompt
includes:
  (i)  general background on mutation testing,
  (ii) the code fragment that was originally present at the placeholder's location,
  (iii) a request to apply mutation testing to the code by replacing the placeholder with a buggy code fragment, and
  (iv) suggestions \textit{how} the code could be mutated.
The LLM is asked to provide three possible replacements for the placeholder, each accompanied by an explanation
how the mutation would change program behavior.  

Figure~\ref{fig:Overview} presents a high-level overview of our approach, which involves three components that work in concert: 
the \textit{prompt generator}, the \textit{mutant generator},
and a version of the \textit{StrykerJS} mutation testing tool that has been modified to apply the mutants created
by \ToolName%
\footnote{
  In particular, we use \StrykerJS' to (i) determine the impact of each mutant on an application's tests and
  classify it as ``killed'', ``survived'', or ``timed-out'' and (ii) generate an interactive web page for
  inspecting results. 
}
We now discuss each of these components.

\begin{figure}
  \centering
  {\footnotesize
  \begin{tabular}{@{\hspace*{-4mm}}c@{\hspace*{-0mm}}|@{\hspace*{-0mm}}c@{\hspace*{-2mm}}}
  \smallcode{if (x === y)\{ ...  \}}
  &
  \smallcode{if (<PLACEHOLDER>)\{ ...  \}}
  \\
   \hline
  \smallcode{switch (x === y)\{ ...  \}}
  &
  \smallcode{switch (<PLACEHOLDER>)\{ ...  \}}
  \\
   \hline
  \smallcode{while (x)\{ ...  \}}
  &
  \smallcode{while (<PLACEHOLDER>)\{ ...  \}}
  \\
   \hline
  \smallcode{do \{ ...  \} while (x)\}}
  &
  \smallcode{do \{ ...  \} while (<PLACEHOLDER>)}
  \\
  \hline
  \begin{lstlisting}[language=JavaScript,numbers=none,basicstyle=\scriptsize]
  for (let i=0; i < x; i++){ 
      ... 
  }
  \end{lstlisting}
  &
  \begin{tabular}{l}
  \smallcode{for (<PLACEHOLDER>; i < x; i++)\{ ... \}}
  \\
  \smallcode{for (let i=0; <PLACEHOLDER>; i++)\{ ... \}}
  \\
  \smallcode{for (let i=0; i < x; <PLACEHOLDER>)\{ ... \}}
  \\
  \smallcode{for (<PLACEHOLDER>)\{ ... \}}
  \\ 
   \end{tabular}
  \\
  \hline
  \smallcode{for (o in obj)\{ ... \}}
  &
  \begin{tabular}{l}
  \smallcode{for (<PLACEHOLDER> in obj)\{ ... \}}
  \\
  \smallcode{for (o in <PLACEHOLDER>)\{ ... \}}
  \\
  \smallcode{for (<PLACEHOLDER>)\{ ... \}}
  \\ 
   \end{tabular}
  \\
   \hline
  \smallcode{for (o of obj)\{ ... \}}
  &
  \begin{tabular}{l}
  \smallcode{for (<PLACEHOLDER> of obj)\{ ... \}}
  \\
  \smallcode{for (o of <PLACEHOLDER>)\{ ... \}}
  \\
  \smallcode{for (<PLACEHOLDER>)\{ ... \}}
  \\ 
   \end{tabular}
  \\
  \hline
  \smallcode{a.m(x,y)}
  &
  \begin{tabular}{l}
  \smallcode{<PLACEHOLDER>(x,y)}
  \\
  \smallcode{a.m(<PLACEHOLDER>,y)}
  \\
  \smallcode{a.m(x,<PLACEHOLDER>)}
  \\ 
  \smallcode{a.m(<PLACEHOLDER>)}
  \end{tabular}
  \\
   \end{tabular}
  }
  \caption{Illustration of the insertion of placeholders  to direct
           the LLM at source locations that need to be mutated.}
    \label{fig:PlaceHolders}
\end{figure}

\paragraph{Prompt generator.} This component takes as input a package and generates a set of prompts.
The prompt generator parses the source files and identifies locations where mutations will be introduced. 
For ease of reference during prompting, the source code fragment corresponding to each location is replaced 
with the text ``\code{<PLACEHOLDER>}''.
\ToolName considers the following locations as candidates for mutation: (i) conditions of  \code{if}, \code{switch}, \code{while},
and \code{do-while} statements, (ii) initializers, updaters, and entire headers of loop statements, and (iii)
receiver, arguments, and entire sequence of arguments for function calls. For each such location, a separate
prompt is created. 
Figure~\ref{fig:PlaceHolders} illustrates where placeholders are introduced into the source code.

The LLM is then given a prompt that is created by instantiating the template shown in Figure~\ref{fig:PromptTemplate}(a),
by replacing \code{\{\{\{code\}\}\}} with the original source code in which  a placeholder has been inserted, 
and  \code{\{\{\{orig\}\}\}} with the code fragment that was replaced by the placeholder.  Figure~\ref{fig:PromptTemplate}(b)
shows the system prompt given to the LLM, which provides background on the role the LLM is expected to play in the conversation
as a mutation testing expert.
As can be seen in Figure~\ref{fig:PromptTemplate}(a), the prompt provides instructions for applying mutation testing to the 
specific source code at hand and details the specific format to which the completion should conform. Specifically, 
we require that the proposed mutants be provided inside ``fenced code blocks'' (i.e., code blocks surrounded by three backquote characters).  
 
\begin{figure}
\centering
\begin{tabular}{c}
\begin{minipage}[t]{0.475\textwidth}
{\scriptsize
\begin{verbatim}
Your task is to apply mutation testing to the following 
code:
```
{{{code}}}
```

by replacing the PLACEHOLDER with a buggy code fragment 
that has different behavior than the original code fragment, 
which was:
```
{{{orig}}}
```
Please consider changes such as using different operators, 
changing constants, referring to different variables, object 
properties, functions, or methods.  

Provide three answers as fenced code blocks containing a 
single line of code, using the following template:

Option 1: The PLACEHOLDER can be replaced with:
```
<code fragment>
```
This would result in different behavior because 
<brief explanation>.

Option 2: The PLACEHOLDER can be replaced with:
```
<code fragment>
```
This would result in different behavior because 
<brief explanation>.

Option 3: The PLACEHOLDER can be replaced with:
```
<code fragment>
```
This would result in different behavior because 
<brief explanation>.

Please conclude your response with "DONE."
\end{verbatim}
}
\end{minipage}
\\[2mm]
{\bf (a)}
\\
\begin{minipage}[t]{0.475\textwidth}
{\scriptsize
\begin{verbatim}
You are an expert in mutation testing. Your job is to 
make small changes to a project's code in order to find 
weaknesses in its test suite. If none of the tests fail 
after you make a change, that indicates that the tests 
may not be as effective as the developers might have 
hoped, and provide them with a starting point for 
improving their test suite.
\end{verbatim}
}
\end{minipage}
\\[3mm]
{\bf (b)}
\end{tabular} 
\caption{Prompt template (a) and system prompt (b) used by \ToolName.}
  \label{fig:PromptTemplate}
\end{figure}

\paragraph{Mutant generator.} This component takes the completions received from the LLM and extracts  
candidate mutants from the instantiated template  by matching a regular expression against
the completion to find the fenced code blocks.
Candidate mutants identical to the original source code
fragment or identical to previously generated mutants are discarded.  The candidate 
mutants are then parsed to check if they are syntactically valid and discarded if this is not the case. 
The resulting mutants are written to a file \code{mutants.json} that is read by a customized version of \StrykerJS
that is described below.  The mutant generator also saves all experimental data to files, including the generated prompts, 
completions received from the LLM, and the configuration options that were used (e.g., the LLM's temperature setting). 

\paragraph{Custom version of \StrykerJS.}  
We modified \StrykerJS to give it an option \code{--usePrecomputed} that, if selected, directs
it to read its set of mutations from a file \code{mutants.json} instead.
\StrykerJS then executes all mutants and determines for each mutant whether it causes tests to fail or time out.
When this analysis is complete, \StrykerJS generates a report as an interactive web page allowing users 
to inspect the generated mutants. The previously shown Figures~\ref{fig:ZipAFolderExample}--\ref{fig:ZipAFolderEventNameMutation} show screenshots 
of our custom version of \StrykerJS.  
   
\paragraph{Pragmatics}

While \ToolName implements a conceptually straightforward technique, considerable engineering effort was required to make it a
practical tool.
We use BabelJS \cite{Babel} for parsing source code to identify locations where placeholders should be inserted,
and to check syntactic validity of candidate mutants. Handlebars \cite{Handlebars}  is used for instantiating
prompt templates. \StrykerJS expects mutants to correspond to a single AST node, so for mutants that do not 
correspond exactly to a single AST node (e.g., loop headers and sequences of arguments passed in function calls), 
it is necessary to expand the mutation to the nearest enclosing AST node, for which we also rely on BabelJS. 

\ToolName has command-line arguments for specifying the prompt template and system template to be used. 
Furthermore, it enables users to specify a number of LLM-specific parameters, such as the maximum length of 
completions that should be generated, the sampling temperature%
\footnote{
  The sampling temperature is a parameter between 0 and 2 that controls the randomness of the completions generated by the LLM.
  Roughly speaking, the higher the temperature the more diverse the completions. At temperature zero, the LLM will always 
  generate the most likely completion, which increases the chance that the same prompt will result in the same completion.
}, and number of lines of source code that should be included in prompts (by default, this is limited to 200
lines surrounding the location of the placeholder).
Since many LLM installations have limited capacity or explicit rate limits, \ToolName provides two command-line options 
to work with such LLMs: 
   \code{--rateLimit <N>}  ensures that least N milliseconds will have elapsed between successive prompts and
   \code{--nrAttempts <N>} will try the same prompt up to N times if a 429 error occurs.  

One possible concern with our approach is that \ToolName relies on a fixed set of 
locations where it introduces placeholders. The current placeholder scheme aims
to strike a balance between creating a number of mutants that is practical and
having mutants that are likely to result in different control flow or data flow
and that are likely to be different than what can be achieved using mutation operators. 
It would be straightforward to modify \ToolName to use a different placeholder scheme.
That said, the examples in Section~\ref{sec:Motivation} show that mutants produced by \ToolName 
(using its current placeholder scheme) involve changing references to variables, properties, and 
functions that cannot be produced using Stryker's mutation operators and that correspond to real-world bugs.

An open-source release of \ToolName can be found at \url{https://github.com/neu-se/llmorpheus} and the customized version of \StrykerJS
that we used for classifying mutants can be found at \url{https://github.com/neu-se/stryker-js}.
 
\begin{table*}
  \centering
  {\scriptsize
\begin{tabular}{l||c|c|r|r|r|r||r|r|r|r|r|r}
    {\bf application} & {\bf description} & {\bf weekly} & {\bf \#LOC} & {\bf \#tests} & \multicolumn{2}{|c||}{\bf coverage} & \multicolumn{6}{|c}{\StrykerJS}\\
                      &                   & {\bf downloads}        &             &               & {\bf stmt} & {\bf branch} & \Total & \Killed & \Survived & \Timeout & \MutScore & \Time\\
    \hline
    \href{https://github.com/infusion/Complex.js/tree/d995ca105e8adef4c38d0ace50643daf84e0dd1c}{\color{blue} \it Complex.js} & \WRAP{complex numbers} & 671K & 1,425 & 216 & 71.82\% & 67.54\% & 1,302 & 763 & 539 & 0 & 58.60 & \ChangedText{405.08} \\   
    \hline  
    \href{https://github.com/manuelmhtr/countries-and-timezones/tree/241dd0f56dfc527bcd87779ae14ed67bd25c1c0e}{\color{blue} \it countries-and-timezones} & \WRAP{accessing countries and timezones data} &  152K & 165 & 58 &  100\% & 92.55\% & 140 & 134 & 6 & 0 & 95.71 & \ChangedText{142.37}\\ 
    \hline
    \href{https://gitlab.com/autokent/crawler-url-parser/tree/202c5b25ad693d284804261e2b3815fe66e0723e}{\color{blue} \it crawler-url-parser} & \WRAP{URL parser for crawling} & 495 & 209 & 185 & 96.39\% & 92.5\% & 226 & 143 & 83 & 0 & 63.27 & \ChangedText{433.53} \\ 
    \hline 
    \href{https://github.com/quilljs/delta/tree/5ffb853d645aa5b4c93e42aa52697e2824afc869}{\color{blue} \it delta} & \WRAP{Format for representing rich text documents and changes} & 1.76M & 806 & 180 & 98.99\% & 95.89\% & 834 & 686 & 88 & 60 & 89.45 & \ChangedText{2747.04} \\ 
    \hline
    \href{https://gitlab.com/demsking/image-downloader/tree/19a53f652824bd0c612cc5bcd3a2eb173a16f938}{\color{blue} \it image-downloader} & \WRAP{downloading image to disk from a given URL} & 17.75K & 64 & 11 & 100\% & 93.75\% & 43 & 28 & 11 & 4 & 74.42 & \ChangedText{284.20} \\
\end{tabular}
  }
  \\[2mm]
  \caption{Subject applications used to evaluate \ToolName.}
  \label{table:SubjectApplications}
\end{table*}

\section{Evaluation}
   \label{sec:Evaluation}

\subsection{Research Questions}

This evaluation aims to answer the following research questions:
\begin{enumerate}
  \item[\bf RQ1] How many mutants does \ToolName create?
  \item[\bf RQ2] How many of the surviving mutants produced by \ToolName are equivalent mutants? 
  \item[\bf RQ3] What is the effect of using different temperature settings?
  \item[\bf RQ4] What is the effect of variations in the prompting strategy used by \ToolName?
  \item[\bf RQ5] How does the effectiveness of \ToolName depend on the LLM that is being used?
  \item[\bf RQ6] What is the cost of running \ToolName?
  \item[\ChangedText{\bf RQ7}] \ChangedText{Is \ToolName capable of producing mutants that resemble existing bugs?}
\end{enumerate}

\subsection{Experimental Setup}

\paragraph{Selecting subject applications}
Our goal is to evaluate \ToolName on real-world JavaScript packages that have test suites. Moreover, we 
want to compare the mutants generated by\ToolName to those generated using traditional mutation testing techniques, so we decided
to focus on projects for which the state-of-the-art StrykerJS mutation testing tool \cite{Stryker} could be applied successfully.
As a starting point for benchmark selection, we considered the 25 subject applications that were used to evaluate
TestPilot \cite{DBLP:journals/tse/SchaferNET24}, a recent LLM-based unit test generation tool. These applications are written in JavaScript or TypeScript, 
cover a range of different domains, and have test suites that can be executed successfully.    

Of these 25 subject applications, 10 could not be used because StrykerJS does not work on them, either because its dependences conflict
with those of the subject application itself%
\footnote{
  Running StrykerJS on an application requires installing it locally among the subject project libraries. Stryker itself 
  depends on various other packages that also need to be installed, and these packages may conflict with packages that the 
  subject application itself depends upon.
}, or because it crashes.  On one package, \textit{simple-statistics}, StrykerJS
requires approximately 10 hours of running time, which makes using it impractical. We excluded another package, \textit{fs-extra}, a utility
library for accessing the file system, because we observed that mutating this application poses a significant security risk, as the
mutated code was corrupting our local file system. This left us with \NrSubjectApplications subject applications for which Table~\ref{table:SubjectApplications} 
provides key characteristics. The first set of columns in the table show, from left to right, the name of the package, a short description of its functionality, the number of 
weekly downloads according to \url{npmjs.com}, the number of lines of source code, the number of tests, and the statement and branch coverage achieved by those 
tests, respectively. 
The second set of columns shows the results of running \StrykerJS on the applications: the total set of mutants, the number of mutants that were killed,
survived, and that timed out, the \textit{mutation score}%
\footnote{
  The mutation score aims to provide a measure of the quality of a test suite by calculating 
  the fraction of the total number of mutants that are detected (i.e., killed or timed out),
  see \url{https://stryker-mutator.io/docs/General/faq/}.
} reported by \StrykerJS, and the time required to run \StrykerJS, respectively.

\begin{sloppypar}
\paragraph{LLM selection}
\ChangedText{RQ5 explores how the effectiveness of the proposed technique depends on the LLM being used.  
We use Meta's  \CodeLlamaThirtyFour model for our main experiments. In addition, we evaluate the technique with Meta's
\CodeLlamaThirteen and \LlamaThreeThree  models, with Mistral's \Mixtral model, and
with OpenAI's \GPTFouroMini model.  The \textit{codellama} models are specifically trained for tasks involving code.
\LlamaThreeThree is a newer and larger model from Meta that supercedes the smaller, specialized \textit{codellama} models.
\Mixtral is a state-of-the-art general-purpose ``mixture-of-experts'' LLM developed by Mistral. 
\GPTFouroMini is a smaller, faster, and lower-cost variant of OpenAI's popular \GPTFouro model
The  \CodeLlamaThirtyFour, \CodeLlamaThirteen, \LlamaThreeThree, and \Mixtral LLMs    
are ``open'' in the sense that their training process is documented.  We relied on several commercial LLM service providers (\url{https://octo.ai}, 
\url{https://openai.com}, and \url{https://openrouter.ai}) for the experiments reported on in this paper.
}    
\end{sloppypar}

\paragraph{LLM Temperature settings}
LLMs have a temperature parameter that reflects the amount of randomness or creativity in their completions. For a task such as
mutation testing, randomness and creativity may determine whether generated mutants are killed or survive.
Therefore, we conduct experiments using several temperature settings.

\paragraph{\ChangedText{Similarity to real-world bugs}}
\ChangedText{
Previous work on evaluating mutation testing techniques has focused on ``coupling'' to determine whether mutants resemble real-world bugs~\cite{Just14AreMutants,Laurent22Revisiting,Gay23HowClosely}. 
This involves determining whether a test suite that detects particular mutants also detects particular real faults, and requires a curated dataset of isolated faults.
While there is a wealth of such datasets constructed from open-source projects written in Java, we found only one JavaScript dataset, the Bugs.js suite~\cite{Gyimesi19BugsJS}.
Unfortunately, we found that most of these subjects could 
not be used at all due to their reliance on outdated versions of various libraries and because of their incompatibility with modern Node.js versions that \StrykerJS  
requires, causing them to be incompatible with \ToolName.  These projects also have flaky tests%
\footnote{
  See \url{https://github.com/BugsJS/bug-dataset/issues/11}.
}, making it particularly challenging to perform mutation analysis~\cite{Shi19Mitigating}.
We, therefore, opted to conduct a case study involving four real-world bugs from the Bugs.js suite that we could reproduce reliably.  
}

\begin{table*}[hbt!]
\centering
{\scriptsize
\begin{tabular}{l||r|r|r|r|r|r|r|r|r|r|r|r|r|r|r}
  {\bf application} & {\bf \#prompts} & \multicolumn{9}{|c|}{\ToolName} & \multicolumn{4}{|c|}{\ChangedText{\StrykerJS}} \\
  & &  {\bf \ChangedText{\Candidates}} & {\bf \ChangedText{\Invalid}} & {\bf \ChangedText{\Identical}} & {\bf \ChangedText{\Duplicate}} & \Total & \Killed & \Survived & \Timeout & \MutScore & \ChangedText{\Total} & \ChangedText{\Killed} & \ChangedText{\Survived} & \ChangedText{\Timeout} & \ChangedText{\MutScore} \\
  \hline
  \hline
\textit{Complex.js} 
      & 490 
      & \ChangedText{1,451} 
      & \ChangedText{194} 
      & \ChangedText{13} 
      & \ChangedText{45} 
      & 1,199 
      & 725 
      & 473 
      & 1 
      & 60.55 
      & \ChangedText{1,302}
      & \ChangedText{763} 
      & \ChangedText{539} 
      & \ChangedText{0} 
      & \ChangedText{58.60} 
      \\ 
\hline
\textit{countries-and-timezones} 
      & 106 
      & \ChangedText{318} 
      & \ChangedText{89} 
      & \ChangedText{0} 
      & \ChangedText{12} 
      & 217 
      & 188 
      & 29 
      & 0 
      & 86.64 
      & \ChangedText{140}
      & \ChangedText{134} 
      & \ChangedText{6} 
      & \ChangedText{0} 
      & \ChangedText{95.71} 
      \\ 
\hline
\textit{crawler-url-parser} 
      & 176 
      & \ChangedText{521} 
      & \ChangedText{205} 
      & \ChangedText{14} 
      & \ChangedText{17} 
      & 285 
      & 157 
      & 128 
      & 0 
      & 55.09 
      & \ChangedText{226}
      & \ChangedText{143} 
      & \ChangedText{83} 
      & \ChangedText{0} 
      & \ChangedText{63.27} 
      \\ 
\hline
\textit{delta} 
      & 462 
      & \ChangedText{1,367} 
      & \ChangedText{565} 
      & \ChangedText{10} 
      & \ChangedText{25} 
      & 767 
      & 634 
      & 101 
      & 32 
      & 86.83 
      & \ChangedText{834}
      & \ChangedText{686} 
      & \ChangedText{88} 
      & \ChangedText{60} 
      & \ChangedText{89.45} 
      \\
\end{tabular}
  }
  \\[2mm]
  \caption{Results from LLMorpheus experiment \ChangedText{(run \#312)}.
    Model: \textit{codellama-34b-instruct}, 
    temperature: 0.0, 
    maxTokens: 250, 
    template: \textit{template-full.hb}, 
    systemPrompt: \textit{SystemPrompt-MutationTestingExpert.txt}. 
  }
  \label{table:CodeLlama34Full0.0}

{\scriptsize
\begin{tabular}{l||rrrr|rrrr|rrrr|rrrr}
    & \multicolumn{4}{|c|}{\bf temp. 0.0 \ChangedText{(run \#312)}} &  \multicolumn{4}{|c|}{\bf temp. 0.25 \ChangedText{(run \#348)}} & \multicolumn{4}{|c|}{\bf temp. 0.50 \ChangedText{(run \#318)}} &  \multicolumn{4}{|c}{\bf temp. 1.0 \ChangedText{(run \#341)}} \\
    &  \Total & \Killed & \Survived & \Timeout
    &  \Total & \Killed & \Survived & \Timeout
    &  \Total & \Killed & \Survived & \Timeout
    &  \Total & \Killed & \Survived & \Timeout  \\
\hline
\hline\textit{Complex.js} & 1,199 & 725 & 473 & 1 & 1,197 & 730 & 466 & 1 & 1,191 & 739 & 452 & 0 & 1,028 & 648 & 379 & 1 \\
\textit{countries-and-timezones} & 217 & 188 & 29 & 0 & 219 & 181 & 38 & 0 & 224 & 194 & 30 & 0 & 186 & 156 & 30 & 0 \\
\textit{crawler-url-parser} & 285 & 157 & 128 & 0 & 260 & 167 & 93 & 0 & 298 & 166 & 108 & 24 & 278 & 202 & 76 & 0 \\
\textit{delta} & 767 & 634 & 101 & 32 & 781 & 642 & 111 & 28 & 769 & 642 & 93 & 34 & 698 & 583 & 83 & 32 \\
\textit{image-downloader} & 89 & 72 & 17 & 0 & 86 & 71 & 15 & 0 & 89 & 68 & 21 & 0 & 75 & 53 & 22 & 0 \\
\textit{node-dirty} & 275 & 163 & 100 & 12 & 279 & 175 & 93 & 11 & 277 & 158 & 107 & 12 & 246 & 150 & 84 & 12 \\
\textit{node-geo-point} & 302 & 223 & 79 & 0 & 293 & 225 & 68 & 0 & 302 & 230 & 72 & 0 & 273 & 213 & 60 & 0 \\
\textit{node-jsonfile} & 154 & 49 & 48 & 57 & 151 & 52 & 41 & 58 & 153 & 51 & 43 & 59 & 132 & 50 & 22 & 60 \\
\textit{plural} & 281 & 205 & 75 & 1 & 273 & 208 & 63 & 2 & 289 & 219 & 69 & 1 & 299 & 229 & 69 & 1 \\
\textit{pull-stream} & 769 & 441 & 271 & 57 & 779 & 452 & 270 & 57 & 796 & 465 & 278 & 53 & 743 & 461 & 235 & 47 \\
\textit{q} & 2,035 & 158 & 1,792 & 85 & 2,050 & 153 & 1,813 & 84 & 2,073 & 163 & 1,823 & 87 & 1,899 & 147 & 1,671 & 81 \\
\textit{spacl-core} & 239 & 199 & 39 & 1 & 223 & 187 & 36 & 0 & 250 & 210 & 39 & 1 & 218 & 180 & 38 & 0 \\
\textit{zip-a-folder} & 100 & 23 & 3 & 74 & 97 & 24 & 4 & 69 & 87 & 48 & 33 & 6 & 96 & 54 & 38 & 4 \\
\end{tabular}
}
\caption{Number of mutants generated using the \CodeLlamaThirtyFour LLM at temperatures 0.0, 0.25, 0.5, and 1.0 \ChangedText{(showing one run of each)}}
\label{table:Temperature}
\end{table*}

\subsection{RQ1: How many mutants does \ToolName create?}

To answer this question, we ran \ToolName on the projects listed in Table~\ref{table:SubjectApplications} using the \CodeLlamaThirtyFour LLM at temperature 0.0 and 
the prompt templates shown in Figure~\ref{fig:PromptTemplate}.  The results, shown in Table~\ref{table:CodeLlama34Full0.0}, 
show that \ToolName produces between \NrPromptsMin and \NrPromptsMax prompts for these projects.
\ChangedText{
   The next 4 columns in  the table show the number of ``candidate mutants'', i.e., code fragments obtained by replacing placeholders with code fragments suggested by
   the LLM. 
     The subcolumn labeled ``candidates'' shows the total number of candidate mutants,
     the subcolumn labeled ``invalid'' shows the number of candidate mutants that were found to be syntactically invalid,
     the subcolumn labeled ``identical'' shows the number of candidate mutants that were found to be identical to the original code, and
     the subcolumn labeled ``duplicate'' shows the number of  candidate mutants that were found to be duplicated.     
From this data, it can be inferred that, on average, 
  \PercentageInvalidMutantCandidates (\NrInvalidMutantCandidates/\TotalMutantCandidates) of candidate mutants are discarded because they are syntactically invalid,
  \PercentageIdenticalMutantCandidates (\NrIdenticalMutantCandidates/\TotalMutantCandidates) are discarded because they are identical to the original code, and
  \PercentageDuplicateMutantCandidates (\NrDuplicateMutantCandidates/\TotalMutantCandidates) are discarded because they are duplicates.
This suggests that LLMs generally do not have too much trouble with generating syntactically correct code, which is consistent with recent findings by others
  \cite{LemieuxICSE2023,DBLP:journals/tse/SchaferNET24}.   
}     
\ChangedText{
The next column, labeled ``mutants'', shows the number of mutants that remain after discarding the useless candidate mutants. Here, the reader can see that
between \FullMutantsMin and \FullMutantsMax mutants are generated for the subject packages.
Of these mutants,  
  between \FullKilledMutantsMin and \FullKilledMutantsMax are killed, 
  between  \FullSurvivingMutantsMin and \FullSurvivingMutantsMax survive, and
  between \FullTimeoutMutantsMin and \FullTimeoutMutantsMax time out.
Aggregating the results over all projects, it can be seen that
  \PercentageKilledMutants (\NrKilledMutants/\TotalMutants) of all mutants are killed,
  \PercentageSurvivingMutants (\NrSurvivingMutants/\TotalMutants) of all mutants survive, and
  \PercentageTimedOutMutants (\NrTimedOutMutants/\TotalMutants) of all mutants time out.
}  

\ChangedText{
The table also shows the \textit{mutation score}%
\footnote{\ChangedText{
  See \url{https://stryker-mutator.io/docs/General/faq/}.
}} as reported by StrykerJS, which aims to provide a measure of the quality of a test suite by calculating 
the fraction of the total number of mutants that are detected (i.e., killed or timed out).    
}

\ChangedText{
  To facilitate a quantitative comparison with \StrykerJS, the last 5 columns in Table~\ref{table:CodeLlama34Full0.0} repeat 
  the results of running \StrykerJS on the subject applications from Table~\ref{table:SubjectApplications}.
  From this data, it can be seen that---in the aggregate for the \NrSubjectApplications projects under consideration---\ToolName produces \NrSurvivingMutants 
  surviving mutants whereas \StrykerJS produces \StrykerJSNrSurvivingMutants surviving mutants.  
  However, it should be noted that the difference in the number of mutants and surviving mutants varies significantly between subject applications.
  For example, for \textit{Complex.js} \StrykerJS produces more mutants (\StrykerComplexJSNrMutants vs. \LLMorpheusComplexJSNrMutants) than \ToolName, of which more 
  survive (\StrykerComplexJSNrSurvivingMutants vs. \LLMorpheusComplexJSNrSurvivingMutants).
  On the other hand, for \textit{q}, the situation is reversed with \StrykerJS producing fewer mutants (\StrykerQNrMutants vs. \LLMorpheusQNrMutants) and 
  fewer surviving mutants (\StrykerQNrSurvivingMutants vs.\LLMorpheusQNrSurvivingMutants) than \ToolName.
  We conjecture that such differences are due to the subject programs having different characteristics, which makes them amenable to different types of mutations.
  Here, \textit{Complex.js} makes heavy use of arithmetic operators to implement mathematical operations on complex numbers, and such operators are prime
  candidates for \StrykerJS's standard mutation operators.  Moreover, \textit{q} makes heavy use of method calls, which are targeted by \ToolName's placeholder-based
  strategy but much less so by \StrykerJS's mutation operators.  
}

\begin{figure*}
\centering
	\includegraphics[width=0.8\textwidth]{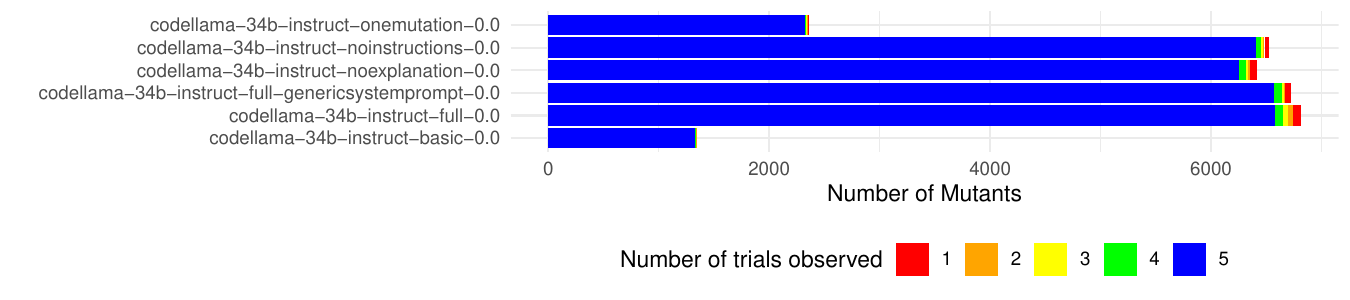}
	\vspace*{-5mm}
	\caption{\begin{revision*}Stability of mutants generated by LLMorpheus with \CodeLlamaThirtyFour at temperature 0.0. For each replacement generated at each position, we count the number of trials (of 5 total) where that replacement was generated.\end{revision*}}
\label{fig:34bstability}

\end{figure*}

LLMs are nondeterministic, even at temperature 0, so a subsequent experiment may produce results that differ from those shown in Table~\ref{table:CodeLlama34Full0.0}.
To determine to what extent this is the case, we repeated the same experiment 4 more times and measured how often the same mutants occur.
We found that, at temperature 0.0, the results of \ToolName are generally stable across runs, with between \MinPercentageOfCommonMutants and 
\MaxPercentageOfCommonMutants of all mutants being observed in all 5 experiments%
\footnote{
  All experimental data associated with this experiment and the other experiments are included with this submission as supplemental materials.
}.
\begin{revision*}
Figure~\ref{fig:34bstability} visualizes the stability of \ToolName when varying the prompt settings, and our supplemental materials include results across all settings of all models.
\end{revision*}

\begin{takeaway}
  Using \CodeLlamaThirtyFour at temperature 0, \ToolName generates between \FullMutantsMin and \FullMutantsMax mutants, of which
  between \FullSurvivingMutantsMin and \FullSurvivingMutantsMax survive. These results are stable across experiments, with between 
  \MinPercentageOfCommonMutants and \MaxPercentageOfCommonMutants of all mutants being observed in all five experiments.
\end{takeaway} 
\vspace*{-3mm}

\begin{table}
\centering
\footnotesize
\begin{tabular}{l||r|r|r|| r|r|r}
& \multicolumn{3}{c ||}{\textbf{\ToolName}} & \multicolumn{3}{c}{\textbf{Stryker.js}} \\
\textbf{Project} & \Equiv & \NotEquiv & \Unknown & \Equiv & \NotEquiv & \Unknown \\
\hline
\textit{Complex.js} & 6 & 44 & 0 & 0 & 50 & 0\\
\textit{countries-and-timezones} & 17 & 11 & 1 & 0 & 6 & 0\\
\textit{crawler-url-parser} & 16 & 33 & 1 & 0 & 47 & 3\\
\textit{delta} & 2 & 48 & 0 & 2 & 48 & 0\\
\textit{image-downloader} & 4 & 12 & 1 & 0 & 4 & 4\\
\textit{node-dirty} & 13 & 36 & 1 & 0 & 46 & 4\\
\textit{node-geo-point} & 1 & 47 & 2 & 1 & 41 & 8\\
\textit{node-jsonfile} & 7 & 25 & 0 & 0 & 3 & 0\\
\textit{plural} & 16 & 34 & 0 & 0 & 36 & 1\\
\textit{pull-stream} & 3 & 44 & 3 & 0 & 48 & 2\\
\textit{q} & 2 & 48 & 0 & 0 & 48 & 2\\
\textit{spacl-core} & 18 & 21 & 0 & 2 & 12 & 6\\
\textit{zip-a-folder} & 0 & 0 & 0 & 0 & 6 & 0\\
\midrule
\textit{Total} & 105 & 403 & 9 & 5 & 395 & 30\\
\end{tabular}	
\\[2mm]
\caption{Number of equivalent surviving mutants generated by \ToolName and \StrykerJS. 
 }
\label{table:EquivalentMutants}
\vspace*{-7mm}
\end{table}
 
\subsection{RQ2: How many of the surviving mutants are equivalent mutants?}

One of the key challenges in mutation testing is the phenomenon of \textit{equivalent mutants}: mutants that have 
equivalent behavior as the original code~\cite{DeMillo78Hints}. Mutants produced by \ToolName may involve arbitrary code changes, so it is entirely possible 
for the LLM to suggest code that is effectively a refactored version of the code that was originally present.
\begin{revision*}
Determining mutant equivalence requires a deep understanding of the program's intended behavior, as a mutant might change the behavior of the program but within bounds that are valid to the program's specification.
To determine to what extent surviving mutants produced by \ToolName are equivalent, we conducted a study in which two authors manually examined 50 surviving mutants%
\footnote{
  For projects with fewer than 50 surviving mutants, we used as many as were available.
} in each project and classified each mutant as ``equivalent'', ``not equivalent'', or ``unknown''.
After independently coding each sampled mutant, the two authors met, discussed disagreement, and came to a consensus on all mutants.
We define these categories as follows.

We labeled a mutant as \emph{equivalent} if we could determine that the change could not cause \emph{any} observable difference in behavior.
For example, mutants that added extra parameters to methods (beyond those accepted by the receiver method) are trivially equivalent, as they are discarded by the runtime.
Other mutants are far from trivial to evaluate, and we manually wrote test code to attempt to discern the impact of, e.g., changing a condition from \code{if(!handler)} to \code{if(handler === undefined}).
Such a mutant is equivalent if \code{handler} is never any other ``falsy'' value (e.g., \code{null}, \code{false}, \code{NaN}, \code{0}, or the empty string \code{''}).

We labeled a mutant as \emph{not equivalent} if it produced a change that could be observed as a behavioral change to a client of the library.
For example, a mutant in the statement \code{const hours=Math.floor(totalMinutes/60)} that changes the call from \code{Math.floor} to \code{Math.round} 
will result in the value of \code{hours} being incorrect.
Of course, if \code{hours} is never used (or it doesn't matter that it is off-by-one), then the mutant could still be equivalent.
Hence, we also found it necessary to trace through code to determine that some mutants were not equivalent.

In some cases, the mutant resulted in a change to the behavior of the program, but it was truly impossible to determine whether or not this change impacted the correctness of the program as intended by its developers.
For example, a mutant that redirects a log statement from \code{console.error} to \code{console.warn} may or may not represent a mutant that developers find to deviate from the application's intended behavior.
Similarly, mutants that change the contents of error messages may or may not represent an equivalent mutant.
We labeled all such cases \emph{unknown}.

The results are shown in Table~\ref{table:EquivalentMutants}.
Of the \NrMutantsInEquivalenceCaseStudy mutants under consideration, the majority \NrNotEquivalentMutants (\PercentageNotEquivalentMutants) are ``not equivalent'', \NrEquivalentMutants (\PercentageEquivalentMutants) are ``equivalent'',  and remaining \NrUnknownMutants (\PercentageUnknownMutants) are classified as ``unknown''. 
To place these findings in perspective, we applied the same manual classification to up to 50 surviving mutants produced by \StrykerJS  
and found that of  \StrykerNrMutantsInEquivalenceCaseStudy surviving mutants, \StrykerNrNotEquivalentMutants (\StrykerPercentageNotEquivalentMutants) 
are ``not equivalent'', \StrykerNrEquivalentMutants (\StykerPercentageEquivalentMutants) are ``equivalent'', and
\StrykerNrUnknownMutants (\StrykerPercentageUnknownMutants) are classified as ``unknown''.  
%

We further examined the \NrEquivalentMutants  equivalent mutants, and observed a number of common patterns, including:
  (i)  checking for \code{null}-ness or \code{undefined}-ness in different ways  (e.g., replacing \code{x != null} with \code{!x} or vice versa),
  (ii) refactoring of calls to the \code{String.substring} method with one of its near-equivalent counterparts \code{String.substr} and \code{String.slice},
 (iii) adding modifiers such as \code{/g} or \code{/m} to a regular expression in cases where this does not have any effect,
  (iv) calls to the \code{Array.slice} method in cases where this does not have any effect, and 
   (v) calling functions with more arguments than are declared.
For the \NrEquivalentMutants equivalent mutants under consideration, approximately 40\% fall into one of these categories.  
We expect that most of these equivalent mutants can be filtered out using an AST-based static analysis. However, further investigation is needed 
because some mutants that cause behavioral differences are syntactically similar to these patterns. This means that any pattern-matching-based 
approach should consider the context in which the mutation occurs to determine whether a mutant is likely to be equivalent. 
Section~\ref{sec:Conclusions} will discuss future work that aims to reduce the number of equivalent mutants.


\end{revision*}

\begin{takeaway}
  The majority (\PercentageNotEquivalentMutants) of the surviving mutants produced by \ToolName are not equivalent to the original code fragments they replace. 
  \ToolName produces significantly more ``equivalent'' mutants than \StrykerJS. However, the number
  of ``not-equivalent'' mutants exceeds the number of equivalent mutants by more than a factor of three\ChangedText{, and preliminary analysis reveals good 
  potential for future work on automatically filtering out  equivalent mutants using static analysis.}
\end{takeaway}
\vspace*{-5mm}

\subsection{RQ3: What is the effect of different temperature settings?}

To explore the impact of an LLM's temperature setting, we repeated the experiment with the \CodeLlamaThirtyFour LLM using temperatures 0.25, 0.50, and 1.0.  
The results of these experiments are summarized in Table~\ref{table:Temperature}.  
As can be seen from the table, the total number of mutants and the number of surviving mutants at temperatures 0.0, 0.25, and 0.50
are generally fairly similar. However, at temperature 1.0, both the total number of mutants and the number of surviving mutants decline noticeably
compared to the results for temperature 0.0. Inspection of the results revealed that this is partly because more of 
the generated mutants are syntactically invalid.  
 
A secondary question is how temperature affects the variability of results. To answer this question, we repeated the
experiment 5 times at each temperature and measured how many distinct mutants occur and how many mutants occur in all five runs. 
We found that, at higher temperatures, the number of distinct mutants increases rapidly and that the number of mutants common to all runs
decreases accordingly.  For example, for \textit{Complex.js}, \ToolName generates \ComplexJSDistinctZero distinct mutants at temperature 0 of which \ComplexJSCommonZero 
(\ComplexJSCommonZeroPercentage) are common to
all five runs. At temperature 0.25, the number number of distinct mutants increases to \ComplexJSDistinctTwentyFive, of which \ComplexJSCommonTwentyFive (\ComplexJSCommonTwentyFivePercentage) 
are common to all five runs.
At temperature 0.5, there are \ComplexJSDistinctFifty distinct mutants of which \ComplexJSCommonFifty (\ComplexJSCommonFiftyPercentage) are common to all runs. At temperature 1.0, there are \ComplexJSDistinctOne distinct mutants, of which \ComplexJSCommonOne (\ComplexJSCommonOnePercentage) are common to all runs, 
meaning that, effectively, at temperature 1.0 each run produces completely different mutants.  The results for the other subject applications are similar.
The supplemental materials associated with this paper include an analysis showing the overall variability in mutants killed and survived across each of the five runs.

\begin{takeaway}
  \ToolName generally produces similar numbers of mutants at temperatures $\leq 0.5$, of which a similar number survives. At temperature 1.0, the
  number of generated and surviving mutants decline noticeably because more candidate mutants are syntactically invalid.
  The variability of results is inversely dependent on the temperature, with mostly the same mutants being produced  at temperature 0,
  and mostly different mutants at temperature 1 in different runs.
\end{takeaway}
\vspace*{-3mm}

\subsection{RQ4: What is the effect of variations in the prompting strategy used by \ToolName?}
  \label{sec:RQ4}

Thus far, we have evaluated the effectiveness of the prompt template of Figure~\ref{fig:PromptTemplate}(a)  (henceforth referred to as \texttt{full}) 
by measuring how many mutants are generated and classifying them as ``killed'', ``survived'', or ``timed-out'' (see Table~\ref{table:CodeLlama34Full0.0}).  
To determine what the effect is of each component of this prompt, we experimented with the following variations%
\footnote{
  All prompt templates are included with the supplemental materials.
}:

\paragraph{\it onemutation} This variant requests just one replacement of the placeholder instead of three possible replacements.

\paragraph{\it noexplanation} This variant omits the phrase ``\code{This would result in different behavior because <brief explanation>.}''.

\paragraph{\it noinstructions} This variant omits the phrase ``\code{Please consider changes such as using different operators, changing constants,
referring to different variables, object properties, functions, or methods.}''

\paragraph{\it genericsystemprompt} In this variant, we replace the system prompt of Figure~\ref{fig:PromptTemplate}(b) with a generic message
 ``\code{You are a programming assistant. You are expected to be concise and precise and avoid any unnecessary examples, tests, and verbosity.}''

\paragraph{\it basic} This minimal template only asks the LLM to provide a code fragment that the placeholder can be replaced with, without any additional context.

\begin{table*}
\centering
{\scriptsize
\begin{tabular}{l@{\hspace*{0.5mm}}||r@{\hspace*{1mm}}r@{\hspace*{1mm}}r@{\hspace*{1mm}}r@{\hspace*{1mm}}|%
  r@{\hspace*{1mm}}r@{\hspace*{1mm}}r@{\hspace*{1mm}}r@{\hspace*{1mm}}|%
  r@{\hspace*{1mm}}r@{\hspace*{1mm}}r@{\hspace*{1mm}}r@{\hspace*{1mm}}|%
  r@{\hspace*{1mm}}r@{\hspace*{1mm}}r@{\hspace*{1mm}}r@{\hspace*{1mm}}|%
  r@{\hspace*{1mm}}r@{\hspace*{1mm}}r@{\hspace*{1mm}}r@{\hspace*{1mm}}|%
   @{\hspace*{1mm}}r@{\hspace*{1mm}}r@{\hspace*{1mm}}r@{\hspace*{1mm}}r%
}
  & \multicolumn{4}{|c|}{\bf full} &  \multicolumn{4}{|c|}{\bf onemutation} &   \multicolumn{4}{|c|}{\bf noexplanation} &  \multicolumn{4}{|c}{\bf noinstructions} &  \multicolumn{4}{|c}{\bf genericsystemprompt} &  \multicolumn{4}{@{\hspace*{-1.05mm}}|c}{\bf basic} \\
  & \multicolumn{4}{|c|}{\ChangedText{(run \#312)}} &  \multicolumn{4}{|c|}{\ChangedText{(run \#365)}} &   \multicolumn{4}{|c|}{\ChangedText{(run \#372)}} &  \multicolumn{4}{|c}{\ChangedText{(run \#378)}} &  \multicolumn{4}{|c}{\ChangedText{(run \#384)}} &  \multicolumn{4}{@{\hspace*{-1.05mm}}|c}{\ChangedText{(run \#390)}} \\
  &  \Total & \Killed & \Survived & \Timeout
  &  \Total & \Killed & \Survived & \Timeout
  &  \Total & \Killed & \Survived & \Timeout
  &  \Total & \Killed & \Survived & \Timeout
  &  \Total & \Killed & \Survived & \Timeout
  &  \Total & \Killed & \Survived & \Timeout  \\
  \hline
  \hline\begin{minipage}[t]{1.2cm}Complex.js\end{minipage} & 1,199 & 725 & 473 & 1 & 406 & 245 & 161 & 0 & 1,125 & 676 & 448 & 1 & 1,137 & 696 & 440 & 1 & 1,199 & 740 & 458 & 1 & 185 & 120 & 65 & 0 \\
\begin{minipage}[t]{1.2cm}countries-and-timezones\end{minipage} & 217 & 188 & 29 & 0 & 79 & 65 & 14 & 0 & 211 & 183 & 28 & 0 & 218 & 174 & 44 & 0 & 217 & 191 & 26 & 0 & 48 & 44 & 4 & 0 \\
\begin{minipage}[t]{1.2cm}crawler-url-parser\end{minipage} & 285 & 157 & 128 & 0 & 86 & 50 & 36 & 0 & 239 & 140 & 99 & 0 & 246 & 134 & 112 & 0 & 246 & 143 & 103 & 0 & 67 & 49 & 18 & 0 \\
\end{tabular}
}
\caption{Number of mutants generated using the \CodeLlamaThirtyFour LLM at temperature 0.0 using templates full, onemutation, noexplanation, noinstructions, gen.system prompt, basic \ChangedText{(showing one run of each)}.}
\label{table:Templates}
\end{table*}

Table~\ref{table:Templates} shows, for each template, the total number of mutants, and the number that were killed, survived, and timed out,
respectively. From these results, it can be seen that:
\begin{itemize}
  \item
    \textit{full} and \textit{genericsystemprompt}  produced the most mutants and performed similarly, demonstrating that the use of a specialized system prompt
   has minimal impact,
 \item
    \textit{noexplanation} and \textit{noinstructions} produce only slightly fewer mutants and surviving mutants than  \textit{full} and \textit{genericsystemprompt}, 
    so including instructions or requesting explanations for suggested mutations has limited impact, 
 \item
   using  \textit{onemutation}  dramatically reduces the number of mutants from \TotalMutants to \TotalMutantsOneMutation, demonstrating that it is helpful to request
   multiple suggestions, and
 \item
   using  \textit{basic}  reduces the number of mutants to \TotalMutantsBasic, suggesting that additional context in prompts is helpful.
\end{itemize}
\begin{revision*}
We separately analyzed the variability of these results (Table~\ref{table:Templates} presents the results from a single trial), and found the number of mutants killed and survived to be quite stable across trials (the supplemental materials provide further detail).	
\end{revision*}

We also investigated how similar mutants produced using the different prompt templates are to the original code fragments they replace.
As manually inspecting sufficient samples of mutants from each of the different configurations would be infeasible, we instead rely on an automated measure.
We calculate the Levenshtein string edit distance for each mutant between the mutated code and the original code.
Table ~\ref{table:StringSimilarityTemplates} reports the average string edit distance scores for each of the prompt templates by project.

\begin{table} 
\centering
\begin{minipage}{0.375\textwidth}
\scriptsize
\setlength{\tabcolsep}{2pt}
\begin{tabular}{l||r|r|r|r|r|r}
& \Rotate{\bf full} & \Rotate{\bf onemutation} & \Rotate{\bf noexplanation} & \Rotate{\bf noinstructions} & \Rotate{\begin{minipage}{1cm}\bf generic sys. prmpt\end{minipage}} & \Rotate{\bf basic}\\
\midrule
\textit{Complex.js} & 4.27 & 3.37 & 5.09 & 4.27 & 4.17 & 11.98\\
\textit{countries-and-timezones} & 11.13 & 7.75 & 11.17 & 10.87 & 10.85 & 11.29\\
\textit{crawler-url-parser} & 9.50 & 6.41 & 9.46 & 9.49 & 9.30 & 20.04\\
\textit{delta} & 9.55 & 7.38 & 9.91 & 9.43 & 9.14 & 19.63\\
\textit{image-downloader} & 12.67 & 8.82 & 12.89 & 11.01 & 11.48 & 21.92\\
\textit{node-dirty} & 7.53 & 6.90 & 7.58 & 7.41 & 7.51 & 17.52\\
\textit{node-geo-point} & 8.86 & 6.10 & 8.79 & 7.75 & 8.66 & 15.66\\
\textit{node-jsonfile} & 9.73 & 6.98 & 9.76 & 7.77 & 8.91 & 11.64\\
\textit{plural} & 8.14 & 5.21 & 8.41 & 7.58 & 7.80 & 23.64\\
\textit{pull-stream} & 6.72 & 4.57 & 7.53 & 7.48 & 7.30 & 11.92\\
\textit{q} & 8.61 & 7.61 & 9.21 & 8.60 & 8.58 & 16.18\\
\textit{spacl-core} & 9.30 & 5.86 & 10.44 & 9.43 & 9.44 & 14.27\\
\textit{zip-a-folder} & 9.85 & 5.33 & 9.02 & 10.05 & 10.10 & 24.60\\
\end{tabular}
\end{minipage}
\\[2mm]
\caption{
    Average string similarity of mutants to the original code fragments that they replace, for mutants generated using each of the prompt templates at temperature 0.0 using \CodeLlamaThirtyFour.
}
\label{table:StringSimilarityTemplates}
\vspace*{-6mm}
\end{table}

Interpreting the results across different projects is challenging, as each project uses different code idioms that might lead to different mutations.
However, we observe several interesting trends by comparing the mutant similarity across prompts (within the same project).
We find the \textit{basic} template to produce the mutants that are \emph{least similar} to the original code.
We examined samples of these mutants and found that many were creative changes that injected large code blocks in place of short, simple values.
For example, in \textit{crawler-url-parser}, the mutant with the largest string edit distance (297) involves replacing a constant \code{TRUE} with an object literal. 
While the \textit{onemutation} template tended to produce mutants most similar to the original code, this is likely due to the more limited sample space.
We infer that prompting for multiple mutants can result in the LLM suggesting more significant code changes than it would otherwise have.

\begin{takeaway}
  The \textit{full} template produces the most mutants and surviving mutants overall. Using a specialized system prompt has a marginal effect. 
  Including instructions on performing mutations and requesting explanations for mutations  only 
  modestly affects the number of mutants and surviving mutants. Requesting only one mutation dramatically reduces the
  number of generated and surviving mutants, and even greater reductions are observed if the LLM is only asked to fill in the placeholder
  without additional guidance.    
\end{takeaway}
\vspace*{-5mm}

\subsection{RQ5: how does the effectiveness of \ToolName depend on the LLM being used?}

\ChangedText{
The results discussed thus far were obtained with the \textit{codellama-} \textit{34b-instruct} LLM. To determine how
the quality of results depends on the particular LLM being used; we also experimented with
\CodeLlamaThirteen, \LlamaThreeThree, \Mixtral, and \GPTFouroMini at temperature 0.0.
}

\begin{figure*}
\centering
\begin{tabular}{c}
\includegraphics[width=0.95\textwidth]{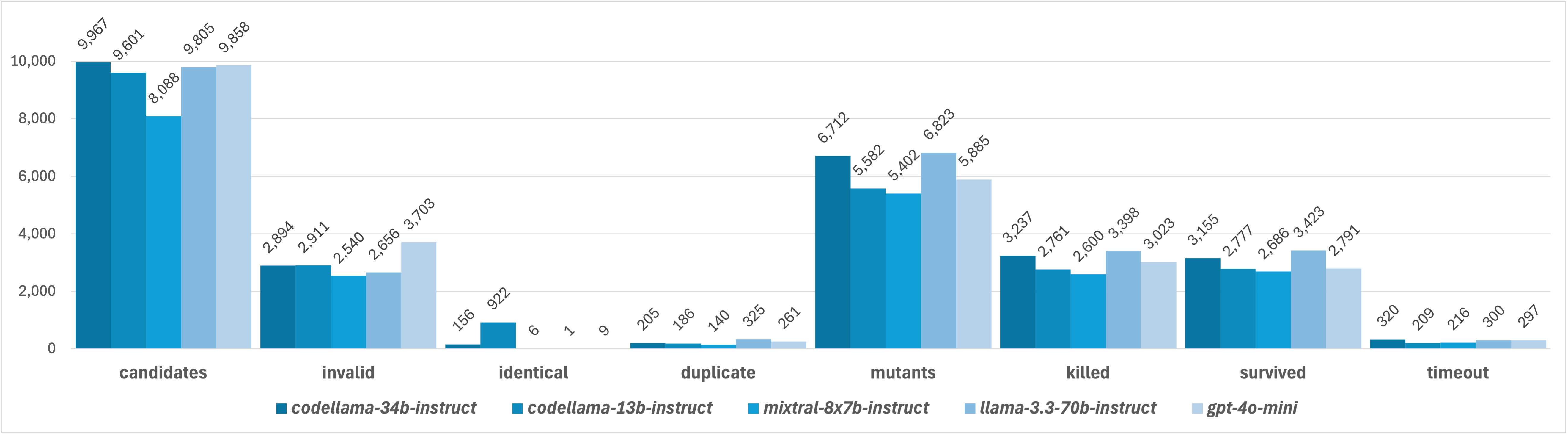}
\end{tabular}
\caption{Comparison of the number of mutant candidates and mutants generated with the \CodeLlamaThirteen, \Mixtral, \LlamaThreeThree, 
         and \GPTFouroMini LLMs at temperature 0.0. This chart was created from the data shown in Tables~\ref{table:CodeLlama34Full0.0}
         and~\ref{table:CompareLLMs}.
}
\label{fig:compareLLMs}
\end{figure*}

\ChangedText{
Table~\ref{table:CompareLLMs} shows the number of mutant candidates produced using each model (along with a breakdown
how many of those candidates are syntactically invalid, identical to the original code, or duplicates), and
the number of mutants produced using each model, classified as killed, surviving, and timed-out.   
Figure~\ref{fig:compareLLMs} shows a visual comparison of the total number of mutant candidates and mutants
produced using each of the five LLMs under consideration, aggregated over all \NrBenchmarks subject applications. 
From these results, it can be seen that:
\begin{itemize}
  \item
    \begin{sloppypar}
    The \CodeLlamaThirtyFour model generates the largest number of mutant candidates (\TotalMutantCandidates), though the 
    number of mutant candidates produced by \CodeLlamaThirteen, \Mixtral, \LlamaThreeThree, and \GPTFouroMini are
    quite similar. \CodeLlamaThirteen produces noticeably fewer mutant candidates (\CodeLlamaThirteenTotalMutantCandidates).
    \end{sloppypar}
  \item
   All models produce a significant number of mutant candidates that is syntactically invalid, ranging from 
   \GPTFouroMiniInvalid in the case of \GPTFouroMini to \MixtralInvalid in the case of \Mixtral.
  \item
   \CodeLlamaThirteen is the only model that produces a significant number of mutant candidates  that are identical
   to the original code fragments that they replace (\CodeLlamaThirteenIdentical).  
 \item
    None of the models produces a significant number of mutant candidates that are duplicates.
 \item 
   The number of mutants that remains after discarding the invalid, identical, and duplicate mutant candidates ranges
   from \MixtralTotalMutants in the case of \Mixtral to \LlamaThreeThreeTotalMutants in the case of \LlamaThreeThree, with \CodeLlamaThirtyFour producing 
   almost as many valid mutants (\TotalMutants). 
 \item 
   \LlamaThreeThree produces the most surviving mutants (\LlamaThreeThreeSurvivingMutants), followed by \CodeLlamaThirtyFour (\NrSurvivingMutants).   
\end{itemize}
}

\begin{table*}
\centering
{\scriptsize
\begin{tabular}{l||rrrrrrrr|rrrrrrrr}
 & \multicolumn{8}{|c|}{\it codellama-13b-instruct (run \#354)} &   \multicolumn{8}{|c}{\it mixtral-8x7b-instruct (run \#360)} \\
 & \Candidates & \Invalid & \Identical & \Duplicate & \Total & \Killed & \Survived & \Timeout 
 & \Candidates & \Invalid & \Identical & \Duplicate & \Total & \Killed & \Survived & \Timeout \\
\hline\textit{Complex.js} & 1,410 & 339 & 116 & 28 & 955 & 553 & 401 & 1 & 1,272 & 310 & 0 & 15 & 962 & 589 & 373 & 0 \\
\textit{countries-and-timezones} & 305 & 83 & 15 & 1 & 207 & 177 & 30 & 0 & 272 & 65 & 2 & 5 & 205 & 166 & 39 & 0 \\
\textit{crawler-url-parser} & 494 & 186 & 51 & 12 & 247 & 129 & 118 & 0 & 411 & 165 & 0 & 3 & 234 & 130 & 104 & 0 \\
\textit{delta} & 1,334 & 530 & 92 & 16 & 712 & 583 & 107 & 22 & 1,132 & 452 & 0 & 24 & 680 & 516 & 128 & 36 \\
\textit{image-downloader} & 122 & 40 & 5 & 2 & 77 & 48 & 29 & 0 & 107 & 38 & 0 & 1 & 69 & 46 & 23 & 0 \\
\textit{node-dirty} & 439 & 161 & 33 & 11 & 245 & 142 & 92 & 11 & 300 & 109 & 0 & 10 & 191 & 111 & 72 & 8 \\
\textit{node-geo-point} & 390 & 64 & 21 & 16 & 304 & 237 & 67 & 0 & 341 & 88 & 0 & 11 & 247 & 166 & 81 & 0 \\
\textit{node-jsonfile} & 191 & 43 & 10 & 7 & 138 & 43 & 45 & 50 & 155 & 23 & 0 & 4 & 132 & 54 & 32 & 46 \\
\textit{plural} & 407 & 100 & 99 & 17 & 208 & 154 & 53 & 1 & 299 & 73 & 0 & 8 & 226 & 166 & 60 & 0 \\
\textit{pull-stream} & 1,002 & 279 & 54 & 13 & 669 & 386 & 237 & 46 & 934 & 255 & 1 & 6 & 678 & 386 & 248 & 44 \\
\textit{q} & 2,993 & 901 & 379 & 55 & 1,713 & 122 & 1,518 & 73 & 2,418 & 772 & 3 & 50 & 1,643 & 112 & 1,460 & 71 \\
\textit{spacl-core} & 377 & 142 & 40 & 7 & 185 & 160 & 25 & 0 & 330 & 152 & 0 & 3 & 157 & 134 & 22 & 1 \\
\textit{zip-a-folder} & 137 & 43 & 7 & 1 & 87 & 27 & 55 & 5 & 117 & 38 & 0 & 0 & 78 & 24 & 44 & 10 \\
\hline\textit{Total} & 9,601 & 2,911 & 922 & 186 & 5,582 & 2,761 & 2,777 & 209 & 8,088 & 2,540 & 6 & 140 & 5,402 & 2,600 & 2,686 & 216 \\
\end{tabular}
}
\\[5mm]
{\scriptsize
\begin{tabular}{l||rrrrrrrr|rrrrrrrr}
 & \multicolumn{8}{|c|}{\it llama-3.3-70b-instruct (run \#23)} &   \multicolumn{8}{|c}{\it gpt-4o-mini (run \#58)} \\
 & \Candidates & \Invalid & \Identical & \Duplicate & \Total & \Killed & \Survived & \Timeout 
 & \Candidates & \Invalid & \Identical & \Duplicate & \Total & \Killed & \Survived & \Timeout \\
\hline\textit{Complex.js} & 1,417 & 279 & 0 & 53 & 1,138 & 690 & 447 & 1 & 1,432 & 446 & 0 & 38 & 986 & 596 & 390 & 0 \\
\textit{countries-and-timezones} & 304 & 64 & 0 & 14 & 240 & 207 & 33 & 0 & 308 & 99 & 0 & 9 & 209 & 171 & 38 & 0 \\
\textit{crawler-url-parser} & 506 & 175 & 0 & 22 & 319 & 208 & 111 & 0 & 516 & 227 & 0 & 11 & 275 & 181 & 94 & 0 \\
\textit{delta} & 1,333 & 539 & 0 & 54 & 794 & 626 & 130 & 38 & 1,345 & 636 & 2 & 40 & 707 & 564 & 108 & 35 \\
\textit{image-downloader} & 122 & 43 & 0 & 5 & 79 & 54 & 25 & 0 & 123 & 58 & 0 & 3 & 65 & 45 & 20 & 0 \\
\textit{node-dirty} & 453 & 121 & 1 & 10 & 331 & 168 & 142 & 21 & 453 & 188 & 0 & 9 & 265 & 154 & 103 & 8 \\
\textit{node-geo-point} & 399 & 39 & 0 & 22 & 358 & 255 & 103 & 0 & 399 & 86 & 0 & 20 & 311 & 225 & 86 & 0 \\
\textit{node-jsonfile} & 198 & 25 & 0 & 6 & 173 & 64 & 37 & 72 & 198 & 44 & 0 & 6 & 154 & 64 & 26 & 64 \\
\textit{plural} & 427 & 96 & 0 & 29 & 331 & 244 & 87 & 0 & 428 & 110 & 5 & 26 & 313 & 257 & 55 & 1 \\
\textit{pull-stream} & 1,044 & 262 & 0 & 10 & 782 & 465 & 265 & 52 & 1,037 & 302 & 0 & 16 & 735 & 420 & 247 & 68 \\
\textit{q} & 3,074 & 855 & 0 & 80 & 2,219 & 127 & 2,006 & 86 & 3,084 & 1,287 & 2 & 69 & 1,795 & 137 & 1,597 & 61 \\
\textit{spacl-core} & 383 & 134 & 0 & 18 & 236 & 203 & 32 & 1 & 392 & 158 & 0 & 10 & 215 & 195 & 20 & 0 \\
\textit{zip-a-folder} & 145 & 24 & 0 & 2 & 121 & 87 & 5 & 29 & 143 & 62 & 0 & 4 & 81 & 14 & 7 & 60 \\
\hline\textit{Total} & 9,805 & 2,656 & 1 & 325 & 6,823 & 3,398 & 3,423 & 300 & 9,858 & 3,703 & 9 & 261 & 5,885 & 3,023 & 2,791 & 297 \\
\end{tabular}
}

  \vspace*{3mm}
  \caption{Mutants generated with the \CodeLlamaThirteen, \Mixtral, \LlamaThreeThree, and \GPTFouroMini LLMs, using
  the following parameters:
   temperature: 0.0, 
    maxTokens: 250, 
    maxNrPrompts: 2000, 
    template: \textit{template-full.hb}, 
    systemPrompt: \textit{SystemPrompt-MutationTestingExpert.txt}, 
    rateLimit: 0, 
    nrAttempts: 3. 
  }
 \label{table:CompareLLMs}
 \end{table*}

\begin{sloppypar}
We also explored the variability of results produced using \CodeLlamaThirteen, \Mixtral, \ChangedText{\LlamaThreeThree and \GPTFouroMini} by conducting
each experiment 5 times, and determined how many distinct mutants are produced and how many 
 mutants occur in all five runs. We found that, at temperature 0, the results
obtained with \CodeLlamaThirteen are very stable across runs, with 
  \CodeLlamaThirteenCommonMutantsMinPercentage--\CodeLlamaThirteenCommonMutantsMaxPercentage of all
  mutants occurring in each of the five runs.
\end{sloppypar}

\begin{sloppypar}
However, with \Mixtral, \LlamaThreeThree and \GPTFouroMini we encountered more variability.
With \Mixtral,  between \MixtralCommonMutantsMinPercentage--\MixtralCommonMutantsMaxPercentage of mutants occur in all 5 runs,
with \LlamaThreeThree, between \LlamaThreeThreeCommonMutantsMinPercentage--\LlamaThreeThreeCommonMutantsMaxPercentage occur in all 5 runs, and
with \GPTFouroMini, between \GPTFouroMiniCommonMutantsMinPercentage--\GPTFouroMiniCommonMutantsMaxPercentage.
We also analyzed the variance of the number of mutants killed and survived, finding that, despite the diversity of mutants across trials, the mutation score was relatively stable.
The supplemental materials include tables showing the average and standard deviation of the number of mutants killed and survived.
\end{sloppypar}

We also examined the string similarity of mutants produced by the \ChangedText{five} LLMs to the original code and
found that the \emph{mixtral-8x7b-instruct} model tends to generate mutants with the greatest string edit distance in the most projects.
%
%
We examined the top 2 mutants with the greatest string edit distance generated by this model for each project, finding several cases of unusual completions.
In \emph{q}, mixtral's most dissimilar mutants (distance 219) replaced a string literal that referred to the function \code{"allResolved"} with a declaration of the same function.
In  \emph{delta}, mixtral's most dissimilar mutants (distance 155) apply a \code{reduce} operation to an object before invoking \code{Object.keys} on it.
We saw similar trends for mixtral across all  projects, with mutants that tended to include large code declarations.
Examining the mutants with the greatest string edit distance for the other four LLMs, we did not find significant trends that held across all targets.
Further details can be found in the supplemental materials.

\begin{takeaway}
\ChangedText{
  All five LLMs under consideration can be used successfully to generate large
  numbers of (surviving) mutants.  \LlamaThreeThree and \CodeLlamaThirtyFour tend to produce
  the largest number of surviving mutants, and \CodeLlamaThirtyFour tends to produce results 
  that are stable across experiments when temperature 0 is used.
  \LlamaThreeThree, \Mixtral, and \GPTFouroMini produce highly variable results, even at temperature 0.
}  
\end{takeaway}

\subsection{RQ6: What is the cost of running \ToolName?}

The primary costs of running \ToolName are the time required to run experiments and the expenses associated with LLM usage. 
Regarding the latter, for the experiments reported on in this paper, we have relied on several  commercial LLM service providers (\url{octo.ai}, \url{openrouter.ai},
and \url{openai.com}).
Such costs tend to vary depending on the provider and the LLM being used, and are typically calculated as a function of the number of ``tokens''
used in the prompt and the completion%
\footnote{
  Depending on the provider, there may also be additional costs associated with the number of requests, though that was not the case for our experiments.
}.  The cost of commercial LLM providers also tends to vary over time, and when a newer version of an LLM is released, it often costs the
same as the older version that it replaced. 
In our experiments, the total number of tokens used for running a full experiment with \ToolName varied by \ChangedText{less than 20\%} 
for the five LLMs that we used%
\footnote{
  Calculated from the total number of tokens reported in the Supplemental Materials associated with this paper, for experiments using the ``full' prompt template
  at temperature 0. 
}, suggesting that token usage is a reasonable proxy for the financial costs incurred.
For these reasons, we use the number of input and output tokens used in our experiments as the primary cost metric for evaluating \ToolName's LLM usage.
For completeness, we also discuss the expense in US dollars at the time of running the experiments below, but the reader should be aware that 
these costs are likely to vary over time.

\begin{table}
\centering
{\scriptsize
\begin{tabular}{l@{\hspace{1mm}}||@{\hspace{1mm}}r@{\hspace{1mm}}|@{\hspace{1mm}}r@{\hspace{1mm}}|@{\hspace{1mm}}r@{\hspace{1mm}}|@{\hspace{1mm}}r@{\hspace{1mm}}|@{\hspace{1mm}}r}
{\bf \hspace*{6mm}project} & \multicolumn{2}{c}{\bf time (sec)} & \multicolumn{3}{@{\hspace{-1mm}}|c}{\bf \#tokens} \\
               & {\it LLMorpheus} & {\it StrykerJS} & {\bf prompt} & {\bf compl.} & {\bf total} \\
\hline
  Complex.js & 3,050.00 & 637.85 & 967,508 & 102,517 & 1,070,025 \\ 
countries-and-timezones & 1,070.89 & 313.86 & 105,828 & 23,441 & 129,269 \\ 
crawler-url-parser & 1,642.70 & 929.43 & 386,223 & 39,175 & 425,398 \\ 
delta & 2,961.66 & 3,839.60 & 890,252 & 98,974 & 989,226 \\ 
image-downloader & 430.53 & 379.25 & 24,655 & 9,134 & 33,789 \\ 
node-dirty & 1,526.20 & 241.81 & 246,248 & 33,070 & 279,318 \\ 
node-geo-point & 1,411.11 & 987.17 & 316,333 & 30,013 & 346,346 \\ 
node-jsonfile & 690.61 & 474.78 & 57,516 & 14,797 & 72,313 \\ 
plural & 1,521.32 & 155.24 & 265,602 & 34,174 & 299,776 \\ 
pull-stream & 2,492.50 & 1,608.97 & 208,130 & 76,513 & 284,643 \\ 
q & 5,241.46 & 14,034.67 & 2,127,655 & 220,215 & 2,347,870 \\ 
spacl-core & 1,351.08 & 798.96 & 162,705 & 29,236 & 191,941 \\ 
zip-a-folder & 500.57 & 1,156.11 & 82,457 & 10,725 & 93,182 \\ 
\hline
  \textit{Total} & 23,890.64 & 25,557.70 & 5,841,112 & 721,984 & 6,563,096 \\
  \end{tabular}
  }
  \\[2mm]
  \caption{Results from LLMorpheus experiment \ChangedText{(run \#312)}.
    Model: \textit{codellama-34b-instruct}, 
    temperature: 0.0, 
    maxTokens: 250, 
    template: \textit{template-full.hb}, 
    systemPrompt: \textit{SystemPrompt-MutationTestingExpert.txt}
  }
  \label{table:Cost}
  \vspace*{-5mm}
\end{table}

The {\bf time} column in Table~\ref{table:Cost} shows the time needed to run \ToolName and the modified version of \StrykerJS
on each subject application. As can be seen in the table, \ToolName requires between \MinLLMorpheusTimeInSeconds 
seconds (about \MinLLMorpheusTimeInMinutes minutes) and \MaxLLMorpheusTimeInSeconds seconds (about \MaxLLMorpheusTimeInMinutes minutes)  
and \StrykerJS  between \MinPrecomputedStrykerTimeInSeconds seconds (about \MinPrecomputedStrykerTimeInMinutes minutes) and 
\MaxPrecomputedStrykerTimeInSeconds seconds (about \MaxPrecomputedStrykerTimeInMinutes minutes).  
  
\ChangedText{The}
last three columns of Table~\ref{table:Cost} show the number of tokens used in prompts and completions for
each subject application and in the aggregate. From these results, it can be seen that running \ToolName required 
between \NrPromptTokensMin and \NrPromptTokensMax prompt tokens, and between \NrCompletionTokensMin and \NrCompletionTokensMax completion tokens.
Hence, in the aggregate, \TotalPromptTokens prompt tokens and \TotalCompletionTokens completion tokens were required.
At the time of \ChangedText{conducting the experiments}, the cost of the \CodeLlamaThirtyFour LLM using \url{octo.ai}'s LLM service was \$0.50 per 
million input tokens and \$1.00 per million output tokens, so for running \ToolName on all \NrSubjectApplications applications, a total cost of approximately \$3.62 was incurred.
\ChangedText{
Moreover, at the time of conducting our experiments, the \LlamaThreeThree model that we used can be accessed from \$0.12 per million input tokens and \$0.30 
per million output tokens at \url{openrouter.ai}, and the \GPTFouroMini model that we used can be accessed from \$0.15 per million input tokens and \$0.60 per
million output tokens from \url{openai.com}. Hence, a full experiment can be run for less than \$1 with \LlamaThreeThree, and for approximately \$1.30
with \GPTFouroMini. 
}   

It should be pointed out that the cost of the LLMs we used is significantly lower than that of \ChangedText{larger} state-of-the-art proprietary LLMs such as 
OpenAI's \ChangedText{\GPTFouro, for which \url{https://openai.com/pricing} quotes a cost of \$2.50 per million input tokens and \$10 per million output tokens}
at the time of writing. While such models might be even more capable of suggesting useful mutants, it is encouraging to see that lower-cost LLMs can achieve good results.

\begin{takeaway}
  \ToolName requires between \MinLLMorpheusTimeInMinutes and \MaxLLMorpheusTimeInMinutes minutes to generate mutants for \NrSubjectApplications subject applications.
  \ChangedText{
    At the time of conducting our experiments, a full experiment with \ToolName on all \NrSubjectApplications applications costs up to \$3.62 
    depending on the LLM being used, suggesting that cost is not a prohibitive limiting factor.
  }
\end{takeaway}

\begin{figure*}
\centering
\begin{tabular}[t]{ll}
\begin{minipage}[t]{0.35\textwidth}
  \includegraphics[width=\textwidth,valign=t]{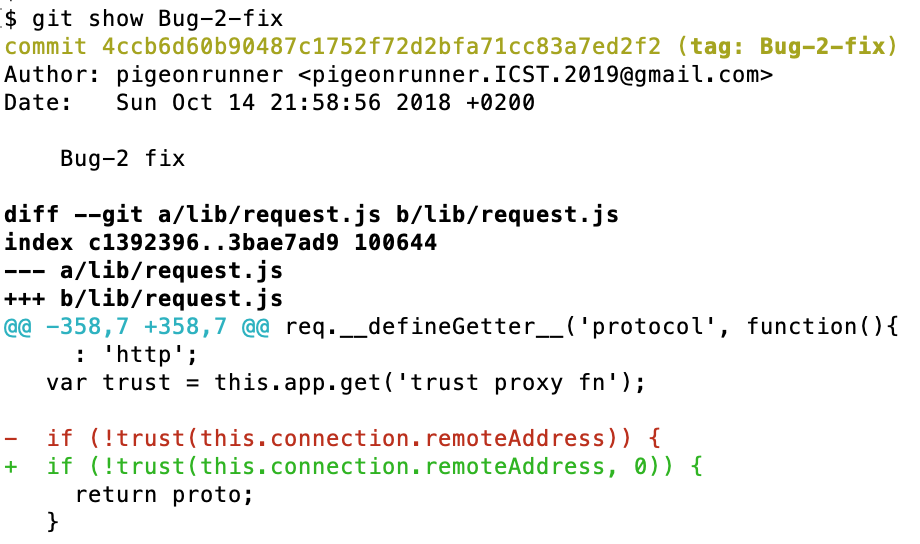} 
\end{minipage} & 
\begin{minipage}[t]{0.35\textwidth}
  \includegraphics[width=\textwidth,valign=t]{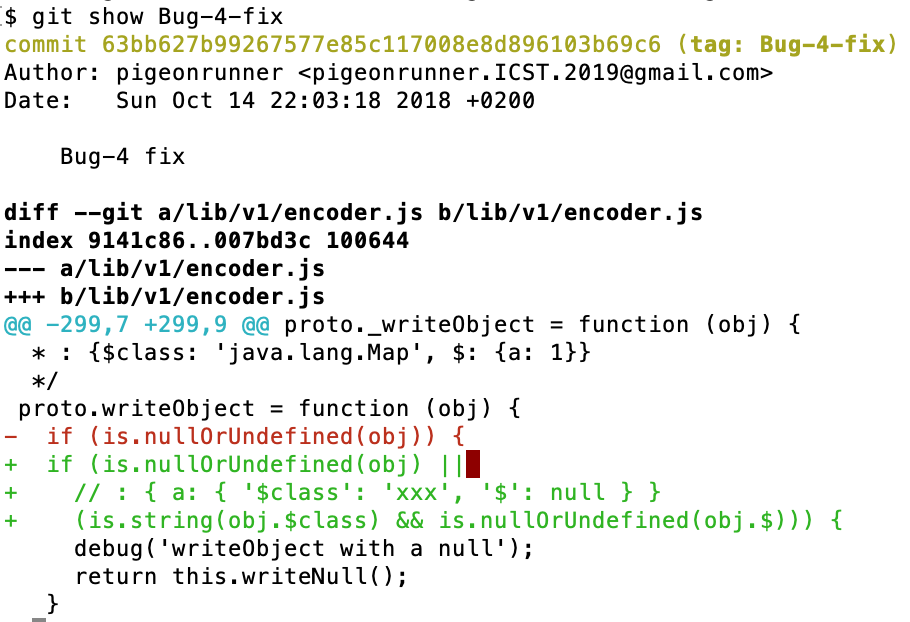} 
\end{minipage}\\[-2mm]
\multicolumn{1}{c}{\textbf{(a) Bug \#2 in Express}} & \multicolumn{1}{c}{\textbf{(b) Bug \#4 in Hessian}}\\
\\[-4mm]
\begin{minipage}[t]{0.485\textwidth}
  \includegraphics[width=\textwidth,valign=t]{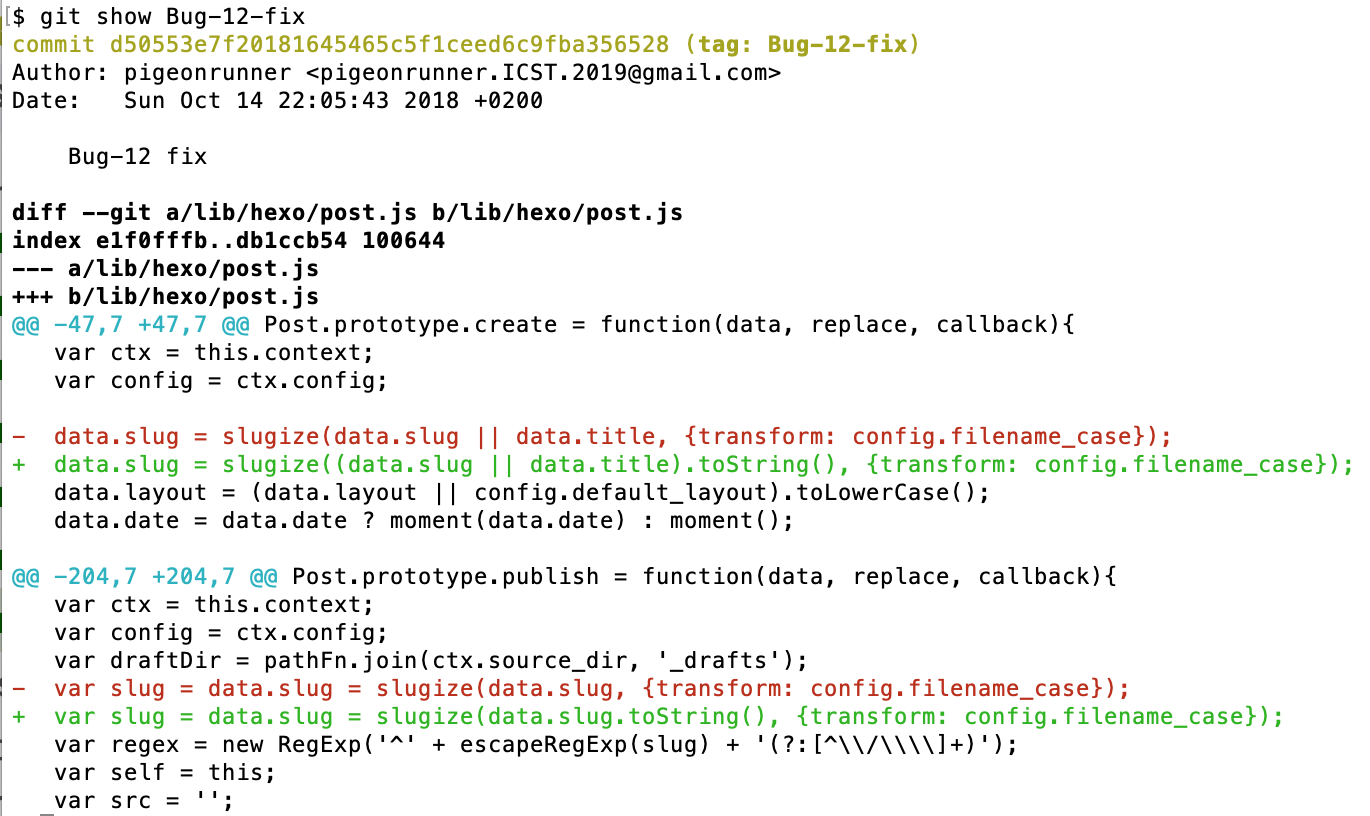} 
\end{minipage}
& 
\begin{minipage}[t]{0.485\textwidth}
  \includegraphics[width=\textwidth,valign=t]{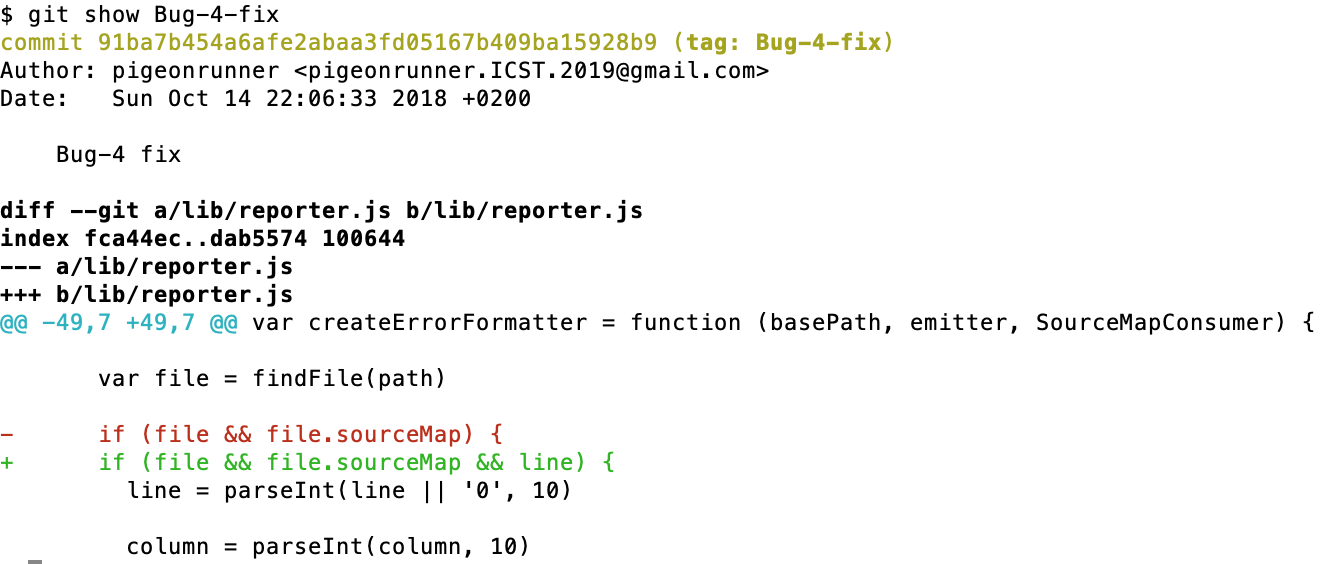} 
\end{minipage}  
\\
\multicolumn{1}{c}{\textbf{(a) Bug \#2 in Hexo}} & \multicolumn{1}{c}{\textbf{(b) Bug \#4 in Karma}}\\
\end{tabular}
\caption{
  Patches corresponding to 4 real-world bugs selected from the Bugs.js suite \cite{Gyimesi19BugsJS}
  that were used in a case study to determine \ToolName's ability to generate mutants that resemble
  real-world bugs.
}
\label{fig:BugsJS}
\end{figure*}

\subsection{\ChangedText{RQ7: Is \ToolName capable of producing mutants that resemble existing bugs?}}

\ChangedText{
  To determine whether \ToolName is capable of producing mutants that resemble existing bugs, we conducted a case study involving
  4 bugs from Bugs.js \cite{Gyimesi19BugsJS}, a collection of 453 bugs from real-world JavaScript applications. For each of these bugs,
  the original faulty version is provided, along with a cleaned patch extracted from the bug fix and instructions on how to execute the test cases.
 In this study, we applied \ToolName to the \textit{fixed} version of a program by introducing placeholders near the location of the fix,
 generating mutants, executing the program's tests for each of these mutants and checking if the observed test failures were identical to 
 those caused by the original bug.        
  Below, we report on our findings for 4 such bugs. 
}

\paragraph{\ChangedText{Express.js Bug\#2}}

\ChangedText{
Figure~\ref{fig:BugsJS}(a) shows bug \#2 in Express, a popular web framework for Node.js. 
This bug occurs at line 361 in the file \code{lib/request.js} and involves the invocation of a function \code{trust} with a single argument 
\code{this.connection.remoteAddress}. Here, the fix involved the addition of a second argument, \code{0}.
Reintroducing this bug in the fixed version causes 2 tests to fail. When applied to the fixed version, \ToolName creates the following 3 mutants:
\begin{itemize}
  \item
    replacing \code{!trust(this.connection.remoteAddress, 0)} with \code{trust(this.connection.remoteAddress, 1)}
  \item
    replacing \code{!trust(this.connection.remoteAddress, 0)} with \code{!trust(this.connection.remoteAddress)}
  \item
    replacing \code{!trust(this.connection.remoteAddress, 0)} with \code{trust(this.connection.localAddress, 0)}
\end{itemize}
The second mutant is identical to the original bug.  The other two mutants cause multiple test failures that differ from those caused by the original bug.
}

\ChangedText{
We repeated the same experiment 4 more times%
\footnote{
  Data for 5 experiments with each of these bugs is included with supplemental materials.
}, and found that in some cases \ToolName produces mutants such as \code{!trust(this.connection.localAddress, 1)} that
differ from the original bug but cause the same test failures. Moreover, in one experiment, \ToolName produced a mutant \code{!this.app.get('trust proxy')} that reproduces one of the
two test failures that were caused by the original bug.  
}

\paragraph{\ChangedText{Hessian.js Bug\#4}}

\ChangedText{
Figure~\ref{fig:BugsJS}(b) shows bug \#4 in Hessian, a serialization framework. This bug occurs at line 302 in file \code{lib/v1/encoder.js} and involves the condition
of an \code{if}-statement. Here, the fix for the bug involves changing the condition from \code{is.nullOrUndefined(obj)} to 
  \code{is.nullOrUndefined(obj) || (is.string(obj.$class) && is.nullOrUndefined(obj.$))}.  
Reintroducing this bug in the fixed version results in a test failure.
}  

\ChangedText{
When applied to the fixed version, \ToolName creates the following 3 mutants:
\begin{itemize}
  \item
    replacing \code{is.nullOrUndefined(obj) || (is.string(obj.$class) &&} \code{is.nullOrUndefined(obj.$))} with \code{obj === null},
  \item
    replacing \code{is.nullOrUndefined(obj) || (is.string(obj.$class) &&} \code{is.nullOrUndefined(obj.$))} with \code{is.nullOrUndefined(obj.$)}
  \item
    replacing \code{is.nullOrUndefined(obj) || (is.string(obj.$class) &&} \code{is.nullOrUndefined(obj.$))} with \code{!obj}
\end{itemize}
In this case, none of the generated mutants are identical to the original bug. However, the first and the third mutants \textit{cause exactly the same test 
failures as the original bug}. The second mutant causes multiple test failures that differ from those produced by the original bug.  
We repeated the same experiment 4 more times and while \ToolName never reproduced the original bug, it produced mutants with the same behavior as the 
original bug on multiple occasions.
}
 
\paragraph{\ChangedText{Hexo Bug\#12}}

\ChangedText{
Figure~\ref{fig:BugsJS}(c) shows bug \#12 in Hexo, a blogging framework for Node.js. The bug involves lines 50 and 207 in file \code{lib/hexo/post.js}.
Here, the fix involves inserting a call to \code{toString} on the expression that is passed as the first argument to a function \code{slugize}
in each of these lines.
This issue causes two test failures, one corresponding to each of the lines that was changed in the fix, so it can be viewed as two similar bugs that can be fixed
independently.  
}

\ChangedText{
We decided to focus on the slightly simpler expression at line~207, where \ToolName creates the following 7 mutants:
\begin{enumerate}
   \item
     replacing \code{slugize((data.slug || data.title).toString(), {transform: config.filename_case})} with \code{slugize((data.slug && data.title).toString(), {transform: config.filename_case})}
   \item
     replacing \code{data.slug.toString()} with \code{data.slug}
   \item
     replacing \code{data.slug.toString()} with \code{data.title}
   \item
     replacing \code{data.slug.toString()} with \code{data.slug + 'test'}
   \item
     replacing \code{slugize(data.slug.toString(), {transform: config.filename_case})} with \code{slugize(data.slug.toUpperCase())}
   \item
     replacing \code{slugize(data.slug.toString(), {transform: config.filename_case})} with \code{slugize(data.title)}
   \item
     replacing  \code{slugize(data.slug.toString(), {transform: config.filename_case})} with \code{slugize(data.slug + Math.random())}  
\end{enumerate}
The second of these mutants is identical to the first part of the original bug and causes one of the two failures associated with the entire buggy code fragment. 
The other mutants cause multiple test failures that differ from the original bug. 
}

\subsection{\ChangedText{Karma Bug \#4}}

\ChangedText{
Figure~\ref{fig:BugsJS}(d) shows bug \#4 in Karma, a testing framework. The bug occurs on line~46 in file \code{lib/reporter.js} 
where a buggy condition \code{file && file.sourceMap} in an if-statement was changed to \code{file && file.sourceMap && line} in the fixed version. 
} 
 
\ChangedText{ 
When applied to the fixed version, \ToolName creates the following 3 mutants at this line:
\begin{itemize}
  \item
    replacing \code{file && file.sourceMap && line} with \code{file && line},
  \item
   replacing \code{file && file.sourceMap && line} with \code{file && file.sourceMap}, and
  \item
   replacing \code{file && file.sourceMap && line} with \code{file &&!file.sourceMap && line}
\end{itemize}  
The first of these mutants has the same behavior as the fixed version, i.e., it is a surviving mutant, indicating that testing may not be sufficiently rigorous.
The second mutant is identical to the original bug. The third mutant causes multiple test failures that differ from those caused by the original bug. 
} 

\medskip 

\ChangedText{
In each of these cases, \ToolName produced either a mutant that was identical to a bug that was originally observed, or one that caused the same test failures 
as a bug that was originally observed. The latter case suggests that \ToolName's ability to generate  mutants that resemble real-world tests is not entirely due 
to training-set leakage.
}
 
\begin{takeaway}
\ChangedText{
  In each of the experiments under consideration, \ToolName was able capable to produce mutants that resemble existing bugs.
}
\end{takeaway}

\subsection{Experimental Data}

All experimental data associated with the experiments reported on in this paper can be found at
\url{https://github.com/neu-se/mutation-testing-data}. 

\section{Threats to Validity}
\label{sec:Threats}
The projects used to evaluate \ToolName may not be representative of the entire ecosystem of JavaScript packages.
To mitigate this risk, we select popular packages used in prior JavaScript testing tool evaluations 
and report results per project, discussing the full range of behaviors we witness.
As in many evaluations of LLM-based tools, the validity of our conclusions may be threatened by including our evaluation subjects in the training data for the models.
If the model were in fact trained on bugs in some of the programs we asked it to create bugs in, one would expect its performance on those programs to vary 
significantly from those that it was not pre-trained on. We mitigate this risk by 
\ChangedText{
  conducting experiments with five  LLMs, four of which are ``open'' in the sense that the training process is documented, thus
  enabling reproducibility and detailed analysis of experimental results.
}

\ChangedText{
  Truly determining if a mutant is equivalent requires significant effort \ChangedText{ and despite the best efforts of two authors to rigorously evaluate 
  them, there may be errors in our categorizing of mutants.}
}

\ChangedText{
  One of the key evaluation criteria used in previous work on mutation testing is ``coupling'', i.e., determining whether a test suite that detects particular mutants also detects particular real faults \cite{Just14AreMutants,Laurent22Revisiting,Gay23HowClosely}.   
  We investigated the feasibility of conducting such a study  using the Bugs.js
  suite~\cite{Gyimesi19BugsJS}, but we found that most of these subjects could not be used at all due to their reliance on outdated versions of various libraries and 
  because of their incompatibility with modern Node.js versions that \StrykerJS  requires, causing them to be incompatible with \ToolName.
%
These projects also have flaky tests, making it particularly challenging to perform mutation analysis~\cite{Shi19Mitigating}.
  We therefore opted for conducting a case study involving 4 real-world bugs from the Bugs.js suite that we were able to reproduce reliably.  
  The results of this case study may be skewed because the code for previous buggy versions of the applications
  may have been included in the training set of the LLM that we used. However, in several cases, we observed that \ToolName produced mutants that 
  \textit{differed from the original bug but that caused the same test failures}. This suggests that \ToolName's ability to produce mutants that resemble real-world
  bugs is not entirely due to training-set leakage.   
}

Evaluating tools that rely on LLMs face significant reproducibility challenges.
We mitigate these risks by 
  (i) evaluating \ToolName using four open LLMs that are version-controlled and permanently archived (in addition to one popular proprietary LLM),
  (ii) repeating each experiment 5 times and
  (iii) making all experimental data available as supplemental materials, and
  (iv) making \ToolName, our evaluation scripts and results publicly available.
  \ChangedText{ Including all results for all experiments in the main body of this paper would significantly decrease the readability of the work. Where we observed significant variability in results, we include data regarding that distribution in the paper directly. In all cases, the supplemental materials associated with this paper include results for all trials of all experiments and summary tables that describe the observed variability for each configuration.}

Lastly, a possible concern is that \ToolName only supports JavaScript and TypeScript, and that applicability beyond these languages may be unclear.
Implementing the same approach for a different language would involve various steps (parsing ASTs, executing tests, etc.) 
that are language-specific and would involve very significant engineering effort, but should otherwise be straightforward.

\section{Related Work}
\label{sec:Related}
Mutation testing, first introduced in the 1970's~\cite{DeMillo78Hints}, has a long history \cite{DBLP:journals/ac/PapadakisK00TH19}.
The era of ``big code'' and software repository mining has enabled the large-scale evaluation of the core hypothesis behind mutation testing, namely, that mutants are coupled to real faults.
Just \etal mined real faults from Java applications and found a statistically significant correlation between mutation detection and real fault detection~\cite{Just14AreMutants}.
This finding has since been replicated on newer, larger datasets consisting of faults from even more Java programs~\cite{Laurent22Revisiting}.
Gay and Salahirad extended this methodology to examine the extent to which individual mutation operators are most coupled to real faults~\cite{Gay23HowClosely}.
While this has demonstrated that test suites that detect more mutants are also likely to detect more bugs, it also underscores the need for new mutation approaches that can generate faults that are coupled to more real bugs.

\textbf{ML for Mutation Testing:}
Several recent projects have considered the use of LLMs and other AI-based techniques for mutation testing.
{\(\mathrm{\mu}\)}Bert  \cite{DBLP:conf/icst/DegiovanniP22,DBLP:journals/corr/abs-2301-03543} resembles \ToolName in that both techniques select some designated code fragments, and query a model what they could be replaced with.  {\(\mathrm{\mu}\)}Bert masks one token at a time, so its mutations involve changes to a single variable 
or operator.  By contrast, \ToolName' placeholders  correspond to (sequences of) AST nodes so it may suggest mutations involving more significant changes to 
complex expressions. A crucial difference between the techniques is that \ToolName utilizes prompts that provide an LLM with additional guidance whereas 
{\(\mathrm{\mu}\)}Bert  provides no way  of guiding the mutations at all, and is therefore completely at the mercy of what the model thinks masked tokens should 
be replaced with. In our experiments with different prompts (Section~\ref{sec:RQ4}), the \textit{basic} prompt is analogous to {\(\mathrm{\mu}\)}Bert in that it merely asks 
the LLM what placeholders should be replaced with. Our results show this to be much less effective at producing interesting mutants, thus demonstrating the usefulness 
of including additional information in prompts.  Our work also differs from {\cite{DBLP:conf/icst/DegiovanniP22} by considering several LLMs, different temperatures 
and by targeting a different language.

\ChangedText{
  In recent work, Garg et al. \cite{DBLP:conf/icst/GargDPT24} explore the coupling between mutants generated using {\(\mathrm{\mu}\)}Bert and 45 reproducible 
  vulnerabilities from the Vul4J dataset. They distinguish between {\it strongly coupled mutants} that fail the same tests for the same reasons as the 
  vulnerabilities and {\it test coupled mutants} that fail the same tests but for some different reason as the vulnerabilities. While they find the majority 
  (32 of 45) of {\(\mathrm{\mu}\)}Bert-generated mutants to be strongly coupled, they also find that strongly coupled mutants are scarce, representing just 1.17\% of killable mutants.
  It would be interesting to explore whether the use of more elaborate prompting strategies such as those employed by \ToolName could be used to increase
  the ratio of strongly coupled mutants.   
}

\ChangedText{
  Tian et al. \cite{ DBLP:conf/issta/TianSWCK024} consider the use of LLMs for determining whether mutants are equivalent and compare 
  their effectiveness to that of traditional techniques for mutant equivalence detection.  Their study considers the detection of equivalent 
  mutants in 19 Java programs from the MutantBench suite \cite{ DBLP:conf/icst/HijfteO21}, from which mutants were derived using standard 
  mutation operators from $\mu$Java \cite{ Ma06MuJava}. Tian et al. experimented with 10 LLMs. They consider 10 state-of-the-art LLMs and 
  several strategies for fine-tuning and prompting, and consider three widely used traditional techniques (compiler-based, ML-based, and 
  Tree-Based Neural NetWork) as the baseline for comparison. Their results indicate that LLMs are significantly better than traditional 
  techniques at equivalent mutant detection, with the fine-tuned code embedding strategy being the most effective. It would be interesting 
  to explore to what extent these results carry over to the detection of mutants that were produced using LLMs using tools such as \ToolName.
}

Similar to our interests, Wang et al.~\cite{Wang24AnExploratory} perform an exploratory study on using large language models to generate mutants.
Unlike our prompting strategy that generates up to three mutants per-AST node, Wang et al. explore a strategy that generates mutants at the granularity of entire methods.
We demonstrate the nuances of prompt engineering in this context by exploring performance under different prompts (RQ4).
These complementary works demonstrate the potential of using LLMs for mutation testing.

Several projects \cite{DBLP:conf/gi-ws/KangY23,DBLP:conf/ssbse/BrownleeCEGHPSS23} have considered the use of LLMs as mutation operators in the context of Genetic 
and Search-Based techniques to improve the efficiency of the search. 
Brownlee et al. \cite{DBLP:conf/ssbse/BrownleeCEGHPSS23} consider the generation of alternate implementations for methods and experimented with prompts exhibiting different levels of detail, 
similar to our experiments reported on in Section~\ref{sec:RQ4}, finding that more detailed prompting generally improves the number of successful patches.

Several other works rely on LLMs to validate the results of mutation testing tools.  Li and Shin
\cite{DBLP:conf/cain/LiS24} use 4 syntactic mutation operators and then observe the change to the natural language description that an LLM generates of the mutated code. 
MuTAP \cite{DBLP:journals/infsof/DakhelNMKD24} uses an off-the-shelf syntactic mutation tool to generate mutants for a Python program, and then prompts an LLM to generate 
a test that can detect those mutants.

\ChangedText{
\textbf{Equivalent Mutants:}
  Kushigian et al. \cite{DBLP:conf/issta/KushigianKFPMJ24} study the types and prevalence of equivalent mutants in Java programs, considering why 
  they are equivalent and how challenging it is to detect that they are. Their study considers 19 Java open-source programs from which mutants are 
  derived using Major \cite{ Just14Major}, a rule-based mutation-testing framework for Java that supports similar mutation operators as \StrykerJS. 
  Their findings indicate that around 3\% of mutants are equivalent, and these equivalent mutants are further classified according to criteria that 
  reflect \textit{why} a mutant is equivalent, and \textit{how} this could be determined. Based on these findings, Kushigian et al. propose Equivalent 
  Mutant Suppression (EMS), a collection of simple static checks for detecting equivalent mutants.    
}

\textbf{Improving Mutation operators:}
Other approaches for mutation testing aim to generate mutants that represent a wider variety of faults.
``Higher-order mutation'' combines multiple mutations concurrently, creating more complex faults, but still limited by the set of operators implemented~\cite{Ghiduk17Higher,Jia09Higher}.
More recently, Brown \etal improve mutation by mining patches for new idioms to use as mutation operators~\cite{Brown17Care}.
Beller \etal design a similar tool and evaluate it at Facebook, with the goal of increasing adoption of mutation testing~\cite{Beller21WhatItWouldTake}
Taking this idea further, Tufano \etal create \emph{DeepMutation}, an approach that learns models for performing mutation from real bugs~\cite{Tufano19Learning}.
This idea was refined by Tian \etal's \emph{LEAM}, which improves the search process by leveraging program grammars~\cite{Tian23Learning}.
Patra and Pradel's \emph{SemSeed} learns to generate mutants from fixes of real-world identifier and literal semantic bugs~\cite{Patra21Semantic}.
Unlike these approaches, \ToolName uses a \emph{pre-trained LLM}, requiring no training to apply it to a new project.

\ChangedText{
\textbf{Mutation Testing Applications and Tools: }
Bel{\'{e}}n~S{\'{a}}nchez et al. \cite{ DBLP:journals/tse/SanchezPSTP24} report on a results of a qualitative study among open-source developers on 
the use of mutation testing. Their findings indicate that developers find mutation testing useful for improving test suite quality, detecting bugs, and 
improving code maintainabiity and that performance considerations are the biggest impediment to adoption.
}
Much of the research advancing the state of mutation testing tooling has targeted Java, such as MuJava~\cite{Ma06MuJava}, Javalanche~\cite{Schuler09Javalanche}, Jumble~\cite{Irvine07Jumble}, Judy~\cite{Madeyski10Judy} and Major~\cite{Just14Major}.
Gopinath \etal empirically compared two of these research-oriented tools tools (Judy~\cite{Madeyski10Judy}, Major~\cite{Just14Major}) with an industry-oriented tool (Pit~\cite{Coles16PIT}), finding that despite the stated similarities between the tools, each produced a somewhat different set of mutants~\cite{Gopinath17DoesChoice}.
Pit is actively maintained, and the open-source tool is also available packaged with professional plugins under the name `ArcMutate'~\cite{Coles24ArcMutate}.
Also aimed at practitioners, the \emph{Stryker} mutation tool is a framework that supports code written in JavaScript, TypeScript, C\#, and Scala~\cite{Stryker}.
We build \ToolName atop Stryker.
Deb \etal examine a new, language-agnostic approach to generating mutants using regular expressions~\cite{Deb24Syntax}.
Future work may examine the feasibility of implementing \ToolName using this approach.


\ChangedText{
\textbf{Mutation and Test Generation:}
There is a long line of research on test generation techniques that specifically target mutated code.
DeMillo and Offutt \cite{DBLP:journals/tse/DeMilloO91} presented a technique that relies on solving systems of algebraic constraints to derive test
cases that target mutated code.
Fraser and Zeller \cite{DBLP:conf/issta/FraserZ10} present $\mu$TEST, an approach that automatically generates unit tests for object-oriented classes 
based on mutation analysis. Their test generation technique uses mutations as the coverage criterion that it aims to maximize and creates tests containing 
oracles that test the mutated value.
Chekam et al. \cite{DBLP:journals/tosem/ChekamPCT21} present a test generation technique based on symbolic execution that systematically searches for situations where program behaviors of the original program
diverges from that of mutated versions.
Lee et al. \cite{DBLP:conf/kbse/LeeVCPB23} present a grey-box fuzzing technique that involves executing both the original and the
mutated code in the same fuzzing driver to direct the generation of test inputs towards those that kill mutants.
Adapting LLM-based test generation techniques \cite{tufano2021unit,bareiss22,DBLP:journals/tse/SchaferNET24} to target mutated code would be an interested topic for future  work.
}

\textbf{LLMs and Testing:}
Beyond mutation testing, LLMs have also been used for test generation. 
Barei{\ss} et al.~\cite{bareiss22} present an approach for test generation that follows a few-shot learning paradigm,
outperforming traditional feedback-directed test generation~\cite{Pacheco:2007}.
Tufano et al. \cite{tufano2021unit} present an approach for test generation using a BART transformer model~\cite{lewis2019bart} that 
is fine-tuned on a training set of functions and corresponding tests.
Lemieux et al.~\cite{LemieuxICSE2023} present an approach where tests generated by Codex are used to assist
search-based testing techniques~\cite{PanichellaMOSA2018} in situations where such techniques get ``stuck'' because 
the generated test cases diverge too far from expected uses of the code under test. 
TestPilot \cite{DBLP:journals/tse/SchaferNET24} produces unit tests for JavaScript programs by prompting an LLM  with 
the start of a test for an API function, with information about that function (signature, body, and usage examples mined from project documentation)
embedded in code comments. In response, the LLM will produce a candidate test, which it executes to determine whether it passes or fails. In
case of failure, TestPilot attempts to fix the failing test by re-prompting the LLM with the error message.   
In principle, \ToolName can be used to evaluate such test generation techniques by providing a means
to assess the quality of the generated tests. 

\section{Conclusions and Future Work}
  \label{sec:Conclusions}
  
We have presented \ToolName, an LLM-based technique for mutation testing. In this approach,
code fragments at designated locations in the program's source code are replaced with the
word ``PLACEHOLDER'', and an   LLM is given a prompt that includes:
   general background on mutation testing, 
  the original code fragment, and  
  instructions directing the LLM to replace the placeholder with a buggy piece of code.
The mutants produced by \ToolName are passed to a modified version of the popular \StrykerJS mutation testing tool,
which runs the tests, classifies mutants, and creates an interactive web page for inspecting the results.
  
An empirical evaluation on \NrSubjectApplications subject applications demonstrates that
\ToolName is capable of producing mutants that resemble real bugs that cannot be produced using standard
mutation operators. We found that the majority (\ChangedText{\PercentageNotEquivalentMutants}) of surviving mutants produced by \ToolName
are behavioral changes and that \ChangedText{\PercentageEquivalentMutants} of them are equivalent mutants.
Experiments with variations on the prompt template reveal that the ``full'' template that includes
all information performs best and that omitting parts of the information from this template matters
to varying degrees. From experiments with \ChangedText{five LLMs, we found that \LlamaThreeThree and \CodeLlamaThirtyFour
generally produced} the largest number of mutants and surviving mutants.  
\ChangedText{
Moreover, a case study involving four real-world bugs from the Bugs.js suite \cite{Gyimesi19BugsJS} revealed that,
in each case, \ToolName produced either a mutant that was identical to a bug that was originally observed, or one that 
caused the same test failures as a bug that was originally observed.
}

The number of mutants produced by \ToolName can become quite large, and executing them can take considerable time.
\ChangedText{In future work, we plan to explore techniques for pruning and prioritizing mutants, focusing
particularly on reducing the number of equivalent mutants.  From a manual investigation of 105 equivalent
mutants, we observed several common patterns, such as replacing an expression “!x”  with “x === null” or “x === undefined”
or replacing call to \code{String.substring} with calls to \code{String.substr} and \code{String.slice}, two methods with similar semantics.
We expect that most of these equivalent mutants can be filtered out using simple AST-based analysis. However, further 
investigation is needed because a few of the mutants that cause behavioral differences are syntactically similar to 
these patterns. This means that any pattern matching based approach should consider the context in which the mutation occurs 
to make a determination whether a mutant is likely to be equivalent. To deal with more challenging cases,
future work could also explore the use of symbolic execution or efficient formal reasoning techniques for 
automatically identifying mutants that are likely to be equivalent.
}   
 
\ChangedText{
In our research we have used LLMs in their default configuration, without any fine-tuning. The strong results obtained with
the relatively small \CodeLlamaThirtyFour LLM that is trained for code-related tasks suggests that fine-tuning an LLM for
the specific task of mutation testing might be worthwhile, particularly with the goal of optimizing the number of mutants 
that are not equivalent.  
}


\pagebreak

\onecolumn
\appendix

\section{Appendix: Experimental Data}

This appendix contains the following experimental results:

\begin{itemize}
  \item 
    Section~\ref{app:CodeLlamaThirtyFour0.0} includes the results for 5 experiments using the \CodeLlamaThirtyFour LLM at temperature 0.0.
  \item
    Section~\ref{app:CodeLlamaThirtyFour0.25} includes the results for 5 experiments using the \CodeLlamaThirtyFour LLM at temperature 0.25.
  \item
    Section~\ref{app:CodeLlamaThirtyFour0.5} includes the results for 5 experiments using the \CodeLlamaThirtyFour LLM at temperature 0.5.
  \item
    Section~\ref{app:CodeLlamaThirtyFour0.5} includes the results for 5 experiments using the \CodeLlamaThirtyFour LLM at temperature 1.0.  
  \item
    Section~\ref{app:CodeLlamaThirtyFourVariability} includes the results of variability analysis of the mutants observed in 5 runs using \CodeLlamaThirtyFour, for each
    of the temperature settings under consideration.
  \item
    Section~\ref{app:CodeLlamaThirteen} includes results for 5 experiments using the \CodeLlamaThirteen LLM at temperature 0.0.
  \item
    Section~\ref{app:Mixtral} includes results for 5 experiments using the \Mixtral LLM at temperature 0.0.
  \item 
    \ChangedText{Section~\ref{app:Gpt4oMini} includes results for 5 experiments using the \GPTFouroMini LLM at temperature 0.0.}
  \item 
    \ChangedText{Section~\ref{app:LlamaThreeThree} includes results for 5 experiments using the \LlamaThreeThree LLM at temperature 0.0.}  
  \item
    Section~\ref{app:OneMutation} includes results for 5 experiments using the {\tt onemutation} template. 
  \item
    Section~\ref{app:NoExplanation} includes results for 5 experiments using the {\tt noexplanation} template. 
  \item  
    Section~\ref{app:NoInstructions} includes results for 5 experiments using the {\tt noinstructions} template.
  \item  
    Section~\ref{app:GenericSystemPrompt} includes results for 5 experiments using a generic system prompt.
  \item  
    Section~\ref{app:Basic} includes results for 5 experiments using the {\tt basic} template.
  \item 
    \ChangedText{Section~\ref{app:allSummary} includes a summary of the variability of the results for all 5 experiments for all configurations of all models in the paper.}  
  \item
    Section~\ref{app:StrykerStandardMutators} includes the results of running the standard mutation operators of \StrykerJS on the subject applications.  
  \item
    Section~\ref{app:StringEditDistance} reports measurements of the average string edit distance observed when using different LLMs.
\end{itemize}

\newpage
 
\FloatBarrier

\subsection{Results for \CodeLlamaThirtyFour, {\tt full} template, temperature 0.0}
\label{app:CodeLlamaThirtyFour0.0}

\ChangedText{
Tables~\ref{table:Mutants:run312:codellama-34b-instruct:template-full.hb:0.0}--\ref{table:Cost:run317:codellama-34b-instruct:template-full.hb:0.0}}
show the results for 5 experiments with the \textit{codellama-34b-instruct} model
at temperature 0.0 using the default prompt and system prompt shown in Figure~\ref{fig:PromptTemplate}.

\begin{table*}[hbt!]
\centering
{\scriptsize
\begin{tabular}{l||r|r|r|r|r|r|r|r|r|r}
  {\bf application} & {\bf \#prompts} & \multicolumn{4}{|c|}{\bf \ChangedText{mutant candidates}} & {\bf \#mutants} & {\bf \#killed} & {\bf \#survived} & {\bf \#timeout} & {\bf mut.} \\
  & &  {\bf \ChangedText{total}} & {\bf \ChangedText{invalid}} & {\bf \ChangedText{identical}} & {\bf \ChangedText{duplicate}}  &  & & & & {\bf score} \\
  \hline
  \hline
\textit{Complex.js} & 490 & \ChangedText{1,451} & \ChangedText{194} & \ChangedText{13} & \ChangedText{45} & 1,199 & 725 & 473 & 1 & 60.55 \\ 
\hline
\textit{countries-and-timezones} & 106 & \ChangedText{318} & \ChangedText{89} & \ChangedText{0} & \ChangedText{12} & 217 & 188 & 29 & 0 & 86.64 \\ 
\hline
\textit{crawler-url-parser} & 176 & \ChangedText{521} & \ChangedText{205} & \ChangedText{14} & \ChangedText{17} & 285 & 157 & 128 & 0 & 55.09 \\ 
\hline
\textit{delta} & 462 & \ChangedText{1,367} & \ChangedText{565} & \ChangedText{10} & \ChangedText{25} & 767 & 634 & 101 & 32 & 86.83 \\ 
\hline
\textit{image-downloader} & 42 & \ChangedText{124} & \ChangedText{33} & \ChangedText{2} & \ChangedText{0} & 89 & 72 & 17 & 0 & 80.90 \\ 
\hline
\textit{node-dirty} & 154 & \ChangedText{450} & \ChangedText{153} & \ChangedText{15} & \ChangedText{7} & 275 & 163 & 100 & 12 & 63.64 \\
\end{tabular}
  }
  \\[2mm]
  \caption{Results from LLMorpheus experiment \ChangedText{(run \#312)}.
    Model: \textit{codellama-34b-instruct}, 
    temperature: 0.0, 
    maxTokens: 250, 
    template: \textit{template-full.hb}, 
    systemPrompt: \textit{SystemPrompt-MutationTestingExpert.txt}. 
  }
  \label{table:Mutants:run312:codellama-34b-instruct:template-full.hb:0.0}
\end{table*}

\begin{table*}[hbt!]
\centering
{\scriptsize
\begin{tabular}{l||r|r|r|r|r|r|r|r|r|r}
  {\bf application} & {\bf \#prompts} & \multicolumn{4}{|c|}{\bf \ChangedText{mutant candidates}} & {\bf \#mutants} & {\bf \#killed} & {\bf \#survived} & {\bf \#timeout} & {\bf mut.} \\
  & &  {\bf \ChangedText{total}} & {\bf \ChangedText{invalid}} & {\bf \ChangedText{identical}} & {\bf \ChangedText{duplicate}}  &  & & & & {\bf score} \\
  \hline
  \hline
\textit{Complex.js} & 490 & \ChangedText{1,450} & \ChangedText{194} & \ChangedText{13} & \ChangedText{45} & 1,198 & 726 & 471 & 1 & 60.68 \\ 
\hline
\textit{countries-and-timezones} & 106 & \ChangedText{318} & \ChangedText{89} & \ChangedText{0} & \ChangedText{12} & 217 & 188 & 29 & 0 & 86.64 \\ 
\hline
\textit{crawler-url-parser} & 176 & \ChangedText{521} & \ChangedText{208} & \ChangedText{14} & \ChangedText{17} & 282 & 156 & 126 & 0 & 55.32 \\ 
\hline
\textit{delta} & 462 & \ChangedText{1,367} & \ChangedText{565} & \ChangedText{10} & \ChangedText{24} & 768 & 636 & 100 & 32 & 86.98 \\ 
\hline
\textit{image-downloader} & 42 & \ChangedText{124} & \ChangedText{34} & \ChangedText{0} & \ChangedText{0} & 90 & 73 & 17 & 0 & 81.11 \\ 
\hline
\textit{node-dirty} & 154 & \ChangedText{450} & \ChangedText{154} & \ChangedText{15} & \ChangedText{7} & 274 & 161 & 101 & 12 & 63.14 \\
\end{tabular}
  }
  \\[2mm]
  \caption{Results from LLMorpheus experiment \ChangedText{(run \#314)}.
    Model: \textit{codellama-34b-instruct}, 
    temperature: 0.0, 
    maxTokens: 250, 
    template: \textit{template-full.hb}, 
    systemPrompt: \textit{SystemPrompt-MutationTestingExpert.txt}. 
  }
  \label{table:Mutants:run314:codellama-34b-instruct:template-full.hb:0.0}
\end{table*}

\begin{table*}[hbt!]
\centering
{\scriptsize
\begin{tabular}{l||r|r|r|r|r|r|r|r|r|r}
  {\bf application} & {\bf \#prompts} & \multicolumn{4}{|c|}{\bf \ChangedText{mutant candidates}} & {\bf \#mutants} & {\bf \#killed} & {\bf \#survived} & {\bf \#timeout} & {\bf mut.} \\
  & &  {\bf \ChangedText{total}} & {\bf \ChangedText{invalid}} & {\bf \ChangedText{identical}} & {\bf \ChangedText{duplicate}}  &  & & & & {\bf score} \\
  \hline
  \hline
\textit{Complex.js} & 490 & \ChangedText{1,452} & \ChangedText{195} & \ChangedText{13} & \ChangedText{45} & 1,199 & 724 & 474 & 1 & 60.47 \\ 
\hline
\textit{countries-and-timezones} & 106 & \ChangedText{318} & \ChangedText{89} & \ChangedText{0} & \ChangedText{12} & 217 & 188 & 29 & 0 & 86.64 \\ 
\hline
\textit{crawler-url-parser} & 176 & \ChangedText{522} & \ChangedText{208} & \ChangedText{14} & \ChangedText{17} & 283 & 157 & 126 & 0 & 55.48 \\ 
\hline
\textit{delta} & 462 & \ChangedText{1,367} & \ChangedText{566} & \ChangedText{10} & \ChangedText{26} & 765 & 633 & 100 & 32 & 86.93 \\ 
\hline
\textit{image-downloader} & 42 & \ChangedText{124} & \ChangedText{35} & \ChangedText{0} & \ChangedText{0} & 89 & 72 & 17 & 0 & 80.90 \\ 
\hline
\textit{node-dirty} & 154 & \ChangedText{450} & \ChangedText{154} & \ChangedText{14} & \ChangedText{7} & 275 & 161 & 102 & 12 & 62.91 \\
\end{tabular}
  }
  \\[2mm]
  \caption{Results from LLMorpheus experiment \ChangedText{(run \#315)}.
    Model: \textit{codellama-34b-instruct}, 
    temperature: 0.0, 
    maxTokens: 250, 
    template: \textit{template-full.hb}, 
    systemPrompt: \textit{SystemPrompt-MutationTestingExpert.txt}. 
  }
  \label{table:Mutants:run315:codellama-34b-instruct:template-full.hb:0.0}
\end{table*}

\begin{table*}[hbt!]
\centering
{\scriptsize
\begin{tabular}{l||r|r|r|r|r|r|r|r|r|r}
  {\bf application} & {\bf \#prompts} & \multicolumn{4}{|c|}{\bf \ChangedText{mutant candidates}} & {\bf \#mutants} & {\bf \#killed} & {\bf \#survived} & {\bf \#timeout} & {\bf mut.} \\
  & &  {\bf \ChangedText{total}} & {\bf \ChangedText{invalid}} & {\bf \ChangedText{identical}} & {\bf \ChangedText{duplicate}}  &  & & & & {\bf score} \\
  \hline
  \hline
\textit{Complex.js} & 490 & \ChangedText{1,452} & \ChangedText{195} & \ChangedText{13} & \ChangedText{45} & 1,199 & 725 & 473 & 1 & 60.55 \\ 
\hline
\textit{countries-and-timezones} & 106 & \ChangedText{318} & \ChangedText{89} & \ChangedText{0} & \ChangedText{12} & 217 & 188 & 29 & 0 & 86.64 \\ 
\hline
\textit{crawler-url-parser} & 176 & \ChangedText{521} & \ChangedText{207} & \ChangedText{14} & \ChangedText{17} & 283 & 157 & 125 & 1 & 55.83 \\ 
\hline
\textit{delta} & 462 & \ChangedText{1,367} & \ChangedText{565} & \ChangedText{10} & \ChangedText{26} & 766 & 634 & 100 & 32 & 86.95 \\ 
\hline
\textit{image-downloader} & 42 & \ChangedText{124} & \ChangedText{35} & \ChangedText{0} & \ChangedText{0} & 89 & 72 & 17 & 0 & 80.90 \\ 
\hline
\textit{node-dirty} & 154 & \ChangedText{450} & \ChangedText{153} & \ChangedText{15} & \ChangedText{8} & 274 & 160 & 102 & 12 & 62.77 \\
\end{tabular}
  }
  \\[2mm]
  \caption{Results from LLMorpheus experiment \ChangedText{(run \#316)}.
    Model: \textit{codellama-34b-instruct}, 
    temperature: 0.0, 
    maxTokens: 250, 
    template: \textit{template-full.hb}, 
    systemPrompt: \textit{SystemPrompt-MutationTestingExpert.txt}. 
  }
  \label{table:Mutants:run316:codellama-34b-instruct:template-full.hb:0.0}
\end{table*}

\begin{table*}[hbt!]
\centering
{\scriptsize
\begin{tabular}{l||r|r|r|r|r|r|r|r|r|r}
  {\bf application} & {\bf \#prompts} & \multicolumn{4}{|c|}{\bf \ChangedText{mutant candidates}} & {\bf \#mutants} & {\bf \#killed} & {\bf \#survived} & {\bf \#timeout} & {\bf mut.} \\
  & &  {\bf \ChangedText{total}} & {\bf \ChangedText{invalid}} & {\bf \ChangedText{identical}} & {\bf \ChangedText{duplicate}}  &  & & & & {\bf score} \\
  \hline
  \hline
\textit{Complex.js} & 490 & \ChangedText{1,450} & \ChangedText{194} & \ChangedText{13} & \ChangedText{44} & 1,199 & 726 & 472 & 1 & 60.63 \\ 
\hline
\textit{countries-and-timezones} & 106 & \ChangedText{317} & \ChangedText{89} & \ChangedText{0} & \ChangedText{12} & 216 & 187 & 29 & 0 & 86.57 \\ 
\hline
\textit{crawler-url-parser} & 176 & \ChangedText{521} & \ChangedText{205} & \ChangedText{14} & \ChangedText{17} & 285 & 157 & 128 & 0 & 55.09 \\ 
\hline
\textit{delta} & 462 & \ChangedText{1,367} & \ChangedText{565} & \ChangedText{10} & \ChangedText{25} & 767 & 636 & 100 & 31 & 86.96 \\ 
\hline
\textit{image-downloader} & 42 & \ChangedText{124} & \ChangedText{32} & \ChangedText{2} & \ChangedText{0} & 90 & 73 & 17 & 0 & 81.11 \\ 
\hline
\textit{node-dirty} & 154 & \ChangedText{450} & \ChangedText{153} & \ChangedText{15} & \ChangedText{7} & 275 & 161 & 102 & 12 & 62.91 \\
\end{tabular}
  }
  \\[2mm]
  \caption{Results from LLMorpheus experiment \ChangedText{(run \#317)}.
    Model: \textit{codellama-34b-instruct}, 
    temperature: 0.0, 
    maxTokens: 250, 
    template: \textit{template-full.hb}, 
    systemPrompt: \textit{SystemPrompt-MutationTestingExpert.txt}. 
  }
  \label{table:Mutants:run317:codellama-34b-instruct:template-full.hb:0.0}
\end{table*}

\begin{table*}[hbt!]
\centering
{\scriptsize
\begin{tabular}{l||r|r|r|r|r}
\multicolumn{1}{c|}{\bf project} & \multicolumn{2}{|c|}{\bf time (sec)} & \multicolumn{3}{|c}{\bf \#tokens} \\
               & {\it LLMorpheus} & {\it StrykerJS} & {\bf prompt} & {\bf compl.} & {\bf total} \\
\hline
  Complex.js & 3,050.00 & 637.85 & 967,508 & 102,517 & 1,070,025 \\ 
countries-and-timezones & 1,070.89 & 313.86 & 105,828 & 23,441 & 129,269 \\ 
crawler-url-parser & 1,642.70 & 929.43 & 386,223 & 39,175 & 425,398 \\ 
delta & 2,961.66 & 3,839.60 & 890,252 & 98,974 & 989,226 \\ 
image-downloader & 430.53 & 379.25 & 24,655 & 9,134 & 33,789 \\ 
node-dirty & 1,526.20 & 241.81 & 246,248 & 33,070 & 279,318 \\ 
node-geo-point & 1,411.11 & 987.17 & 316,333 & 30,013 & 346,346 \\ 
node-jsonfile & 690.61 & 474.78 & 57,516 & 14,797 & 72,313 \\ 
plural & 1,521.32 & 155.24 & 265,602 & 34,174 & 299,776 \\ 
pull-stream & 2,492.50 & 1,608.97 & 208,130 & 76,513 & 284,643 \\ 
q & 5,241.46 & 14,034.67 & 2,127,655 & 220,215 & 2,347,870 \\ 
spacl-core & 1,351.08 & 798.96 & 162,705 & 29,236 & 191,941 \\ 
zip-a-folder & 500.57 & 1,156.11 & 82,457 & 10,725 & 93,182 \\ 
\hline
  \textit{Total} & 23,890.64 & 25,557.70 & 5,841,112 & 721,984 & 6,563,096 \\
  \end{tabular}
  }
  \\[2mm]
  \caption{Results from LLMorpheus experiment \ChangedText{(run \#312)}.
    Model: \textit{codellama-34b-instruct}, 
    temperature: 0.0, 
    maxTokens: 250, 
    template: \textit{template-full.hb}, 
    systemPrompt: \textit{SystemPrompt-MutationTestingExpert.txt}
  }
  \label{table:Cost:run312:codellama-34b-instruct:template-full.hb:0.0}
\end{table*}

\begin{table*}[hbt!]
\centering
{\scriptsize
\begin{tabular}{l||r|r|r|r|r}
\multicolumn{1}{c|}{\bf project} & \multicolumn{2}{|c|}{\bf time (sec)} & \multicolumn{3}{|c}{\bf \#tokens} \\
               & {\it LLMorpheus} & {\it StrykerJS} & {\bf prompt} & {\bf compl.} & {\bf total} \\
\hline
  Complex.js & 3,087.58 & 637.10 & 967,508 & 102,428 & 1,069,936 \\ 
countries-and-timezones & 1,070.89 & 313.12 & 105,828 & 23,427 & 129,255 \\ 
crawler-url-parser & 1,645.11 & 679.89 & 386,223 & 39,210 & 425,433 \\ 
delta & 2,941.55 & 3,838.23 & 890,252 & 98,951 & 989,203 \\ 
image-downloader & 430.54 & 377.12 & 24,655 & 9,175 & 33,830 \\ 
node-dirty & 1,526.11 & 248.57 & 246,248 & 33,038 & 279,286 \\ 
node-geo-point & 1,411.06 & 999.59 & 316,333 & 29,959 & 346,292 \\ 
node-jsonfile & 690.69 & 478.32 & 57,516 & 14,829 & 72,345 \\ 
plural & 1,521.04 & 147.85 & 265,602 & 34,164 & 299,766 \\ 
pull-stream & 2,477.99 & 1,400.76 & 208,130 & 76,398 & 284,528 \\ 
q & 5,231.88 & 14,004.86 & 2,127,655 & 220,252 & 2,347,907 \\ 
spacl-core & 1,351.05 & 805.30 & 162,705 & 29,283 & 191,988 \\ 
zip-a-folder & 500.57 & 1,152.62 & 82,457 & 10,705 & 93,162 \\ 
\hline
  \textit{Total} & 23,886.05 & 25,083.32 & 5,841,112 & 721,819 & 6,562,931 \\
  \end{tabular}
  }
  \\[2mm]
  \caption{Results from LLMorpheus experiment \ChangedText{(run \#314)}.
    Model: \textit{codellama-34b-instruct}, 
    temperature: 0.0, 
    maxTokens: 250, 
    template: \textit{template-full.hb}, 
    systemPrompt: \textit{SystemPrompt-MutationTestingExpert.txt}
  }
  \label{table:Cost:run314:codellama-34b-instruct:template-full.hb:0.0}
\end{table*}

\begin{table*}[hbt!]
\centering
{\scriptsize
\begin{tabular}{l||r|r|r|r|r}
\multicolumn{1}{c|}{\bf project} & \multicolumn{2}{|c|}{\bf time (sec)} & \multicolumn{3}{|c}{\bf \#tokens} \\
               & {\it LLMorpheus} & {\it StrykerJS} & {\bf prompt} & {\bf compl.} & {\bf total} \\
\hline
  Complex.js & 3,000.46 & 630.00 & 967,508 & 102,314 & 1,069,822 \\ 
countries-and-timezones & 1,070.90 & 314.35 & 105,828 & 23,438 & 129,266 \\ 
crawler-url-parser & 1,644.58 & 1,051.08 & 386,223 & 39,105 & 425,328 \\ 
delta & 3,006.51 & 3,795.29 & 890,252 & 98,978 & 989,230 \\ 
image-downloader & 430.54 & 376.08 & 24,655 & 9,186 & 33,841 \\ 
node-dirty & 1,526.69 & 247.49 & 246,248 & 33,089 & 279,337 \\ 
node-geo-point & 1,411.05 & 1,003.93 & 316,333 & 30,010 & 346,343 \\ 
node-jsonfile & 690.67 & 478.93 & 57,516 & 14,803 & 72,319 \\ 
plural & 1,521.23 & 148.33 & 265,602 & 34,082 & 299,684 \\ 
pull-stream & 2,492.02 & 1,392.68 & 208,130 & 76,599 & 284,729 \\ 
q & 5,296.20 & 14,072.49 & 2,127,655 & 220,395 & 2,348,050 \\ 
spacl-core & 1,351.09 & 802.30 & 162,705 & 29,334 & 192,039 \\ 
zip-a-folder & 500.56 & 1,155.04 & 82,457 & 10,764 & 93,221 \\ 
\hline
  \textit{Total} & 23,942.51 & 25,467.99 & 5,841,112 & 722,097 & 6,563,209 \\
  \end{tabular}
  }
  \\[2mm]
  \caption{Results from LLMorpheus experiment \ChangedText{(run \#315)}.
    Model: \textit{codellama-34b-instruct}, 
    temperature: 0.0, 
    maxTokens: 250, 
    template: \textit{template-full.hb}, 
    systemPrompt: \textit{SystemPrompt-MutationTestingExpert.txt}
  }
  \label{table:Cost:run315:codellama-34b-instruct:template-full.hb:0.0}
\end{table*}

\begin{table*}[hbt!]
\centering
{\scriptsize
\begin{tabular}{l||r|r|r|r|r}
\multicolumn{1}{c|}{\bf project} & \multicolumn{2}{|c|}{\bf time (sec)} & \multicolumn{3}{|c}{\bf \#tokens} \\
               & {\it LLMorpheus} & {\it StrykerJS} & {\bf prompt} & {\bf compl.} & {\bf total} \\
\hline
  Complex.js & 3,034.16 & 631.23 & 967,508 & 102,497 & 1,070,005 \\ 
countries-and-timezones & 1,070.91 & 309.40 & 105,828 & 23,444 & 129,272 \\ 
crawler-url-parser & 1,644.09 & 1,022.89 & 386,223 & 39,174 & 425,397 \\ 
delta & 2,983.03 & 3,969.11 & 890,252 & 99,003 & 989,255 \\ 
image-downloader & 430.51 & 374.45 & 24,655 & 9,148 & 33,803 \\ 
node-dirty & 1,563.30 & 251.17 & 246,248 & 33,068 & 279,316 \\ 
node-geo-point & 1,411.00 & 1,001.75 & 316,333 & 30,041 & 346,374 \\ 
node-jsonfile & 690.69 & 474.83 & 57,516 & 14,750 & 72,266 \\ 
plural & 1,521.09 & 151.19 & 265,602 & 34,132 & 299,734 \\ 
pull-stream & 2,541.67 & 1,398.89 & 208,130 & 76,567 & 284,697 \\ 
q & 5,399.09 & 13,959.40 & 2,127,655 & 220,191 & 2,347,846 \\ 
spacl-core & 1,351.08 & 959.37 & 162,705 & 29,287 & 191,992 \\ 
zip-a-folder & 510.58 & 1,154.14 & 82,457 & 10,725 & 93,182 \\ 
\hline
  \textit{Total} & 24,151.21 & 25,657.83 & 5,841,112 & 722,027 & 6,563,139 \\
  \end{tabular}
  }
  \\[2mm]
  \caption{Results from LLMorpheus experiment \ChangedText{(run \#316)}.
    Model: \textit{codellama-34b-instruct}, 
    temperature: 0.0, 
    maxTokens: 250, 
    template: \textit{template-full.hb}, 
    systemPrompt: \textit{SystemPrompt-MutationTestingExpert.txt}
  }
  \label{table:Cost:run316:codellama-34b-instruct:template-full.hb:0.0}
\end{table*}

\begin{table*}[hbt!]
\centering
{\scriptsize
\begin{tabular}{l||r|r|r|r|r}
\multicolumn{1}{c|}{\bf project} & \multicolumn{2}{|c|}{\bf time (sec)} & \multicolumn{3}{|c}{\bf \#tokens} \\
               & {\it LLMorpheus} & {\it StrykerJS} & {\bf prompt} & {\bf compl.} & {\bf total} \\
\hline
  Complex.js & 3,019.69 & 669.52 & 967,508 & 102,524 & 1,070,032 \\ 
countries-and-timezones & 1,070.82 & 313.99 & 105,828 & 23,425 & 129,253 \\ 
crawler-url-parser & 1,641.20 & 958.26 & 386,223 & 39,160 & 425,383 \\ 
delta & 3,013.74 & 3,867.52 & 890,252 & 99,031 & 989,283 \\ 
image-downloader & 430.56 & 378.47 & 24,655 & 9,117 & 33,772 \\ 
node-dirty & 1,527.49 & 251.95 & 246,248 & 33,113 & 279,361 \\ 
node-geo-point & 1,451.19 & 1,042.75 & 316,333 & 29,894 & 346,227 \\ 
node-jsonfile & 690.66 & 479.95 & 57,516 & 14,803 & 72,319 \\ 
plural & 1,521.18 & 150.87 & 265,602 & 34,163 & 299,765 \\ 
pull-stream & 2,486.78 & 1,384.03 & 208,130 & 76,520 & 284,650 \\ 
q & 5,250.40 & 14,006.76 & 2,127,655 & 220,193 & 2,347,848 \\ 
spacl-core & 1,350.99 & 808.62 & 162,705 & 29,297 & 192,002 \\ 
zip-a-folder & 500.63 & 1,171.49 & 82,457 & 10,705 & 93,162 \\ 
\hline
  \textit{Total} & 23,955.33 & 25,484.18 & 5,841,112 & 721,945 & 6,563,057 \\
  \end{tabular}
  }
  \\[2mm]
  \caption{Results from LLMorpheus experiment \ChangedText{(run \#317)}.
    Model: \textit{codellama-34b-instruct}, 
    temperature: 0.0, 
    maxTokens: 250, 
    template: \textit{template-full.hb}, 
    systemPrompt: \textit{SystemPrompt-MutationTestingExpert.txt}
  }
  \label{table:Cost:run317:codellama-34b-instruct:template-full.hb:0.0}
\end{table*}

\FloatBarrier

\subsection{Results for \CodeLlamaThirtyFour, {\tt full} template, temperature 0.25}
 \label{app:CodeLlamaThirtyFour0.25}

\ChangedText{
Tables~\ref{table:Mutants:run348:codellama-34b-instruct:template-full.hb:0.25}--\ref{table:Cost:run353:codellama-34b-instruct:template-full.hb:0.25}}
show the results for 5 experiments with the \textit{codellama-34b-instruct} model
at temperature 0.25 using the default prompt and system prompt shown in Figure~\ref{fig:PromptTemplate}.

\begin{table*}[hbt!]
\centering
{\scriptsize
\begin{tabular}{l||r|r|r|r|r|r|r|r|r|r}
  {\bf application} & {\bf \#prompts} & \multicolumn{4}{|c|}{\bf \ChangedText{mutant candidates}} & {\bf \#mutants} & {\bf \#killed} & {\bf \#survived} & {\bf \#timeout} & {\bf mut.} \\
  & &  {\bf \ChangedText{total}} & {\bf \ChangedText{invalid}} & {\bf \ChangedText{identical}} & {\bf \ChangedText{duplicate}}  &  & & & & {\bf score} \\
  \hline
  \hline
\textit{Complex.js} & 490 & \ChangedText{1,460} & \ChangedText{210} & \ChangedText{9} & \ChangedText{44} & 1,197 & 730 & 466 & 1 & 61.07 \\ 
\hline
\textit{countries-and-timezones} & 106 & \ChangedText{315} & \ChangedText{86} & \ChangedText{0} & \ChangedText{10} & 219 & 181 & 38 & 0 & 82.65 \\ 
\hline
\textit{crawler-url-parser} & 176 & \ChangedText{515} & \ChangedText{216} & \ChangedText{7} & \ChangedText{12} & 260 & 167 & 93 & 0 & 64.23 \\ 
\hline
\textit{delta} & 462 & \ChangedText{1,366} & \ChangedText{550} & \ChangedText{10} & \ChangedText{25} & 781 & 642 & 111 & 28 & 85.79 \\ 
\hline
\textit{image-downloader} & 42 & \ChangedText{125} & \ChangedText{38} & \ChangedText{1} & \ChangedText{0} & 86 & 71 & 15 & 0 & 82.56 \\ 
\hline
\textit{node-dirty} & 154 & \ChangedText{457} & \ChangedText{157} & \ChangedText{9} & \ChangedText{12} & 279 & 175 & 93 & 11 & 66.67 \\
\end{tabular}
  }
  \\[2mm]
  \caption{Results from LLMorpheus experiment \ChangedText{(run \#348)}.
    Model: \textit{codellama-34b-instruct}, 
    temperature: 0.25, 
    maxTokens: 250, 
    template: \textit{template-full.hb}, 
    systemPrompt: \textit{SystemPrompt-MutationTestingExpert.txt}. 
  }
  \label{table:Mutants:run348:codellama-34b-instruct:template-full.hb:0.25}
\end{table*}

\begin{table*}[hbt!]
\centering
{\scriptsize
\begin{tabular}{l||r|r|r|r|r|r|r|r|r|r}
  {\bf application} & {\bf \#prompts} & \multicolumn{4}{|c|}{\bf \ChangedText{mutant candidates}} & {\bf \#mutants} & {\bf \#killed} & {\bf \#survived} & {\bf \#timeout} & {\bf mut.} \\
  & &  {\bf \ChangedText{total}} & {\bf \ChangedText{invalid}} & {\bf \ChangedText{identical}} & {\bf \ChangedText{duplicate}}  &  & & & & {\bf score} \\
  \hline
  \hline
\textit{Complex.js} & 490 & \ChangedText{1,460} & \ChangedText{202} & \ChangedText{13} & \ChangedText{45} & 1,200 & 729 & 471 & 0 & 60.75 \\ 
\hline
\textit{countries-and-timezones} & 106 & \ChangedText{316} & \ChangedText{87} & \ChangedText{1} & \ChangedText{5} & 223 & 198 & 25 & 0 & 88.79 \\ 
\hline
\textit{crawler-url-parser} & 176 & \ChangedText{518} & \ChangedText{206} & \ChangedText{13} & \ChangedText{15} & 269 & 163 & 106 & 0 & 60.59 \\ 
\hline
\textit{delta} & 462 & \ChangedText{1,362} & \ChangedText{576} & \ChangedText{5} & \ChangedText{29} & 752 & 608 & 110 & 34 & 85.37 \\ 
\hline
\textit{image-downloader} & 42 & \ChangedText{123} & \ChangedText{35} & \ChangedText{3} & \ChangedText{0} & 85 & 67 & 18 & 0 & 78.82 \\ 
\hline
\textit{node-dirty} & 154 & \ChangedText{457} & \ChangedText{165} & \ChangedText{17} & \ChangedText{4} & 271 & 163 & 96 & 12 & 64.58 \\
\end{tabular}
  }
  \\[2mm]
  \caption{Results from LLMorpheus experiment \ChangedText{(run \#350)}.
    Model: \textit{codellama-34b-instruct}, 
    temperature: 0.25, 
    maxTokens: 250, 
    template: \textit{template-full.hb}, 
    systemPrompt: \textit{SystemPrompt-MutationTestingExpert.txt}. 
  }
  \label{table:Mutants:run350:codellama-34b-instruct:template-full.hb:0.25}
\end{table*}

\begin{table*}[hbt!]
\centering
{\scriptsize
\begin{tabular}{l||r|r|r|r|r|r|r|r|r|r}
  {\bf application} & {\bf \#prompts} & \multicolumn{4}{|c|}{\bf \ChangedText{mutant candidates}} & {\bf \#mutants} & {\bf \#killed} & {\bf \#survived} & {\bf \#timeout} & {\bf mut.} \\
  & &  {\bf \ChangedText{total}} & {\bf \ChangedText{invalid}} & {\bf \ChangedText{identical}} & {\bf \ChangedText{duplicate}}  &  & & & & {\bf score} \\
  \hline
  \hline
\textit{Complex.js} & 490 & \ChangedText{1,451} & \ChangedText{199} & \ChangedText{12} & \ChangedText{45} & 1,195 & 737 & 457 & 1 & 61.76 \\ 
\hline
\textit{countries-and-timezones} & 106 & \ChangedText{315} & \ChangedText{88} & \ChangedText{0} & \ChangedText{12} & 215 & 189 & 26 & 0 & 87.91 \\ 
\hline
\textit{crawler-url-parser} & 176 & \ChangedText{512} & \ChangedText{211} & \ChangedText{16} & \ChangedText{19} & 247 & 144 & 103 & 0 & 58.30 \\ 
\hline
\textit{delta} & 462 & \ChangedText{1,362} & \ChangedText{554} & \ChangedText{8} & \ChangedText{24} & 776 & 647 & 100 & 29 & 87.11 \\ 
\hline
\textit{image-downloader} & 42 & \ChangedText{123} & \ChangedText{36} & \ChangedText{0} & \ChangedText{0} & 87 & 71 & 16 & 0 & 81.61 \\ 
\hline
\textit{node-dirty} & 154 & \ChangedText{454} & \ChangedText{156} & \ChangedText{14} & \ChangedText{10} & 274 & 162 & 100 & 12 & 63.50 \\
\end{tabular}
  }
  \\[2mm]
  \caption{Results from LLMorpheus experiment \ChangedText{(run \#351)}.
    Model: \textit{codellama-34b-instruct}, 
    temperature: 0.25, 
    maxTokens: 250, 
    template: \textit{template-full.hb}, 
    systemPrompt: \textit{SystemPrompt-MutationTestingExpert.txt}. 
  }
  \label{table:Mutants:run351:codellama-34b-instruct:template-full.hb:0.25}
\end{table*}

\begin{table*}[hbt!]
\centering
{\scriptsize
\begin{tabular}{l||r|r|r|r|r|r|r|r|r|r}
  {\bf application} & {\bf \#prompts} & \multicolumn{4}{|c|}{\bf \ChangedText{mutant candidates}} & {\bf \#mutants} & {\bf \#killed} & {\bf \#survived} & {\bf \#timeout} & {\bf mut.} \\
  & &  {\bf \ChangedText{total}} & {\bf \ChangedText{invalid}} & {\bf \ChangedText{identical}} & {\bf \ChangedText{duplicate}}  &  & & & & {\bf score} \\
  \hline
  \hline
\textit{Complex.js} & 490 & \ChangedText{1,457} & \ChangedText{222} & \ChangedText{7} & \ChangedText{36} & 1,192 & 721 & 471 & 0 & 60.49 \\ 
\hline
\textit{countries-and-timezones} & 106 & \ChangedText{317} & \ChangedText{86} & \ChangedText{2} & \ChangedText{2} & 227 & 200 & 27 & 0 & 88.11 \\ 
\hline
\textit{crawler-url-parser} & 176 & \ChangedText{513} & \ChangedText{213} & \ChangedText{15} & \ChangedText{12} & 260 & 151 & 109 & 0 & 58.08 \\ 
\hline
\textit{delta} & 462 & \ChangedText{1,370} & \ChangedText{576} & \ChangedText{7} & \ChangedText{22} & 765 & 619 & 109 & 37 & 85.75 \\ 
\hline
\textit{image-downloader} & 42 & \ChangedText{122} & \ChangedText{37} & \ChangedText{0} & \ChangedText{3} & 82 & 62 & 20 & 0 & 75.61 \\ 
\hline
\textit{node-dirty} & 154 & \ChangedText{454} & \ChangedText{160} & \ChangedText{12} & \ChangedText{15} & 267 & 158 & 97 & 12 & 63.67 \\
\end{tabular}
  }
  \\[2mm]
  \caption{Results from LLMorpheus experiment \ChangedText{(run \#352)}.
    Model: \textit{codellama-34b-instruct}, 
    temperature: 0.25, 
    maxTokens: 250, 
    template: \textit{template-full.hb}, 
    systemPrompt: \textit{SystemPrompt-MutationTestingExpert.txt}. 
  }
  \label{table:Mutants:run352:codellama-34b-instruct:template-full.hb:0.25}
\end{table*}

\begin{table*}[hbt!]
\centering
{\scriptsize
\begin{tabular}{l||r|r|r|r|r|r|r|r|r|r}
  {\bf application} & {\bf \#prompts} & \multicolumn{4}{|c|}{\bf \ChangedText{mutant candidates}} & {\bf \#mutants} & {\bf \#killed} & {\bf \#survived} & {\bf \#timeout} & {\bf mut.} \\
  & &  {\bf \ChangedText{total}} & {\bf \ChangedText{invalid}} & {\bf \ChangedText{identical}} & {\bf \ChangedText{duplicate}}  &  & & & & {\bf score} \\
  \hline
  \hline
\textit{Complex.js} & 490 & \ChangedText{1,451} & \ChangedText{226} & \ChangedText{11} & \ChangedText{46} & 1,168 & 717 & 451 & 0 & 61.39 \\ 
\hline
\textit{countries-and-timezones} & 106 & \ChangedText{313} & \ChangedText{80} & \ChangedText{0} & \ChangedText{8} & 225 & 199 & 26 & 0 & 88.44 \\ 
\hline
\textit{crawler-url-parser} & 176 & \ChangedText{523} & \ChangedText{204} & \ChangedText{15} & \ChangedText{19} & 268 & 161 & 107 & 0 & 60.07 \\ 
\hline
\textit{delta} & 462 & \ChangedText{1,368} & \ChangedText{574} & \ChangedText{11} & \ChangedText{23} & 760 & 620 & 106 & 34 & 86.05 \\ 
\hline
\textit{image-downloader} & 42 & \ChangedText{126} & \ChangedText{34} & \ChangedText{3} & \ChangedText{1} & 85 & 71 & 14 & 0 & 83.53 \\ 
\hline
\textit{node-dirty} & 154 & \ChangedText{450} & \ChangedText{161} & \ChangedText{17} & \ChangedText{7} & 265 & 167 & 86 & 12 & 67.55 \\
\end{tabular}
  }
  \\[2mm]
  \caption{Results from LLMorpheus experiment \ChangedText{(run \#353)}.
    Model: \textit{codellama-34b-instruct}, 
    temperature: 0.25, 
    maxTokens: 250, 
    template: \textit{template-full.hb}, 
    systemPrompt: \textit{SystemPrompt-MutationTestingExpert.txt}. 
  }
  \label{table:Mutants:run353:codellama-34b-instruct:template-full.hb:0.25}
\end{table*}

\begin{table*}[hbt!]
\centering
{\scriptsize
\begin{tabular}{l||r|r|r|r|r}
\multicolumn{1}{c|}{\bf project} & \multicolumn{2}{|c|}{\bf time (sec)} & \multicolumn{3}{|c}{\bf \#tokens} \\
               & {\it LLMorpheus} & {\it StrykerJS} & {\bf prompt} & {\bf compl.} & {\bf total} \\
\hline
  Complex.js & 3,044.19 & 631.79 & 967,508 & 101,588 & 1,069,096 \\ 
countries-and-timezones & 1,070.90 & 327.25 & 105,828 & 23,471 & 129,299 \\ 
crawler-url-parser & 1,646.30 & 818.05 & 386,223 & 39,000 & 425,223 \\ 
delta & 2,954.37 & 3,844.00 & 890,252 & 99,341 & 989,593 \\ 
image-downloader & 430.54 & 348.86 & 24,655 & 9,217 & 33,872 \\ 
node-dirty & 1,532.06 & 232.35 & 246,248 & 32,400 & 278,648 \\ 
node-geo-point & 1,708.63 & 961.35 & 291,061 & 26,301 & 317,362 \\ 
node-jsonfile & 740.73 & 502.52 & 57,516 & 14,400 & 71,916 \\ 
plural & 1,537.89 & 149.38 & 265,602 & 33,182 & 298,784 \\ 
pull-stream & 2,522.38 & 1,395.14 & 208,130 & 76,091 & 284,221 \\ 
q & 5,294.80 & 14,085.10 & 2,127,655 & 218,620 & 2,346,275 \\ 
spacl-core & 1,351.05 & 728.32 & 162,705 & 29,167 & 191,872 \\ 
zip-a-folder & 500.56 & 1,086.06 & 82,457 & 10,557 & 93,014 \\ 
\hline
  \textit{Total} & 24,334.42 & 25,110.17 & 5,815,840 & 713,335 & 6,529,175 \\
  \end{tabular}
  }
  \\[2mm]
  \caption{Results from LLMorpheus experiment \ChangedText{(run \#348)}.
    Model: \textit{codellama-34b-instruct}, 
    temperature: 0.25, 
    maxTokens: 250, 
    template: \textit{template-full.hb}, 
    systemPrompt: \textit{SystemPrompt-MutationTestingExpert.txt}
  }
  \label{table:Cost:run348:codellama-34b-instruct:template-full.hb:0.25}
\end{table*}

\begin{table*}[hbt!]
\centering
{\scriptsize
\begin{tabular}{l||r|r|r|r|r}
\multicolumn{1}{c|}{\bf project} & \multicolumn{2}{|c|}{\bf time (sec)} & \multicolumn{3}{|c}{\bf \#tokens} \\
               & {\it LLMorpheus} & {\it StrykerJS} & {\bf prompt} & {\bf compl.} & {\bf total} \\
\hline
  Complex.js & 3,029.62 & 629.66 & 967,508 & 100,540 & 1,068,048 \\ 
countries-and-timezones & 1,070.85 & 314.98 & 105,828 & 23,186 & 129,014 \\ 
crawler-url-parser & 1,777.23 & 867.78 & 386,223 & 38,916 & 425,139 \\ 
delta & 2,978.88 & 3,758.68 & 890,252 & 99,176 & 989,428 \\ 
image-downloader & 430.55 & 366.05 & 24,655 & 9,223 & 33,878 \\ 
node-dirty & 1,528.73 & 237.16 & 246,248 & 32,776 & 279,024 \\ 
node-geo-point & 1,411.08 & 1,021.61 & 316,333 & 29,301 & 345,634 \\ 
node-jsonfile & 690.69 & 485.24 & 57,516 & 14,071 & 71,587 \\ 
plural & 1,521.09 & 154.20 & 265,602 & 33,560 & 299,162 \\ 
pull-stream & 2,503.60 & 1,383.12 & 208,130 & 76,551 & 284,681 \\ 
q & 5,379.31 & 13,584.78 & 2,127,655 & 217,699 & 2,345,354 \\ 
spacl-core & 1,350.98 & 739.03 & 162,705 & 29,184 & 191,889 \\ 
zip-a-folder & 500.57 & 531.10 & 82,457 & 10,753 & 93,210 \\ 
\hline
  \textit{Total} & 24,173.19 & 24,073.41 & 5,841,112 & 714,936 & 6,556,048 \\
  \end{tabular}
  }
  \\[2mm]
  \caption{Results from LLMorpheus experiment \ChangedText{(run \#350)}.
    Model: \textit{codellama-34b-instruct}, 
    temperature: 0.25, 
    maxTokens: 250, 
    template: \textit{template-full.hb}, 
    systemPrompt: \textit{SystemPrompt-MutationTestingExpert.txt}
  }
  \label{table:Cost:run350:codellama-34b-instruct:template-full.hb:0.25}
\end{table*}

\begin{table*}[hbt!]
\centering
{\scriptsize
\begin{tabular}{l||r|r|r|r|r}
\multicolumn{1}{c|}{\bf project} & \multicolumn{2}{|c|}{\bf time (sec)} & \multicolumn{3}{|c}{\bf \#tokens} \\
               & {\it LLMorpheus} & {\it StrykerJS} & {\bf prompt} & {\bf compl.} & {\bf total} \\
\hline
  Complex.js & 3,026.34 & 650.01 & 967,508 & 101,118 & 1,068,626 \\ 
countries-and-timezones & 1,070.91 & 309.33 & 105,828 & 23,331 & 129,159 \\ 
crawler-url-parser & 1,644.86 & 811.88 & 386,223 & 39,215 & 425,438 \\ 
delta & 2,962.02 & 3,896.15 & 890,252 & 99,274 & 989,526 \\ 
image-downloader & 430.54 & 373.38 & 24,655 & 9,163 & 33,818 \\ 
node-dirty & 1,526.45 & 247.44 & 246,248 & 32,894 & 279,142 \\ 
node-geo-point & 1,421.09 & 935.39 & 316,333 & 29,830 & 346,163 \\ 
node-jsonfile & 690.68 & 504.57 & 57,516 & 14,702 & 72,218 \\ 
plural & 1,521.32 & 151.42 & 265,602 & 33,298 & 298,900 \\ 
pull-stream & 2,497.56 & 1,351.06 & 208,130 & 76,100 & 284,230 \\ 
q & 5,341.32 & 13,704.14 & 2,127,655 & 218,805 & 2,346,460 \\ 
spacl-core & 1,351.03 & 756.66 & 162,705 & 28,939 & 191,644 \\ 
zip-a-folder & 500.57 & 1,119.70 & 82,457 & 10,786 & 93,243 \\ 
\hline
  \textit{Total} & 23,984.67 & 24,811.13 & 5,841,112 & 717,455 & 6,558,567 \\
  \end{tabular}
  }
  \\[2mm]
  \caption{Results from LLMorpheus experiment \ChangedText{(run \#351)}.
    Model: \textit{codellama-34b-instruct}, 
    temperature: 0.25, 
    maxTokens: 250, 
    template: \textit{template-full.hb}, 
    systemPrompt: \textit{SystemPrompt-MutationTestingExpert.txt}
  }
  \label{table:Cost:run351:codellama-34b-instruct:template-full.hb:0.25}
\end{table*}

\begin{table*}[hbt!]
\centering
{\scriptsize
\begin{tabular}{l||r|r|r|r|r}
\multicolumn{1}{c|}{\bf project} & \multicolumn{2}{|c|}{\bf time (sec)} & \multicolumn{3}{|c}{\bf \#tokens} \\
               & {\it LLMorpheus} & {\it StrykerJS} & {\bf prompt} & {\bf compl.} & {\bf total} \\
\hline
  Complex.js & 3,031.72 & 631.91 & 967,508 & 102,075 & 1,069,583 \\ 
countries-and-timezones & 1,070.83 & 324.27 & 105,828 & 23,502 & 129,330 \\ 
crawler-url-parser & 1,642.15 & 835.28 & 386,223 & 39,240 & 425,463 \\ 
delta & 2,969.01 & 3,912.89 & 890,252 & 99,383 & 989,635 \\ 
image-downloader & 430.57 & 486.28 & 24,655 & 9,228 & 33,883 \\ 
node-dirty & 1,527.16 & 237.65 & 246,248 & 32,850 & 279,098 \\ 
node-geo-point & 1,411.10 & 1,015.42 & 316,333 & 28,895 & 345,228 \\ 
node-jsonfile & 690.64 & 496.41 & 57,516 & 14,557 & 72,073 \\ 
plural & 1,521.03 & 148.10 & 265,602 & 33,162 & 298,764 \\ 
pull-stream & 2,509.05 & 1,370.63 & 208,130 & 75,917 & 284,047 \\ 
q & 5,300.14 & 14,131.12 & 2,127,655 & 218,921 & 2,346,576 \\ 
spacl-core & 1,351.01 & 712.33 & 162,705 & 28,809 & 191,514 \\ 
zip-a-folder & 500.63 & 1,266.61 & 82,457 & 10,707 & 93,164 \\ 
\hline
  \textit{Total} & 23,955.04 & 25,568.93 & 5,841,112 & 717,246 & 6,558,358 \\
  \end{tabular}
  }
  \\[2mm]
  \caption{Results from LLMorpheus experiment \ChangedText{(run \#352)}.
    Model: \textit{codellama-34b-instruct}, 
    temperature: 0.25, 
    maxTokens: 250, 
    template: \textit{template-full.hb}, 
    systemPrompt: \textit{SystemPrompt-MutationTestingExpert.txt}
  }
  \label{table:Cost:run352:codellama-34b-instruct:template-full.hb:0.25}
\end{table*}

\begin{table*}[hbt!]
\centering
{\scriptsize
\begin{tabular}{l||r|r|r|r|r}
\multicolumn{1}{c|}{\bf project} & \multicolumn{2}{|c|}{\bf time (sec)} & \multicolumn{3}{|c}{\bf \#tokens} \\
               & {\it LLMorpheus} & {\it StrykerJS} & {\bf prompt} & {\bf compl.} & {\bf total} \\
\hline
  Complex.js & 3,082.97 & 613.65 & 967,508 & 101,316 & 1,068,824 \\ 
countries-and-timezones & 1,070.86 & 326.66 & 105,828 & 22,979 & 128,807 \\ 
crawler-url-parser & 1,641.17 & 853.01 & 386,223 & 38,790 & 425,013 \\ 
delta & 2,952.50 & 3,773.99 & 890,252 & 99,524 & 989,776 \\ 
image-downloader & 470.58 & 359.76 & 24,655 & 8,898 & 33,553 \\ 
node-dirty & 1,526.98 & 239.71 & 246,248 & 32,476 & 278,724 \\ 
node-geo-point & 1,411.01 & 1,001.71 & 316,333 & 29,427 & 345,760 \\ 
node-jsonfile & 690.74 & 500.76 & 57,516 & 14,495 & 72,011 \\ 
plural & 1,521.11 & 158.00 & 265,602 & 33,838 & 299,440 \\ 
pull-stream & 2,510.21 & 1,355.46 & 208,130 & 75,432 & 283,562 \\ 
q & 5,390.06 & 14,133.05 & 2,127,655 & 217,855 & 2,345,510 \\ 
spacl-core & 1,350.98 & 752.72 & 162,705 & 29,399 & 192,104 \\ 
zip-a-folder & 520.57 & 564.11 & 82,457 & 10,749 & 93,206 \\ 
\hline
  \textit{Total} & 24,139.74 & 24,632.57 & 5,841,112 & 715,178 & 6,556,290 \\
  \end{tabular}
  }
  \\[2mm]
  \caption{Results from LLMorpheus experiment \ChangedText{(run \#353)}.
    Model: \textit{codellama-34b-instruct}, 
    temperature: 0.25, 
    maxTokens: 250, 
    template: \textit{template-full.hb}, 
    systemPrompt: \textit{SystemPrompt-MutationTestingExpert.txt}
  }
  \label{table:Cost:run353:codellama-34b-instruct:template-full.hb:0.25}
\end{table*}

\FloatBarrier

\subsection{Results for \CodeLlamaThirtyFour, {\tt full} template, temperature 0.5}
 \label{app:CodeLlamaThirtyFour0.5}
 
Tables~\ref{table:Mutants:run318:codellama-34b-instruct:template-full.hb:0.5}--\ref{table:Cost:run322:codellama-34b-instruct:template-full.hb:0.5}}
show the results for 5 experiments with the \textit{codellama-34b-instruct} model
at temperature 0.5 using the default prompt and system prompt shown in Figure~\ref{fig:PromptTemplate}.

\begin{table*}[hbt!]
\centering
{\scriptsize
\begin{tabular}{l||r|r|r|r|r|r|r|r|r|r}
  {\bf application} & {\bf \#prompts} & \multicolumn{4}{|c|}{\bf \ChangedText{mutant candidates}} & {\bf \#mutants} & {\bf \#killed} & {\bf \#survived} & {\bf \#timeout} & {\bf mut.} \\
  & &  {\bf \ChangedText{total}} & {\bf \ChangedText{invalid}} & {\bf \ChangedText{identical}} & {\bf \ChangedText{duplicate}}  &  & & & & {\bf score} \\
  \hline
  \hline
\textit{Complex.js} & 490 & \ChangedText{1,448} & \ChangedText{211} & \ChangedText{7} & \ChangedText{39} & 1,191 & 739 & 452 & 0 & 62.05 \\ 
\hline
\textit{countries-and-timezones} & 106 & \ChangedText{311} & \ChangedText{81} & \ChangedText{3} & \ChangedText{3} & 224 & 194 & 30 & 0 & 86.61 \\ 
\hline
\textit{crawler-url-parser} & 176 & \ChangedText{510} & \ChangedText{185} & \ChangedText{14} & \ChangedText{13} & 298 & 166 & 108 & 24 & 63.76 \\ 
\hline
\textit{delta} & 462 & \ChangedText{1,369} & \ChangedText{560} & \ChangedText{13} & \ChangedText{27} & 769 & 642 & 93 & 34 & 87.91 \\ 
\hline
\textit{image-downloader} & 42 & \ChangedText{126} & \ChangedText{35} & \ChangedText{1} & \ChangedText{1} & 89 & 68 & 21 & 0 & 76.40 \\ 
\hline
\textit{node-dirty} & 154 & \ChangedText{454} & \ChangedText{162} & \ChangedText{8} & \ChangedText{7} & 277 & 158 & 107 & 12 & 61.37 \\
\end{tabular}
  }
  \\[2mm]
  \caption{Results from LLMorpheus experiment \ChangedText{(run \#318)}.
    Model: \textit{codellama-34b-instruct}, 
    temperature: 0.5, 
    maxTokens: 250, 
    template: \textit{template-full.hb}, 
    systemPrompt: \textit{SystemPrompt-MutationTestingExpert.txt}. 
  }
  \label{table:Mutants:run318:codellama-34b-instruct:template-full.hb:0.5}
\end{table*}

\begin{table*}[hbt!]
\centering
{\scriptsize
\begin{tabular}{l||r|r|r|r|r|r|r|r|r|r}
  {\bf application} & {\bf \#prompts} & \multicolumn{4}{|c|}{\bf \ChangedText{mutant candidates}} & {\bf \#mutants} & {\bf \#killed} & {\bf \#survived} & {\bf \#timeout} & {\bf mut.} \\
  & &  {\bf \ChangedText{total}} & {\bf \ChangedText{invalid}} & {\bf \ChangedText{identical}} & {\bf \ChangedText{duplicate}}  &  & & & & {\bf score} \\
  \hline
  \hline
\textit{Complex.js} & 490 & \ChangedText{1,456} & \ChangedText{250} & \ChangedText{7} & \ChangedText{33} & 1,166 & 714 & 452 & 0 & 61.23 \\ 
\hline
\textit{countries-and-timezones} & 106 & \ChangedText{316} & \ChangedText{81} & \ChangedText{4} & \ChangedText{10} & 221 & 192 & 29 & 0 & 86.88 \\ 
\hline
\textit{crawler-url-parser} & 176 & \ChangedText{515} & \ChangedText{176} & \ChangedText{11} & \ChangedText{18} & 310 & 173 & 137 & 0 & 55.81 \\ 
\hline
\textit{delta} & 462 & \ChangedText{1,366} & \ChangedText{548} & \ChangedText{8} & \ChangedText{18} & 792 & 668 & 89 & 35 & 88.76 \\ 
\hline
\textit{image-downloader} & 42 & \ChangedText{126} & \ChangedText{36} & \ChangedText{3} & \ChangedText{5} & 82 & 54 & 28 & 0 & 65.85 \\ 
\hline
\textit{node-dirty} & 154 & \ChangedText{456} & \ChangedText{147} & \ChangedText{8} & \ChangedText{13} & 288 & 173 & 104 & 11 & 63.89 \\
\end{tabular}
  }
  \\[2mm]
  \caption{Results from LLMorpheus experiment \ChangedText{(run \#319)}.
    Model: \textit{codellama-34b-instruct}, 
    temperature: 0.5, 
    maxTokens: 250, 
    template: \textit{template-full.hb}, 
    systemPrompt: \textit{SystemPrompt-MutationTestingExpert.txt}. 
  }
  \label{table:Mutants:run319:codellama-34b-instruct:template-full.hb:0.5}
\end{table*}

\begin{table*}[hbt!]
\centering
{\scriptsize
\begin{tabular}{l||r|r|r|r|r|r|r|r|r|r}
  {\bf application} & {\bf \#prompts} & \multicolumn{4}{|c|}{\bf \ChangedText{mutant candidates}} & {\bf \#mutants} & {\bf \#killed} & {\bf \#survived} & {\bf \#timeout} & {\bf mut.} \\
  & &  {\bf \ChangedText{total}} & {\bf \ChangedText{invalid}} & {\bf \ChangedText{identical}} & {\bf \ChangedText{duplicate}}  &  & & & & {\bf score} \\
  \hline
  \hline
\textit{Complex.js} & 490 & \ChangedText{1,459} & \ChangedText{232} & \ChangedText{10} & \ChangedText{35} & 1,182 & 730 & 452 & 0 & 61.76 \\ 
\hline
\textit{countries-and-timezones} & 106 & \ChangedText{315} & \ChangedText{86} & \ChangedText{4} & \ChangedText{3} & 222 & 199 & 23 & 0 & 89.64 \\ 
\hline
\textit{crawler-url-parser} & 176 & \ChangedText{516} & \ChangedText{201} & \ChangedText{15} & \ChangedText{16} & 284 & 166 & 117 & 1 & 58.80 \\ 
\hline
\textit{delta} & 462 & \ChangedText{1,357} & \ChangedText{567} & \ChangedText{9} & \ChangedText{21} & 760 & 621 & 104 & 35 & 86.32 \\ 
\hline
\textit{image-downloader} & 42 & \ChangedText{125} & \ChangedText{33} & \ChangedText{3} & \ChangedText{4} & 85 & 60 & 25 & 0 & 70.59 \\ 
\hline
\textit{node-dirty} & 154 & \ChangedText{457} & \ChangedText{156} & \ChangedText{9} & \ChangedText{13} & 279 & 166 & 101 & 12 & 63.80 \\
\end{tabular}
  }
  \\[2mm]
  \caption{Results from LLMorpheus experiment \ChangedText{(run \#320)}.
    Model: \textit{codellama-34b-instruct}, 
    temperature: 0.5, 
    maxTokens: 250, 
    template: \textit{template-full.hb}, 
    systemPrompt: \textit{SystemPrompt-MutationTestingExpert.txt}. 
  }
  \label{table:Mutants:run320:codellama-34b-instruct:template-full.hb:0.5}
\end{table*}

\begin{table*}[hbt!]
\centering
{\scriptsize
\begin{tabular}{l||r|r|r|r|r|r|r|r|r|r}
  {\bf application} & {\bf \#prompts} & \multicolumn{4}{|c|}{\bf \ChangedText{mutant candidates}} & {\bf \#mutants} & {\bf \#killed} & {\bf \#survived} & {\bf \#timeout} & {\bf mut.} \\
  & &  {\bf \ChangedText{total}} & {\bf \ChangedText{invalid}} & {\bf \ChangedText{identical}} & {\bf \ChangedText{duplicate}}  &  & & & & {\bf score} \\
  \hline
  \hline
\textit{Complex.js} & 490 & \ChangedText{1,456} & \ChangedText{216} & \ChangedText{9} & \ChangedText{41} & 1,190 & 716 & 474 & 0 & 60.17 \\ 
\hline
\textit{countries-and-timezones} & 106 & \ChangedText{306} & \ChangedText{70} & \ChangedText{6} & \ChangedText{9} & 221 & 191 & 29 & 1 & 86.88 \\ 
\hline
\textit{crawler-url-parser} & 176 & \ChangedText{512} & \ChangedText{202} & \ChangedText{18} & \ChangedText{13} & 279 & 158 & 121 & 0 & 56.63 \\ 
\hline
\textit{delta} & 462 & \ChangedText{1,358} & \ChangedText{554} & \ChangedText{12} & \ChangedText{17} & 775 & 634 & 105 & 36 & 86.45 \\ 
\hline
\textit{image-downloader} & 42 & \ChangedText{125} & \ChangedText{36} & \ChangedText{2} & \ChangedText{2} & 85 & 68 & 17 & 0 & 80.00 \\ 
\hline
\textit{node-dirty} & 154 & \ChangedText{454} & \ChangedText{166} & \ChangedText{12} & \ChangedText{7} & 269 & 152 & 105 & 12 & 60.97 \\
\end{tabular}
  }
  \\[2mm]
  \caption{Results from LLMorpheus experiment \ChangedText{(run \#321)}.
    Model: \textit{codellama-34b-instruct}, 
    temperature: 0.5, 
    maxTokens: 250, 
    template: \textit{template-full.hb}, 
    systemPrompt: \textit{SystemPrompt-MutationTestingExpert.txt}. 
  }
  \label{table:Mutants:run321:codellama-34b-instruct:template-full.hb:0.5}
\end{table*}

\begin{table*}[hbt!]
\centering
{\scriptsize
\begin{tabular}{l||r|r|r|r|r|r|r|r|r|r}
  {\bf application} & {\bf \#prompts} & \multicolumn{4}{|c|}{\bf \ChangedText{mutant candidates}} & {\bf \#mutants} & {\bf \#killed} & {\bf \#survived} & {\bf \#timeout} & {\bf mut.} \\
  & &  {\bf \ChangedText{total}} & {\bf \ChangedText{invalid}} & {\bf \ChangedText{identical}} & {\bf \ChangedText{duplicate}}  &  & & & & {\bf score} \\
  \hline
  \hline
\textit{Complex.js} & 490 & \ChangedText{1,451} & \ChangedText{219} & \ChangedText{7} & \ChangedText{39} & 1,186 & 725 & 461 & 0 & 61.13 \\ 
\hline
\textit{countries-and-timezones} & 106 & \ChangedText{311} & \ChangedText{78} & \ChangedText{1} & \ChangedText{8} & 224 & 197 & 27 & 0 & 87.95 \\ 
\hline
\textit{crawler-url-parser} & 176 & \ChangedText{518} & \ChangedText{200} & \ChangedText{8} & \ChangedText{13} & 297 & 173 & 124 & 0 & 58.25 \\ 
\hline
\textit{delta} & 462 & \ChangedText{1,362} & \ChangedText{535} & \ChangedText{8} & \ChangedText{28} & 791 & 660 & 92 & 39 & 88.37 \\ 
\hline
\textit{image-downloader} & 42 & \ChangedText{125} & \ChangedText{40} & \ChangedText{1} & \ChangedText{4} & 80 & 68 & 12 & 0 & 85.00 \\ 
\hline
\textit{node-dirty} & 154 & \ChangedText{455} & \ChangedText{159} & \ChangedText{8} & \ChangedText{5} & 283 & 161 & 110 & 12 & 61.13 \\
\end{tabular}
  }
  \\[2mm]
  \caption{Results from LLMorpheus experiment \ChangedText{(run \#322)}.
    Model: \textit{codellama-34b-instruct}, 
    temperature: 0.5, 
    maxTokens: 250, 
    template: \textit{template-full.hb}, 
    systemPrompt: \textit{SystemPrompt-MutationTestingExpert.txt}. 
  }
  \label{table:Mutants:run322:codellama-34b-instruct:template-full.hb:0.5}
\end{table*}

\begin{table*}[hbt!]
\centering
{\scriptsize
\begin{tabular}{l||r|r|r|r|r}
\multicolumn{1}{c|}{\bf project} & \multicolumn{2}{|c|}{\bf time (sec)} & \multicolumn{3}{|c}{\bf \#tokens} \\
               & {\it LLMorpheus} & {\it StrykerJS} & {\bf prompt} & {\bf compl.} & {\bf total} \\
\hline
  Complex.js & 3,016.85 & 592.07 & 967,508 & 100,627 & 1,068,135 \\ 
countries-and-timezones & 1,070.93 & 319.44 & 105,828 & 22,932 & 128,760 \\ 
crawler-url-parser & 1,668.66 & 1,244.64 & 386,223 & 38,920 & 425,143 \\ 
delta & 2,980.99 & 3,869.42 & 890,252 & 98,347 & 988,599 \\ 
image-downloader & 430.55 & 365.90 & 24,655 & 8,925 & 33,580 \\ 
node-dirty & 1,532.57 & 247.29 & 246,248 & 32,995 & 279,243 \\ 
node-geo-point & 1,411.06 & 1,006.90 & 316,333 & 29,454 & 345,787 \\ 
node-jsonfile & 720.71 & 494.61 & 57,516 & 15,051 & 72,567 \\ 
plural & 1,522.56 & 152.03 & 265,602 & 33,550 & 299,152 \\ 
pull-stream & 2,506.81 & 1,378.63 & 208,130 & 75,239 & 283,369 \\ 
q & 5,332.46 & 14,311.02 & 2,127,655 & 217,886 & 2,345,541 \\ 
spacl-core & 1,351.01 & 850.39 & 162,705 & 28,919 & 191,624 \\ 
zip-a-folder & 900.98 & 471.62 & 72,362 & 9,438 & 81,800 \\ 
\hline
  \textit{Total} & 24,446.14 & 25,303.97 & 5,831,017 & 712,283 & 6,543,300 \\
  \end{tabular}
  }
  \\[2mm]
  \caption{Results from LLMorpheus experiment \ChangedText{(run \#318)}.
    Model: \textit{codellama-34b-instruct}, 
    temperature: 0.5, 
    maxTokens: 250, 
    template: \textit{template-full.hb}, 
    systemPrompt: \textit{SystemPrompt-MutationTestingExpert.txt}
  }
  \label{table:Cost:run318:codellama-34b-instruct:template-full.hb:0.5}
\end{table*}

\begin{table*}[hbt!]
\centering
{\scriptsize
\begin{tabular}{l||r|r|r|r|r}
\multicolumn{1}{c|}{\bf project} & \multicolumn{2}{|c|}{\bf time (sec)} & \multicolumn{3}{|c}{\bf \#tokens} \\
               & {\it LLMorpheus} & {\it StrykerJS} & {\bf prompt} & {\bf compl.} & {\bf total} \\
\hline
  Complex.js & 3,065.41 & 611.71 & 967,508 & 100,190 & 1,067,698 \\ 
countries-and-timezones & 1,070.87 & 313.43 & 105,828 & 22,897 & 128,725 \\ 
crawler-url-parser & 1,643.05 & 1,045.88 & 386,223 & 39,148 & 425,371 \\ 
delta & 3,028.61 & 4,054.65 & 890,252 & 99,121 & 989,373 \\ 
image-downloader & 430.56 & 496.08 & 24,655 & 8,793 & 33,448 \\ 
node-dirty & 1,527.72 & 252.26 & 246,248 & 33,054 & 279,302 \\ 
node-geo-point & 1,411.09 & 1,060.50 & 316,333 & 28,836 & 345,169 \\ 
node-jsonfile & 690.68 & 520.06 & 57,516 & 14,997 & 72,513 \\ 
plural & 1,522.17 & 146.72 & 265,602 & 33,944 & 299,546 \\ 
pull-stream & 2,517.82 & 1,365.08 & 208,130 & 75,400 & 283,530 \\ 
q & 5,232.53 & 13,865.94 & 2,127,655 & 217,305 & 2,344,960 \\ 
spacl-core & 1,351.05 & 848.71 & 162,705 & 28,593 & 191,298 \\ 
zip-a-folder & 500.59 & 1,105.81 & 82,457 & 10,544 & 93,001 \\ 
\hline
  \textit{Total} & 23,992.16 & 25,686.82 & 5,841,112 & 712,822 & 6,553,934 \\
  \end{tabular}
  }
  \\[2mm]
  \caption{Results from LLMorpheus experiment \ChangedText{(run \#319)}.
    Model: \textit{codellama-34b-instruct}, 
    temperature: 0.5, 
    maxTokens: 250, 
    template: \textit{template-full.hb}, 
    systemPrompt: \textit{SystemPrompt-MutationTestingExpert.txt}
  }
  \label{table:Cost:run319:codellama-34b-instruct:template-full.hb:0.5}
\end{table*}

\begin{table*}[hbt!]
\centering
{\scriptsize
\begin{tabular}{l||r|r|r|r|r}
\multicolumn{1}{c|}{\bf project} & \multicolumn{2}{|c|}{\bf time (sec)} & \multicolumn{3}{|c}{\bf \#tokens} \\
               & {\it LLMorpheus} & {\it StrykerJS} & {\bf prompt} & {\bf compl.} & {\bf total} \\
\hline
  Complex.js & 3,080.04 & 618.08 & 967,508 & 102,242 & 1,069,750 \\ 
countries-and-timezones & 1,070.88 & 311.67 & 105,828 & 22,556 & 128,384 \\ 
crawler-url-parser & 1,645.90 & 965.22 & 386,223 & 38,423 & 424,646 \\ 
delta & 2,992.70 & 3,820.99 & 890,252 & 99,184 & 989,436 \\ 
image-downloader & 430.53 & 500.27 & 24,655 & 9,240 & 33,895 \\ 
node-dirty & 1,527.92 & 252.60 & 246,248 & 32,911 & 279,159 \\ 
node-geo-point & 1,740.11 & 890.54 & 289,389 & 26,285 & 315,674 \\ 
node-jsonfile & 690.70 & 487.19 & 57,516 & 14,355 & 71,871 \\ 
plural & 1,677.14 & 158.69 & 249,979 & 30,944 & 280,923 \\ 
pull-stream & 2,510.30 & 1,455.39 & 208,130 & 75,369 & 283,499 \\ 
q & 5,453.21 & 14,010.80 & 2,127,655 & 217,999 & 2,345,654 \\ 
spacl-core & 1,351.08 & 861.29 & 162,705 & 28,654 & 191,359 \\ 
zip-a-folder & 500.56 & 1,278.26 & 82,457 & 10,677 & 93,134 \\ 
\hline
  \textit{Total} & 24,671.06 & 25,610.98 & 5,798,545 & 708,839 & 6,507,384 \\
  \end{tabular}
  }
  \\[2mm]
  \caption{Results from LLMorpheus experiment \ChangedText{(run \#320)}.
    Model: \textit{codellama-34b-instruct}, 
    temperature: 0.5, 
    maxTokens: 250, 
    template: \textit{template-full.hb}, 
    systemPrompt: \textit{SystemPrompt-MutationTestingExpert.txt}
  }
  \label{table:Cost:run320:codellama-34b-instruct:template-full.hb:0.5}
\end{table*}

\begin{table*}[hbt!]
\centering
{\scriptsize
\begin{tabular}{l||r|r|r|r|r}
\multicolumn{1}{c|}{\bf project} & \multicolumn{2}{|c|}{\bf time (sec)} & \multicolumn{3}{|c}{\bf \#tokens} \\
               & {\it LLMorpheus} & {\it StrykerJS} & {\bf prompt} & {\bf compl.} & {\bf total} \\
\hline
  Complex.js & 3,025.70 & 617.91 & 967,508 & 101,251 & 1,068,759 \\ 
countries-and-timezones & 1,070.86 & 333.07 & 105,828 & 23,224 & 129,052 \\ 
crawler-url-parser & 1,686.57 & 939.88 & 386,223 & 39,014 & 425,237 \\ 
delta & 3,020.99 & 3,992.69 & 890,252 & 99,683 & 989,935 \\ 
image-downloader & 430.56 & 347.98 & 24,655 & 9,059 & 33,714 \\ 
node-dirty & 1,526.16 & 227.18 & 246,248 & 32,693 & 278,941 \\ 
node-geo-point & 1,411.04 & 1,057.16 & 316,333 & 29,723 & 346,056 \\ 
node-jsonfile & 690.68 & 565.63 & 57,516 & 14,528 & 72,044 \\ 
plural & 1,521.10 & 156.75 & 265,602 & 34,049 & 299,651 \\ 
pull-stream & 2,516.94 & 1,385.89 & 208,130 & 75,071 & 283,201 \\ 
q & 5,280.14 & 14,131.09 & 2,127,655 & 217,911 & 2,345,566 \\ 
spacl-core & 1,350.99 & 879.83 & 162,705 & 28,767 & 191,472 \\ 
zip-a-folder & 510.57 & 542.42 & 82,457 & 10,709 & 93,166 \\ 
\hline
  \textit{Total} & 24,042.30 & 25,177.49 & 5,841,112 & 715,682 & 6,556,794 \\
  \end{tabular}
  }
  \\[2mm]
  \caption{Results from LLMorpheus experiment \ChangedText{(run \#321)}.
    Model: \textit{codellama-34b-instruct}, 
    temperature: 0.5, 
    maxTokens: 250, 
    template: \textit{template-full.hb}, 
    systemPrompt: \textit{SystemPrompt-MutationTestingExpert.txt}
  }
  \label{table:Cost:run321:codellama-34b-instruct:template-full.hb:0.5}
\end{table*}

\begin{table*}[hbt!]
\centering
{\scriptsize
\begin{tabular}{l||r|r|r|r|r}
\multicolumn{1}{c|}{\bf project} & \multicolumn{2}{|c|}{\bf time (sec)} & \multicolumn{3}{|c}{\bf \#tokens} \\
               & {\it LLMorpheus} & {\it StrykerJS} & {\bf prompt} & {\bf compl.} & {\bf total} \\
\hline
  Complex.js & 3,058.33 & 648.14 & 967,508 & 100,962 & 1,068,470 \\ 
countries-and-timezones & 1,070.88 & 328.85 & 105,828 & 23,165 & 128,993 \\ 
crawler-url-parser & 1,645.45 & 966.36 & 386,223 & 38,649 & 424,872 \\ 
delta & 2,954.29 & 4,010.15 & 890,252 & 97,995 & 988,247 \\ 
image-downloader & 430.57 & 330.02 & 24,655 & 9,182 & 33,837 \\ 
node-dirty & 1,528.13 & 239.77 & 246,248 & 32,568 & 278,816 \\ 
node-geo-point & 1,411.08 & 1,024.88 & 316,333 & 29,330 & 345,663 \\ 
node-jsonfile & 690.67 & 543.04 & 57,516 & 14,833 & 72,349 \\ 
plural & 1,522.09 & 148.56 & 265,602 & 33,906 & 299,508 \\ 
pull-stream & 2,509.37 & 1,433.54 & 208,130 & 75,574 & 283,704 \\ 
q & 5,254.98 & 13,947.69 & 2,127,655 & 216,579 & 2,344,234 \\ 
spacl-core & 1,351.03 & 834.35 & 162,705 & 29,103 & 191,808 \\ 
zip-a-folder & 500.60 & 1,144.50 & 82,457 & 10,347 & 92,804 \\ 
\hline
  \textit{Total} & 23,927.46 & 25,599.86 & 5,841,112 & 712,193 & 6,553,305 \\
  \end{tabular}
  }
  \\[2mm]
  \caption{Results from LLMorpheus experiment \ChangedText{(run \#322)}.
    Model: \textit{codellama-34b-instruct}, 
    temperature: 0.5, 
    maxTokens: 250, 
    template: \textit{template-full.hb}, 
    systemPrompt: \textit{SystemPrompt-MutationTestingExpert.txt}
  }
  \label{table:Cost:run322:codellama-34b-instruct:template-full.hb:0.5}
\end{table*}

\FloatBarrier

\subsection{Results for \CodeLlamaThirtyFour, {\tt full} template, temperature 1.0}
 \label{app:CodeLlamaThirtyFour1.0}

Tables~\ref{table:Mutants:run341:codellama-34b-instruct:template-full.hb:1.0}--\ref{table:Cost:run347:codellama-34b-instruct:template-full.hb:1.0}
show the results for 5 experiments with the \textit{codellama-34b-instruct} model
at temperature 1.0 using the default prompt and system prompt shown in Figure~\ref{fig:PromptTemplate}.

\begin{table*}[hbt!]
\centering
{\scriptsize
\begin{tabular}{l||r|r|r|r|r|r|r|r|r|r}
  {\bf application} & {\bf \#prompts} & \multicolumn{4}{|c|}{\bf \ChangedText{mutant candidates}} & {\bf \#mutants} & {\bf \#killed} & {\bf \#survived} & {\bf \#timeout} & {\bf mut.} \\
  & &  {\bf \ChangedText{total}} & {\bf \ChangedText{invalid}} & {\bf \ChangedText{identical}} & {\bf \ChangedText{duplicate}}  &  & & & & {\bf score} \\
  \hline
  \hline
\textit{Complex.js} & 490 & \ChangedText{1,405} & \ChangedText{345} & \ChangedText{5} & \ChangedText{27} & 1,028 & 648 & 379 & 1 & 63.13 \\ 
\hline
\textit{countries-and-timezones} & 106 & \ChangedText{300} & \ChangedText{108} & \ChangedText{2} & \ChangedText{4} & 186 & 156 & 30 & 0 & 83.87 \\ 
\hline
\textit{crawler-url-parser} & 176 & \ChangedText{490} & \ChangedText{202} & \ChangedText{4} & \ChangedText{6} & 278 & 202 & 76 & 0 & 72.66 \\ 
\hline
\textit{delta} & 462 & \ChangedText{1,250} & \ChangedText{519} & \ChangedText{11} & \ChangedText{22} & 698 & 583 & 83 & 32 & 88.11 \\ 
\hline
\textit{image-downloader} & 42 & \ChangedText{118} & \ChangedText{42} & \ChangedText{1} & \ChangedText{0} & 75 & 53 & 22 & 0 & 70.67 \\ 
\hline
\textit{node-dirty} & 154 & \ChangedText{438} & \ChangedText{175} & \ChangedText{9} & \ChangedText{8} & 246 & 150 & 84 & 12 & 65.85 \\
\end{tabular}
  }
  \\[2mm]
  \caption{Results from LLMorpheus experiment \ChangedText{(run \#341)}.
    Model: \textit{codellama-34b-instruct}, 
    temperature: 1.0, 
    maxTokens: 250, 
    template: \textit{template-full.hb}, 
    systemPrompt: \textit{SystemPrompt-MutationTestingExpert.txt}. 
  }
  \label{table:Mutants:run341:codellama-34b-instruct:template-full.hb:1.0}
\end{table*}

\begin{table*}[hbt!]
\centering
{\scriptsize
\begin{tabular}{l||r|r|r|r|r|r|r|r|r|r}
  {\bf application} & {\bf \#prompts} & \multicolumn{4}{|c|}{\bf \ChangedText{mutant candidates}} & {\bf \#mutants} & {\bf \#killed} & {\bf \#survived} & {\bf \#timeout} & {\bf mut.} \\
  & &  {\bf \ChangedText{total}} & {\bf \ChangedText{invalid}} & {\bf \ChangedText{identical}} & {\bf \ChangedText{duplicate}}  &  & & & & {\bf score} \\
  \hline
  \hline
\textit{Complex.js} & 490 & \ChangedText{1,382} & \ChangedText{363} & \ChangedText{9} & \ChangedText{23} & 987 & 611 & 376 & 0 & 61.90 \\ 
\hline
\textit{countries-and-timezones} & 106 & \ChangedText{254} & \ChangedText{79} & \ChangedText{7} & \ChangedText{4} & 164 & 145 & 19 & 0 & 88.41 \\ 
\hline
\textit{crawler-url-parser} & 176 & \ChangedText{497} & \ChangedText{228} & \ChangedText{6} & \ChangedText{6} & 257 & 177 & 80 & 0 & 68.87 \\ 
\hline
\textit{delta} & 462 & \ChangedText{1,307} & \ChangedText{580} & \ChangedText{7} & \ChangedText{19} & 701 & 596 & 75 & 30 & 89.30 \\ 
\hline
\textit{image-downloader} & 42 & \ChangedText{115} & \ChangedText{52} & \ChangedText{1} & \ChangedText{3} & 59 & 46 & 13 & 0 & 77.97 \\ 
\hline
\textit{node-dirty} & 154 & \ChangedText{438} & \ChangedText{159} & \ChangedText{4} & \ChangedText{7} & 268 & 164 & 96 & 8 & 64.18 \\
\end{tabular}
  }
  \\[2mm]
  \caption{Results from LLMorpheus experiment \ChangedText{(run \#342)}.
    Model: \textit{codellama-34b-instruct}, 
    temperature: 1.0, 
    maxTokens: 250, 
    template: \textit{template-full.hb}, 
    systemPrompt: \textit{SystemPrompt-MutationTestingExpert.txt}. 
  }
  \label{table:Mutants:run342:codellama-34b-instruct:template-full.hb:1.0}
\end{table*}

\begin{table*}[hbt!]
\centering
{\scriptsize
\begin{tabular}{l||r|r|r|r|r|r|r|r|r|r}
  {\bf application} & {\bf \#prompts} & \multicolumn{4}{|c|}{\bf \ChangedText{mutant candidates}} & {\bf \#mutants} & {\bf \#killed} & {\bf \#survived} & {\bf \#timeout} & {\bf mut.} \\
  & &  {\bf \ChangedText{total}} & {\bf \ChangedText{invalid}} & {\bf \ChangedText{identical}} & {\bf \ChangedText{duplicate}}  &  & & & & {\bf score} \\
  \hline
  \hline
\textit{Complex.js} & 490 & \ChangedText{1,378} & \ChangedText{346} & \ChangedText{7} & \ChangedText{23} & 1,002 & 637 & 365 & 0 & 63.57 \\ 
\hline
\textit{countries-and-timezones} & 106 & \ChangedText{297} & \ChangedText{94} & \ChangedText{1} & \ChangedText{4} & 198 & 177 & 21 & 0 & 89.39 \\ 
\hline
\textit{crawler-url-parser} & 176 & \ChangedText{491} & \ChangedText{207} & \ChangedText{6} & \ChangedText{9} & 269 & 173 & 96 & 0 & 64.31 \\ 
\hline
\textit{delta} & 462 & \ChangedText{1,296} & \ChangedText{535} & \ChangedText{11} & \ChangedText{23} & 727 & 611 & 85 & 31 & 88.31 \\ 
\hline
\textit{image-downloader} & 42 & \ChangedText{118} & \ChangedText{39} & \ChangedText{1} & \ChangedText{0} & 78 & 70 & 8 & 0 & 89.74 \\ 
\hline
\textit{node-dirty} & 154 & \ChangedText{437} & \ChangedText{180} & \ChangedText{4} & \ChangedText{8} & 245 & 136 & 97 & 12 & 60.41 \\
\end{tabular}
  }
  \\[2mm]
  \caption{Results from LLMorpheus experiment \ChangedText{(run \#343)}.
    Model: \textit{codellama-34b-instruct}, 
    temperature: 1.0, 
    maxTokens: 250, 
    template: \textit{template-full.hb}, 
    systemPrompt: \textit{SystemPrompt-MutationTestingExpert.txt}. 
  }
  \label{table:Mutants:run343:codellama-34b-instruct:template-full.hb:1.0}
\end{table*}

\begin{table*}[hbt!]
\centering
{\scriptsize
\begin{tabular}{l||r|r|r|r|r|r|r|r|r|r}
  {\bf application} & {\bf \#prompts} & \multicolumn{4}{|c|}{\bf \ChangedText{mutant candidates}} & {\bf \#mutants} & {\bf \#killed} & {\bf \#survived} & {\bf \#timeout} & {\bf mut.} \\
  & &  {\bf \ChangedText{total}} & {\bf \ChangedText{invalid}} & {\bf \ChangedText{identical}} & {\bf \ChangedText{duplicate}}  &  & & & & {\bf score} \\
  \hline
  \hline
\textit{Complex.js} & 490 & \ChangedText{1,385} & \ChangedText{318} & \ChangedText{3} & \ChangedText{22} & 1,042 & 667 & 374 & 1 & 64.11 \\ 
\hline
\textit{countries-and-timezones} & 106 & \ChangedText{290} & \ChangedText{84} & \ChangedText{1} & \ChangedText{3} & 202 & 183 & 19 & 0 & 90.59 \\ 
\hline
\textit{crawler-url-parser} & 176 & \ChangedText{504} & \ChangedText{221} & \ChangedText{4} & \ChangedText{3} & 276 & 200 & 76 & 0 & 72.46 \\ 
\hline
\textit{delta} & 462 & \ChangedText{1,296} & \ChangedText{531} & \ChangedText{5} & \ChangedText{16} & 744 & 639 & 71 & 34 & 90.46 \\ 
\hline
\textit{image-downloader} & 42 & \ChangedText{116} & \ChangedText{38} & \ChangedText{2} & \ChangedText{1} & 75 & 49 & 26 & 0 & 65.33 \\ 
\hline
\textit{node-dirty} & 154 & \ChangedText{412} & \ChangedText{142} & \ChangedText{7} & \ChangedText{3} & 260 & 145 & 100 & 15 & 61.54 \\
\end{tabular}
  }
  \\[2mm]
  \caption{Results from LLMorpheus experiment \ChangedText{(run \#345)}.
    Model: \textit{codellama-34b-instruct}, 
    temperature: 1.0, 
    maxTokens: 250, 
    template: \textit{template-full.hb}, 
    systemPrompt: \textit{SystemPrompt-MutationTestingExpert.txt}. 
  }
  \label{table:Mutants:run345:codellama-34b-instruct:template-full.hb:1.0}
\end{table*}

\begin{table*}[hbt!]
\centering
{\scriptsize
\begin{tabular}{l||r|r|r|r|r|r|r|r|r|r}
  {\bf application} & {\bf \#prompts} & \multicolumn{4}{|c|}{\bf \ChangedText{mutant candidates}} & {\bf \#mutants} & {\bf \#killed} & {\bf \#survived} & {\bf \#timeout} & {\bf mut.} \\
  & &  {\bf \ChangedText{total}} & {\bf \ChangedText{invalid}} & {\bf \ChangedText{identical}} & {\bf \ChangedText{duplicate}}  &  & & & & {\bf score} \\
  \hline
  \hline
\textit{Complex.js} & 490 & \ChangedText{1,366} & \ChangedText{318} & \ChangedText{6} & \ChangedText{24} & 1,018 & 640 & 378 & 0 & 62.87 \\ 
\hline
\textit{countries-and-timezones} & 106 & \ChangedText{298} & \ChangedText{88} & \ChangedText{1} & \ChangedText{2} & 207 & 177 & 30 & 0 & 85.51 \\ 
\hline
\textit{crawler-url-parser} & 176 & \ChangedText{494} & \ChangedText{232} & \ChangedText{4} & \ChangedText{6} & 252 & 178 & 74 & 0 & 70.63 \\ 
\hline
\textit{delta} & 462 & \ChangedText{1,315} & \ChangedText{617} & \ChangedText{4} & \ChangedText{13} & 679 & 564 & 83 & 32 & 87.78 \\ 
\hline
\textit{image-downloader} & 42 & \ChangedText{122} & \ChangedText{41} & \ChangedText{0} & \ChangedText{6} & 75 & 56 & 17 & 2 & 77.33 \\ 
\hline
\textit{node-dirty} & 154 & \ChangedText{435} & \ChangedText{165} & \ChangedText{2} & \ChangedText{1} & 267 & 155 & 103 & 9 & 61.42 \\
\end{tabular}
  }
  \\[2mm]
  \caption{Results from LLMorpheus experiment \ChangedText{(run \#347)}.
    Model: \textit{codellama-34b-instruct}, 
    temperature: 1.0, 
    maxTokens: 250, 
    template: \textit{template-full.hb}, 
    systemPrompt: \textit{SystemPrompt-MutationTestingExpert.txt}. 
  }
  \label{table:Mutants:run347:codellama-34b-instruct:template-full.hb:1.0}
\end{table*}

\begin{table*}[hbt!]
\centering
{\scriptsize
\begin{tabular}{l||r|r|r|r|r}
\multicolumn{1}{c|}{\bf project} & \multicolumn{2}{|c|}{\bf time (sec)} & \multicolumn{3}{|c}{\bf \#tokens} \\
               & {\it LLMorpheus} & {\it StrykerJS} & {\bf prompt} & {\bf compl.} & {\bf total} \\
\hline
  Complex.js & 3,090.47 & 545.31 & 967,508 & 101,516 & 1,069,024 \\ 
countries-and-timezones & 1,070.88 & 268.04 & 105,828 & 23,041 & 128,869 \\ 
crawler-url-parser & 1,786.52 & 902.86 & 377,644 & 36,825 & 414,469 \\ 
delta & 3,134.68 & 3,531.25 & 877,266 & 93,814 & 971,080 \\ 
image-downloader & 430.55 & 457.54 & 24,655 & 9,002 & 33,657 \\ 
node-dirty & 1,526.84 & 215.86 & 246,248 & 32,721 & 278,969 \\ 
node-geo-point & 1,411.05 & 897.39 & 316,333 & 29,494 & 345,827 \\ 
node-jsonfile & 710.73 & 477.34 & 57,516 & 14,447 & 71,963 \\ 
plural & 1,533.27 & 157.43 & 265,602 & 31,993 & 297,595 \\ 
pull-stream & 2,504.20 & 1,268.62 & 208,130 & 73,625 & 281,755 \\ 
q & 5,355.10 & 13,120.09 & 2,127,655 & 213,824 & 2,341,479 \\ 
spacl-core & 1,351.04 & 679.82 & 162,705 & 28,574 & 191,279 \\ 
zip-a-folder & 500.58 & 493.11 & 82,457 & 10,747 & 93,204 \\ 
\hline
  \textit{Total} & 24,405.89 & 23,014.65 & 5,819,547 & 699,623 & 6,519,170 \\
  \end{tabular}
  }
  \\[2mm]
  \caption{Results from LLMorpheus experiment \ChangedText{(run \#341)}.
    Model: \textit{codellama-34b-instruct}, 
    temperature: 1.0, 
    maxTokens: 250, 
    template: \textit{template-full.hb}, 
    systemPrompt: \textit{SystemPrompt-MutationTestingExpert.txt}
  }
  \label{table:Cost:run341:codellama-34b-instruct:template-full.hb:1.0}
\end{table*}

\begin{table*}[hbt!]
\centering
{\scriptsize
\begin{tabular}{l||r|r|r|r|r}
\multicolumn{1}{c|}{\bf project} & \multicolumn{2}{|c|}{\bf time (sec)} & \multicolumn{3}{|c}{\bf \#tokens} \\
               & {\it LLMorpheus} & {\it StrykerJS} & {\bf prompt} & {\bf compl.} & {\bf total} \\
\hline
  Complex.js & 3,099.95 & 515.00 & 967,508 & 99,831 & 1,067,339 \\ 
countries-and-timezones & 1,537.50 & 234.01 & 96,352 & 19,743 & 116,095 \\ 
crawler-url-parser & 1,643.69 & 843.29 & 386,223 & 36,746 & 422,969 \\ 
delta & 3,119.28 & 3,448.69 & 890,252 & 95,930 & 986,182 \\ 
image-downloader & 430.53 & 351.02 & 24,655 & 8,785 & 33,440 \\ 
node-dirty & 1,526.21 & 216.70 & 246,248 & 31,856 & 278,104 \\ 
node-geo-point & 1,422.53 & 877.31 & 316,333 & 30,037 & 346,370 \\ 
node-jsonfile & 690.70 & 537.75 & 57,516 & 14,738 & 72,254 \\ 
plural & 1,521.12 & 159.03 & 265,602 & 32,288 & 297,890 \\ 
pull-stream & 2,659.93 & 1,222.06 & 208,130 & 73,902 & 282,032 \\ 
q & 5,264.59 & 12,818.11 & 2,127,655 & 212,677 & 2,340,332 \\ 
spacl-core & 1,361.03 & 682.26 & 162,705 & 27,970 & 190,675 \\ 
zip-a-folder & 500.58 & 908.74 & 82,457 & 9,890 & 92,347 \\ 
\hline
  \textit{Total} & 24,777.63 & 22,813.96 & 5,831,636 & 694,393 & 6,526,029 \\
  \end{tabular}
  }
  \\[2mm]
  \caption{Results from LLMorpheus experiment \ChangedText{(run \#342)}.
    Model: \textit{codellama-34b-instruct}, 
    temperature: 1.0, 
    maxTokens: 250, 
    template: \textit{template-full.hb}, 
    systemPrompt: \textit{SystemPrompt-MutationTestingExpert.txt}
  }
  \label{table:Cost:run342:codellama-34b-instruct:template-full.hb:1.0}
\end{table*}

\begin{table*}[hbt!]
\centering
{\scriptsize
\begin{tabular}{l||r|r|r|r|r}
\multicolumn{1}{c|}{\bf project} & \multicolumn{2}{|c|}{\bf time (sec)} & \multicolumn{3}{|c}{\bf \#tokens} \\
               & {\it LLMorpheus} & {\it StrykerJS} & {\bf prompt} & {\bf compl.} & {\bf total} \\
\hline
  Complex.js & 3,024.06 & 525.74 & 967,508 & 98,687 & 1,066,195 \\ 
countries-and-timezones & 1,070.86 & 293.27 & 105,828 & 23,318 & 129,146 \\ 
crawler-url-parser & 1,644.85 & 845.43 & 386,223 & 36,638 & 422,861 \\ 
delta & 2,948.43 & 3,663.87 & 890,252 & 96,381 & 986,633 \\ 
image-downloader & 430.52 & 274.68 & 24,655 & 9,228 & 33,883 \\ 
node-dirty & 1,526.11 & 216.05 & 246,248 & 31,882 & 278,130 \\ 
node-geo-point & 1,411.07 & 852.26 & 316,333 & 29,152 & 345,485 \\ 
node-jsonfile & 690.73 & 549.73 & 57,516 & 13,670 & 71,186 \\ 
plural & 1,521.17 & 153.33 & 265,602 & 31,946 & 297,548 \\ 
pull-stream & 2,506.18 & 1,294.61 & 208,130 & 72,667 & 280,797 \\ 
q & 5,178.82 & 13,300.53 & 2,127,655 & 213,258 & 2,340,913 \\ 
spacl-core & 1,351.08 & 743.18 & 162,705 & 27,865 & 190,570 \\ 
zip-a-folder & 500.55 & 1,038.21 & 82,457 & 10,518 & 92,975 \\ 
\hline
  \textit{Total} & 23,804.42 & 23,750.90 & 5,841,112 & 695,210 & 6,536,322 \\
  \end{tabular}
  }
  \\[2mm]
  \caption{Results from LLMorpheus experiment \ChangedText{(run \#343)}.
    Model: \textit{codellama-34b-instruct}, 
    temperature: 1.0, 
    maxTokens: 250, 
    template: \textit{template-full.hb}, 
    systemPrompt: \textit{SystemPrompt-MutationTestingExpert.txt}
  }
  \label{table:Cost:run343:codellama-34b-instruct:template-full.hb:1.0}
\end{table*}

\begin{table*}[hbt!]
\centering
{\scriptsize
\begin{tabular}{l||r|r|r|r|r}
\multicolumn{1}{c|}{\bf project} & \multicolumn{2}{|c|}{\bf time (sec)} & \multicolumn{3}{|c}{\bf \#tokens} \\
               & {\it LLMorpheus} & {\it StrykerJS} & {\bf prompt} & {\bf compl.} & {\bf total} \\
\hline
  Complex.js & 3,070.12 & 559.72 & 967,508 & 100,998 & 1,068,506 \\ 
countries-and-timezones & 1,070.86 & 285.32 & 105,828 & 22,109 & 127,937 \\ 
crawler-url-parser & 1,644.31 & 872.80 & 386,223 & 38,267 & 424,490 \\ 
delta & 2,974.80 & 3,703.16 & 890,252 & 96,594 & 986,846 \\ 
image-downloader & 430.54 & 438.80 & 24,655 & 8,660 & 33,315 \\ 
node-dirty & 1,527.36 & 248.76 & 246,248 & 31,479 & 277,727 \\ 
node-geo-point & 1,411.07 & 843.77 & 316,333 & 29,660 & 345,993 \\ 
node-jsonfile & 690.69 & 512.58 & 57,516 & 14,276 & 71,792 \\ 
plural & 1,521.47 & 156.91 & 265,602 & 32,805 & 298,407 \\ 
pull-stream & 2,495.10 & 1,354.27 & 208,130 & 73,802 & 281,932 \\ 
q & 5,350.94 & 13,107.50 & 2,127,655 & 213,504 & 2,341,159 \\ 
spacl-core & 1,351.11 & 811.62 & 162,705 & 28,193 & 190,898 \\ 
zip-a-folder & 500.61 & 968.74 & 82,457 & 10,354 & 92,811 \\ 
\hline
  \textit{Total} & 24,038.96 & 23,863.96 & 5,841,112 & 700,701 & 6,541,813 \\
  \end{tabular}
  }
  \\[2mm]
  \caption{Results from LLMorpheus experiment \ChangedText{(run \#345)}.
    Model: \textit{codellama-34b-instruct}, 
    temperature: 1.0, 
    maxTokens: 250, 
    template: \textit{template-full.hb}, 
    systemPrompt: \textit{SystemPrompt-MutationTestingExpert.txt}
  }
  \label{table:Cost:run345:codellama-34b-instruct:template-full.hb:1.0}
\end{table*}

\begin{table*}[hbt!]
\centering
{\scriptsize
\begin{tabular}{l||r|r|r|r|r}
\multicolumn{1}{c|}{\bf project} & \multicolumn{2}{|c|}{\bf time (sec)} & \multicolumn{3}{|c}{\bf \#tokens} \\
               & {\it LLMorpheus} & {\it StrykerJS} & {\bf prompt} & {\bf compl.} & {\bf total} \\
\hline
  Complex.js & 3,021.09 & 531.51 & 967,508 & 100,183 & 1,067,691 \\ 
countries-and-timezones & 1,070.89 & 297.61 & 105,828 & 23,229 & 129,057 \\ 
crawler-url-parser & 1,642.94 & 776.31 & 386,223 & 36,920 & 423,143 \\ 
delta & 2,984.22 & 3,374.88 & 890,252 & 96,698 & 986,950 \\ 
image-downloader & 430.54 & 435.23 & 24,655 & 9,024 & 33,679 \\ 
node-dirty & 1,528.77 & 227.23 & 246,248 & 32,731 & 278,979 \\ 
node-geo-point & 1,411.05 & 858.36 & 316,333 & 29,774 & 346,107 \\ 
node-jsonfile & 690.66 & 494.30 & 57,516 & 14,127 & 71,643 \\ 
plural & 1,521.30 & 172.76 & 265,602 & 32,844 & 298,446 \\ 
pull-stream & 2,481.70 & 1,251.69 & 208,130 & 73,022 & 281,152 \\ 
q & 5,205.03 & 13,013.51 & 2,127,655 & 214,495 & 2,342,150 \\ 
spacl-core & 1,351.07 & 731.40 & 162,705 & 27,723 & 190,428 \\ 
zip-a-folder & 500.58 & 1,168.69 & 82,457 & 10,656 & 93,113 \\ 
\hline
  \textit{Total} & 23,839.84 & 23,333.49 & 5,841,112 & 701,426 & 6,542,538 \\
  \end{tabular}
  }
  \\[2mm]
  \caption{Results from LLMorpheus experiment \ChangedText{(run \#347)}.
    Model: \textit{codellama-34b-instruct}, 
    temperature: 1.0, 
    maxTokens: 250, 
    template: \textit{template-full.hb}, 
    systemPrompt: \textit{SystemPrompt-MutationTestingExpert.txt}
  }
  \label{table:Cost:run347:codellama-34b-instruct:template-full.hb:1.0}
\end{table*}

\FloatBarrier

\subsection{Results for variability analysis for \CodeLlamaThirtyFour}
\label{app:CodeLlamaThirtyFourVariability}

\ChangedText{
Tables~\ref{table:Variability_codellama-34b-instruct_0.0}--\ref{table:Variability_codellama-34b-instruct_1.0} show the variability of the mutants observed in 5 runs 
using \CodeLlamaThirtyFour, for each of the temperature settings 
that were previously discussed in Sections~\ref{app:CodeLlamaThirtyFour0.0}--\ref{app:CodeLlamaThirtyFour1.0}.}

\begin{table}[hbt!]
\centering
{\footnotesize
\begin{tabular}{l|r|r|r|r}

{\bf application}  & {\bf \#min} &  {\bf \#max} &  {\bf \#distinct} & {\bf \#common}\\
\hline
Complex.js & 1,198 & 1,199 & 1,217 & 1,181 (97.04\%) \\ 
countries-and-timezones & 216 & 217 & 218 & 215 (98.62\%) \\ 
crawler-url-parser & 282 & 285 & 289 & 278 (96.19\%) \\ 
delta & 765 & 768 & 787 & 746 (94.79\%) \\ 
image-downloader & 89 & 90 & 90 & 89 (98.89\%) \\ 
node-dirty & 274 & 275 & 283 & 266 (93.99\%) \\ 
node-geo-point & 302 & 302 & 307 & 297 (96.74\%) \\ 
node-jsonfile & 153 & 154 & 158 & 149 (94.30\%) \\ 
plural & 279 & 281 & 288 & 274 (95.14\%) \\ 
pull-stream & 758 & 760 & 770 & 746 (96.88\%) \\ 
q & 2,031 & 2,035 & 2,053 & 2,014 (98.10\%) \\ 
spacl-core & 238 & 239 & 252 & 225 (89.29\%) \\ 
zip-a-folder & 100 & 100 & 102 & 98 (96.08\%) \\ 
\end{tabular}
}
\caption{
  Variability of the mutants generated in 5 runs of \ToolName using the \textit{codellama-34b-instruct} LLM
       at temperature 0.0 \ChangedText{(run \#312,\#314,\#315,\#316,\#317)}. The columns of the table show, from left to right:
    (i) the minimum number of mutants observed in any of the runs,
    (ii) the maximum number of mutants observed in any of the runs,
    (iii) the total number of distinct mutants observed in all runs, and
    (iv) the number (percentage) of mutants are observed in all runs.
}
\label{table:Variability_codellama-34b-instruct_0.0}
\end{table}

\begin{table}[hbt!]
\centering
{\footnotesize
\begin{tabular}{l|r|r|r|r}

{\bf application}  & {\bf \#min} &  {\bf \#max} &  {\bf \#distinct} & {\bf \#common}\\
\hline
Complex.js & 1,168 & 1,200 & 2,354 & 447 (18.99\%) \\ 
countries-and-timezones & 215 & 227 & 551 & 53 (9.62\%) \\ 
crawler-url-parser & 266 & 285 & 700 & 67 (9.57\%) \\ 
delta & 752 & 781 & 1,706 & 247 (14.48\%) \\ 
image-downloader & 82 & 88 & 202 & 27 (13.37\%) \\ 
node-dirty & 265 & 279 & 600 & 82 (13.67\%) \\ 
node-geo-point & 296 & 314 & 660 & 99 (15.00\%) \\ 
node-jsonfile & 151 & 158 & 369 & 42 (11.38\%) \\ 
plural & 273 & 291 & 678 & 79 (11.65\%) \\ 
pull-stream & 764 & 770 & 1,664 & 280 (16.83\%) \\ 
q & 1,980 & 2,050 & 4,491 & 714 (15.90\%) \\ 
spacl-core & 230 & 244 & 613 & 58 (9.46\%) \\ 
zip-a-folder & 98 & 105 & 233 & 36 (15.45\%) \\ 
\end{tabular}
}
\caption{
  Variability of the mutants generated in 5 runs of \ToolName using the \textit{codellama-34b-instruct} LLM
       at temperature 0.25 \ChangedText{(run \#348,\#350,\#351,\#352,\#353)}. The columns of the table show, from left to right:
    (i) the minimum number of mutants observed in any of the runs,
    (ii) the maximum number of mutants observed in any of the runs,
    (iii) the total number of distinct mutants observed in all runs, and
    (iv) the number (percentage) of mutants are observed in all runs.
}
\label{table:Variability_codellama-34b-instruct_0.25}
\end{table}

\begin{table}[hbt!]
\centering
{\footnotesize
\begin{tabular}{l|r|r|r|r}

{\bf application}  & {\bf \#min} &  {\bf \#max} &  {\bf \#distinct} & {\bf \#common}\\
\hline
Complex.js & 1,166 & 1,191 & 3,196 & 205 (6.41\%) \\ 
countries-and-timezones & 221 & 224 & 735 & 13 (1.77\%) \\ 
crawler-url-parser & 279 & 310 & 966 & 31 (3.21\%) \\ 
delta & 760 & 792 & 2,322 & 94 (4.05\%) \\ 
image-downloader & 80 & 89 & 273 & 13 (4.76\%) \\ 
node-dirty & 269 & 288 & 831 & 33 (3.97\%) \\ 
node-geo-point & 269 & 312 & 905 & 44 (4.86\%) \\ 
node-jsonfile & 150 & 160 & 487 & 15 (3.08\%) \\ 
plural & 285 & 296 & 893 & 30 (3.36\%) \\ 
pull-stream & 779 & 809 & 2,183 & 161 (7.38\%) \\ 
q & 2,020 & 2,073 & 5,964 & 327 (5.48\%) \\ 
spacl-core & 250 & 262 & 782 & 23 (2.94\%) \\ 
zip-a-folder & 87 & 114 & 313 & 17 (5.43\%) \\ 
\end{tabular}
}
\caption{
  Variability of the mutants generated in 5 runs of \ToolName using the \textit{codellama-34b-instruct} LLM
       at temperature 0.5 \ChangedText{(run \#318,\#319,\#320,\#321,\#322)}. The columns of the table show, from left to right:
    (i) the minimum number of mutants observed in any of the runs,
    (ii) the maximum number of mutants observed in any of the runs,
    (iii) the total number of distinct mutants observed in all runs, and
    (iv) the number (percentage) of mutants are observed in all runs.
}
\label{table:Variability_codellama-34b-instruct_0.5}
\end{table}

\begin{table}[hbt!]
\centering
{\footnotesize
\begin{tabular}{l|r|r|r|r}

{\bf application}  & {\bf \#min} &  {\bf \#max} &  {\bf \#distinct} & {\bf \#common}\\
\hline
Complex.js & 987 & 1,042 & 4,200 & 17 (0.40\%) \\ 
countries-and-timezones & 164 & 207 & 835 & 2 (0.24\%) \\ 
crawler-url-parser & 252 & 278 & 1,171 & 1 (0.09\%) \\ 
delta & 681 & 744 & 3,013 & 2 (0.07\%) \\ 
image-downloader & 59 & 78 & 298 & 1 (0.34\%) \\ 
node-dirty & 245 & 268 & 1,075 & 2 (0.19\%) \\ 
node-geo-point & 259 & 273 & 1,111 & 2 (0.18\%) \\ 
node-jsonfile & 132 & 150 & 606 & 1 (0.17\%) \\ 
plural & 295 & 308 & 1,316 & 5 (0.38\%) \\ 
pull-stream & 736 & 749 & 2,814 & 30 (1.07\%) \\ 
q & 1,868 & 1,942 & 7,695 & 43 (0.56\%) \\ 
spacl-core & 205 & 240 & 958 & 2 (0.21\%) \\ 
zip-a-folder & 84 & 102 & 366 & 6 (1.64\%) \\ 
\end{tabular}
}
\caption{
  Variability of the mutants generated in 5 runs of \ToolName using the \textit{codellama-34b-instruct} LLM
       at temperature 1.0 \ChangedText{(run \#341,\#342,\#343,\#345,\#347)}. The columns of the table show, from left to right:
    (i) the minimum number of mutants observed in any of the runs,
    (ii) the maximum number of mutants observed in any of the runs,
    (iii) the total number of distinct mutants observed in all runs, and
    (iv) the number (percentage) of mutants are observed in all runs.
}
\label{table:Variability_codellama-34b-instruct_1.0}
\end{table}

\begin{figure*}
\includegraphics[width=\textwidth]{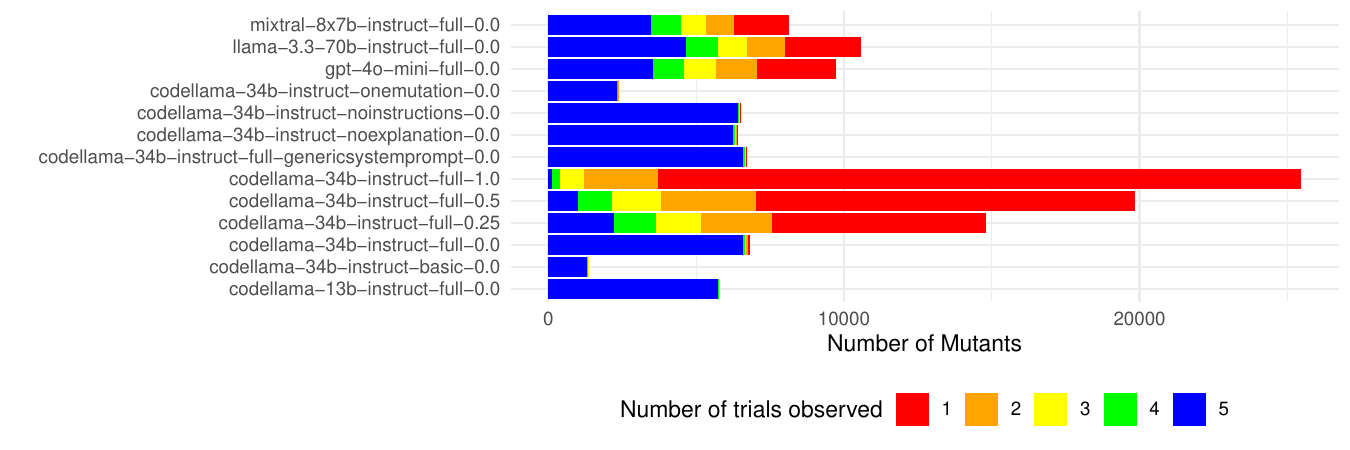}
	\caption{\begin{revision*}Variability of mutants generated by \ToolName. For each replacement generated at each position, we count the number of trials (of 5 total) where that replacement was generated.\end{revision*}}
	\label{fig:allstability}
\end{figure*}

\FloatBarrier
\begin{revision*}
\subsection{Visualization of variability across all configurations}	
Figure~\ref{fig:allstability} visualizes the variability of \ToolName across all configurations. The figure shows, in the aggregate for all \NrSubjectApplications subject applications, in how many runs each mutant was observed.
\end{revision*}

\FloatBarrier
 
\subsection{Results for \CodeLlamaThirteen}
     \label{app:CodeLlamaThirteen}

Tables~\ref{table:Mutants:run354:codellama-13b-instruct:template-full.hb:0.0}--\ref{table:Cost:run359:codellama-13b-instruct:template-full.hb:0.0}
show the results for 5 experiments with the \textit{codellama-13b-instruct} model
at temperature 0.0 using the default prompt and system prompt shown in Figure~\ref{fig:PromptTemplate}.
Table~\ref{table:Variability_codellama-13b-instruct_0.0} shows the variability of the mutants observed in 5 runs using \CodeLlamaThirteen at temperature 0.0.

\begin{table*}[hbt!]
\centering
{\scriptsize
\begin{tabular}{l||r|r|r|r|r|r|r|r|r|r}
  {\bf application} & {\bf \#prompts} & \multicolumn{4}{|c|}{\bf \ChangedText{mutant candidates}} & {\bf \#mutants} & {\bf \#killed} & {\bf \#survived} & {\bf \#timeout} & {\bf mut.} \\
  & &  {\bf \ChangedText{total}} & {\bf \ChangedText{invalid}} & {\bf \ChangedText{identical}} & {\bf \ChangedText{duplicate}}  &  & & & & {\bf score} \\
  \hline
  \hline
\textit{Complex.js} & 490 & \ChangedText{1,438} & \ChangedText{339} & \ChangedText{116} & \ChangedText{28} & 955 & 553 & 401 & 1 & 58.01 \\ 
\hline
\textit{countries-and-timezones} & 106 & \ChangedText{306} & \ChangedText{83} & \ChangedText{15} & \ChangedText{1} & 207 & 177 & 30 & 0 & 85.51 \\ 
\hline
\textit{crawler-url-parser} & 176 & \ChangedText{506} & \ChangedText{186} & \ChangedText{51} & \ChangedText{12} & 247 & 129 & 118 & 0 & 52.23 \\ 
\hline
\textit{delta} & 462 & \ChangedText{1,350} & \ChangedText{530} & \ChangedText{92} & \ChangedText{16} & 712 & 583 & 107 & 22 & 84.97 \\ 
\hline
\textit{image-downloader} & 42 & \ChangedText{124} & \ChangedText{40} & \ChangedText{5} & \ChangedText{2} & 77 & 48 & 29 & 0 & 62.34 \\ 
\hline
\textit{node-dirty} & 154 & \ChangedText{450} & \ChangedText{161} & \ChangedText{33} & \ChangedText{11} & 245 & 142 & 92 & 11 & 62.45 \\
\end{tabular}
  }
  \\[2mm]
  \caption{Results from LLMorpheus experiment \ChangedText{(run \#354)}.
    Model: \textit{codellama-13b-instruct}, 
    temperature: 0.0, 
    maxTokens: 250, 
    template: \textit{template-full.hb}, 
    systemPrompt: \textit{SystemPrompt-MutationTestingExpert.txt}. 
  }
  \label{table:Mutants:run354:codellama-13b-instruct:template-full.hb:0.0}
\end{table*}

\begin{table*}[hbt!]
\centering
{\scriptsize
\begin{tabular}{l||r|r|r|r|r|r|r|r|r|r}
  {\bf application} & {\bf \#prompts} & \multicolumn{4}{|c|}{\bf \ChangedText{mutant candidates}} & {\bf \#mutants} & {\bf \#killed} & {\bf \#survived} & {\bf \#timeout} & {\bf mut.} \\
  & &  {\bf \ChangedText{total}} & {\bf \ChangedText{invalid}} & {\bf \ChangedText{identical}} & {\bf \ChangedText{duplicate}}  &  & & & & {\bf score} \\
  \hline
  \hline
\textit{Complex.js} & 490 & \ChangedText{1,439} & \ChangedText{340} & \ChangedText{116} & \ChangedText{28} & 955 & 553 & 401 & 1 & 58.01 \\ 
\hline
\textit{countries-and-timezones} & 106 & \ChangedText{306} & \ChangedText{83} & \ChangedText{15} & \ChangedText{1} & 207 & 177 & 30 & 0 & 85.51 \\ 
\hline
\textit{crawler-url-parser} & 176 & \ChangedText{506} & \ChangedText{186} & \ChangedText{51} & \ChangedText{12} & 247 & 129 & 118 & 0 & 52.23 \\ 
\hline
\textit{delta} & 462 & \ChangedText{1,350} & \ChangedText{530} & \ChangedText{92} & \ChangedText{16} & 712 & 583 & 107 & 22 & 84.97 \\ 
\hline
\textit{image-downloader} & 42 & \ChangedText{124} & \ChangedText{40} & \ChangedText{5} & \ChangedText{2} & 77 & 48 & 29 & 0 & 62.34 \\ 
\hline
\textit{node-dirty} & 154 & \ChangedText{450} & \ChangedText{161} & \ChangedText{33} & \ChangedText{11} & 245 & 142 & 92 & 11 & 62.45 \\
\end{tabular}
  }
  \\[2mm]
  \caption{Results from LLMorpheus experiment \ChangedText{(run \#355)}.
    Model: \textit{codellama-13b-instruct}, 
    temperature: 0.0, 
    maxTokens: 250, 
    template: \textit{template-full.hb}, 
    systemPrompt: \textit{SystemPrompt-MutationTestingExpert.txt}. 
  }
  \label{table:Mutants:run355:codellama-13b-instruct:template-full.hb:0.0}
\end{table*}

\begin{table*}[hbt!]
\centering
{\scriptsize
\begin{tabular}{l||r|r|r|r|r|r|r|r|r|r}
  {\bf application} & {\bf \#prompts} & \multicolumn{4}{|c|}{\bf \ChangedText{mutant candidates}} & {\bf \#mutants} & {\bf \#killed} & {\bf \#survived} & {\bf \#timeout} & {\bf mut.} \\
  & &  {\bf \ChangedText{total}} & {\bf \ChangedText{invalid}} & {\bf \ChangedText{identical}} & {\bf \ChangedText{duplicate}}  &  & & & & {\bf score} \\
  \hline
  \hline
\textit{Complex.js} & 490 & \ChangedText{1,439} & \ChangedText{340} & \ChangedText{116} & \ChangedText{28} & 955 & 553 & 401 & 1 & 58.01 \\ 
\hline
\textit{countries-and-timezones} & 106 & \ChangedText{306} & \ChangedText{83} & \ChangedText{15} & \ChangedText{1} & 207 & 177 & 30 & 0 & 85.51 \\ 
\hline
\textit{crawler-url-parser} & 176 & \ChangedText{506} & \ChangedText{186} & \ChangedText{51} & \ChangedText{12} & 247 & 129 & 118 & 0 & 52.23 \\ 
\hline
\textit{delta} & 462 & \ChangedText{1,350} & \ChangedText{530} & \ChangedText{92} & \ChangedText{16} & 712 & 583 & 107 & 22 & 84.97 \\ 
\hline
\textit{image-downloader} & 42 & \ChangedText{124} & \ChangedText{40} & \ChangedText{5} & \ChangedText{2} & 77 & 48 & 29 & 0 & 62.34 \\ 
\hline
\textit{node-dirty} & 154 & \ChangedText{450} & \ChangedText{161} & \ChangedText{33} & \ChangedText{11} & 245 & 142 & 92 & 11 & 62.45 \\
\end{tabular}
  }
  \\[2mm]
  \caption{Results from LLMorpheus experiment \ChangedText{(run \#356)}.
    Model: \textit{codellama-13b-instruct}, 
    temperature: 0.0, 
    maxTokens: 250, 
    template: \textit{template-full.hb}, 
    systemPrompt: \textit{SystemPrompt-MutationTestingExpert.txt}. 
  }
  \label{table:Mutants:run356:codellama-13b-instruct:template-full.hb:0.0}
\end{table*}

\begin{table*}[hbt!]
\centering
{\scriptsize
\begin{tabular}{l||r|r|r|r|r|r|r|r|r|r}
  {\bf application} & {\bf \#prompts} & \multicolumn{4}{|c|}{\bf \ChangedText{mutant candidates}} & {\bf \#mutants} & {\bf \#killed} & {\bf \#survived} & {\bf \#timeout} & {\bf mut.} \\
  & &  {\bf \ChangedText{total}} & {\bf \ChangedText{invalid}} & {\bf \ChangedText{identical}} & {\bf \ChangedText{duplicate}}  &  & & & & {\bf score} \\
  \hline
  \hline
\textit{Complex.js} & 490 & \ChangedText{1,439} & \ChangedText{340} & \ChangedText{116} & \ChangedText{28} & 955 & 553 & 401 & 1 & 58.01 \\ 
\hline
\textit{countries-and-timezones} & 106 & \ChangedText{306} & \ChangedText{83} & \ChangedText{15} & \ChangedText{1} & 207 & 177 & 30 & 0 & 85.51 \\ 
\hline
\textit{crawler-url-parser} & 176 & \ChangedText{506} & \ChangedText{186} & \ChangedText{51} & \ChangedText{12} & 247 & 129 & 118 & 0 & 52.23 \\ 
\hline
\textit{delta} & 462 & \ChangedText{1,350} & \ChangedText{530} & \ChangedText{92} & \ChangedText{16} & 712 & 583 & 107 & 22 & 84.97 \\ 
\hline
\textit{image-downloader} & 42 & \ChangedText{124} & \ChangedText{40} & \ChangedText{5} & \ChangedText{2} & 77 & 48 & 29 & 0 & 62.34 \\ 
\hline
\textit{node-dirty} & 154 & \ChangedText{450} & \ChangedText{161} & \ChangedText{33} & \ChangedText{11} & 245 & 142 & 92 & 11 & 62.45 \\
\end{tabular}
  }
  \\[2mm]
  \caption{Results from LLMorpheus experiment \ChangedText{(run \#358)}.
    Model: \textit{codellama-13b-instruct}, 
    temperature: 0.0, 
    maxTokens: 250, 
    template: \textit{template-full.hb}, 
    systemPrompt: \textit{SystemPrompt-MutationTestingExpert.txt}. 
  }
  \label{table:Mutants:run358:codellama-13b-instruct:template-full.hb:0.0}
\end{table*}

\begin{table*}[hbt!]
\centering
{\scriptsize
\begin{tabular}{l||r|r|r|r|r|r|r|r|r|r}
  {\bf application} & {\bf \#prompts} & \multicolumn{4}{|c|}{\bf \ChangedText{mutant candidates}} & {\bf \#mutants} & {\bf \#killed} & {\bf \#survived} & {\bf \#timeout} & {\bf mut.} \\
  & &  {\bf \ChangedText{total}} & {\bf \ChangedText{invalid}} & {\bf \ChangedText{identical}} & {\bf \ChangedText{duplicate}}  &  & & & & {\bf score} \\
  \hline
  \hline
\textit{Complex.js} & 490 & \ChangedText{1,439} & \ChangedText{340} & \ChangedText{116} & \ChangedText{28} & 955 & 553 & 401 & 1 & 58.01 \\ 
\hline
\textit{countries-and-timezones} & 106 & \ChangedText{306} & \ChangedText{83} & \ChangedText{15} & \ChangedText{1} & 207 & 177 & 30 & 0 & 85.51 \\ 
\hline
\textit{crawler-url-parser} & 176 & \ChangedText{506} & \ChangedText{186} & \ChangedText{51} & \ChangedText{12} & 247 & 124 & 118 & 5 & 52.23 \\ 
\hline
\textit{delta} & 462 & \ChangedText{1,349} & \ChangedText{530} & \ChangedText{91} & \ChangedText{16} & 712 & 583 & 107 & 22 & 84.97 \\ 
\hline
\textit{image-downloader} & 42 & \ChangedText{124} & \ChangedText{40} & \ChangedText{5} & \ChangedText{2} & 77 & 48 & 29 & 0 & 62.34 \\ 
\hline
\textit{node-dirty} & 154 & \ChangedText{450} & \ChangedText{161} & \ChangedText{33} & \ChangedText{11} & 245 & 142 & 92 & 11 & 62.45 \\
\end{tabular}
  }
  \\[2mm]
  \caption{Results from LLMorpheus experiment \ChangedText{(run \#359)}.
    Model: \textit{codellama-13b-instruct}, 
    temperature: 0.0, 
    maxTokens: 250, 
    template: \textit{template-full.hb}, 
    systemPrompt: \textit{SystemPrompt-MutationTestingExpert.txt}. 
  }
  \label{table:Mutants:run359:codellama-13b-instruct:template-full.hb:0.0}
\end{table*}

\begin{table*}[hbt!]
\centering
{\scriptsize
\begin{tabular}{l||r|r|r|r|r}
\multicolumn{1}{c|}{\bf project} & \multicolumn{2}{|c|}{\bf time (sec)} & \multicolumn{3}{|c}{\bf \#tokens} \\
               & {\it LLMorpheus} & {\it StrykerJS} & {\bf prompt} & {\bf compl.} & {\bf total} \\
\hline
  Complex.js & 3,041.69 & 525.25 & 967,508 & 104,246 & 1,071,754 \\ 
countries-and-timezones & 1,070.79 & 310.80 & 105,828 & 23,971 & 129,799 \\ 
crawler-url-parser & 1,637.98 & 803.17 & 386,223 & 39,906 & 426,129 \\ 
delta & 2,993.70 & 3,529.05 & 890,252 & 103,085 & 993,337 \\ 
image-downloader & 430.56 & 460.64 & 24,655 & 9,339 & 33,994 \\ 
node-dirty & 1,526.96 & 211.68 & 246,248 & 34,892 & 281,140 \\ 
node-geo-point & 1,411.04 & 1,000.74 & 316,333 & 30,715 & 347,048 \\ 
node-jsonfile & 690.65 & 425.39 & 57,516 & 15,398 & 72,914 \\ 
plural & 1,521.00 & 111.98 & 265,602 & 34,926 & 300,528 \\ 
pull-stream & 2,489.93 & 1,188.59 & 208,130 & 77,308 & 285,438 \\ 
q & 5,187.67 & 11,850.22 & 2,127,655 & 231,175 & 2,358,830 \\ 
spacl-core & 1,350.97 & 616.43 & 162,705 & 30,694 & 193,399 \\ 
zip-a-folder & 500.51 & 496.32 & 82,457 & 11,494 & 93,951 \\ 
\hline
  \textit{Total} & 23,853.45 & 21,530.25 & 5,841,112 & 747,149 & 6,588,261 \\
  \end{tabular}
  }
  \\[2mm]
  \caption{Results from LLMorpheus experiment \ChangedText{(run \#354)}.
    Model: \textit{codellama-13b-instruct}, 
    temperature: 0.0, 
    maxTokens: 250, 
    template: \textit{template-full.hb}, 
    systemPrompt: \textit{SystemPrompt-MutationTestingExpert.txt}
  }
  \label{table:Cost:run354:codellama-13b-instruct:template-full.hb:0.0}
\end{table*}

\begin{table*}[hbt!]
\centering
{\scriptsize
\begin{tabular}{l||r|r|r|r|r}
\multicolumn{1}{c|}{\bf project} & \multicolumn{2}{|c|}{\bf time (sec)} & \multicolumn{3}{|c}{\bf \#tokens} \\
               & {\it LLMorpheus} & {\it StrykerJS} & {\bf prompt} & {\bf compl.} & {\bf total} \\
\hline
  Complex.js & 3,065.29 & 504.22 & 967,508 & 104,246 & 1,071,754 \\ 
countries-and-timezones & 1,070.80 & 296.35 & 105,828 & 23,971 & 129,799 \\ 
crawler-url-parser & 1,646.62 & 799.72 & 386,223 & 39,938 & 426,161 \\ 
delta & 2,972.99 & 3,511.68 & 890,252 & 103,085 & 993,337 \\ 
image-downloader & 430.51 & 461.29 & 24,655 & 9,339 & 33,994 \\ 
node-dirty & 1,527.57 & 208.98 & 246,248 & 34,892 & 281,140 \\ 
node-geo-point & 1,411.07 & 1,010.61 & 316,333 & 30,715 & 347,048 \\ 
node-jsonfile & 690.63 & 424.92 & 57,516 & 15,398 & 72,914 \\ 
plural & 1,696.97 & 108.86 & 255,187 & 33,552 & 288,739 \\ 
pull-stream & 2,488.59 & 1,182.03 & 208,130 & 77,307 & 285,437 \\ 
q & 5,141.59 & 11,793.14 & 2,127,655 & 231,269 & 2,358,924 \\ 
spacl-core & 1,350.98 & 624.29 & 162,705 & 30,694 & 193,399 \\ 
zip-a-folder & 500.55 & 488.31 & 82,457 & 11,494 & 93,951 \\ 
\hline
  \textit{Total} & 23,994.15 & 21,414.40 & 5,830,697 & 745,900 & 6,576,597 \\
  \end{tabular}
  }
  \\[2mm]
  \caption{Results from LLMorpheus experiment \ChangedText{(run \#355)}.
    Model: \textit{codellama-13b-instruct}, 
    temperature: 0.0, 
    maxTokens: 250, 
    template: \textit{template-full.hb}, 
    systemPrompt: \textit{SystemPrompt-MutationTestingExpert.txt}
  }
  \label{table:Cost:run355:codellama-13b-instruct:template-full.hb:0.0}
\end{table*}

\begin{table*}[hbt!]
\centering
{\scriptsize
\begin{tabular}{l||r|r|r|r|r}
\multicolumn{1}{c|}{\bf project} & \multicolumn{2}{|c|}{\bf time (sec)} & \multicolumn{3}{|c}{\bf \#tokens} \\
               & {\it LLMorpheus} & {\it StrykerJS} & {\bf prompt} & {\bf compl.} & {\bf total} \\
\hline
  Complex.js & 3,041.58 & 506.22 & 967,508 & 104,246 & 1,071,754 \\ 
countries-and-timezones & 1,070.84 & 308.67 & 105,828 & 23,971 & 129,799 \\ 
crawler-url-parser & 1,639.49 & 858.45 & 386,223 & 39,915 & 426,138 \\ 
delta & 2,978.12 & 3,447.64 & 890,252 & 103,085 & 993,337 \\ 
image-downloader & 430.52 & 459.07 & 24,655 & 9,339 & 33,994 \\ 
node-dirty & 1,527.40 & 208.98 & 246,248 & 34,892 & 281,140 \\ 
node-geo-point & 1,411.02 & 1,011.58 & 316,333 & 30,715 & 347,048 \\ 
node-jsonfile & 690.68 & 420.13 & 57,516 & 15,398 & 72,914 \\ 
plural & 1,521.00 & 112.81 & 265,602 & 34,926 & 300,528 \\ 
pull-stream & 2,481.30 & 1,179.99 & 208,130 & 77,302 & 285,432 \\ 
q & 5,249.72 & 11,806.54 & 2,127,655 & 231,355 & 2,359,010 \\ 
spacl-core & 1,351.04 & 617.86 & 162,705 & 30,694 & 193,399 \\ 
zip-a-folder & 500.58 & 495.97 & 82,457 & 11,494 & 93,951 \\ 
\hline
  \textit{Total} & 23,893.29 & 21,433.91 & 5,841,112 & 747,332 & 6,588,444 \\
  \end{tabular}
  }
  \\[2mm]
  \caption{Results from LLMorpheus experiment \ChangedText{(run \#356)}.
    Model: \textit{codellama-13b-instruct}, 
    temperature: 0.0, 
    maxTokens: 250, 
    template: \textit{template-full.hb}, 
    systemPrompt: \textit{SystemPrompt-MutationTestingExpert.txt}
  }
  \label{table:Cost:run356:codellama-13b-instruct:template-full.hb:0.0}
\end{table*}

\begin{table*}[hbt!]
\centering
{\scriptsize
\begin{tabular}{l||r|r|r|r|r}
\multicolumn{1}{c|}{\bf project} & \multicolumn{2}{|c|}{\bf time (sec)} & \multicolumn{3}{|c}{\bf \#tokens} \\
               & {\it LLMorpheus} & {\it StrykerJS} & {\bf prompt} & {\bf compl.} & {\bf total} \\
\hline
  Complex.js & 3,045.49 & 511.92 & 967,508 & 104,246 & 1,071,754 \\ 
countries-and-timezones & 1,070.84 & 308.11 & 105,828 & 23,971 & 129,799 \\ 
crawler-url-parser & 1,639.50 & 877.27 & 386,223 & 39,906 & 426,129 \\ 
delta & 2,975.67 & 3,533.28 & 890,252 & 103,025 & 993,277 \\ 
image-downloader & 430.50 & 458.40 & 24,655 & 9,339 & 33,994 \\ 
node-dirty & 1,527.47 & 209.67 & 246,248 & 34,892 & 281,140 \\ 
node-geo-point & 1,411.06 & 994.07 & 316,333 & 30,715 & 347,048 \\ 
node-jsonfile & 690.64 & 421.39 & 57,516 & 15,398 & 72,914 \\ 
plural & 1,521.06 & 112.22 & 265,602 & 34,926 & 300,528 \\ 
pull-stream & 2,493.00 & 1,182.21 & 208,130 & 77,307 & 285,437 \\ 
q & 5,177.80 & 11,862.47 & 2,127,655 & 231,214 & 2,358,869 \\ 
spacl-core & 1,350.98 & 606.49 & 162,705 & 30,694 & 193,399 \\ 
zip-a-folder & 500.52 & 486.21 & 82,457 & 11,494 & 93,951 \\ 
\hline
  \textit{Total} & 23,834.54 & 21,563.71 & 5,841,112 & 747,127 & 6,588,239 \\
  \end{tabular}
  }
  \\[2mm]
  \caption{Results from LLMorpheus experiment \ChangedText{(run \#358)}.
    Model: \textit{codellama-13b-instruct}, 
    temperature: 0.0, 
    maxTokens: 250, 
    template: \textit{template-full.hb}, 
    systemPrompt: \textit{SystemPrompt-MutationTestingExpert.txt}
  }
  \label{table:Cost:run358:codellama-13b-instruct:template-full.hb:0.0}
\end{table*}

\begin{table*}[hbt!]
\centering
{\scriptsize
\begin{tabular}{l||r|r|r|r|r}
\multicolumn{1}{c|}{\bf project} & \multicolumn{2}{|c|}{\bf time (sec)} & \multicolumn{3}{|c}{\bf \#tokens} \\
               & {\it LLMorpheus} & {\it StrykerJS} & {\bf prompt} & {\bf compl.} & {\bf total} \\
\hline
  Complex.js & 3,065.97 & 514.43 & 967,508 & 104,246 & 1,071,754 \\ 
countries-and-timezones & 1,070.82 & 328.45 & 105,828 & 23,951 & 129,779 \\ 
crawler-url-parser & 1,638.27 & 905.63 & 386,223 & 39,906 & 426,129 \\ 
delta & 2,954.39 & 3,471.08 & 890,252 & 103,085 & 993,337 \\ 
image-downloader & 430.50 & 458.13 & 24,655 & 9,339 & 33,994 \\ 
node-dirty & 1,526.96 & 207.80 & 246,248 & 34,892 & 281,140 \\ 
node-geo-point & 1,411.06 & 1,003.76 & 316,333 & 30,715 & 347,048 \\ 
node-jsonfile & 690.67 & 421.74 & 57,516 & 15,398 & 72,914 \\ 
plural & 1,521.03 & 111.45 & 265,602 & 34,926 & 300,528 \\ 
pull-stream & 2,483.02 & 1,186.04 & 208,130 & 77,307 & 285,437 \\ 
q & 5,276.52 & 11,824.34 & 2,127,655 & 231,254 & 2,358,909 \\ 
spacl-core & 1,350.99 & 617.51 & 162,705 & 30,694 & 193,399 \\ 
zip-a-folder & 500.53 & 490.95 & 82,457 & 11,494 & 93,951 \\ 
\hline
  \textit{Total} & 23,920.75 & 21,541.32 & 5,841,112 & 747,207 & 6,588,319 \\
  \end{tabular}
  }
  \\[2mm]
  \caption{Results from LLMorpheus experiment \ChangedText{(run \#359)}.
    Model: \textit{codellama-13b-instruct}, 
    temperature: 0.0, 
    maxTokens: 250, 
    template: \textit{template-full.hb}, 
    systemPrompt: \textit{SystemPrompt-MutationTestingExpert.txt}
  }
  \label{table:Cost:run359:codellama-13b-instruct:template-full.hb:0.0}
\end{table*}

\begin{table}[hbt!]
\centering
{\footnotesize
\begin{tabular}{l|r|r|r|r}

{\bf application}  & {\bf \#min} &  {\bf \#max} &  {\bf \#distinct} & {\bf \#common}\\
\hline
Complex.js & 955 & 955 & 955 & 955 (100.00\%) \\ 
countries-and-timezones & 207 & 207 & 207 & 207 (100.00\%) \\ 
crawler-url-parser & 257 & 257 & 257 & 257 (100.00\%) \\ 
delta & 712 & 712 & 713 & 711 (99.72\%) \\ 
image-downloader & 77 & 77 & 77 & 77 (100.00\%) \\ 
node-dirty & 245 & 245 & 245 & 245 (100.00\%) \\ 
node-geo-point & 305 & 305 & 305 & 305 (100.00\%) \\ 
node-jsonfile & 137 & 138 & 138 & 137 (99.28\%) \\ 
plural & 200 & 208 & 208 & 200 (96.15\%) \\ 
pull-stream & 663 & 664 & 666 & 661 (99.25\%) \\ 
q & 1,713 & 1,715 & 1,725 & 1,702 (98.67\%) \\ 
spacl-core & 195 & 195 & 197 & 193 (97.97\%) \\ 
zip-a-folder & 87 & 87 & 87 & 87 (100.00\%) \\ 
\end{tabular}
}
\caption{
  Variability of the mutants generated in 5 runs of \ToolName using the \textit{codellama-13b-instruct} LLM
       at temperature 0.0 \ChangedText{(run \#354,\#355,\#356,\#358,\#359)}. The columns of the table show, from left to right:
    (i) the minimum number of mutants observed in any of the runs,
    (ii) the maximum number of mutants observed in any of the runs,
    (iii) the total number of distinct mutants observed in all runs, and
    (iv) the number (percentage) of mutants are observed in all runs.
}
\label{table:Variability_codellama-13b-instruct_0.0}
\end{table}

\FloatBarrier
 
\subsection{Results for \Mixtral}
   \label{app:Mixtral}

Tables~\ref{table:Mutants:run360:mixtral-8x7b-instruct:template-full.hb:0.0}--\ref{table:Cost:run364:mixtral-8x7b-instruct:template-full.hb:0.0}
show the results for 5 experiments with the \textit{mixtral-8x7b-instruct} model
at temperature 0.0 using the default prompt and system prompt shown in Figure~\ref{fig:PromptTemplate}.
Table~\ref{table:Variability_mixtral-8x7b-instruct_0.0} shows the variability of the mutants observed in 5 runs using \Mixtral at temperature 0.0.

\begin{table*}[hbt!]
\centering
{\scriptsize
\begin{tabular}{l||r|r|r|r|r|r|r|r|r|r}
  {\bf application} & {\bf \#prompts} & \multicolumn{4}{|c|}{\bf \ChangedText{mutant candidates}} & {\bf \#mutants} & {\bf \#killed} & {\bf \#survived} & {\bf \#timeout} & {\bf mut.} \\
  & &  {\bf \ChangedText{total}} & {\bf \ChangedText{invalid}} & {\bf \ChangedText{identical}} & {\bf \ChangedText{duplicate}}  &  & & & & {\bf score} \\
  \hline
  \hline
\textit{Complex.js} & 490 & \ChangedText{1,287} & \ChangedText{310} & \ChangedText{0} & \ChangedText{15} & 962 & 589 & 373 & 0 & 61.23 \\ 
\hline
\textit{countries-and-timezones} & 106 & \ChangedText{277} & \ChangedText{65} & \ChangedText{2} & \ChangedText{5} & 205 & 166 & 39 & 0 & 80.98 \\ 
\hline
\textit{crawler-url-parser} & 176 & \ChangedText{414} & \ChangedText{165} & \ChangedText{0} & \ChangedText{3} & 234 & 130 & 104 & 0 & 55.56 \\ 
\hline
\textit{delta} & 462 & \ChangedText{1,156} & \ChangedText{452} & \ChangedText{0} & \ChangedText{24} & 680 & 516 & 128 & 36 & 81.18 \\ 
\hline
\textit{image-downloader} & 42 & \ChangedText{108} & \ChangedText{38} & \ChangedText{0} & \ChangedText{1} & 69 & 46 & 23 & 0 & 66.67 \\ 
\hline
\textit{node-dirty} & 154 & \ChangedText{310} & \ChangedText{109} & \ChangedText{0} & \ChangedText{10} & 191 & 111 & 72 & 8 & 62.30 \\
\end{tabular}
  }
  \\[2mm]
  \caption{Results from LLMorpheus experiment \ChangedText{(run \#360)}.
    Model: \textit{mixtral-8x7b-instruct}, 
    temperature: 0.0, 
    maxTokens: 250, 
    template: \textit{template-full.hb}, 
    systemPrompt: \textit{SystemPrompt-MutationTestingExpert.txt}. 
  }
  \label{table:Mutants:run360:mixtral-8x7b-instruct:template-full.hb:0.0}
\end{table*}

\begin{table*}[hbt!]
\centering
{\scriptsize
\begin{tabular}{l||r|r|r|r|r|r|r|r|r|r}
  {\bf application} & {\bf \#prompts} & \multicolumn{4}{|c|}{\bf \ChangedText{mutant candidates}} & {\bf \#mutants} & {\bf \#killed} & {\bf \#survived} & {\bf \#timeout} & {\bf mut.} \\
  & &  {\bf \ChangedText{total}} & {\bf \ChangedText{invalid}} & {\bf \ChangedText{identical}} & {\bf \ChangedText{duplicate}}  &  & & & & {\bf score} \\
  \hline
  \hline
\textit{Complex.js} & 490 & \ChangedText{1,313} & \ChangedText{303} & \ChangedText{0} & \ChangedText{21} & 989 & 608 & 381 & 0 & 61.48 \\ 
\hline
\textit{countries-and-timezones} & 106 & \ChangedText{270} & \ChangedText{67} & \ChangedText{2} & \ChangedText{5} & 196 & 157 & 39 & 0 & 80.10 \\ 
\hline
\textit{crawler-url-parser} & 176 & \ChangedText{403} & \ChangedText{177} & \ChangedText{0} & \ChangedText{1} & 211 & 114 & 97 & 0 & 54.03 \\ 
\hline
\textit{delta} & 462 & \ChangedText{1,136} & \ChangedText{446} & \ChangedText{0} & \ChangedText{24} & 666 & 509 & 123 & 34 & 81.53 \\ 
\hline
\textit{image-downloader} & 42 & \ChangedText{103} & \ChangedText{37} & \ChangedText{0} & \ChangedText{1} & 65 & 42 & 23 & 0 & 64.62 \\ 
\hline
\textit{node-dirty} & 154 & \ChangedText{342} & \ChangedText{124} & \ChangedText{0} & \ChangedText{9} & 209 & 125 & 76 & 8 & 63.64 \\
\end{tabular}
  }
  \\[2mm]
  \caption{Results from LLMorpheus experiment \ChangedText{(run \#361)}.
    Model: \textit{mixtral-8x7b-instruct}, 
    temperature: 0.0, 
    maxTokens: 250, 
    template: \textit{template-full.hb}, 
    systemPrompt: \textit{SystemPrompt-MutationTestingExpert.txt}. 
  }
  \label{table:Mutants:run361:mixtral-8x7b-instruct:template-full.hb:0.0}
\end{table*}

\begin{table*}[hbt!]
\centering
{\scriptsize
\begin{tabular}{l||r|r|r|r|r|r|r|r|r|r}
  {\bf application} & {\bf \#prompts} & \multicolumn{4}{|c|}{\bf \ChangedText{mutant candidates}} & {\bf \#mutants} & {\bf \#killed} & {\bf \#survived} & {\bf \#timeout} & {\bf mut.} \\
  & &  {\bf \ChangedText{total}} & {\bf \ChangedText{invalid}} & {\bf \ChangedText{identical}} & {\bf \ChangedText{duplicate}}  &  & & & & {\bf score} \\
  \hline
  \hline
\textit{Complex.js} & 490 & \ChangedText{1,302} & \ChangedText{306} & \ChangedText{0} & \ChangedText{16} & 980 & 594 & 386 & 0 & 60.61 \\ 
\hline
\textit{countries-and-timezones} & 106 & \ChangedText{272} & \ChangedText{70} & \ChangedText{2} & \ChangedText{3} & 197 & 154 & 43 & 0 & 78.17 \\ 
\hline
\textit{crawler-url-parser} & 176 & \ChangedText{404} & \ChangedText{168} & \ChangedText{0} & \ChangedText{1} & 223 & 121 & 102 & 0 & 54.26 \\ 
\hline
\textit{delta} & 462 & \ChangedText{1,136} & \ChangedText{455} & \ChangedText{0} & \ChangedText{21} & 660 & 497 & 129 & 34 & 80.45 \\ 
\hline
\textit{image-downloader} & 42 & \ChangedText{103} & \ChangedText{35} & \ChangedText{0} & \ChangedText{1} & 67 & 41 & 26 & 0 & 61.19 \\ 
\hline
\textit{node-dirty} & 154 & \ChangedText{349} & \ChangedText{126} & \ChangedText{0} & \ChangedText{10} & 213 & 123 & 81 & 9 & 61.97 \\
\end{tabular}
  }
  \\[2mm]
  \caption{Results from LLMorpheus experiment \ChangedText{(run \#362)}.
    Model: \textit{mixtral-8x7b-instruct}, 
    temperature: 0.0, 
    maxTokens: 250, 
    template: \textit{template-full.hb}, 
    systemPrompt: \textit{SystemPrompt-MutationTestingExpert.txt}. 
  }
  \label{table:Mutants:run362:mixtral-8x7b-instruct:template-full.hb:0.0}
\end{table*}

\begin{table*}[hbt!]
\centering
{\scriptsize
\begin{tabular}{l||r|r|r|r|r|r|r|r|r|r}
  {\bf application} & {\bf \#prompts} & \multicolumn{4}{|c|}{\bf \ChangedText{mutant candidates}} & {\bf \#mutants} & {\bf \#killed} & {\bf \#survived} & {\bf \#timeout} & {\bf mut.} \\
  & &  {\bf \ChangedText{total}} & {\bf \ChangedText{invalid}} & {\bf \ChangedText{identical}} & {\bf \ChangedText{duplicate}}  &  & & & & {\bf score} \\
  \hline
  \hline
\textit{Complex.js} & 490 & \ChangedText{1,299} & \ChangedText{305} & \ChangedText{0} & \ChangedText{25} & 969 & 587 & 382 & 0 & 60.58 \\ 
\hline
\textit{countries-and-timezones} & 106 & \ChangedText{270} & \ChangedText{61} & \ChangedText{2} & \ChangedText{4} & 203 & 160 & 43 & 0 & 78.82 \\ 
\hline
\textit{crawler-url-parser} & 176 & \ChangedText{401} & \ChangedText{167} & \ChangedText{0} & \ChangedText{1} & 220 & 118 & 102 & 0 & 53.64 \\ 
\hline
\textit{delta} & 462 & \ChangedText{1,122} & \ChangedText{440} & \ChangedText{0} & \ChangedText{22} & 660 & 500 & 126 & 34 & 80.91 \\ 
\hline
\textit{image-downloader} & 42 & \ChangedText{103} & \ChangedText{39} & \ChangedText{0} & \ChangedText{0} & 64 & 41 & 23 & 0 & 64.06 \\ 
\hline
\textit{node-dirty} & 154 & \ChangedText{338} & \ChangedText{125} & \ChangedText{0} & \ChangedText{10} & 203 & 120 & 77 & 6 & 62.07 \\
\end{tabular}
  }
  \\[2mm]
  \caption{Results from LLMorpheus experiment \ChangedText{(run \#363)}.
    Model: \textit{mixtral-8x7b-instruct}, 
    temperature: 0.0, 
    maxTokens: 250, 
    template: \textit{template-full.hb}, 
    systemPrompt: \textit{SystemPrompt-MutationTestingExpert.txt}. 
  }
  \label{table:Mutants:run363:mixtral-8x7b-instruct:template-full.hb:0.0}
\end{table*}

\begin{table*}[hbt!]
\centering
{\scriptsize
\begin{tabular}{l||r|r|r|r|r|r|r|r|r|r}
  {\bf application} & {\bf \#prompts} & \multicolumn{4}{|c|}{\bf \ChangedText{mutant candidates}} & {\bf \#mutants} & {\bf \#killed} & {\bf \#survived} & {\bf \#timeout} & {\bf mut.} \\
  & &  {\bf \ChangedText{total}} & {\bf \ChangedText{invalid}} & {\bf \ChangedText{identical}} & {\bf \ChangedText{duplicate}}  &  & & & & {\bf score} \\
  \hline
  \hline
\textit{Complex.js} & 490 & \ChangedText{1,313} & \ChangedText{328} & \ChangedText{0} & \ChangedText{16} & 969 & 580 & 389 & 0 & 59.86 \\ 
\hline
\textit{countries-and-timezones} & 106 & \ChangedText{269} & \ChangedText{64} & \ChangedText{2} & \ChangedText{4} & 199 & 156 & 43 & 0 & 78.39 \\ 
\hline
\textit{crawler-url-parser} & 176 & \ChangedText{404} & \ChangedText{178} & \ChangedText{0} & \ChangedText{1} & 212 & 117 & 95 & 0 & 55.19 \\ 
\hline
\textit{delta} & 462 & \ChangedText{1,148} & \ChangedText{458} & \ChangedText{0} & \ChangedText{16} & 674 & 521 & 116 & 37 & 82.79 \\ 
\hline
\textit{image-downloader} & 42 & \ChangedText{107} & \ChangedText{40} & \ChangedText{0} & \ChangedText{1} & 66 & 42 & 24 & 0 & 63.64 \\ 
\hline
\textit{node-dirty} & 154 & \ChangedText{339} & \ChangedText{121} & \ChangedText{0} & \ChangedText{10} & 208 & 124 & 73 & 11 & 64.90 \\
\end{tabular}
  }
  \\[2mm]
  \caption{Results from LLMorpheus experiment \ChangedText{(run \#364)}.
    Model: \textit{mixtral-8x7b-instruct}, 
    temperature: 0.0, 
    maxTokens: 250, 
    template: \textit{template-full.hb}, 
    systemPrompt: \textit{SystemPrompt-MutationTestingExpert.txt}. 
  }
  \label{table:Mutants:run364:mixtral-8x7b-instruct:template-full.hb:0.0}
\end{table*}

\begin{table*}[hbt!]
\centering
{\scriptsize
\begin{tabular}{l||r|r|r|r|r}
\multicolumn{1}{c|}{\bf project} & \multicolumn{2}{|c|}{\bf time (sec)} & \multicolumn{3}{|c}{\bf \#tokens} \\
               & {\it LLMorpheus} & {\it StrykerJS} & {\bf prompt} & {\bf compl.} & {\bf total} \\
\hline
  Complex.js & 3,756.38 & 507.41 & 960,545 & 96,072 & 1,056,617 \\ 
countries-and-timezones & 1,080.09 & 296.25 & 104,291 & 22,693 & 126,984 \\ 
crawler-url-parser & 1,697.04 & 762.88 & 384,404 & 33,495 & 417,899 \\ 
delta & 3,636.95 & 3,476.06 & 882,477 & 90,094 & 972,571 \\ 
image-downloader & 430.46 & 387.36 & 24,140 & 8,238 & 32,378 \\ 
node-dirty & 1,665.69 & 159.16 & 234,503 & 24,705 & 259,208 \\ 
node-geo-point & 1,497.29 & 833.16 & 315,891 & 27,864 & 343,755 \\ 
node-jsonfile & 691.90 & 425.95 & 56,273 & 12,371 & 68,644 \\ 
plural & 1,583.06 & 115.03 & 259,916 & 25,067 & 284,983 \\ 
pull-stream & 2,802.21 & 1,201.97 & 204,431 & 70,423 & 274,854 \\ 
q & 7,003.28 & 11,371.39 & 2,103,232 & 192,284 & 2,295,516 \\ 
spacl-core & 1,369.93 & 534.67 & 162,695 & 26,484 & 189,179 \\ 
zip-a-folder & 506.29 & 470.28 & 81,279 & 9,124 & 90,403 \\ 
\hline
  \textit{Total} & 27,720.59 & 20,541.58 & 5,774,077 & 638,914 & 6,412,991 \\
  \end{tabular}
  }
  \\[2mm]
  \caption{Results from LLMorpheus experiment \ChangedText{(run \#360)}.
    Model: \textit{mixtral-8x7b-instruct}, 
    temperature: 0.0, 
    maxTokens: 250, 
    template: \textit{template-full.hb}, 
    systemPrompt: \textit{SystemPrompt-MutationTestingExpert.txt}
  }
  \label{table:Cost:run360:mixtral-8x7b-instruct:template-full.hb:0.0}
\end{table*}

\begin{table*}[hbt!]
\centering
{\scriptsize
\begin{tabular}{l||r|r|r|r|r}
\multicolumn{1}{c|}{\bf project} & \multicolumn{2}{|c|}{\bf time (sec)} & \multicolumn{3}{|c}{\bf \#tokens} \\
               & {\it LLMorpheus} & {\it StrykerJS} & {\bf prompt} & {\bf compl.} & {\bf total} \\
\hline
  Complex.js & 3,372.45 & 520.01 & 960,545 & 98,260 & 1,058,805 \\ 
countries-and-timezones & 1,071.57 & 285.14 & 104,291 & 22,228 & 126,519 \\ 
crawler-url-parser & 1,669.78 & 698.95 & 384,404 & 32,771 & 417,175 \\ 
delta & 3,199.74 & 3,398.17 & 882,477 & 89,050 & 971,527 \\ 
image-downloader & 432.08 & 378.08 & 24,140 & 8,005 & 32,145 \\ 
node-dirty & 1,554.61 & 185.18 & 242,671 & 26,837 & 269,508 \\ 
node-geo-point & 1,416.99 & 850.15 & 318,251 & 28,316 & 346,567 \\ 
node-jsonfile & 740.68 & 431.76 & 56,273 & 12,101 & 68,374 \\ 
plural & 1,546.99 & 120.70 & 261,626 & 25,293 & 286,919 \\ 
pull-stream & 2,660.01 & 1,259.31 & 204,431 & 70,142 & 274,573 \\ 
q & 6,149.24 & 11,173.93 & 2,103,232 & 188,223 & 2,291,455 \\ 
spacl-core & 1,416.02 & 534.70 & 162,695 & 26,751 & 189,446 \\ 
zip-a-folder & 500.53 & 489.08 & 81,279 & 9,372 & 90,651 \\ 
\hline
  \textit{Total} & 25,730.68 & 20,325.15 & 5,786,315 & 637,349 & 6,423,664 \\
  \end{tabular}
  }
  \\[2mm]
  \caption{Results from LLMorpheus experiment \ChangedText{(run \#361)}.
    Model: \textit{mixtral-8x7b-instruct}, 
    temperature: 0.0, 
    maxTokens: 250, 
    template: \textit{template-full.hb}, 
    systemPrompt: \textit{SystemPrompt-MutationTestingExpert.txt}
  }
  \label{table:Cost:run361:mixtral-8x7b-instruct:template-full.hb:0.0}
\end{table*}

\begin{table*}[hbt!]
\centering
{\scriptsize
\begin{tabular}{l||r|r|r|r|r}
\multicolumn{1}{c|}{\bf project} & \multicolumn{2}{|c|}{\bf time (sec)} & \multicolumn{3}{|c}{\bf \#tokens} \\
               & {\it LLMorpheus} & {\it StrykerJS} & {\bf prompt} & {\bf compl.} & {\bf total} \\
\hline
  Complex.js & 3,359.29 & 519.50 & 960,545 & 96,727 & 1,057,272 \\ 
countries-and-timezones & 1,074.42 & 277.16 & 104,291 & 22,353 & 126,644 \\ 
crawler-url-parser & 1,661.26 & 731.82 & 384,404 & 32,772 & 417,176 \\ 
delta & 3,170.76 & 3,384.81 & 882,477 & 89,334 & 971,811 \\ 
image-downloader & 430.53 & 386.39 & 24,140 & 7,934 & 32,074 \\ 
node-dirty & 1,530.96 & 182.12 & 244,297 & 27,524 & 271,821 \\ 
node-geo-point & 1,413.38 & 829.76 & 318,251 & 27,995 & 346,246 \\ 
node-jsonfile & 690.64 & 444.51 & 56,273 & 11,970 & 68,243 \\ 
plural & 1,524.56 & 117.41 & 261,626 & 25,277 & 286,903 \\ 
pull-stream & 2,644.38 & 1,205.37 & 204,431 & 69,081 & 273,512 \\ 
q & 6,079.84 & 11,465.34 & 2,103,232 & 192,672 & 2,295,904 \\ 
spacl-core & 1,354.60 & 540.52 & 162,695 & 26,151 & 188,846 \\ 
zip-a-folder & 500.60 & 464.06 & 81,279 & 9,340 & 90,619 \\ 
\hline
  \textit{Total} & 25,435.21 & 20,548.77 & 5,787,941 & 639,130 & 6,427,071 \\
  \end{tabular}
  }
  \\[2mm]
  \caption{Results from LLMorpheus experiment \ChangedText{(run \#362)}.
    Model: \textit{mixtral-8x7b-instruct}, 
    temperature: 0.0, 
    maxTokens: 250, 
    template: \textit{template-full.hb}, 
    systemPrompt: \textit{SystemPrompt-MutationTestingExpert.txt}
  }
  \label{table:Cost:run362:mixtral-8x7b-instruct:template-full.hb:0.0}
\end{table*}

\begin{table*}[hbt!]
\centering
{\scriptsize
\begin{tabular}{l||r|r|r|r|r}
\multicolumn{1}{c|}{\bf project} & \multicolumn{2}{|c|}{\bf time (sec)} & \multicolumn{3}{|c}{\bf \#tokens} \\
               & {\it LLMorpheus} & {\it StrykerJS} & {\bf prompt} & {\bf compl.} & {\bf total} \\
\hline
  Complex.js & 3,511.86 & 522.92 & 960,545 & 96,846 & 1,057,391 \\ 
countries-and-timezones & 1,073.78 & 299.00 & 104,291 & 22,090 & 126,381 \\ 
crawler-url-parser & 1,664.52 & 748.35 & 384,404 & 32,721 & 417,125 \\ 
delta & 3,343.71 & 3,332.02 & 882,477 & 88,421 & 970,898 \\ 
image-downloader & 440.52 & 362.20 & 24,140 & 7,972 & 32,112 \\ 
node-dirty & 1,532.40 & 164.86 & 244,297 & 26,801 & 271,098 \\ 
node-geo-point & 1,436.26 & 842.23 & 318,251 & 28,074 & 346,325 \\ 
node-jsonfile & 692.85 & 440.06 & 56,273 & 12,731 & 69,004 \\ 
plural & 1,525.78 & 117.87 & 261,626 & 25,198 & 286,824 \\ 
pull-stream & 2,725.42 & 1,227.21 & 204,431 & 70,751 & 275,182 \\ 
q & 6,622.34 & 11,507.75 & 2,103,232 & 194,705 & 2,297,937 \\ 
spacl-core & 1,359.32 & 532.62 & 162,695 & 26,100 & 188,795 \\ 
zip-a-folder & 500.56 & 463.12 & 81,279 & 9,118 & 90,397 \\ 
\hline
  \textit{Total} & 26,429.31 & 20,560.22 & 5,787,941 & 641,528 & 6,429,469 \\
  \end{tabular}
  }
  \\[2mm]
  \caption{Results from LLMorpheus experiment \ChangedText{(run \#363)}.
    Model: \textit{mixtral-8x7b-instruct}, 
    temperature: 0.0, 
    maxTokens: 250, 
    template: \textit{template-full.hb}, 
    systemPrompt: \textit{SystemPrompt-MutationTestingExpert.txt}
  }
  \label{table:Cost:run363:mixtral-8x7b-instruct:template-full.hb:0.0}
\end{table*}

\begin{table*}[hbt!]
\centering
{\scriptsize
\begin{tabular}{l||r|r|r|r|r}
\multicolumn{1}{c|}{\bf project} & \multicolumn{2}{|c|}{\bf time (sec)} & \multicolumn{3}{|c}{\bf \#tokens} \\
               & {\it LLMorpheus} & {\it StrykerJS} & {\bf prompt} & {\bf compl.} & {\bf total} \\
\hline
  Complex.js & 3,649.24 & 528.79 & 960,545 & 98,038 & 1,058,583 \\ 
countries-and-timezones & 1,087.49 & 295.42 & 104,291 & 22,259 & 126,550 \\ 
crawler-url-parser & 1,672.78 & 705.16 & 384,404 & 32,640 & 417,044 \\ 
delta & 3,494.16 & 3,421.43 & 882,477 & 90,244 & 972,721 \\ 
image-downloader & 430.50 & 378.73 & 24,140 & 8,213 & 32,353 \\ 
node-dirty & 1,530.34 & 197.97 & 244,297 & 26,982 & 271,279 \\ 
node-geo-point & 1,432.72 & 865.00 & 318,251 & 28,015 & 346,266 \\ 
node-jsonfile & 702.02 & 419.40 & 56,273 & 11,348 & 67,621 \\ 
plural & 1,533.26 & 118.22 & 261,626 & 25,664 & 287,290 \\ 
pull-stream & 2,763.15 & 1,206.16 & 204,431 & 69,471 & 273,902 \\ 
q & 6,879.90 & 11,339.61 & 2,103,232 & 191,046 & 2,294,278 \\ 
spacl-core & 1,409.48 & 527.44 & 162,695 & 26,807 & 189,502 \\ 
zip-a-folder & 500.54 & 444.67 & 81,279 & 9,009 & 90,288 \\ 
\hline
  \textit{Total} & 27,085.58 & 20,447.99 & 5,787,941 & 639,736 & 6,427,677 \\
  \end{tabular}
  }
  \\[2mm]
  \caption{Results from LLMorpheus experiment \ChangedText{(run \#364)}.
    Model: \textit{mixtral-8x7b-instruct}, 
    temperature: 0.0, 
    maxTokens: 250, 
    template: \textit{template-full.hb}, 
    systemPrompt: \textit{SystemPrompt-MutationTestingExpert.txt}
  }
  \label{table:Cost:run364:mixtral-8x7b-instruct:template-full.hb:0.0}
\end{table*}

\begin{table}[hbt!]
\centering
{\footnotesize
\begin{tabular}{l|r|r|r|r}

{\bf application}  & {\bf \#min} &  {\bf \#max} &  {\bf \#distinct} & {\bf \#common}\\
\hline
Complex.js & 962 & 989 & 1,425 & 604 (42.39\%) \\ 
countries-and-timezones & 196 & 205 & 287 & 132 (45.99\%) \\ 
crawler-url-parser & 225 & 246 & 349 & 144 (41.26\%) \\ 
delta & 660 & 680 & 958 & 431 (44.99\%) \\ 
image-downloader & 64 & 69 & 95 & 42 (44.21\%) \\ 
node-dirty & 191 & 213 & 307 & 120 (39.09\%) \\ 
node-geo-point & 253 & 263 & 368 & 169 (45.92\%) \\ 
node-jsonfile & 121 & 135 & 187 & 81 (43.32\%) \\ 
plural & 226 & 232 & 374 & 128 (34.22\%) \\ 
pull-stream & 669 & 692 & 981 & 451 (45.97\%) \\ 
q & 1,609 & 1,660 & 2,438 & 999 (40.98\%) \\ 
spacl-core & 178 & 181 & 261 & 113 (43.30\%) \\ 
zip-a-folder & 75 & 86 & 112 & 56 (50.00\%) \\ 
\end{tabular}
}
\caption{
  Variability of the mutants generated in 5 runs of \ToolName using the \textit{mixtral-8x7b-instruct} LLM
       at temperature 0.0 \ChangedText{(run \#360,\#361,\#362,\#363,\#364)}. The columns of the table show, from left to right:
    (i) the minimum number of mutants observed in any of the runs,
    (ii) the maximum number of mutants observed in any of the runs,
    (iii) the total number of distinct mutants observed in all runs, and
    (iv) the number (percentage) of mutants are observed in all runs.
}
\label{table:Variability_mixtral-8x7b-instruct_0.0}
\end{table}

\FloatBarrier

\subsection{\ChangedText{Results for \GPTFouroMini}}
   \label{app:Gpt4oMini}

\ChangedText{
Tables~\ref{table:Mutants:run58:gpt-4o-mini:template-full.hb:0.0}--\ref{table:Cost:run63:gpt-4o-mini:template-full.hb:0.0}
show the results for 5 experiments with the \GPTFouroMini model
at temperature 0.0 using the default prompt and system prompt shown in Figure~\ref{fig:PromptTemplate}.
Table~\ref{table:Variability_gpt-4o-mini_0.0} shows the variability of the mutants observed in 5 runs using \GPTFouroMini at temperature 0.0.
}

\begin{table*}[hbt!]
\centering
{\scriptsize
\begin{tabular}{l||r|r|r|r|r|r|r|r|r|r}
  {\bf application} & {\bf \#prompts} & \multicolumn{4}{|c|}{\bf \ChangedText{mutant candidates}} & {\bf \#mutants} & {\bf \#killed} & {\bf \#survived} & {\bf \#timeout} & {\bf mut.} \\
  & &  {\bf \ChangedText{total}} & {\bf \ChangedText{invalid}} & {\bf \ChangedText{identical}} & {\bf \ChangedText{duplicate}}  &  & & & & {\bf score} \\
  \hline
  \hline
\textit{Complex.js} & 490 & \ChangedText{1,470} & \ChangedText{446} & \ChangedText{0} & \ChangedText{38} & 986 & 596 & 390 & 0 & 60.45 \\ 
\hline
\textit{countries-and-timezones} & 106 & \ChangedText{317} & \ChangedText{99} & \ChangedText{0} & \ChangedText{9} & 209 & 171 & 38 & 0 & 81.82 \\ 
\hline
\textit{crawler-url-parser} & 176 & \ChangedText{527} & \ChangedText{227} & \ChangedText{0} & \ChangedText{11} & 275 & 181 & 94 & 0 & 65.82 \\ 
\hline
\textit{delta} & 462 & \ChangedText{1,385} & \ChangedText{636} & \ChangedText{2} & \ChangedText{40} & 707 & 564 & 108 & 35 & 84.72 \\ 
\hline
\textit{image-downloader} & 42 & \ChangedText{126} & \ChangedText{58} & \ChangedText{0} & \ChangedText{3} & 65 & 45 & 20 & 0 & 69.23 \\ 
\hline
\textit{node-dirty} & 154 & \ChangedText{462} & \ChangedText{188} & \ChangedText{0} & \ChangedText{9} & 265 & 154 & 103 & 8 & 61.13 \\
\end{tabular}
  }
  \\[2mm]
  \caption{Results from LLMorpheus experiment \ChangedText{(run \#58)}.
    Model: \textit{gpt-4o-mini}, 
    temperature: 0.0, 
    maxTokens: 250, 
    template: \textit{template-full.hb}, 
    systemPrompt: \textit{SystemPrompt-MutationTestingExpert.txt}. 
  }
  \label{table:Mutants:run58:gpt-4o-mini:template-full.hb:0.0}
\end{table*}

\begin{table*}[hbt!]
\centering
{\scriptsize
%
  }
  \\[2mm]
  \caption{Results from LLMorpheus experiment \ChangedText{(run \#59)}.
    Model: \textit{gpt-4o-mini}, 
    temperature: 0.0, 
    maxTokens: 250, 
    template: \textit{template-full.hb}, 
    systemPrompt: \textit{SystemPrompt-MutationTestingExpert.txt}. 
  }
  \label{table:Mutants:run59:gpt-4o-mini:template-full.hb:0.0}
\end{table*}

\begin{table*}[hbt!]
\centering
{\scriptsize
%
  }
  \\[2mm]
  \caption{Results from LLMorpheus experiment \ChangedText{(run \#60)}.
    Model: \textit{gpt-4o-mini}, 
    temperature: 0.0, 
    maxTokens: 250, 
    template: \textit{template-full.hb}, 
    systemPrompt: \textit{SystemPrompt-MutationTestingExpert.txt}. 
  }
  \label{table:Mutants:run60:gpt-4o-mini:template-full.hb:0.0}
\end{table*}

\begin{table*}[hbt!]
\centering
{\scriptsize
%
  }
  \\[2mm]
  \caption{Results from LLMorpheus experiment \ChangedText{(run \#63)}.
    Model: \textit{gpt-4o-mini}, 
    temperature: 0.0, 
    maxTokens: 250, 
    template: \textit{template-full.hb}, 
    systemPrompt: \textit{SystemPrompt-MutationTestingExpert.txt}. 
  }
  \label{table:Mutants:run63:gpt-4o-mini:template-full.hb:0.0}
\end{table*}

\begin{table*}[hbt!]
\centering
{\scriptsize
%
  }
  \\[2mm]
  \caption{Results from LLMorpheus experiment \ChangedText{(run \#58)}.
    Model: \textit{gpt-4o-mini}, 
    temperature: 0.0, 
    maxTokens: 250, 
    template: \textit{template-full.hb}, 
    systemPrompt: \textit{SystemPrompt-MutationTestingExpert.txt}
  }
  \label{table:Cost:run58:gpt-4o-mini:template-full.hb:0.0}
\end{table*}

\begin{table*}[hbt!]
\centering
{\scriptsize
%
  }
  \\[2mm]
  \caption{Results from LLMorpheus experiment \ChangedText{(run \#59)}.
    Model: \textit{gpt-4o-mini}, 
    temperature: 0.0, 
    maxTokens: 250, 
    template: \textit{template-full.hb}, 
    systemPrompt: \textit{SystemPrompt-MutationTestingExpert.txt}
  }
  \label{table:Cost:run59:gpt-4o-mini:template-full.hb:0.0}
\end{table*}

\begin{table*}[hbt!]
\centering
{\scriptsize
%
  }
  \\[2mm]
  \caption{Results from LLMorpheus experiment \ChangedText{(run \#60)}.
    Model: \textit{gpt-4o-mini}, 
    temperature: 0.0, 
    maxTokens: 250, 
    template: \textit{template-full.hb}, 
    systemPrompt: \textit{SystemPrompt-MutationTestingExpert.txt}
  }
  \label{table:Cost:run60:gpt-4o-mini:template-full.hb:0.0}
\end{table*}

\begin{table*}[hbt!]
\centering
{\scriptsize
%
  }
  \\[2mm]
  \caption{Results from LLMorpheus experiment \ChangedText{(run \#63)}.
    Model: \textit{gpt-4o-mini}, 
    temperature: 0.0, 
    maxTokens: 250, 
    template: \textit{template-full.hb}, 
    systemPrompt: \textit{SystemPrompt-MutationTestingExpert.txt}
  }
  \label{table:Cost:run63:gpt-4o-mini:template-full.hb:0.0}
\end{table*}

\begin{table}[hbt!]
\centering
{\footnotesize
\begin{tabular}{l|r|r|r|r}

{\bf application}  & {\bf \#min} &  {\bf \#max} &  {\bf \#distinct} & {\bf \#common}\\
\hline
Complex.js & 982 & 991 & 1,484 & 613 (41.31\%) \\ 
countries-and-timezones & 204 & 209 & 362 & 95 (26.24\%) \\ 
crawler-url-parser & 280 & 289 & 472 & 153 (32.42\%) \\ 
delta & 698 & 707 & 1,094 & 413 (37.75\%) \\ 
image-downloader & 65 & 68 & 127 & 31 (24.41\%) \\ 
node-dirty & 257 & 265 & 417 & 147 (35.25\%) \\ 
node-geo-point & 307 & 313 & 492 & 189 (38.41\%) \\ 
node-jsonfile & 151 & 155 & 250 & 80 (32.00\%) \\ 
plural & 311 & 315 & 585 & 145 (24.79\%) \\ 
pull-stream & 728 & 734 & 1,176 & 411 (34.95\%) \\ 
q & 1,773 & 1,795 & 2,740 & 1,093 (39.89\%) \\ 
spacl-core & 229 & 236 & 408 & 109 (26.72\%) \\ 
zip-a-folder & 81 & 84 & 131 & 50 (38.17\%) \\ 
\end{tabular}
}
\caption{
  Variability of the mutants generated in 5 runs of \ToolName using the \textit{gpt-4o-mini} LLM
       at temperature 0.0 \ChangedText{(run \#58,\#59,\#60,\#61,\#63)}. The columns of the table show, from left to right:
    (i) the minimum number of mutants observed in any of the runs,
    (ii) the maximum number of mutants observed in any of the runs,
    (iii) the total number of distinct mutants observed in all runs, and
    (iv) the number (percentage) of mutants are observed in all runs.
}
\label{table:Variability_gpt-4o-mini_0.0}
\end{table}

\FloatBarrier

\subsection{\ChangedText{Results for \LlamaThreeThree}}
    \label{app:LlamaThreeThree}
 
\ChangedText{
Tables~\ref{table:Mutants:run23:meta-llama/llama-3.3-70b-instruct:template-full.hb:0.0}--\ref{table:Cost:run27:meta-llama/llama-3.3-70b-instruct:template-full.hb:0.0}
show the results for 5 experiments with the \LlamaThreeThree model at temperature 0.0 using the default prompt and system prompt shown in Figure~\ref{fig:PromptTemplate}.    
Table~\ref{table:Variability_meta-llama/llama-3.3-70b-instruct_0.0} shows the variability of the mutants observed in 5 runs using \GPTFouroMini at temperature 0.0.
}

\begin{table*}[hbt!]
\centering
{\scriptsize

  }
  \\[2mm]
  \caption{Results from LLMorpheus experiment \ChangedText{(run \#23)}.
    Model: \textit{meta-llama/llama-3.3-70b-instruct}, 
    temperature: 0.0, 
    maxTokens: 250, 
    template: \textit{template-full.hb}, 
    systemPrompt: \textit{SystemPrompt-MutationTestingExpert.txt}. 
  }
  \label{table:Mutants:run23:meta-llama/llama-3.3-70b-instruct:template-full.hb:0.0}
\end{table*}

\begin{table*}[hbt!]
\centering
{\scriptsize

  }
  \\[2mm]
  \caption{Results from LLMorpheus experiment \ChangedText{(run \#24)}.
    Model: \textit{meta-llama/llama-3.3-70b-instruct}, 
    temperature: 0.0, 
    maxTokens: 250, 
    template: \textit{template-full.hb}, 
    systemPrompt: \textit{SystemPrompt-MutationTestingExpert.txt}. 
  }
  \label{table:Mutants:run24:meta-llama/llama-3.3-70b-instruct:template-full.hb:0.0}
\end{table*}

\begin{table*}[hbt!]
\centering
{\scriptsize

  }
  \\[2mm]
  \caption{Results from LLMorpheus experiment \ChangedText{(run \#25)}.
    Model: \textit{meta-llama/llama-3.3-70b-instruct}, 
    temperature: 0.0, 
    maxTokens: 250, 
    template: \textit{template-full.hb}, 
    systemPrompt: \textit{SystemPrompt-MutationTestingExpert.txt}. 
  }
  \label{table:Mutants:run25:meta-llama/llama-3.3-70b-instruct:template-full.hb:0.0}
\end{table*}

\begin{table*}[hbt!]
\centering
{\scriptsize

  }
  \\[2mm]
  \caption{Results from LLMorpheus experiment \ChangedText{(run \#26)}.
    Model: \textit{meta-llama/llama-3.3-70b-instruct}, 
    temperature: 0.0, 
    maxTokens: 250, 
    template: \textit{template-full.hb}, 
    systemPrompt: \textit{SystemPrompt-MutationTestingExpert.txt}. 
  }
  \label{table:Mutants:run26:meta-llama/llama-3.3-70b-instruct:template-full.hb:0.0}
\end{table*}

\begin{table*}[hbt!]
\centering
{\scriptsize

  }
  \\[2mm]
  \caption{Results from LLMorpheus experiment \ChangedText{(run \#27)}.
    Model: \textit{meta-llama/llama-3.3-70b-instruct}, 
    temperature: 0.0, 
    maxTokens: 250, 
    template: \textit{template-full.hb}, 
    systemPrompt: \textit{SystemPrompt-MutationTestingExpert.txt}. 
  }
  \label{table:Mutants:run27:meta-llama/llama-3.3-70b-instruct:template-full.hb:0.0}
\end{table*}

\begin{table*}[hbt!]
\centering
{\scriptsize
\begin{tabular}{l||r|r|r|r|r}
\multicolumn{1}{c|}{\bf project} & \multicolumn{2}{|c|}{\bf time (sec)} & \multicolumn{3}{|c}{\bf \#tokens} \\
               & {\it LLMorpheus} & {\it StrykerJS} & {\bf prompt} & {\bf compl.} & {\bf total} \\
\hline
  Complex.js & 2,035.43 & 359.18 & 785,856 & 99,437 & 885,293 \\ 
countries-and-timezones & 503.27 & 250.43 & 82,111 & 20,809 & 102,920 \\ 
crawler-url-parser & 810.25 & 606.98 & 283,056 & 35,138 & 318,194 \\ 
delta & 1,881.70 & 2,576.74 & 700,458 & 90,055 & 790,513 \\ 
image-downloader & 199.98 & 362.58 & 19,515 & 7,900 & 27,415 \\ 
node-dirty & 739.35 & 257.29 & 197,447 & 30,494 & 227,941 \\ 
node-geo-point & 688.13 & 840.31 & 252,643 & 27,340 & 279,983 \\ 
node-jsonfile & 334.39 & 527.30 & 46,029 & 13,579 & 59,608 \\ 
plural & 748.74 & 93.78 & 218,344 & 32,058 & 250,402 \\ 
pull-stream & 1,483.02 & 1,141.94 & 170,717 & 68,203 & 238,920 \\ 
q & 3,563.55 & 14,535.02 & 1,690,971 & 205,818 & 1,896,789 \\ 
spacl-core & 631.83 & 898.43 & 136,724 & 26,473 & 163,197 \\ 
zip-a-folder & 260.98 & 451.69 & 64,841 & 10,049 & 74,890 \\ 
\hline
  \textit{Total} & 13,880.61 & 22,901.67 & 4,648,712 & 667,353 & 5,316,065 \\
  \end{tabular}
  }
  \\[2mm]
  \caption{Results from LLMorpheus experiment \ChangedText{(run \#23)}.
    Model: \textit{meta-llama/llama-3.3-70b-instruct}, 
    temperature: 0.0, 
    maxTokens: 250, 
    template: \textit{template-full.hb}, 
    systemPrompt: \textit{SystemPrompt-MutationTestingExpert.txt}
  }
  \label{table:Cost:run23:meta-llama/llama-3.3-70b-instruct:template-full.hb:0.0}
\end{table*}

\begin{table*}[hbt!]
\centering
{\scriptsize
\begin{tabular}{l||r|r|r|r|r}
\multicolumn{1}{c|}{\bf project} & \multicolumn{2}{|c|}{\bf time (sec)} & \multicolumn{3}{|c}{\bf \#tokens} \\
               & {\it LLMorpheus} & {\it StrykerJS} & {\bf prompt} & {\bf compl.} & {\bf total} \\
\hline
  Complex.js & 1,935.55 & 354.56 & 785,902 & 98,758 & 884,660 \\ 
countries-and-timezones & 517.60 & 248.81 & 82,111 & 21,228 & 103,339 \\ 
crawler-url-parser & 862.45 & 606.56 & 283,056 & 35,246 & 318,302 \\ 
delta & 1,894.76 & 2,611.62 & 700,458 & 90,310 & 790,768 \\ 
image-downloader & 197.05 & 453.31 & 19,515 & 7,908 & 27,423 \\ 
node-dirty & 714.96 & 253.07 & 197,447 & 30,659 & 228,106 \\ 
node-geo-point & 700.66 & 831.74 & 252,643 & 27,406 & 280,049 \\ 
node-jsonfile & 325.93 & 518.65 & 46,029 & 13,248 & 59,277 \\ 
plural & 791.02 & 95.56 & 218,344 & 32,292 & 250,636 \\ 
pull-stream & 1,417.60 & 1,132.73 & 170,717 & 68,708 & 239,425 \\ 
q & 3,506.09 & 14,435.66 & 1,690,971 & 205,518 & 1,896,489 \\ 
spacl-core & 665.00 & 997.34 & 136,724 & 26,424 & 163,148 \\ 
zip-a-folder & 262.84 & 1,296.03 & 64,841 & 10,226 & 75,067 \\ 
\hline
  \textit{Total} & 13,791.51 & 23,835.63 & 4,648,758 & 667,931 & 5,316,689 \\
  \end{tabular}
  }
  \\[2mm]
  \caption{Results from LLMorpheus experiment \ChangedText{(run \#24)}.
    Model: \textit{meta-llama/llama-3.3-70b-instruct}, 
    temperature: 0.0, 
    maxTokens: 250, 
    template: \textit{template-full.hb}, 
    systemPrompt: \textit{SystemPrompt-MutationTestingExpert.txt}
  }
  \label{table:Cost:run24:meta-llama/llama-3.3-70b-instruct:template-full.hb:0.0}
\end{table*}

\begin{table*}[hbt!]
\centering
{\scriptsize
\begin{tabular}{l||r|r|r|r|r}
\multicolumn{1}{c|}{\bf project} & \multicolumn{2}{|c|}{\bf time (sec)} & \multicolumn{3}{|c}{\bf \#tokens} \\
               & {\it LLMorpheus} & {\it StrykerJS} & {\bf prompt} & {\bf compl.} & {\bf total} \\
\hline
  Complex.js & 1,951.37 & 356.92 & 785,856 & 99,172 & 885,028 \\ 
countries-and-timezones & 522.56 & 246.91 & 82,111 & 21,238 & 103,349 \\ 
crawler-url-parser & 804.63 & 604.74 & 283,056 & 35,252 & 318,308 \\ 
delta & 1,862.31 & 2,566.04 & 700,458 & 90,059 & 790,517 \\ 
image-downloader & 195.17 & 449.66 & 19,515 & 7,845 & 27,360 \\ 
node-dirty & 724.87 & 254.24 & 197,447 & 31,135 & 228,582 \\ 
node-geo-point & 664.03 & 860.50 & 252,643 & 27,500 & 280,143 \\ 
node-jsonfile & 353.83 & 501.73 & 46,029 & 13,505 & 59,534 \\ 
plural & 742.45 & 96.06 & 218,344 & 32,077 & 250,421 \\ 
pull-stream & 1,454.40 & 1,137.23 & 170,717 & 68,279 & 238,996 \\ 
q & 3,808.92 & 14,578.83 & 1,690,971 & 204,821 & 1,895,792 \\ 
spacl-core & 632.50 & 974.79 & 136,724 & 26,240 & 162,964 \\ 
zip-a-folder & 261.77 & 1,272.87 & 64,841 & 10,055 & 74,896 \\ 
\hline
  \textit{Total} & 13,978.80 & 23,900.50 & 4,648,712 & 667,178 & 5,315,890 \\
  \end{tabular}
  }
  \\[2mm]
  \caption{Results from LLMorpheus experiment \ChangedText{(run \#25)}.
    Model: \textit{meta-llama/llama-3.3-70b-instruct}, 
    temperature: 0.0, 
    maxTokens: 250, 
    template: \textit{template-full.hb}, 
    systemPrompt: \textit{SystemPrompt-MutationTestingExpert.txt}
  }
  \label{table:Cost:run25:meta-llama/llama-3.3-70b-instruct:template-full.hb:0.0}
\end{table*}

\begin{table*}[hbt!]
\centering
{\scriptsize
\begin{tabular}{l||r|r|r|r|r}
\multicolumn{1}{c|}{\bf project} & \multicolumn{2}{|c|}{\bf time (sec)} & \multicolumn{3}{|c}{\bf \#tokens} \\
               & {\it LLMorpheus} & {\it StrykerJS} & {\bf prompt} & {\bf compl.} & {\bf total} \\
\hline
  Complex.js & 1,965.89 & 356.55 & 785,856 & 99,239 & 885,095 \\ 
countries-and-timezones & 520.83 & 252.88 & 82,111 & 21,050 & 103,161 \\ 
crawler-url-parser & 845.23 & 415.43 & 283,056 & 35,204 & 318,260 \\ 
delta & 1,820.88 & 2,526.25 & 700,458 & 90,336 & 790,794 \\ 
image-downloader & 199.63 & 376.57 & 19,515 & 8,115 & 27,630 \\ 
node-dirty & 732.92 & 248.37 & 197,447 & 30,614 & 228,061 \\ 
node-geo-point & 648.30 & 852.65 & 252,643 & 27,289 & 279,932 \\ 
node-jsonfile & 337.73 & 504.46 & 46,029 & 13,500 & 59,529 \\ 
plural & 771.86 & 96.38 & 218,344 & 32,276 & 250,620 \\ 
pull-stream & 1,482.72 & 1,117.82 & 170,717 & 68,160 & 238,877 \\ 
q & 3,818.36 & 14,500.22 & 1,690,971 & 205,244 & 1,896,215 \\ 
spacl-core & 661.68 & 1,006.28 & 136,724 & 26,127 & 162,851 \\ 
zip-a-folder & 261.75 & 423.42 & 64,841 & 10,119 & 74,960 \\ 
\hline
  \textit{Total} & 14,067.79 & 22,677.29 & 4,648,712 & 667,273 & 5,315,985 \\
  \end{tabular}
  }
  \\[2mm]
  \caption{Results from LLMorpheus experiment \ChangedText{(run \#26)}.
    Model: \textit{meta-llama/llama-3.3-70b-instruct}, 
    temperature: 0.0, 
    maxTokens: 250, 
    template: \textit{template-full.hb}, 
    systemPrompt: \textit{SystemPrompt-MutationTestingExpert.txt}
  }
  \label{table:Cost:run26:meta-llama/llama-3.3-70b-instruct:template-full.hb:0.0}
\end{table*}

\begin{table*}[hbt!]
\centering
{\scriptsize
\begin{tabular}{l||r|r|r|r|r}
\multicolumn{1}{c|}{\bf project} & \multicolumn{2}{|c|}{\bf time (sec)} & \multicolumn{3}{|c}{\bf \#tokens} \\
               & {\it LLMorpheus} & {\it StrykerJS} & {\bf prompt} & {\bf compl.} & {\bf total} \\
\hline
  Complex.js & 1,949.16 & 359.45 & 785,856 & 98,758 & 884,614 \\ 
countries-and-timezones & 518.42 & 249.03 & 82,111 & 21,290 & 103,401 \\ 
crawler-url-parser & 843.83 & 610.56 & 283,056 & 35,543 & 318,599 \\ 
delta & 1,930.29 & 2,578.87 & 700,458 & 90,277 & 790,735 \\ 
image-downloader & 199.97 & 357.91 & 19,515 & 7,985 & 27,500 \\ 
node-dirty & 711.46 & 241.03 & 197,447 & 30,092 & 227,539 \\ 
node-geo-point & 664.09 & 878.84 & 252,643 & 27,234 & 279,877 \\ 
node-jsonfile & 334.95 & 526.44 & 46,029 & 13,288 & 59,317 \\ 
plural & 756.46 & 95.20 & 218,344 & 31,877 & 250,221 \\ 
pull-stream & 1,444.48 & 1,125.62 & 170,717 & 68,975 & 239,692 \\ 
q & 3,636.06 & 14,547.79 & 1,690,971 & 205,379 & 1,896,350 \\ 
spacl-core & 641.07 & 1,006.59 & 136,724 & 26,304 & 163,028 \\ 
zip-a-folder & 255.10 & 1,275.79 & 64,841 & 9,904 & 74,745 \\ 
\hline
  \textit{Total} & 13,885.34 & 23,853.13 & 4,648,712 & 666,906 & 5,315,618 \\
  \end{tabular}
  }
  \\[2mm]
  \caption{Results from LLMorpheus experiment \ChangedText{(run \#27)}.
    Model: \textit{meta-llama/llama-3.3-70b-instruct}, 
    temperature: 0.0, 
    maxTokens: 250, 
    template: \textit{template-full.hb}, 
    systemPrompt: \textit{SystemPrompt-MutationTestingExpert.txt}
  }
  \label{table:Cost:run27:meta-llama/llama-3.3-70b-instruct:template-full.hb:0.0}
\end{table*}

\begin{table}[hbt!]
\centering
{\footnotesize
\begin{tabular}{l|r|r|r|r}

{\bf application}  & {\bf \#min} &  {\bf \#max} &  {\bf \#distinct} & {\bf \#common}\\
\hline
Complex.js & 1,123 & 1,138 & 1,712 & 698 (40.77\%) \\ 
countries-and-timezones & 239 & 243 & 374 & 139 (37.17\%) \\ 
crawler-url-parser & 325 & 331 & 559 & 186 (33.27\%) \\ 
delta & 793 & 803 & 1,222 & 492 (40.26\%) \\ 
image-downloader & 76 & 81 & 131 & 40 (30.53\%) \\ 
node-dirty & 320 & 332 & 511 & 189 (36.99\%) \\ 
node-geo-point & 354 & 364 & 533 & 232 (43.53\%) \\ 
node-jsonfile & 171 & 175 & 273 & 97 (35.53\%) \\ 
plural & 331 & 339 & 598 & 169 (28.26\%) \\ 
pull-stream & 768 & 775 & 1,158 & 500 (43.18\%) \\ 
q & 2,205 & 2,237 & 2,898 & 1,708 (58.94\%) \\ 
spacl-core & 249 & 263 & 445 & 131 (29.44\%) \\ 
zip-a-folder & 117 & 121 & 172 & 78 (45.35\%) \\ 
\end{tabular}
}
\caption{
  Variability of the mutants generated in 5 runs of \ToolName using the \textit{meta-llama/llama-3.3-70b-instruct} LLM
       at temperature 0.0 \ChangedText{(run \#23,\#24,\#25,\#26,\#27)}. The columns of the table show, from left to right:
    (i) the minimum number of mutants observed in any of the runs,
    (ii) the maximum number of mutants observed in any of the runs,
    (iii) the total number of distinct mutants observed in all runs, and
    (iv) the number (percentage) of mutants are observed in all runs.
}
\label{table:Variability_meta-llama/llama-3.3-70b-instruct_0.0}
\end{table}

\FloatBarrier 

\subsection{Results for \texttt{template-onemutation-0.0}}
\label{app:OneMutation}

Tables~\ref{table:Mutants:run365:codellama-34b-instruct:template-onemutation.hb:0.0}--\ref{table:Cost:run371:codellama-34b-instruct:template-onemutation.hb:0.0}
show the results for 5 experiments with the \textit{codellama-34b-instruct} model
at temperature 0.0 using the prompt template of Figure~\ref{fig:PromptTemplateOneMutation} and using the system prompt shown in Figure~\ref{fig:PromptTemplate}.

\begin{table*}[hbt!]
\centering
{\scriptsize

  }
  \\[2mm]
  \caption{Results from LLMorpheus experiment \ChangedText{(run \#365)}.
    Model: \textit{codellama-34b-instruct}, 
    temperature: 0.0, 
    maxTokens: 250, 
    template: \textit{template-onemutation.hb}, 
    systemPrompt: \textit{SystemPrompt-MutationTestingExpert.txt}. 
  }
  \label{table:Mutants:run365:codellama-34b-instruct:template-onemutation.hb:0.0}
\end{table*}

\begin{table*}[hbt!]
\centering
{\scriptsize

  }
  \\[2mm]
  \caption{Results from LLMorpheus experiment \ChangedText{(run \#366)}.
    Model: \textit{codellama-34b-instruct}, 
    temperature: 0.0, 
    maxTokens: 250, 
    template: \textit{template-onemutation.hb}, 
    systemPrompt: \textit{SystemPrompt-MutationTestingExpert.txt}. 
  }
  \label{table:Mutants:run366:codellama-34b-instruct:template-onemutation.hb:0.0}
\end{table*}

\begin{table*}[hbt!]
\centering
{\scriptsize

  }
  \\[2mm]
  \caption{Results from LLMorpheus experiment \ChangedText{(run \#369)}.
    Model: \textit{codellama-34b-instruct}, 
    temperature: 0.0, 
    maxTokens: 250, 
    template: \textit{template-onemutation.hb}, 
    systemPrompt: \textit{SystemPrompt-MutationTestingExpert.txt}. 
  }
  \label{table:Mutants:run369:codellama-34b-instruct:template-onemutation.hb:0.0}
\end{table*}

\begin{table*}[hbt!]
\centering
{\scriptsize

  }
  \\[2mm]
  \caption{Results from LLMorpheus experiment \ChangedText{(run \#370)}.
    Model: \textit{codellama-34b-instruct}, 
    temperature: 0.0, 
    maxTokens: 250, 
    template: \textit{template-onemutation.hb}, 
    systemPrompt: \textit{SystemPrompt-MutationTestingExpert.txt}. 
  }
  \label{table:Mutants:run370:codellama-34b-instruct:template-onemutation.hb:0.0}
\end{table*}

\begin{table*}[hbt!]
\centering
{\scriptsize

  }
  \\[2mm]
  \caption{Results from LLMorpheus experiment \ChangedText{(run \#371)}.
    Model: \textit{codellama-34b-instruct}, 
    temperature: 0.0, 
    maxTokens: 250, 
    template: \textit{template-onemutation.hb}, 
    systemPrompt: \textit{SystemPrompt-MutationTestingExpert.txt}. 
  }
  \label{table:Mutants:run371:codellama-34b-instruct:template-onemutation.hb:0.0}
\end{table*}

\begin{table*}[hbt!]
\centering
{\scriptsize
\begin{tabular}{l||r|r|r|r|r}
\multicolumn{1}{c|}{\bf project} & \multicolumn{2}{|c|}{\bf time (sec)} & \multicolumn{3}{|c}{\bf \#tokens} \\
               & {\it LLMorpheus} & {\it StrykerJS} & {\bf prompt} & {\bf compl.} & {\bf total} \\
\hline
  Complex.js & 2,784.11 & 210.86 & 927,818 & 39,567 & 967,385 \\ 
countries-and-timezones & 1,071.07 & 117.59 & 97,242 & 8,518 & 105,760 \\ 
crawler-url-parser & 1,636.44 & 292.18 & 371,967 & 15,504 & 387,471 \\ 
delta & 2,676.03 & 1,251.08 & 852,830 & 37,401 & 890,231 \\ 
image-downloader & 430.61 & 139.23 & 21,253 & 3,459 & 24,712 \\ 
node-dirty & 1,526.39 & 77.95 & 233,774 & 12,906 & 246,680 \\ 
node-geo-point & 1,411.29 & 330.28 & 304,993 & 11,192 & 316,185 \\ 
node-jsonfile & 690.81 & 183.43 & 52,008 & 5,846 & 57,854 \\ 
plural & 1,521.37 & 54.03 & 253,209 & 13,450 & 266,659 \\ 
pull-stream & 2,400.61 & 499.06 & 179,699 & 30,228 & 209,927 \\ 
q & 4,195.04 & 4,866.38 & 2,042,524 & 82,318 & 2,124,842 \\ 
spacl-core & 1,351.30 & 271.81 & 151,851 & 10,803 & 162,654 \\ 
zip-a-folder & 500.63 & 219.06 & 78,488 & 4,405 & 82,893 \\ 
\hline
  \textit{Total} & 22,195.69 & 8,512.94 & 5,567,656 & 275,597 & 5,843,253 \\
  \end{tabular}
  }
  \\[2mm]
  \caption{Results from LLMorpheus experiment \ChangedText{(run \#365)}.
    Model: \textit{codellama-34b-instruct}, 
    temperature: 0.0, 
    maxTokens: 250, 
    template: \textit{template-onemutation.hb}, 
    systemPrompt: \textit{SystemPrompt-MutationTestingExpert.txt}
  }
  \label{table:Cost:run365:codellama-34b-instruct:template-onemutation.hb:0.0}
\end{table*}

\begin{table*}[hbt!]
\centering
{\scriptsize
\begin{tabular}{l||r|r|r|r|r}
\multicolumn{1}{c|}{\bf project} & \multicolumn{2}{|c|}{\bf time (sec)} & \multicolumn{3}{|c}{\bf \#tokens} \\
               & {\it LLMorpheus} & {\it StrykerJS} & {\bf prompt} & {\bf compl.} & {\bf total} \\
\hline
  Complex.js & 2,757.00 & 215.00 & 927,818 & 39,486 & 967,304 \\ 
countries-and-timezones & 1,091.07 & 116.98 & 97,242 & 8,527 & 105,769 \\ 
crawler-url-parser & 1,636.38 & 300.85 & 371,967 & 15,532 & 387,499 \\ 
delta & 2,681.00 & 1,235.39 & 852,830 & 37,383 & 890,213 \\ 
image-downloader & 460.67 & 139.42 & 21,253 & 3,476 & 24,729 \\ 
node-dirty & 1,526.36 & 75.52 & 233,774 & 12,907 & 246,681 \\ 
node-geo-point & 1,411.28 & 327.69 & 304,993 & 11,211 & 316,204 \\ 
node-jsonfile & 730.85 & 184.62 & 52,008 & 5,779 & 57,787 \\ 
plural & 1,521.30 & 53.73 & 253,209 & 13,418 & 266,627 \\ 
pull-stream & 2,397.86 & 497.98 & 179,699 & 30,310 & 210,009 \\ 
q & 4,204.99 & 4,839.22 & 2,042,524 & 82,262 & 2,124,786 \\ 
spacl-core & 1,351.25 & 273.58 & 151,851 & 10,809 & 162,660 \\ 
zip-a-folder & 500.62 & 219.57 & 78,488 & 4,403 & 82,891 \\ 
\hline
  \textit{Total} & 22,270.63 & 8,479.55 & 5,567,656 & 275,503 & 5,843,159 \\
  \end{tabular}
  }
  \\[2mm]
  \caption{Results from LLMorpheus experiment \ChangedText{(run \#366)}.
    Model: \textit{codellama-34b-instruct}, 
    temperature: 0.0, 
    maxTokens: 250, 
    template: \textit{template-onemutation.hb}, 
    systemPrompt: \textit{SystemPrompt-MutationTestingExpert.txt}
  }
  \label{table:Cost:run366:codellama-34b-instruct:template-onemutation.hb:0.0}
\end{table*}

\begin{table*}[hbt!]
\centering
{\scriptsize
\begin{tabular}{l||r|r|r|r|r}
\multicolumn{1}{c|}{\bf project} & \multicolumn{2}{|c|}{\bf time (sec)} & \multicolumn{3}{|c}{\bf \#tokens} \\
               & {\it LLMorpheus} & {\it StrykerJS} & {\bf prompt} & {\bf compl.} & {\bf total} \\
\hline
  Complex.js & 2,820.16 & 211.48 & 916,945 & 39,061 & 956,006 \\ 
countries-and-timezones & 1,071.10 & 117.43 & 97,242 & 8,548 & 105,790 \\ 
crawler-url-parser & 1,636.46 & 288.17 & 371,967 & 15,519 & 387,486 \\ 
delta & 2,686.73 & 1,239.88 & 852,830 & 37,432 & 890,262 \\ 
image-downloader & 430.69 & 142.46 & 21,253 & 3,475 & 24,728 \\ 
node-dirty & 1,536.37 & 75.72 & 233,774 & 12,869 & 246,643 \\ 
node-geo-point & 1,411.33 & 338.80 & 304,993 & 11,209 & 316,202 \\ 
node-jsonfile & 690.78 & 183.73 & 52,008 & 5,845 & 57,853 \\ 
plural & 1,521.35 & 54.18 & 253,209 & 13,392 & 266,601 \\ 
pull-stream & 2,398.19 & 499.73 & 179,699 & 30,182 & 209,881 \\ 
q & 4,188.82 & 4,807.59 & 2,042,524 & 82,120 & 2,124,644 \\ 
spacl-core & 1,351.23 & 272.94 & 151,851 & 10,813 & 162,664 \\ 
zip-a-folder & 500.64 & 222.35 & 78,488 & 4,405 & 82,893 \\ 
\hline
  \textit{Total} & 22,243.85 & 8,454.47 & 5,556,783 & 274,870 & 5,831,653 \\
  \end{tabular}
  }
  \\[2mm]
  \caption{Results from LLMorpheus experiment \ChangedText{(run \#369)}.
    Model: \textit{codellama-34b-instruct}, 
    temperature: 0.0, 
    maxTokens: 250, 
    template: \textit{template-onemutation.hb}, 
    systemPrompt: \textit{SystemPrompt-MutationTestingExpert.txt}
  }
  \label{table:Cost:run369:codellama-34b-instruct:template-onemutation.hb:0.0}
\end{table*}

\begin{table*}[hbt!]
\centering
{\scriptsize
\begin{tabular}{l||r|r|r|r|r}
\multicolumn{1}{c|}{\bf project} & \multicolumn{2}{|c|}{\bf time (sec)} & \multicolumn{3}{|c}{\bf \#tokens} \\
               & {\it LLMorpheus} & {\it StrykerJS} & {\bf prompt} & {\bf compl.} & {\bf total} \\
\hline
  Complex.js & 2,754.08 & 218.66 & 927,818 & 39,566 & 967,384 \\ 
countries-and-timezones & 1,071.03 & 119.64 & 97,242 & 8,579 & 105,821 \\ 
crawler-url-parser & 1,636.41 & 281.15 & 371,967 & 15,525 & 387,492 \\ 
delta & 2,676.11 & 1,249.77 & 852,830 & 37,449 & 890,279 \\ 
image-downloader & 430.60 & 138.45 & 21,253 & 3,475 & 24,728 \\ 
node-dirty & 1,526.42 & 73.92 & 233,774 & 12,859 & 246,633 \\ 
node-geo-point & 1,411.28 & 329.40 & 304,993 & 11,210 & 316,203 \\ 
node-jsonfile & 690.77 & 183.71 & 52,008 & 5,787 & 57,795 \\ 
plural & 1,521.36 & 53.81 & 253,209 & 13,434 & 266,643 \\ 
pull-stream & 2,397.58 & 499.25 & 179,699 & 30,160 & 209,859 \\ 
q & 4,211.46 & 4,819.77 & 2,042,524 & 82,203 & 2,124,727 \\ 
spacl-core & 1,361.19 & 269.34 & 151,851 & 10,818 & 162,669 \\ 
zip-a-folder & 500.61 & 217.85 & 78,488 & 4,400 & 82,888 \\ 
\hline
  \textit{Total} & 22,188.89 & 8,454.73 & 5,567,656 & 275,465 & 5,843,121 \\
  \end{tabular}
  }
  \\[2mm]
  \caption{Results from LLMorpheus experiment \ChangedText{(run \#370)}.
    Model: \textit{codellama-34b-instruct}, 
    temperature: 0.0, 
    maxTokens: 250, 
    template: \textit{template-onemutation.hb}, 
    systemPrompt: \textit{SystemPrompt-MutationTestingExpert.txt}
  }
  \label{table:Cost:run370:codellama-34b-instruct:template-onemutation.hb:0.0}
\end{table*}

\begin{table*}[hbt!]
\centering
{\scriptsize
\begin{tabular}{l||r|r|r|r|r}
\multicolumn{1}{c|}{\bf project} & \multicolumn{2}{|c|}{\bf time (sec)} & \multicolumn{3}{|c}{\bf \#tokens} \\
               & {\it LLMorpheus} & {\it StrykerJS} & {\bf prompt} & {\bf compl.} & {\bf total} \\
\hline
  Complex.js & 2,763.09 & 214.80 & 927,818 & 39,505 & 967,323 \\ 
countries-and-timezones & 1,071.04 & 115.36 & 97,242 & 8,565 & 105,807 \\ 
crawler-url-parser & 1,636.32 & 285.19 & 371,967 & 15,616 & 387,583 \\ 
delta & 2,676.35 & 1,249.33 & 852,830 & 37,349 & 890,179 \\ 
image-downloader & 430.63 & 137.29 & 21,253 & 3,461 & 24,714 \\ 
node-dirty & 1,526.34 & 74.50 & 233,774 & 12,868 & 246,642 \\ 
node-geo-point & 1,411.31 & 337.02 & 304,993 & 11,183 & 316,176 \\ 
node-jsonfile & 690.79 & 183.54 & 52,008 & 5,774 & 57,782 \\ 
plural & 1,521.35 & 52.33 & 253,209 & 13,401 & 266,610 \\ 
pull-stream & 2,403.01 & 497.08 & 179,699 & 30,238 & 209,937 \\ 
q & 4,196.27 & 4,832.34 & 2,042,524 & 82,120 & 2,124,644 \\ 
spacl-core & 1,351.29 & 269.13 & 151,851 & 10,793 & 162,644 \\ 
zip-a-folder & 500.65 & 220.66 & 78,488 & 4,403 & 82,891 \\ 
\hline
  \textit{Total} & 22,178.44 & 8,468.57 & 5,567,656 & 275,276 & 5,842,932 \\
  \end{tabular}
  }
  \\[2mm]
  \caption{Results from LLMorpheus experiment \ChangedText{(run \#371)}.
    Model: \textit{codellama-34b-instruct}, 
    temperature: 0.0, 
    maxTokens: 250, 
    template: \textit{template-onemutation.hb}, 
    systemPrompt: \textit{SystemPrompt-MutationTestingExpert.txt}
  }
  \label{table:Cost:run371:codellama-34b-instruct:template-onemutation.hb:0.0}
\end{table*}

\begin{figure*}
\centering
\begin{minipage}{0.6\textwidth}
\vspace*{2cm}
{\footnotesize
\begin{verbatim}
Your task is to apply mutation testing to the following code:
```
{{{code}}}
```

by replacing the PLACEHOLDER with a buggy code fragment that has different
behavior than the original code fragment, which was:
```
{{{orig}}}
```
Please consider changes such as using different operators, changing constants,
referring to different variables, object properties, functions, or methods.  

Provide your answer as a fenced code block containing a single line of code,
using the following template:

The PLACEHOLDER can be replaced with:
```
<code fragment>
```
This would result in different behavior because <brief explanation>.

Please conclude your response with "DONE."
\end{verbatim}
}
\end{minipage}
\caption{Variation on the template of Figure~\ref{fig:PromptTemplate} that requests only one mutation.}
\label{fig:PromptTemplateOneMutation}
\end{figure*}

\FloatBarrier

\subsection{Results for \texttt{template-noexplanation-0.0}}
     \label{app:NoExplanation}

Tables~\ref{table:Mutants:run372:codellama-34b-instruct:template-noexplanation.hb:0.0}--\ref{table:Cost:run372:codellama-34b-instruct:template-noexplanation.hb:0.0}
show the results for 5 experiments with the \textit{codellama-34b-instruct} model
at temperature 0.0 using the prompt template of Figure~\ref{fig:PromptTemplateNoExplanation} and using the system prompt shown in Figure~\ref{fig:PromptTemplate}.

\begin{table*}[hbt!]
\centering
{\scriptsize

  }
  \\[2mm]
  \caption{Results from LLMorpheus experiment \ChangedText{(run \#372)}.
    Model: \textit{codellama-34b-instruct}, 
    temperature: 0.0, 
    maxTokens: 250, 
    template: \textit{template-noexplanation.hb}, 
    systemPrompt: \textit{SystemPrompt-MutationTestingExpert.txt}. 
  }
  \label{table:Mutants:run372:codellama-34b-instruct:template-noexplanation.hb:0.0}
\end{table*}

\begin{table*}[hbt!]
\centering
{\scriptsize

  }
  \\[2mm]
  \caption{Results from LLMorpheus experiment \ChangedText{(run \#374)}.
    Model: \textit{codellama-34b-instruct}, 
    temperature: 0.0, 
    maxTokens: 250, 
    template: \textit{template-noexplanation.hb}, 
    systemPrompt: \textit{SystemPrompt-MutationTestingExpert.txt}. 
  }
  \label{table:Mutants:run374:codellama-34b-instruct:template-noexplanation.hb:0.0}
\end{table*}

\begin{table*}[hbt!]
\centering
{\scriptsize

  }
  \\[2mm]
  \caption{Results from LLMorpheus experiment \ChangedText{(run \#375)}.
    Model: \textit{codellama-34b-instruct}, 
    temperature: 0.0, 
    maxTokens: 250, 
    template: \textit{template-noexplanation.hb}, 
    systemPrompt: \textit{SystemPrompt-MutationTestingExpert.txt}. 
  }
  \label{table:Mutants:run375:codellama-34b-instruct:template-noexplanation.hb:0.0}
\end{table*}

\begin{table*}[hbt!]
\centering
{\scriptsize

  }
  \\[2mm]
  \caption{Results from LLMorpheus experiment \ChangedText{(run \#376)}.
    Model: \textit{codellama-34b-instruct}, 
    temperature: 0.0, 
    maxTokens: 250, 
    template: \textit{template-noexplanation.hb}, 
    systemPrompt: \textit{SystemPrompt-MutationTestingExpert.txt}. 
  }
  \label{table:Mutants:run376:codellama-34b-instruct:template-noexplanation.hb:0.0}
\end{table*}

\begin{table*}[hbt!]
\centering
{\scriptsize

  }
  \\[2mm]
  \caption{Results from LLMorpheus experiment \ChangedText{(run \#377)}.
    Model: \textit{codellama-34b-instruct}, 
    temperature: 0.0, 
    maxTokens: 250, 
    template: \textit{template-noexplanation.hb}, 
    systemPrompt: \textit{SystemPrompt-MutationTestingExpert.txt}. 
  }
  \label{table:Mutants:run377:codellama-34b-instruct:template-noexplanation.hb:0.0}
\end{table*}

\begin{table*}[hbt!]
\centering
{\scriptsize
\begin{tabular}{l||r|r|r|r|r}
\multicolumn{1}{c|}{\bf project} & \multicolumn{2}{|c|}{\bf time (sec)} & \multicolumn{3}{|c}{\bf \#tokens} \\
               & {\it LLMorpheus} & {\it StrykerJS} & {\bf prompt} & {\bf compl.} & {\bf total} \\
\hline
  Complex.js & 3,056.33 & 600.05 & 948,398 & 75,551 & 1,023,949 \\ 
countries-and-timezones & 1,070.68 & 309.71 & 101,694 & 23,742 & 125,436 \\ 
crawler-url-parser & 1,656.21 & 778.34 & 379,359 & 31,115 & 410,474 \\ 
delta & 2,870.28 & 3,625.25 & 872,234 & 64,880 & 937,114 \\ 
image-downloader & 430.47 & 329.74 & 23,017 & 9,110 & 32,127 \\ 
node-dirty & 1,526.57 & 243.81 & 240,242 & 24,279 & 264,521 \\ 
node-geo-point & 1,411.01 & 1,009.62 & 310,873 & 26,100 & 336,973 \\ 
node-jsonfile & 690.59 & 482.67 & 54,864 & 15,154 & 70,018 \\ 
plural & 1,522.48 & 146.41 & 259,635 & 26,465 & 286,100 \\ 
pull-stream & 2,632.74 & 1,388.06 & 194,441 & 73,821 & 268,262 \\ 
q & 4,694.78 & 12,908.57 & 2,086,666 & 127,647 & 2,214,313 \\ 
spacl-core & 1,350.87 & 711.26 & 157,479 & 28,201 & 185,680 \\ 
zip-a-folder & 500.54 & 1,097.23 & 80,546 & 10,243 & 90,789 \\ 
\hline
  \textit{Total} & 23,413.56 & 23,630.72 & 5,709,448 & 536,308 & 6,245,756 \\
  \end{tabular}
  }
  \\[2mm]
  \caption{Results from LLMorpheus experiment \ChangedText{(run \#372)}.
    Model: \textit{codellama-34b-instruct}, 
    temperature: 0.0, 
    maxTokens: 250, 
    template: \textit{template-noexplanation.hb}, 
    systemPrompt: \textit{SystemPrompt-MutationTestingExpert.txt}
  }
  \label{table:Cost:run372:codellama-34b-instruct:template-noexplanation.hb:0.0}
\end{table*}

\begin{table*}[hbt!]
\centering
{\scriptsize
\begin{tabular}{l||r|r|r|r|r}
\multicolumn{1}{c|}{\bf project} & \multicolumn{2}{|c|}{\bf time (sec)} & \multicolumn{3}{|c}{\bf \#tokens} \\
               & {\it LLMorpheus} & {\it StrykerJS} & {\bf prompt} & {\bf compl.} & {\bf total} \\
\hline
  Complex.js & 3,088.22 & 620.07 & 948,398 & 75,464 & 1,023,862 \\ 
countries-and-timezones & 1,070.66 & 302.71 & 101,694 & 23,766 & 125,460 \\ 
crawler-url-parser & 1,653.80 & 799.42 & 379,359 & 31,089 & 410,448 \\ 
delta & 2,871.99 & 3,718.30 & 872,234 & 65,148 & 937,382 \\ 
image-downloader & 430.48 & 328.19 & 23,017 & 9,096 & 32,113 \\ 
node-dirty & 1,526.65 & 242.49 & 240,242 & 24,129 & 264,371 \\ 
node-geo-point & 1,410.97 & 985.29 & 310,873 & 26,143 & 337,016 \\ 
node-jsonfile & 690.54 & 481.86 & 54,864 & 15,125 & 69,989 \\ 
plural & 1,522.07 & 145.74 & 259,635 & 26,527 & 286,162 \\ 
pull-stream & 2,643.33 & 1,393.56 & 194,441 & 73,922 & 268,363 \\ 
q & 4,623.16 & 12,860.56 & 2,086,666 & 127,954 & 2,214,620 \\ 
spacl-core & 1,360.90 & 649.12 & 157,479 & 28,174 & 185,653 \\ 
zip-a-folder & 500.51 & 1,117.51 & 80,546 & 10,267 & 90,813 \\ 
\hline
  \textit{Total} & 23,393.28 & 23,644.84 & 5,709,448 & 536,804 & 6,246,252 \\
  \end{tabular}
  }
  \\[2mm]
  \caption{Results from LLMorpheus experiment \ChangedText{(run \#374)}.
    Model: \textit{codellama-34b-instruct}, 
    temperature: 0.0, 
    maxTokens: 250, 
    template: \textit{template-noexplanation.hb}, 
    systemPrompt: \textit{SystemPrompt-MutationTestingExpert.txt}
  }
  \label{table:Cost:run374:codellama-34b-instruct:template-noexplanation.hb:0.0}
\end{table*}

\begin{table*}[hbt!]
\centering
{\scriptsize
\begin{tabular}{l||r|r|r|r|r}
\multicolumn{1}{c|}{\bf project} & \multicolumn{2}{|c|}{\bf time (sec)} & \multicolumn{3}{|c}{\bf \#tokens} \\
               & {\it LLMorpheus} & {\it StrykerJS} & {\bf prompt} & {\bf compl.} & {\bf total} \\
\hline
  Complex.js & 3,054.59 & 596.41 & 948,398 & 75,593 & 1,023,991 \\ 
countries-and-timezones & 1,070.75 & 306.48 & 101,694 & 23,740 & 125,434 \\ 
crawler-url-parser & 1,653.25 & 791.38 & 379,359 & 31,096 & 410,455 \\ 
delta & 2,869.41 & 3,656.14 & 872,234 & 64,872 & 937,106 \\ 
image-downloader & 430.47 & 328.44 & 23,017 & 9,109 & 32,126 \\ 
node-dirty & 1,526.59 & 236.98 & 240,242 & 24,096 & 264,338 \\ 
node-geo-point & 1,410.99 & 992.09 & 310,873 & 26,143 & 337,016 \\ 
node-jsonfile & 690.55 & 485.04 & 54,864 & 15,130 & 69,994 \\ 
plural & 1,521.62 & 145.62 & 259,635 & 26,482 & 286,117 \\ 
pull-stream & 2,631.99 & 1,391.93 & 194,441 & 73,754 & 268,195 \\ 
q & 4,695.78 & 12,866.72 & 2,086,666 & 127,918 & 2,214,584 \\ 
spacl-core & 1,350.86 & 706.03 & 157,479 & 28,159 & 185,638 \\ 
zip-a-folder & 500.51 & 1,109.02 & 80,546 & 10,227 & 90,773 \\ 
\hline
  \textit{Total} & 23,407.38 & 23,612.27 & 5,709,448 & 536,319 & 6,245,767 \\
  \end{tabular}
  }
  \\[2mm]
  \caption{Results from LLMorpheus experiment \ChangedText{(run \#375)}.
    Model: \textit{codellama-34b-instruct}, 
    temperature: 0.0, 
    maxTokens: 250, 
    template: \textit{template-noexplanation.hb}, 
    systemPrompt: \textit{SystemPrompt-MutationTestingExpert.txt}
  }
  \label{table:Cost:run375:codellama-34b-instruct:template-noexplanation.hb:0.0}
\end{table*}

\begin{table*}[hbt!]
\centering
{\scriptsize
\begin{tabular}{l||r|r|r|r|r}
\multicolumn{1}{c|}{\bf project} & \multicolumn{2}{|c|}{\bf time (sec)} & \multicolumn{3}{|c}{\bf \#tokens} \\
               & {\it LLMorpheus} & {\it StrykerJS} & {\bf prompt} & {\bf compl.} & {\bf total} \\
\hline
  Complex.js & 3,053.84 & 617.90 & 948,398 & 75,377 & 1,023,775 \\ 
countries-and-timezones & 1,070.72 & 306.03 & 101,694 & 23,805 & 125,499 \\ 
crawler-url-parser & 1,656.36 & 838.63 & 379,359 & 31,102 & 410,461 \\ 
delta & 2,886.73 & 3,644.21 & 872,234 & 64,947 & 937,181 \\ 
image-downloader & 430.48 & 330.08 & 23,017 & 9,107 & 32,124 \\ 
node-dirty & 1,526.82 & 243.25 & 240,242 & 24,153 & 264,395 \\ 
node-geo-point & 1,410.93 & 999.44 & 310,873 & 26,143 & 337,016 \\ 
node-jsonfile & 690.54 & 480.47 & 54,864 & 15,130 & 69,994 \\ 
plural & 1,522.37 & 144.72 & 259,635 & 26,473 & 286,108 \\ 
pull-stream & 2,649.32 & 1,395.27 & 194,441 & 73,826 & 268,267 \\ 
q & 4,627.96 & 12,851.25 & 2,086,666 & 127,807 & 2,214,473 \\ 
spacl-core & 1,350.90 & 680.52 & 157,479 & 28,203 & 185,682 \\ 
zip-a-folder & 500.52 & 1,098.96 & 80,546 & 10,244 & 90,790 \\ 
\hline
  \textit{Total} & 23,377.51 & 23,630.74 & 5,709,448 & 536,317 & 6,245,765 \\
  \end{tabular}
  }
  \\[2mm]
  \caption{Results from LLMorpheus experiment \ChangedText{(run \#376)}.
    Model: \textit{codellama-34b-instruct}, 
    temperature: 0.0, 
    maxTokens: 250, 
    template: \textit{template-noexplanation.hb}, 
    systemPrompt: \textit{SystemPrompt-MutationTestingExpert.txt}
  }
  \label{table:Cost:run376:codellama-34b-instruct:template-noexplanation.hb:0.0}
\end{table*}

\begin{table*}[hbt!]
\centering
{\scriptsize
\begin{tabular}{l||r|r|r|r|r}
\multicolumn{1}{c|}{\bf project} & \multicolumn{2}{|c|}{\bf time (sec)} & \multicolumn{3}{|c}{\bf \#tokens} \\
               & {\it LLMorpheus} & {\it StrykerJS} & {\bf prompt} & {\bf compl.} & {\bf total} \\
\hline
  Complex.js & 3,048.61 & 622.31 & 948,398 & 75,411 & 1,023,809 \\ 
countries-and-timezones & 1,070.74 & 302.32 & 101,694 & 23,740 & 125,434 \\ 
crawler-url-parser & 1,653.97 & 835.85 & 379,359 & 30,947 & 410,306 \\ 
delta & 2,880.16 & 3,648.84 & 872,234 & 65,086 & 937,320 \\ 
image-downloader & 430.49 & 326.72 & 23,017 & 9,110 & 32,127 \\ 
node-dirty & 1,526.71 & 230.04 & 240,242 & 24,142 & 264,384 \\ 
node-geo-point & 1,410.98 & 963.40 & 310,873 & 26,313 & 337,186 \\ 
node-jsonfile & 690.67 & 530.41 & 54,864 & 15,130 & 69,994 \\ 
plural & 1,522.04 & 148.93 & 259,635 & 26,465 & 286,100 \\ 
pull-stream & 2,630.34 & 1,397.17 & 194,441 & 73,763 & 268,204 \\ 
q & 4,627.86 & 12,869.01 & 2,086,666 & 127,790 & 2,214,456 \\ 
spacl-core & 1,350.91 & 714.31 & 157,479 & 28,174 & 185,653 \\ 
zip-a-folder & 500.51 & 1,073.84 & 80,546 & 10,244 & 90,790 \\ 
\hline
  \textit{Total} & 23,344.00 & 23,663.14 & 5,709,448 & 536,315 & 6,245,763 \\
  \end{tabular}
  }
  \\[2mm]
  \caption{Results from LLMorpheus experiment \ChangedText{(run \#377)}.
    Model: \textit{codellama-34b-instruct}, 
    temperature: 0.0, 
    maxTokens: 250, 
    template: \textit{template-noexplanation.hb}, 
    systemPrompt: \textit{SystemPrompt-MutationTestingExpert.txt}
  }
  \label{table:Cost:run377:codellama-34b-instruct:template-noexplanation.hb:0.0}
\end{table*}

\begin{figure*}[h]
\centering
\begin{minipage}{0.6\textwidth}
{\footnotesize
\begin{verbatim}
Your task is to apply mutation testing to the following code:
```
{{{code}}}
```

by replacing the PLACEHOLDER with a buggy code fragment that has different
behavior than the original code fragment, which was:
```
{{{orig}}}
```
Please consider changes such as using different operators, changing constants,
referring to different variables, object properties, functions, or methods.  

Provide three answers as fenced code blocks containing a single line of code,
using the following template:

Option 1: The PLACEHOLDER can be replaced with:
```
<code fragment>
```

Option 2: The PLACEHOLDER can be replaced with:
```
<code fragment>
```

Option 3: The PLACEHOLDER can be replaced with:
```
<code fragment>
```

Please conclude your response with "DONE."
\end{verbatim}
}
\end{minipage}
\caption{Variation on the template of Figure~\ref{fig:PromptTemplate} that does not request explanations for the suggested mutations.}
\label{fig:PromptTemplateNoExplanation}
\end{figure*}

\FloatBarrier

\subsection{Results for \texttt{template-noinstructions-0.0}}
\label{app:NoInstructions}

Tables~\ref{table:Mutants:run378:codellama-34b-instruct:template-noinstructions.hb:0.0}--\ref{table:Cost:run382:codellama-34b-instruct:template-noinstructions.hb:0.0}
show the results for 5 experiments with the \textit{codellama-34b-instruct} model
at temperature 0.0 using the prompt template of Figure~\ref{fig:PromptTemplateNoInstructions} and using the system prompt shown in Figure~\ref{fig:PromptTemplate}.

\begin{table*}[hbt!]
\centering
{\scriptsize

  }
  \\[2mm]
  \caption{Results from LLMorpheus experiment \ChangedText{(run \#378)}.
    Model: \textit{codellama-34b-instruct}, 
    temperature: 0.0, 
    maxTokens: 250, 
    template: \textit{template-noinstructions.hb}, 
    systemPrompt: \textit{SystemPrompt-MutationTestingExpert.txt}. 
  }
  \label{table:Mutants:run378:codellama-34b-instruct:template-noinstructions.hb:0.0}
\end{table*}

\begin{table*}[hbt!]
\centering
{\scriptsize

  }
  \\[2mm]
  \caption{Results from LLMorpheus experiment \ChangedText{(run \#379)}.
    Model: \textit{codellama-34b-instruct}, 
    temperature: 0.0, 
    maxTokens: 250, 
    template: \textit{template-noinstructions.hb}, 
    systemPrompt: \textit{SystemPrompt-MutationTestingExpert.txt}. 
  }
  \label{table:Mutants:run379:codellama-34b-instruct:template-noinstructions.hb:0.0}
\end{table*}

\begin{table*}[hbt!]
\centering
{\scriptsize

  }
  \\[2mm]
  \caption{Results from LLMorpheus experiment \ChangedText{(run \#380)}.
    Model: \textit{codellama-34b-instruct}, 
    temperature: 0.0, 
    maxTokens: 250, 
    template: \textit{template-noinstructions.hb}, 
    systemPrompt: \textit{SystemPrompt-MutationTestingExpert.txt}. 
  }
  \label{table:Mutants:run380:codellama-34b-instruct:template-noinstructions.hb:0.0}
\end{table*}

\begin{table*}[hbt!]
\centering
{\scriptsize

  }
  \\[2mm]
  \caption{Results from LLMorpheus experiment \ChangedText{(run \#381)}.
    Model: \textit{codellama-34b-instruct}, 
    temperature: 0.0, 
    maxTokens: 250, 
    template: \textit{template-noinstructions.hb}, 
    systemPrompt: \textit{SystemPrompt-MutationTestingExpert.txt}. 
  }
  \label{table:Mutants:run381:codellama-34b-instruct:template-noinstructions.hb:0.0}
\end{table*}

\begin{table*}[hbt!]
\centering
{\scriptsize

  }
  \\[2mm]
  \caption{Results from LLMorpheus experiment \ChangedText{(run \#382)}.
    Model: \textit{codellama-34b-instruct}, 
    temperature: 0.0, 
    maxTokens: 250, 
    template: \textit{template-noinstructions.hb}, 
    systemPrompt: \textit{SystemPrompt-MutationTestingExpert.txt}. 
  }
  \label{table:Mutants:run382:codellama-34b-instruct:template-noinstructions.hb:0.0}
\end{table*}

\begin{table*}[hbt!]
\centering
{\scriptsize
\begin{tabular}{l||r|r|r|r|r}
\multicolumn{1}{c|}{\bf project} & \multicolumn{2}{|c|}{\bf time (sec)} & \multicolumn{3}{|c}{\bf \#tokens} \\
               & {\it LLMorpheus} & {\it StrykerJS} & {\bf prompt} & {\bf compl.} & {\bf total} \\
\hline
  Complex.js & 3,363.13 & 610.09 & 953,788 & 104,944 & 1,058,732 \\ 
countries-and-timezones & 1,070.70 & 333.60 & 102,860 & 23,506 & 126,366 \\ 
crawler-url-parser & 1,668.12 & 865.68 & 381,295 & 38,817 & 420,112 \\ 
delta & 3,242.91 & 4,060.50 & 877,316 & 99,522 & 976,838 \\ 
image-downloader & 430.47 & 361.71 & 23,479 & 8,905 & 32,384 \\ 
node-dirty & 1,530.58 & 228.43 & 241,936 & 33,033 & 274,969 \\ 
node-geo-point & 1,410.89 & 1,019.95 & 312,413 & 28,975 & 341,388 \\ 
node-jsonfile & 690.55 & 469.01 & 55,612 & 14,598 & 70,210 \\ 
plural & 1,523.05 & 141.16 & 261,318 & 34,491 & 295,809 \\ 
pull-stream & 2,608.43 & 1,433.25 & 198,302 & 74,144 & 272,446 \\ 
q & 5,802.92 & 13,526.38 & 2,098,227 & 218,277 & 2,316,504 \\ 
spacl-core & 1,350.79 & 626.46 & 158,953 & 29,519 & 188,472 \\ 
zip-a-folder & 500.51 & 1,087.03 & 81,085 & 10,694 & 91,779 \\ 
\hline
  \textit{Total} & 25,193.04 & 24,763.25 & 5,746,584 & 719,425 & 6,466,009 \\
  \end{tabular}
  }
  \\[2mm]
  \caption{Results from LLMorpheus experiment \ChangedText{(run \#378)}.
    Model: \textit{codellama-34b-instruct}, 
    temperature: 0.0, 
    maxTokens: 250, 
    template: \textit{template-noinstructions.hb}, 
    systemPrompt: \textit{SystemPrompt-MutationTestingExpert.txt}
  }
  \label{table:Cost:run378:codellama-34b-instruct:template-noinstructions.hb:0.0}
\end{table*}

\begin{table*}[hbt!]
\centering
{\scriptsize
\begin{tabular}{l||r|r|r|r|r}
\multicolumn{1}{c|}{\bf project} & \multicolumn{2}{|c|}{\bf time (sec)} & \multicolumn{3}{|c}{\bf \#tokens} \\
               & {\it LLMorpheus} & {\it StrykerJS} & {\bf prompt} & {\bf compl.} & {\bf total} \\
\hline
  Complex.js & 3,324.82 & 606.13 & 953,788 & 104,909 & 1,058,697 \\ 
countries-and-timezones & 1,070.79 & 314.67 & 102,860 & 23,502 & 126,362 \\ 
crawler-url-parser & 1,673.87 & 789.35 & 381,295 & 38,816 & 420,111 \\ 
delta & 3,186.56 & 3,835.89 & 877,316 & 99,463 & 976,779 \\ 
image-downloader & 430.46 & 362.97 & 23,479 & 8,961 & 32,440 \\ 
node-dirty & 1,530.82 & 229.51 & 241,936 & 33,036 & 274,972 \\ 
node-geo-point & 1,410.92 & 1,035.06 & 312,413 & 28,975 & 341,388 \\ 
node-jsonfile & 691.16 & 471.51 & 55,612 & 14,598 & 70,210 \\ 
plural & 1,523.31 & 142.93 & 261,318 & 34,484 & 295,802 \\ 
pull-stream & 2,610.16 & 1,428.74 & 198,302 & 74,195 & 272,497 \\ 
q & 5,806.23 & 13,570.79 & 2,098,227 & 218,057 & 2,316,284 \\ 
spacl-core & 1,350.78 & 598.84 & 158,953 & 29,457 & 188,410 \\ 
zip-a-folder & 500.50 & 1,065.03 & 81,085 & 10,694 & 91,779 \\ 
\hline
  \textit{Total} & 25,110.38 & 24,451.42 & 5,746,584 & 719,147 & 6,465,731 \\
  \end{tabular}
  }
  \\[2mm]
  \caption{Results from LLMorpheus experiment \ChangedText{(run \#379)}.
    Model: \textit{codellama-34b-instruct}, 
    temperature: 0.0, 
    maxTokens: 250, 
    template: \textit{template-noinstructions.hb}, 
    systemPrompt: \textit{SystemPrompt-MutationTestingExpert.txt}
  }
  \label{table:Cost:run379:codellama-34b-instruct:template-noinstructions.hb:0.0}
\end{table*}

\begin{table*}[hbt!]
\centering
{\scriptsize
\begin{tabular}{l||r|r|r|r|r}
\multicolumn{1}{c|}{\bf project} & \multicolumn{2}{|c|}{\bf time (sec)} & \multicolumn{3}{|c}{\bf \#tokens} \\
               & {\it LLMorpheus} & {\it StrykerJS} & {\bf prompt} & {\bf compl.} & {\bf total} \\
\hline
  Complex.js & 3,340.83 & 597.34 & 953,788 & 105,056 & 1,058,844 \\ 
countries-and-timezones & 1,070.75 & 323.54 & 102,860 & 23,502 & 126,362 \\ 
crawler-url-parser & 1,659.88 & 803.01 & 381,295 & 38,801 & 420,096 \\ 
delta & 3,211.07 & 3,830.91 & 877,316 & 99,521 & 976,837 \\ 
image-downloader & 430.47 & 362.10 & 23,479 & 8,905 & 32,384 \\ 
node-dirty & 1,530.67 & 228.98 & 241,936 & 33,033 & 274,969 \\ 
node-geo-point & 1,410.92 & 997.90 & 312,413 & 29,053 & 341,466 \\ 
node-jsonfile & 690.56 & 466.32 & 55,612 & 14,598 & 70,210 \\ 
plural & 1,522.82 & 141.80 & 261,318 & 34,484 & 295,802 \\ 
pull-stream & 2,609.79 & 1,444.02 & 198,302 & 74,220 & 272,522 \\ 
q & 6,945.19 & 13,543.90 & 2,098,227 & 218,309 & 2,316,536 \\ 
spacl-core & 1,350.87 & 605.57 & 158,953 & 29,527 & 188,480 \\ 
zip-a-folder & 500.50 & 1,075.42 & 81,085 & 10,694 & 91,779 \\ 
\hline
  \textit{Total} & 26,274.31 & 24,420.80 & 5,746,584 & 719,703 & 6,466,287 \\
  \end{tabular}
  }
  \\[2mm]
  \caption{Results from LLMorpheus experiment \ChangedText{(run \#380)}.
    Model: \textit{codellama-34b-instruct}, 
    temperature: 0.0, 
    maxTokens: 250, 
    template: \textit{template-noinstructions.hb}, 
    systemPrompt: \textit{SystemPrompt-MutationTestingExpert.txt}
  }
  \label{table:Cost:run380:codellama-34b-instruct:template-noinstructions.hb:0.0}
\end{table*}

\begin{table*}[hbt!]
\centering
{\scriptsize
\begin{tabular}{l||r|r|r|r|r}
\multicolumn{1}{c|}{\bf project} & \multicolumn{2}{|c|}{\bf time (sec)} & \multicolumn{3}{|c}{\bf \#tokens} \\
               & {\it LLMorpheus} & {\it StrykerJS} & {\bf prompt} & {\bf compl.} & {\bf total} \\
\hline
  Complex.js & 3,492.66 & 629.00 & 953,788 & 104,886 & 1,058,674 \\ 
countries-and-timezones & 1,177.99 & 323.34 & 102,860 & 23,502 & 126,362 \\ 
crawler-url-parser & 1,751.37 & 786.06 & 381,295 & 38,800 & 420,095 \\ 
delta & 3,365.20 & 3,872.79 & 877,316 & 99,562 & 976,878 \\ 
image-downloader & 441.26 & 358.96 & 23,479 & 8,961 & 32,440 \\ 
node-dirty & 1,608.97 & 228.53 & 241,936 & 33,053 & 274,989 \\ 
node-geo-point & 1,492.54 & 1,035.93 & 312,413 & 28,984 & 341,397 \\ 
node-jsonfile & 717.49 & 465.56 & 55,612 & 14,548 & 70,160 \\ 
plural & 1,630.97 & 140.39 & 261,318 & 34,444 & 295,762 \\ 
pull-stream & 2,738.37 & 1,433.35 & 198,302 & 74,222 & 272,524 \\ 
q & 6,155.10 & 13,521.99 & 2,098,227 & 218,163 & 2,316,390 \\ 
spacl-core & 1,440.32 & 585.59 & 158,953 & 29,512 & 188,465 \\ 
zip-a-folder & 509.75 & 1,063.09 & 81,085 & 10,694 & 91,779 \\ 
\hline
  \textit{Total} & 26,522.01 & 24,444.59 & 5,746,584 & 719,331 & 6,465,915 \\
  \end{tabular}
  }
  \\[2mm]
  \caption{Results from LLMorpheus experiment \ChangedText{(run \#381)}.
    Model: \textit{codellama-34b-instruct}, 
    temperature: 0.0, 
    maxTokens: 250, 
    template: \textit{template-noinstructions.hb}, 
    systemPrompt: \textit{SystemPrompt-MutationTestingExpert.txt}
  }
  \label{table:Cost:run381:codellama-34b-instruct:template-noinstructions.hb:0.0}
\end{table*}

\begin{table*}[hbt!]
\centering
{\scriptsize
\begin{tabular}{l||r|r|r|r|r}
\multicolumn{1}{c|}{\bf project} & \multicolumn{2}{|c|}{\bf time (sec)} & \multicolumn{3}{|c}{\bf \#tokens} \\
               & {\it LLMorpheus} & {\it StrykerJS} & {\bf prompt} & {\bf compl.} & {\bf total} \\
\hline
  Complex.js & 3,378.96 & 601.89 & 953,788 & 104,940 & 1,058,728 \\ 
countries-and-timezones & 1,080.75 & 325.94 & 102,860 & 23,502 & 126,362 \\ 
crawler-url-parser & 1,660.11 & 741.99 & 381,295 & 38,801 & 420,096 \\ 
delta & 3,233.82 & 3,810.62 & 877,316 & 99,525 & 976,841 \\ 
image-downloader & 430.46 & 361.68 & 23,479 & 8,905 & 32,384 \\ 
node-dirty & 1,531.35 & 234.89 & 241,936 & 33,044 & 274,980 \\ 
node-geo-point & 1,410.90 & 1,014.28 & 312,413 & 28,969 & 341,382 \\ 
node-jsonfile & 690.54 & 466.32 & 55,612 & 14,598 & 70,210 \\ 
plural & 1,523.06 & 136.58 & 261,318 & 34,492 & 295,810 \\ 
pull-stream & 2,606.22 & 1,436.07 & 198,302 & 74,135 & 272,437 \\ 
q & 5,921.20 & 13,597.41 & 2,098,227 & 218,214 & 2,316,441 \\ 
spacl-core & 1,350.81 & 585.65 & 158,953 & 29,520 & 188,473 \\ 
zip-a-folder & 500.48 & 1,078.08 & 81,085 & 10,745 & 91,830 \\ 
\hline
  \textit{Total} & 25,318.67 & 24,391.40 & 5,746,584 & 719,390 & 6,465,974 \\
  \end{tabular}
  }
  \\[2mm]
  \caption{Results from LLMorpheus experiment \ChangedText{(run \#382)}.
    Model: \textit{codellama-34b-instruct}, 
    temperature: 0.0, 
    maxTokens: 250, 
    template: \textit{template-noinstructions.hb}, 
    systemPrompt: \textit{SystemPrompt-MutationTestingExpert.txt}
  }
  \label{table:Cost:run382:codellama-34b-instruct:template-noinstructions.hb:0.0}
\end{table*}

\begin{figure*}[b]
\centering
\begin{minipage}{0.6\textwidth}
{\footnotesize
\begin{verbatim}
Your task is to apply mutation testing to the following code:
```
{{{code}}}
```

by replacing the PLACEHOLDER with a buggy code fragment that has different
behavior than the original code fragment, which was:
```
{{{orig}}}
```  

Provide three answers as fenced code blocks containing a single line of code,
using the following template:

Option 1: The PLACEHOLDER can be replaced with:
```
<code fragment>
```
This would result in different behavior because <brief explanation>.

Option 2: The PLACEHOLDER can be replaced with:
```
<code fragment>
```
This would result in different behavior because <brief explanation>.

Option 3: The PLACEHOLDER can be replaced with:
```
<code fragment>
```
This would result in different behavior because <brief explanation>.

Please conclude your response with "DONE."
\end{verbatim}
}
\end{minipage}
\caption{Variation on the template of Figure~\ref{fig:PromptTemplate} that does not provide instructions on how to create mutants.}
\label{fig:PromptTemplateNoInstructions}
\end{figure*}

\FloatBarrier

\subsection{Results~for~\texttt{template-full-genericsystemprompt-0.0}}
\label{app:GenericSystemPrompt}

Tables~\ref{table:Mutants:run384:codellama-34b-instruct:template-full.hb:0.0}--\ref{table:Cost:run388:codellama-34b-instruct:template-full.hb:0.0} 
show the results for 5 experiments with the \CodeLlamaThirtyFour model
at temperature 0.0 using the prompt template of Figure~\ref{fig:PromptTemplate} and the generic system prompt of Figure~\ref{fig:GenericSystemPrompt}.

\begin{table*}[hbt!]
\centering
{\scriptsize

  }
  \\[2mm]
  \caption{Results from LLMorpheus experiment \ChangedText{(run \#384)}.
    Model: \textit{codellama-34b-instruct}, 
    temperature: 0.0, 
    maxTokens: 250, 
    template: \textit{template-full.hb}, 
    systemPrompt: \textit{SystemPrompt-Generic.txt}. 
  }
  \label{table:Mutants:run384:codellama-34b-instruct:template-full.hb:0.0}
\end{table*}

\begin{table*}[hbt!]
\centering
{\scriptsize

  }
  \\[2mm]
  \caption{Results from LLMorpheus experiment \ChangedText{(run \#385)}.
    Model: \textit{codellama-34b-instruct}, 
    temperature: 0.0, 
    maxTokens: 250, 
    template: \textit{template-full.hb}, 
    systemPrompt: \textit{SystemPrompt-Generic.txt}. 
  }
  \label{table:Mutants:run385:codellama-34b-instruct:template-full.hb:0.0}
\end{table*}

\begin{table*}[hbt!]
\centering
{\scriptsize

  }
  \\[2mm]
  \caption{Results from LLMorpheus experiment \ChangedText{(run \#386)}.
    Model: \textit{codellama-34b-instruct}, 
    temperature: 0.0, 
    maxTokens: 250, 
    template: \textit{template-full.hb}, 
    systemPrompt: \textit{SystemPrompt-Generic.txt}. 
  }
  \label{table:Mutants:run386:codellama-34b-instruct:template-full.hb:0.0}
\end{table*}

\begin{table*}[hbt!]
\centering
{\scriptsize

  }
  \\[2mm]
  \caption{Results from LLMorpheus experiment \ChangedText{(run \#387)}.
    Model: \textit{codellama-34b-instruct}, 
    temperature: 0.0, 
    maxTokens: 250, 
    template: \textit{template-full.hb}, 
    systemPrompt: \textit{SystemPrompt-Generic.txt}. 
  }
  \label{table:Mutants:run387:codellama-34b-instruct:template-full.hb:0.0}
\end{table*}

\begin{table*}[hbt!]
\centering
{\scriptsize

  }
  \\[2mm]
  \caption{Results from LLMorpheus experiment \ChangedText{(run \#388)}.
    Model: \textit{codellama-34b-instruct}, 
    temperature: 0.0, 
    maxTokens: 250, 
    template: \textit{template-full.hb}, 
    systemPrompt: \textit{SystemPrompt-Generic.txt}. 
  }
  \label{table:Mutants:run388:codellama-34b-instruct:template-full.hb:0.0}
\end{table*}

\begin{table*}[hbt!]
\centering
{\scriptsize
\begin{tabular}{l||r|r|r|r|r}
\multicolumn{1}{c|}{\bf project} & \multicolumn{2}{|c|}{\bf time (sec)} & \multicolumn{3}{|c}{\bf \#tokens} \\
               & {\it LLMorpheus} & {\it StrykerJS} & {\bf prompt} & {\bf compl.} & {\bf total} \\
\hline
  Complex.js & 3,198.91 & 631.03 & 943,498 & 97,397 & 1,040,895 \\ 
countries-and-timezones & 1,070.69 & 306.50 & 100,634 & 22,822 & 123,456 \\ 
crawler-url-parser & 1,666.72 & 773.61 & 377,599 & 38,968 & 416,567 \\ 
delta & 3,188.39 & 3,973.51 & 867,614 & 96,702 & 964,316 \\ 
image-downloader & 430.51 & 376.36 & 22,597 & 8,748 & 31,345 \\ 
node-dirty & 1,531.02 & 243.07 & 238,702 & 32,642 & 271,344 \\ 
node-geo-point & 1,410.94 & 993.80 & 309,473 & 28,703 & 338,176 \\ 
node-jsonfile & 690.59 & 470.19 & 54,184 & 13,966 & 68,150 \\ 
plural & 1,523.29 & 146.19 & 258,105 & 33,232 & 291,337 \\ 
pull-stream & 2,621.42 & 1,374.37 & 190,931 & 73,130 & 264,061 \\ 
q & 5,911.30 & 13,892.03 & 2,076,156 & 216,002 & 2,292,158 \\ 
spacl-core & 1,350.90 & 688.46 & 156,139 & 28,052 & 184,191 \\ 
zip-a-folder & 500.50 & 1,139.30 & 80,056 & 10,370 & 90,426 \\ 
\hline
  \textit{Total} & 25,095.20 & 25,008.45 & 5,675,688 & 700,734 & 6,376,422 \\
  \end{tabular}
  }
  \\[2mm]
  \caption{Results from LLMorpheus experiment \ChangedText{(run \#384)}.
    Model: \textit{codellama-34b-instruct}, 
    temperature: 0.0, 
    maxTokens: 250, 
    template: \textit{template-full.hb}, 
    systemPrompt: \textit{SystemPrompt-Generic.txt}
  }
  \label{table:Cost:run384:codellama-34b-instruct:template-full.hb:0.0}
\end{table*}

\begin{table*}[hbt!]
\centering
{\scriptsize
\begin{tabular}{l||r|r|r|r|r}
\multicolumn{1}{c|}{\bf project} & \multicolumn{2}{|c|}{\bf time (sec)} & \multicolumn{3}{|c}{\bf \#tokens} \\
               & {\it LLMorpheus} & {\it StrykerJS} & {\bf prompt} & {\bf compl.} & {\bf total} \\
\hline
  Complex.js & 3,198.28 & 628.16 & 943,498 & 97,360 & 1,040,858 \\ 
countries-and-timezones & 1,070.75 & 313.88 & 100,634 & 22,817 & 123,451 \\ 
crawler-url-parser & 1,667.00 & 785.35 & 377,599 & 38,968 & 416,567 \\ 
delta & 3,193.53 & 4,061.32 & 867,614 & 96,672 & 964,286 \\ 
image-downloader & 430.50 & 373.49 & 22,597 & 8,748 & 31,345 \\ 
node-dirty & 1,530.73 & 243.51 & 238,702 & 32,641 & 271,343 \\ 
node-geo-point & 1,411.00 & 1,029.33 & 309,473 & 28,703 & 338,176 \\ 
node-jsonfile & 690.57 & 473.38 & 54,184 & 13,976 & 68,160 \\ 
plural & 1,523.28 & 142.88 & 258,105 & 33,183 & 291,288 \\ 
pull-stream & 2,622.56 & 1,368.80 & 190,931 & 73,002 & 263,933 \\ 
q & 5,761.33 & 13,943.75 & 2,076,156 & 216,075 & 2,292,231 \\ 
spacl-core & 1,350.88 & 668.94 & 156,139 & 28,048 & 184,187 \\ 
zip-a-folder & 500.52 & 1,170.72 & 80,056 & 10,370 & 90,426 \\ 
\hline
  \textit{Total} & 24,950.92 & 25,203.51 & 5,675,688 & 700,563 & 6,376,251 \\
  \end{tabular}
  }
  \\[2mm]
  \caption{Results from LLMorpheus experiment \ChangedText{(run \#385)}.
    Model: \textit{codellama-34b-instruct}, 
    temperature: 0.0, 
    maxTokens: 250, 
    template: \textit{template-full.hb}, 
    systemPrompt: \textit{SystemPrompt-Generic.txt}
  }
  \label{table:Cost:run385:codellama-34b-instruct:template-full.hb:0.0}
\end{table*}

\begin{table*}[hbt!]
\centering
{\scriptsize
\begin{tabular}{l||r|r|r|r|r}
\multicolumn{1}{c|}{\bf project} & \multicolumn{2}{|c|}{\bf time (sec)} & \multicolumn{3}{|c}{\bf \#tokens} \\
               & {\it LLMorpheus} & {\it StrykerJS} & {\bf prompt} & {\bf compl.} & {\bf total} \\
\hline
  Complex.js & 3,246.58 & 637.01 & 943,498 & 97,351 & 1,040,849 \\ 
countries-and-timezones & 1,070.74 & 317.35 & 100,634 & 22,822 & 123,456 \\ 
crawler-url-parser & 1,666.54 & 828.48 & 377,599 & 38,968 & 416,567 \\ 
delta & 3,141.81 & 3,972.35 & 867,614 & 96,648 & 964,262 \\ 
image-downloader & 430.53 & 375.91 & 22,597 & 8,740 & 31,337 \\ 
node-dirty & 1,531.64 & 237.14 & 238,702 & 32,632 & 271,334 \\ 
node-geo-point & 1,410.80 & 1,007.52 & 309,473 & 28,670 & 338,143 \\ 
node-jsonfile & 690.60 & 479.57 & 54,184 & 13,982 & 68,166 \\ 
plural & 1,523.32 & 141.62 & 258,105 & 33,221 & 291,326 \\ 
pull-stream & 2,590.59 & 1,371.20 & 190,931 & 73,097 & 264,028 \\ 
q & 5,912.17 & 13,970.49 & 2,076,156 & 216,015 & 2,292,171 \\ 
spacl-core & 1,461.05 & 690.80 & 156,139 & 28,074 & 184,213 \\ 
zip-a-folder & 500.51 & 1,128.17 & 80,056 & 10,370 & 90,426 \\ 
\hline
  \textit{Total} & 25,176.88 & 25,157.62 & 5,675,688 & 700,590 & 6,376,278 \\
  \end{tabular}
  }
  \\[2mm]
  \caption{Results from LLMorpheus experiment \ChangedText{(run \#386)}.
    Model: \textit{codellama-34b-instruct}, 
    temperature: 0.0, 
    maxTokens: 250, 
    template: \textit{template-full.hb}, 
    systemPrompt: \textit{SystemPrompt-Generic.txt}
  }
  \label{table:Cost:run386:codellama-34b-instruct:template-full.hb:0.0}
\end{table*}

\begin{table*}[hbt!]
\centering
{\scriptsize
\begin{tabular}{l||r|r|r|r|r}
\multicolumn{1}{c|}{\bf project} & \multicolumn{2}{|c|}{\bf time (sec)} & \multicolumn{3}{|c}{\bf \#tokens} \\
               & {\it LLMorpheus} & {\it StrykerJS} & {\bf prompt} & {\bf compl.} & {\bf total} \\
\hline
  Complex.js & 3,229.79 & 627.53 & 943,498 & 97,331 & 1,040,829 \\ 
countries-and-timezones & 1,070.66 & 315.55 & 100,634 & 22,813 & 123,447 \\ 
crawler-url-parser & 1,677.85 & 830.94 & 377,599 & 39,015 & 416,614 \\ 
delta & 3,135.61 & 3,928.08 & 867,614 & 96,647 & 964,261 \\ 
image-downloader & 430.47 & 374.42 & 22,597 & 8,748 & 31,345 \\ 
node-dirty & 1,530.49 & 236.58 & 238,702 & 32,642 & 271,344 \\ 
node-geo-point & 1,410.83 & 1,012.70 & 309,473 & 28,703 & 338,176 \\ 
node-jsonfile & 690.54 & 473.16 & 54,184 & 13,999 & 68,183 \\ 
plural & 1,523.30 & 142.61 & 258,105 & 33,221 & 291,326 \\ 
pull-stream & 2,590.76 & 1,370.63 & 190,931 & 73,109 & 264,040 \\ 
q & 5,913.75 & 13,931.58 & 2,076,156 & 216,172 & 2,292,328 \\ 
spacl-core & 1,350.92 & 677.55 & 156,139 & 28,048 & 184,187 \\ 
zip-a-folder & 500.55 & 1,132.09 & 80,056 & 10,370 & 90,426 \\ 
\hline
  \textit{Total} & 25,055.51 & 25,053.41 & 5,675,688 & 700,818 & 6,376,506 \\
  \end{tabular}
  }
  \\[2mm]
  \caption{Results from LLMorpheus experiment \ChangedText{(run \#387)}.
    Model: \textit{codellama-34b-instruct}, 
    temperature: 0.0, 
    maxTokens: 250, 
    template: \textit{template-full.hb}, 
    systemPrompt: \textit{SystemPrompt-Generic.txt}
  }
  \label{table:Cost:run387:codellama-34b-instruct:template-full.hb:0.0}
\end{table*}

\begin{table*}[hbt!]
\centering
{\scriptsize
\begin{tabular}{l||r|r|r|r|r}
\multicolumn{1}{c|}{\bf project} & \multicolumn{2}{|c|}{\bf time (sec)} & \multicolumn{3}{|c}{\bf \#tokens} \\
               & {\it LLMorpheus} & {\it StrykerJS} & {\bf prompt} & {\bf compl.} & {\bf total} \\
\hline
  Complex.js & 3,200.91 & 614.84 & 943,498 & 97,375 & 1,040,873 \\ 
countries-and-timezones & 1,501.26 & 297.33 & 95,530 & 21,102 & 116,632 \\ 
crawler-url-parser & 1,664.69 & 819.62 & 377,599 & 38,970 & 416,569 \\ 
delta & 3,159.72 & 4,099.01 & 867,614 & 96,624 & 964,238 \\ 
image-downloader & 430.49 & 373.69 & 22,597 & 8,748 & 31,345 \\ 
node-dirty & 1,555.91 & 239.75 & 238,702 & 32,632 & 271,334 \\ 
node-geo-point & 1,410.92 & 994.48 & 309,473 & 28,709 & 338,182 \\ 
node-jsonfile & 690.59 & 467.78 & 54,184 & 13,996 & 68,180 \\ 
plural & 1,523.33 & 146.45 & 258,105 & 33,232 & 291,337 \\ 
pull-stream & 2,596.08 & 1,354.67 & 190,931 & 73,098 & 264,029 \\ 
q & 5,912.23 & 13,931.75 & 2,076,156 & 216,129 & 2,292,285 \\ 
spacl-core & 1,350.85 & 689.90 & 156,139 & 28,037 & 184,176 \\ 
zip-a-folder & 530.56 & 1,147.59 & 80,056 & 10,370 & 90,426 \\ 
\hline
  \textit{Total} & 25,527.53 & 25,176.85 & 5,670,584 & 699,022 & 6,369,606 \\
  \end{tabular}
  }
  \\[2mm]
  \caption{Results from LLMorpheus experiment \ChangedText{(run \#388)}.
    Model: \textit{codellama-34b-instruct}, 
    temperature: 0.0, 
    maxTokens: 250, 
    template: \textit{template-full.hb}, 
    systemPrompt: \textit{SystemPrompt-Generic.txt}
  }
  \label{table:Cost:run388:codellama-34b-instruct:template-full.hb:0.0}
\end{table*}

\begin{figure*}[b]
\centering
\begin{minipage}{0.8\textwidth}
{\footnotesize
\begin{verbatim}
You are a programming assistant. You are expected to be concise and precise and 
avoid any unnecessary examples, tests, and verbosity.
\end{verbatim}
}
\end{minipage}
\caption{Generic system prompt.}
\label{fig:GenericSystemPrompt}
\end{figure*}

\FloatBarrier

\subsection{Results~for~\texttt{template-basic-0.0}}
\label{app:Basic}

Tables~\ref{table:Mutants:run390:codellama-34b-instruct:template-basic.hb:0.0}--\ref{table:Cost:run394:codellama-34b-instruct:template-basic.hb:0.0} show the results for 5 experiments with the \CodeLlamaThirtyFour model
at temperature 0.0 using the prompt template of Figure~\ref{fig:PromptTemplateBasic} and using the system prompt shown in Figure~\ref{fig:PromptTemplate}.

\begin{table*}[hbt!]
\centering
{\scriptsize

  }
  \\[2mm]
  \caption{Results from LLMorpheus experiment \ChangedText{(run \#390)}.
    Model: \textit{codellama-34b-instruct}, 
    temperature: 0.0, 
    maxTokens: 250, 
    template: \textit{template-basic.hb}, 
    systemPrompt: \textit{SystemPrompt-MutationTestingExpert.txt}. 
  }
  \label{table:Mutants:run390:codellama-34b-instruct:template-basic.hb:0.0}
\end{table*}

\begin{table*}[hbt!]
\centering
{\scriptsize

  }
  \\[2mm]
  \caption{Results from LLMorpheus experiment \ChangedText{(run \#391)}.
    Model: \textit{codellama-34b-instruct}, 
    temperature: 0.0, 
    maxTokens: 250, 
    template: \textit{template-basic.hb}, 
    systemPrompt: \textit{SystemPrompt-MutationTestingExpert.txt}. 
  }
  \label{table:Mutants:run391:codellama-34b-instruct:template-basic.hb:0.0}
\end{table*}

\begin{table*}[hbt!]
\centering
{\scriptsize

  }
  \\[2mm]
  \caption{Results from LLMorpheus experiment \ChangedText{(run \#392)}.
    Model: \textit{codellama-34b-instruct}, 
    temperature: 0.0, 
    maxTokens: 250, 
    template: \textit{template-basic.hb}, 
    systemPrompt: \textit{SystemPrompt-MutationTestingExpert.txt}. 
  }
  \label{table:Mutants:run392:codellama-34b-instruct:template-basic.hb:0.0}
\end{table*}

\begin{table*}[hbt!]
\centering
{\scriptsize

  }
  \\[2mm]
  \caption{Results from LLMorpheus experiment \ChangedText{(run \#393)}.
    Model: \textit{codellama-34b-instruct}, 
    temperature: 0.0, 
    maxTokens: 250, 
    template: \textit{template-basic.hb}, 
    systemPrompt: \textit{SystemPrompt-MutationTestingExpert.txt}. 
  }
  \label{table:Mutants:run393:codellama-34b-instruct:template-basic.hb:0.0}
\end{table*}

\begin{table*}[hbt!]
\centering
{\scriptsize

  }
  \\[2mm]
  \caption{Results from LLMorpheus experiment \ChangedText{(run \#394)}.
    Model: \textit{codellama-34b-instruct}, 
    temperature: 0.0, 
    maxTokens: 250, 
    template: \textit{template-basic.hb}, 
    systemPrompt: \textit{SystemPrompt-MutationTestingExpert.txt}. 
  }
  \label{table:Mutants:run394:codellama-34b-instruct:template-basic.hb:0.0}
\end{table*}

\begin{table*}[hbt!]
\centering
{\scriptsize
\begin{tabular}{l||r|r|r|r|r}
\multicolumn{1}{c|}{\bf project} & \multicolumn{2}{|c|}{\bf time (sec)} & \multicolumn{3}{|c}{\bf \#tokens} \\
               & {\it LLMorpheus} & {\it StrykerJS} & {\bf prompt} & {\bf compl.} & {\bf total} \\
\hline
  Complex.js & 2,731.54 & 97.72 & 893,966 & 14,460 & 908,426 \\ 
countries-and-timezones & 1,071.22 & 73.08 & 89,939 & 3,087 & 93,026 \\ 
crawler-url-parser & 1,636.67 & 215.52 & 359,498 & 5,557 & 365,055 \\ 
delta & 2,659.93 & 910.49 & 820,541 & 13,472 & 834,013 \\ 
image-downloader & 430.67 & 65.98 & 18,348 & 1,448 & 19,796 \\ 
node-dirty & 1,526.59 & 39.53 & 223,071 & 4,425 & 227,496 \\ 
node-geo-point & 1,411.43 & 204.76 & 295,321 & 4,217 & 299,538 \\ 
node-jsonfile & 690.87 & 77.82 & 47,346 & 1,831 & 49,177 \\ 
plural & 1,521.52 & 48.57 & 241,953 & 5,075 & 247,028 \\ 
pull-stream & 2,382.28 & 245.31 & 156,016 & 9,287 & 165,303 \\ 
q & 4,158.17 & 2,697.98 & 1,970,359 & 30,059 & 2,000,418 \\ 
spacl-core & 1,351.47 & 85.42 & 142,466 & 4,013 & 146,479 \\ 
zip-a-folder & 500.75 & 235.31 & 75,033 & 1,594 & 76,627 \\ 
\hline
  \textit{Total} & 22,073.11 & 4,997.50 & 5,333,857 & 98,525 & 5,432,382 \\
  \end{tabular}
  }
  \\[2mm]
  \caption{Results from LLMorpheus experiment \ChangedText{(run \#390)}.
    Model: \textit{codellama-34b-instruct}, 
    temperature: 0.0, 
    maxTokens: 250, 
    template: \textit{template-basic.hb}, 
    systemPrompt: \textit{SystemPrompt-MutationTestingExpert.txt}
  }
  \label{table:Cost:run390:codellama-34b-instruct:template-basic.hb:0.0}
\end{table*}

\begin{table*}[hbt!]
\centering
{\scriptsize
\begin{tabular}{l||r|r|r|r|r}
\multicolumn{1}{c|}{\bf project} & \multicolumn{2}{|c|}{\bf time (sec)} & \multicolumn{3}{|c}{\bf \#tokens} \\
               & {\it LLMorpheus} & {\it StrykerJS} & {\bf prompt} & {\bf compl.} & {\bf total} \\
\hline
  Complex.js & 2,732.79 & 102.55 & 893,966 & 14,472 & 908,438 \\ 
countries-and-timezones & 1,071.27 & 72.88 & 89,939 & 3,113 & 93,052 \\ 
crawler-url-parser & 1,636.61 & 211.77 & 359,498 & 5,557 & 365,055 \\ 
delta & 2,667.43 & 897.69 & 820,541 & 13,458 & 833,999 \\ 
image-downloader & 430.64 & 65.54 & 18,348 & 1,448 & 19,796 \\ 
node-dirty & 1,526.57 & 38.93 & 223,071 & 4,422 & 227,493 \\ 
node-geo-point & 1,411.51 & 203.69 & 295,321 & 4,218 & 299,539 \\ 
node-jsonfile & 690.88 & 77.42 & 47,346 & 1,831 & 49,177 \\ 
plural & 1,521.47 & 47.92 & 241,953 & 5,075 & 247,028 \\ 
pull-stream & 2,382.21 & 245.17 & 156,016 & 9,288 & 165,304 \\ 
q & 4,159.06 & 2,700.88 & 1,970,359 & 30,059 & 2,000,418 \\ 
spacl-core & 1,351.31 & 85.36 & 142,466 & 4,007 & 146,473 \\ 
zip-a-folder & 500.73 & 227.68 & 75,033 & 1,594 & 76,627 \\ 
\hline
  \textit{Total} & 22,082.48 & 4,977.48 & 5,333,857 & 98,542 & 5,432,399 \\
  \end{tabular}
  }
  \\[2mm]
  \caption{Results from LLMorpheus experiment \ChangedText{(run \#391)}.
    Model: \textit{codellama-34b-instruct}, 
    temperature: 0.0, 
    maxTokens: 250, 
    template: \textit{template-basic.hb}, 
    systemPrompt: \textit{SystemPrompt-MutationTestingExpert.txt}
  }
  \label{table:Cost:run391:codellama-34b-instruct:template-basic.hb:0.0}
\end{table*}

\begin{table*}[hbt!]
\centering
{\scriptsize
\begin{tabular}{l||r|r|r|r|r}
\multicolumn{1}{c|}{\bf project} & \multicolumn{2}{|c|}{\bf time (sec)} & \multicolumn{3}{|c}{\bf \#tokens} \\
               & {\it LLMorpheus} & {\it StrykerJS} & {\bf prompt} & {\bf compl.} & {\bf total} \\
\hline
  Complex.js & 2,730.84 & 97.59 & 893,966 & 14,461 & 908,427 \\ 
countries-and-timezones & 1,071.16 & 72.83 & 89,939 & 3,113 & 93,052 \\ 
crawler-url-parser & 1,636.68 & 220.83 & 359,498 & 5,576 & 365,074 \\ 
delta & 2,659.86 & 887.50 & 820,541 & 13,473 & 834,014 \\ 
image-downloader & 430.65 & 65.72 & 18,348 & 1,449 & 19,797 \\ 
node-dirty & 1,526.53 & 37.30 & 223,071 & 4,496 & 227,567 \\ 
node-geo-point & 1,411.50 & 195.42 & 295,321 & 4,230 & 299,551 \\ 
node-jsonfile & 690.89 & 77.90 & 47,346 & 1,831 & 49,177 \\ 
plural & 1,521.54 & 49.08 & 241,953 & 5,075 & 247,028 \\ 
pull-stream & 2,382.22 & 248.75 & 156,016 & 9,288 & 165,304 \\ 
q & 4,158.01 & 2,694.11 & 1,970,359 & 30,070 & 2,000,429 \\ 
spacl-core & 1,351.43 & 85.92 & 142,466 & 4,013 & 146,479 \\ 
zip-a-folder & 500.71 & 229.75 & 75,033 & 1,594 & 76,627 \\ 
\hline
  \textit{Total} & 22,072.01 & 4,962.71 & 5,333,857 & 98,669 & 5,432,526 \\
  \end{tabular}
  }
  \\[2mm]
  \caption{Results from LLMorpheus experiment \ChangedText{(run \#392)}.
    Model: \textit{codellama-34b-instruct}, 
    temperature: 0.0, 
    maxTokens: 250, 
    template: \textit{template-basic.hb}, 
    systemPrompt: \textit{SystemPrompt-MutationTestingExpert.txt}
  }
  \label{table:Cost:run392:codellama-34b-instruct:template-basic.hb:0.0}
\end{table*}

\begin{table*}[hbt!]
\centering
{\scriptsize
\begin{tabular}{l||r|r|r|r|r}
\multicolumn{1}{c|}{\bf project} & \multicolumn{2}{|c|}{\bf time (sec)} & \multicolumn{3}{|c}{\bf \#tokens} \\
               & {\it LLMorpheus} & {\it StrykerJS} & {\bf prompt} & {\bf compl.} & {\bf total} \\
\hline
  Complex.js & 2,730.22 & 96.64 & 893,966 & 14,459 & 908,425 \\ 
countries-and-timezones & 1,071.13 & 74.16 & 89,939 & 3,112 & 93,051 \\ 
crawler-url-parser & 1,636.64 & 206.03 & 359,498 & 5,556 & 365,054 \\ 
delta & 2,659.88 & 897.06 & 820,541 & 13,471 & 834,012 \\ 
image-downloader & 430.66 & 65.63 & 18,348 & 1,449 & 19,797 \\ 
node-dirty & 1,526.57 & 38.82 & 223,071 & 4,425 & 227,496 \\ 
node-geo-point & 1,411.45 & 200.23 & 295,321 & 4,217 & 299,538 \\ 
node-jsonfile & 690.84 & 77.61 & 47,346 & 1,831 & 49,177 \\ 
plural & 1,556.64 & 47.70 & 238,779 & 5,029 & 243,808 \\ 
pull-stream & 2,382.19 & 247.28 & 156,016 & 9,288 & 165,304 \\ 
q & 4,156.62 & 2,695.05 & 1,970,359 & 30,071 & 2,000,430 \\ 
spacl-core & 1,351.45 & 84.84 & 142,466 & 4,008 & 146,474 \\ 
zip-a-folder & 500.75 & 235.51 & 75,033 & 1,594 & 76,627 \\ 
\hline
  \textit{Total} & 22,105.05 & 4,966.57 & 5,330,683 & 98,510 & 5,429,193 \\
  \end{tabular}
  }
  \\[2mm]
  \caption{Results from LLMorpheus experiment \ChangedText{(run \#393)}.
    Model: \textit{codellama-34b-instruct}, 
    temperature: 0.0, 
    maxTokens: 250, 
    template: \textit{template-basic.hb}, 
    systemPrompt: \textit{SystemPrompt-MutationTestingExpert.txt}
  }
  \label{table:Cost:run393:codellama-34b-instruct:template-basic.hb:0.0}
\end{table*}

\begin{table*}[hbt!]
\centering
{\scriptsize
\begin{tabular}{l||r|r|r|r|r}
\multicolumn{1}{c|}{\bf project} & \multicolumn{2}{|c|}{\bf time (sec)} & \multicolumn{3}{|c}{\bf \#tokens} \\
               & {\it LLMorpheus} & {\it StrykerJS} & {\bf prompt} & {\bf compl.} & {\bf total} \\
\hline
  Complex.js & 2,730.17 & 96.67 & 893,966 & 14,460 & 908,426 \\ 
countries-and-timezones & 1,071.19 & 73.83 & 89,939 & 3,086 & 93,025 \\ 
crawler-url-parser & 1,636.65 & 205.98 & 359,498 & 5,577 & 365,075 \\ 
delta & 2,659.91 & 887.25 & 820,541 & 13,475 & 834,016 \\ 
image-downloader & 430.71 & 65.63 & 18,348 & 1,449 & 19,797 \\ 
node-dirty & 1,526.60 & 38.47 & 223,071 & 4,500 & 227,571 \\ 
node-geo-point & 1,411.44 & 197.24 & 295,321 & 4,217 & 299,538 \\ 
node-jsonfile & 690.88 & 78.16 & 47,346 & 1,831 & 49,177 \\ 
plural & 1,522.37 & 47.69 & 241,953 & 5,075 & 247,028 \\ 
pull-stream & 2,382.20 & 245.09 & 156,016 & 9,290 & 165,306 \\ 
q & 4,154.13 & 2,691.66 & 1,970,359 & 30,055 & 2,000,414 \\ 
spacl-core & 1,351.41 & 83.75 & 142,466 & 4,013 & 146,479 \\ 
zip-a-folder & 500.72 & 232.15 & 75,033 & 1,594 & 76,627 \\ 
\hline
  \textit{Total} & 22,068.39 & 4,943.55 & 5,333,857 & 98,622 & 5,432,479 \\
  \end{tabular}
  }
  \\[2mm]
  \caption{Results from LLMorpheus experiment \ChangedText{(run \#394)}.
    Model: \textit{codellama-34b-instruct}, 
    temperature: 0.0, 
    maxTokens: 250, 
    template: \textit{template-basic.hb}, 
    systemPrompt: \textit{SystemPrompt-MutationTestingExpert.txt}
  }
  \label{table:Cost:run394:codellama-34b-instruct:template-basic.hb:0.0}
\end{table*}

\begin{figure*}[!htb]
\centering
\begin{minipage}{0.6\textwidth}
{\footnotesize
\begin{verbatim}
Consider the following code fragment:
```
{{{code}}}
```

Please provide a code fragment that PLACEHOLDER can be replaced with.  

Provide your answer as a fenced code block containing a single line of code,
using the following template:

The PLACEHOLDER can be replaced with:
```
<code fragment>
```

Please conclude your response with "DONE."
\end{verbatim}
}
\end{minipage}
\caption{A minimal template for requesting a replacement for the PLACEHOLDER.}
\label{fig:PromptTemplateBasic}
\end{figure*}

\newpage
\FloatBarrier

\begin{revision*}
\subsection{Summary of variability of results for all models examined in the paper}
\label{app:allSummary}
For each configuration of \ToolName, this section provides a summary of the mutation results, showing the average of each value across all five trials, along with the standard deviation for that value.
\begin{itemize}
\item Model codellama-13b-instruct with template full at temperature 0.0: Table \ref{tab:appendix-variability-codellama-13b-instruct-full-0.0}
\item Model codellama-34b-instruct with template basic at temperature 0.0: Table \ref{tab:appendix-variability-codellama-34b-instruct-basic-0.0}
\item Model codellama-34b-instruct with template full at temperature 0.0: Table \ref{tab:appendix-variability-codellama-34b-instruct-full-0.0}
\item Model codellama-34b-instruct with template full at temperature 0.25: Table \ref{tab:appendix-variability-codellama-34b-instruct-full-0.25}
\item Model codellama-34b-instruct with template full at temperature 0.5: Table \ref{tab:appendix-variability-codellama-34b-instruct-full-0.5}
\item Model codellama-34b-instruct with template full at temperature 1.0: Table \ref{tab:appendix-variability-codellama-34b-instruct-full-1.0}
\item Model codellama-34b-instruct with template full at temperature genericsystemprompt-0.0: Table \ref{tab:appendix-variability-codellama-34b-instruct-full-genericsystemprompt-0.0}
\item Model codellama-34b-instruct with template noexplanation at temperature 0.0: Table \ref{tab:appendix-variability-codellama-34b-instruct-noexplanation-0.0}
\item Model codellama-34b-instruct with template noinstructions at temperature 0.0: Table \ref{tab:appendix-variability-codellama-34b-instruct-noinstructions-0.0}
\item Model codellama-34b-instruct with template onemutation at temperature 0.0: Table \ref{tab:appendix-variability-codellama-34b-instruct-onemutation-0.0}
\item Model gpt-4o-mini with template full at temperature 0.0: Table \ref{tab:appendix-variability-gpt-4o-mini-full-0.0}
\item Model llama-3.3-70b-instruct with template full at temperature 0.0: Table \ref{tab:appendix-variability-llama-3.3-70b-instruct-full-0.0}
\item Model mixtral-8x7b-instruct with template full at temperature 0.0: Table \ref{tab:appendix-variability-mixtral-8x7b-instruct-full-0.0}
\end{itemize}

\begin{table}
\centering
\caption{\label{tab:appendix-variability-codellama-13b-instruct-full-0.0}Summary of mutants for codellama-13b-instruct-full-0.0 (runs 354, 355, 356, 358, 359). Each column shows the average number of mutants from all runs, $\pm$ the standard deviation.}
\centering
\resizebox{\ifdim\width>\linewidth\linewidth\else\width\fi}{!}{
}
\end{table}

\begin{table}
\centering
\caption{\label{tab:appendix-variability-codellama-34b-instruct-basic-0.0}Summary of mutants for codellama-34b-instruct-basic-0.0 (runs 390, 391, 392, 393, 394). Each column shows the average number of mutants from all runs, $\pm$ the standard deviation.}
\centering
\resizebox{\ifdim\width>\linewidth\linewidth\else\width\fi}{!}{
%
}
\end{table}

\begin{table}
\centering
\caption{\label{tab:appendix-variability-codellama-34b-instruct-full-0.0}Summary of mutants for codellama-34b-instruct-full-0.0 (runs 312, 314, 315, 316, 317). Each column shows the average number of mutants from all runs, $\pm$ the standard deviation.}
\centering
\resizebox{\ifdim\width>\linewidth\linewidth\else\width\fi}{!}{
%
}
\end{table}

\begin{table}
\centering
\caption{\label{tab:appendix-variability-codellama-34b-instruct-full-0.25}Summary of mutants for codellama-34b-instruct-full-0.25 (runs 348, 350, 351, 352, 353). Each column shows the average number of mutants from all runs, $\pm$ the standard deviation.}
\centering
\resizebox{\ifdim\width>\linewidth\linewidth\else\width\fi}{!}{
%
}
\end{table}

\begin{table}
\centering
\caption{\label{tab:appendix-variability-codellama-34b-instruct-full-0.5}Summary of mutants for codellama-34b-instruct-full-0.5 (runs 318, 319, 320, 321, 322). Each column shows the average number of mutants from all runs, $\pm$ the standard deviation.}
\centering
\resizebox{\ifdim\width>\linewidth\linewidth\else\width\fi}{!}{
%
}
\end{table}

\begin{table}
\centering
\caption{\label{tab:appendix-variability-codellama-34b-instruct-full-genericsystemprompt-0.0}Summary of mutants for codellama-34b-instruct-full-genericsystemprompt-0.0 (runs 384, 385, 386, 387, 388). Each column shows the average number of mutants from all runs, $\pm$ the standard deviation.}
\centering
\resizebox{\ifdim\width>\linewidth\linewidth\else\width\fi}{!}{
%
}
\end{table}

\begin{table}
\centering
\caption{\label{tab:appendix-variability-codellama-34b-instruct-noexplanation-0.0}Summary of mutants for codellama-34b-instruct-noexplanation-0.0 (runs 372, 374, 375, 376, 377). Each column shows the average number of mutants from all runs, $\pm$ the standard deviation.}
\centering
\resizebox{\ifdim\width>\linewidth\linewidth\else\width\fi}{!}{
%
}
\end{table}

\begin{table}
\centering
\caption{\label{tab:appendix-variability-codellama-34b-instruct-noinstructions-0.0}Summary of mutants for codellama-34b-instruct-noinstructions-0.0 (runs 378, 379, 380, 381, 382). Each column shows the average number of mutants from all runs, $\pm$ the standard deviation.}
\centering
\resizebox{\ifdim\width>\linewidth\linewidth\else\width\fi}{!}{
%
}
\end{table}

\begin{table}
\centering
\caption{\label{tab:appendix-variability-codellama-34b-instruct-onemutation-0.0}Summary of mutants for codellama-34b-instruct-onemutation-0.0 (runs 365, 366, 369, 370, 371). Each column shows the average number of mutants from all runs, $\pm$ the standard deviation.}
\centering
\resizebox{\ifdim\width>\linewidth\linewidth\else\width\fi}{!}{
%
}
\end{table}

\begin{table}
\centering
\caption{\label{tab:appendix-variability-gpt-4o-mini-full-0.0}Summary of mutants for gpt-4o-mini-full-0.0 (runs 58, 59, 60, 61, 63). Each column shows the average number of mutants from all runs, $\pm$ the standard deviation.}
\centering
\resizebox{\ifdim\width>\linewidth\linewidth\else\width\fi}{!}{
%
}
\end{table}

\begin{table}
\centering
\caption{\label{tab:appendix-variability-llama-3.3-70b-instruct-full-0.0}Summary of mutants for llama-3.3-70b-instruct-full-0.0 (runs 23, 24, 25, 26, 27). Each column shows the average number of mutants from all runs, $\pm$ the standard deviation.}
\centering
\resizebox{\ifdim\width>\linewidth\linewidth\else\width\fi}{!}{
%
}
\end{table}

\begin{table}
\centering
\caption{\label{tab:appendix-variability-mixtral-8x7b-instruct-full-0.0}Summary of mutants for mixtral-8x7b-instruct-full-0.0 (runs 360, 361, 362, 363, 364). Each column shows the average number of mutants from all runs, $\pm$ the standard deviation.}
\centering
\resizebox{\ifdim\width>\linewidth\linewidth\else\width\fi}{!}{
%
}
\end{table}

\end{revision*}

 \FloatBarrier

\subsection{StrykerJS standard mutation operators}
\label{app:StrykerStandardMutators}

Table~\ref{table:StrykerJSResults} shows the results of running the standard mutation operators of \StrykerJS on the subject applications.

\begin{table*}[htb!]
  \centering
  {scriptsize
  %
  }
  \caption{Results of applying the standard mutation operators of StrykerJS.}
  \label{table:StrykerJSResults}
\end{table*}


\subsection{String edit distance measurements for different LLMs}
\label{app:StringEditDistance}

Table~\ref{table:StringSimilarityLLMs} shows average string similarity for each project, for each of the five LLMs under consideration.

\begin{table*}[hbt!]
\centering
%
\end{minipage}
\end{tabular}
\\[2mm]
\caption{
  Average string similarity to the original code fragments that they replace, for mutants generated using five LLMs at temperature 0.0.
}
  \label{table:StringSimilarityLLMs}
\end{table*}

\end{document}